\documentclass[a4paper,11pt,times,custombib,index]{PhDThesisPSnPDF}

\input{preamble}

\newcommand{\aap}{    {\it A \& A}}
\newcommand{\aaps}{   {\it A \& A Suppl.}}

\newcommand{\actaa}{  {\it Acta Astron.}}

\newcommand{\aj}{     {\it AJ}}
\newcommand{\apj}{    {\it ApJ}}
\newcommand{\apjl}{   {\it ApJL}}
\newcommand{\apjs}{   {\it ApJ Suppl.}}
\newcommand{\apss}{   {\it Astrophys. Space Sci.}}
\newcommand{\azh}{    {\it Astro. Zhurnal}}

\newcommand{\mnras}{  {\it MNRAS}}
\newcommand{\nat}{    {\it Nature}}
\newcommand{\pasp}{   {\it Pub. Astron. Soc. Pac.}}

\newcommand{\ssr}{    {\it Space Sci. Rev.}}
\newcommand{\araa}{   {\it Anu Rev. in Astronomy and Astrophysics.}}
\newcommand{\aplett}{ {\it Astrophysical Letters}}
\newcommand{\memsai}{ {\it MmSAI}}
\newcommand{\na}{     {\it NewA}}

\begin{document}
\title{Panchromatic study of star clusters: binaries, blue lurkers, blue stragglers and membership}

\submitdate{9 March 2022}
\degree{Doctor of Philosophy}
\dept{Joint Astronomy Programme\\
Department of Physics}
\faculty{Faculty of Science}
\author{Vikrant Vinayak Jadhav}

\begin{titlepage}
\maketitle
\end{titlepage}
\null
\thispagestyle{empty} 
\newpage

 \vspace*{\fill}
 \begin{center}
\large\bf \textcopyright \ Vikrant Vinayak Jadhav\\

 \large\bf All rights reserved
 \end{center}
 \vspace*{\fill}
 \thispagestyle{empty}
%

\setcounter{secnumdepth}{3}
\setcounter{tocdepth}{3}

\pagenumbering{roman}

\begin{declaration}
I hereby declare that the work reported in this doctoral thesis titled ``Panchromatic study of star clusters: binaries, blue lurkers, blue stragglers and membership'' is entirely original and is the result of
investigations carried out by me in the Department of Physics, Indian Institute of Science, Bangalore, under the supervision of Prof. Annapurni Subramaniam at the Indian Institute of Astrophysics, Bangalore and Dr. Rajeev Kumar Jain at the Indian Institute of Science, Bangalore.

I further declare that this work has not formed the basis for the award of any degree, diploma, fellowship, associateship or similar title of any University or Institution.

\end{declaration}

\begin{dedication} 
\centering

\LARGE{\it To,\\
My Family and Friends}

\end{dedication}

\begin{acknowledgements}

Here, I want to express my gratitude to my supervisor, Prof. Annapurni Subramaniam, for constant motivation and guidance. I am thankful for the friendly environment, informative discussions, constructive criticism and the long and exhaustive reviews of the manuscripts. I would also like to thank Prof. Ram Sagar, Prof. Kaushar Vaidya, Prof. Smitha Subramanian and Prof. Sudhanshu Barway for exciting and thought-provoking discussions. 

I also want to thank Prof. Rajeev Kumar Jain, Prof. Nirupam Roy, Prof. Maheswar Gopinathan, Prof. Aruna Goswami, Prof. Prateek Sharma, Prof. Banibrata Mukhopadhyay, Prof. Sivarani Thirupathi, Prof. Piyali Chatterjee, Prof. Bacham Eswar Reddy, Prof. Gajendra Pandey and Prof. Tarun Deep Saini for their support. 
I also thank all the JAP instructors who taught us in our coursework.
I would also like to thank all the staff members at IIA and office staff of the Department of Physics, IISc, for their help in administrative works.

I thank my friends, contemporaries, and seniors in IISc, Bhaskara and various institutes - Sahel, Prerna, Jyoti, Anirban, Snehalata, Prasanta, Chayan, Deepthi, Dhanush, Ankit, Manika, Sharmila, Anju, Khushboo, Raghu, Sipra, Samyaday, Gaurav, Ranjan, Suchira, Atanu, Shubham, Sudeb, Surajit, Rashid, Samriddhi for making my life memorable. Special thanks to Sindhu for the discussions and help at the beginning of my PhD.

Finally, I want to thank my family members, without whom I would not have been here.

\end{acknowledgements}

\begin{publication}
\textbf{Publications in Refereed Journals}
\begin{enumerate}
    \item {\it UVIT Open Cluster Study. I. Detection of a White Dwarf Companion to a Blue Straggler in M67: Evidence of Formation through Mass Transfer},
        \\Sindhu, N., Subramaniam, A., \textbf{Jadhav, V. V.}, Chatterjee, S., Geller, A. M., Knigge, C., Leigh, N., Puzia, T. H.,   Shara, M., \& Simunovic, M. (2019), 
        \\\href{https://ui.adsabs.harvard.edu/abs/2019ApJ...882...43S}{The Astrophysical Journal, 882, 43}
    \item {\it UVIT Open Cluster Study. II. Detection of Extremely Low Mass White Dwarfs and Post-Mass Transfer Binaries in M67}, 
        \\ \textbf{Jadhav, V. V.}, Sindhu, N., \& Subramaniam, A. (2019),
        \\\href{https://ui.adsabs.harvard.edu/abs/2019ApJ...886...13J}{The Astrophysical Journal, 886, 13}\footnote{presented in Chapter~\ref{ch:UOCS2}}
    \item {\it UVIT/ASTROSAT studies of Blue Straggler stars and post-mass transfer systems in star clusters: Detection of one more blue lurker in M67},
        \\ Subramaniam, A., Pandey Sindhu, \textbf{Jadhav V.} and Sahu Snehalata (2020),
        \\ \href{https://ui.adsabs.harvard.edu/abs/2020JApA...41...45S/abstract}{Journal of Astrophysics and Astronomy, 41,45}
    \item {\it UOCS. III. UVIT catalogue of open clusters with machine learning based membership using Gaia DR2 astrometry},
        \\ \textbf{Jadhav, V. V.}, Pennock, C. M., Subramaniam, A., Sagar R., \& Nayak P. K. (2020),
        \\ \href{https://ui.adsabs.harvard.edu/abs/2021MNRAS.503..236J/abstract}{Monthly Notices of the Royal Astronomical Society, 503, 236}\footnote{presented in Chapter~\ref{ch:UOCS3}}
    \item {\em UOCS. IV. Characterising blue straggler stars in old open cluster King 2 with ASTROSAT},
        \\ \textbf{Vikrant V. Jadhav}, Sindhu Pandey, Annapurni Subramaniam \& Ram Sagar (2021),
        \\ \href{https://ui.adsabs.harvard.edu/abs/2021JApA...42...89J}{Journal of Astrophysics and Astronomy, 42, 89}  \footnote{presented in Chapter~\ref{ch:UOCS4}}      
    \item \textit{Blue Straggler Stars in Open Clusters using Gaia: Dependence to Cluster Parameters and Possible Formation Pathways},
        \\ \textbf{Vikrant V. Jadhav} \& Annapurni Subramaniam (2021),
        \\ \href{https://ui.adsabs.harvard.edu/abs/2021MNRAS.507.1699J/abstract}{Monthly Notices of the Royal Astronomical Society, 507, 1699} \footnote{presented in Chapter~\ref{ch:BSS_catalogue}}        
    \item \textit{High Mass-ratio Binary Population of Open Clusters and their radial segregation},
        \\ \textbf{Vikrant V. Jadhav}, Kaustubh Roy, Naman Joshi \& Annapurni Subramaniam (2021),
        \\ \href{https://ui.adsabs.harvard.edu/abs/2021AJ....162..264J/abstract}{The Astronomical Journal, 162, 264}
    \item \textit{UOCS.VI. UVIT/AstroSat detection of low-mass white dwarf companions to 4 more blue stragglers in M67},
        \\ Sindhu Pandey, Annapurni Subramaniam \& \textbf{Vikrant V. Jadhav} (2021),
        \\ \href{https://ui.adsabs.harvard.edu/abs/2021MNRAS.507.2373P/abstract}{Monthly Notices of the Royal Astronomical Society, 507, 2373}
    \item \textit{UOCS - VII. Blue Straggler Populations of Open Cluster NGC 7789 with UVIT/AstroSat},
        \\ Kaushar Vaidya, Anju Panthi, Manan Agarwal, Sindhu Pandey, Khushboo K. Rao, \textbf{Vikrant Jadhav} \& Annapurni Subramaniam (2022),
        \\ \href{https://ui.adsabs.harvard.edu/abs/2022arXiv220108773V}{Monthly Notices of the Royal Astronomical Society, 511, 2274}
    \item \textit{Characterization of hot populations of Melotte 66 open cluster using Swift/UVOT},
        \\ Khushboo K. Rao, Kaushar Vaidya, Manan Agarwal, Anju Panthi, \textbf{Vikrant Jadhav} \& Annapurni Subramaniam (2022, submitted)
        \\ {Monthly Notices of the Royal Astronomical Society}
        
    \item \textit{UOCS--VIII. UV Study of the open cluster NGC 2506 using ASTROSAT},
        \\ Anju Panthi, Kaushar Vaidya, \textbf{Vikrant Jadhav}, Khushboo K. Rao,
Annapurni Subramaniam, Manan Agarwal \& Sindhu Pandey (2022 submitted),
        \\ {Monthly Notices of the Royal Astronomical Society}
\end{enumerate}

\textbf{Proceedings}
\begin{enumerate}

    \item {\it Detection of White Dwarf Companions to Blue Straggler Stars from UVIT Observations of M67}, 
        \\N, Sindhu., Subramaniam, A., Geller, A. M., \textbf{Jadhav, V.}, Knigge, C., Simunovic, M., Leigh, N., Shara, M., \& Puzia, T. H. (2019), 
        \\\href{https://ui.adsabs.harvard.edu/abs/2020IAUS..351..482S/abstract}{IAUS, 351, 482}
\end{enumerate}


\end{publication}

\begin{abstract}
Binary systems can evolve into immensely different exotic systems such as blue straggler stars (BSSs), yellow straggler stars, cataclysmic variables, type Ia supernovae depending on their initial mass, the orbital parameters and how the two components evolve. The formation and evolution scenarios for some of these exotic objects are still ambiguous, as they differ significantly from the standard single stellar evolution theory. The UV to infrared panchromatic study of binary stars can characterise them, determine evolutionary history and forecast their evolution. 

The aim of this thesis is to study the demographics of post-mass-transfer systems (BSSs, white dwarfs (WDs) and blue lurkers) in the open clusters and their formation pathways. There are multiple formation pathways theorised for the BSSs, but which pathways are favoured and where is not entirely known. We also need a homogeneous catalogue of such systems to analyse the pathways. The \textit{Gaia} DR2 release in 2018 has provided deep and precise all-sky data required to determine cluster membership and find post-mass-transfer systems. Similarly, UV imaging from UVIT/\textit{AstroSat} telescope is crucial to study the hot components in the interacting systems such as WDs and BSSs. 
In this thesis, I have utilised UVIT, \textit{Gaia} and other archival data to do a comprehensive panchromatic study of open cluster BSSs and post-mass-transfer systems.

This thesis consists of 7 chapters. \textbf{Chapter \ref{ch:intro}} contains the basics of star formation, star clusters and evolution of low mass stars. The chapter also contains the overview of binary evolution and the resulting stellar exotica: BSSs are the most massive stars in a cluster formed via binary or higher-order stellar interactions; blue lurkers are lower mass stars formed through such interactions; extremely low mass (ELM) WDs and hot subdwarfs are results of mass loss by the progenitor. 
\textbf{Chapter \ref{ch:data_and_methods}} contains the details about observational facilities and telescopes used in this thesis. The chapter also contains the details of the research methods and models: photometry, spectral energy distributions of single and binary sources, the isochrone models and evolutionary tracks.  

In \textbf{chapter \ref{ch:UOCS3}}, I present the study of six open clusters (Berkeley 67, King 2, NGC 2420, NGC 2477, NGC 2682 and NGC 6940) using the UVIT and \textit{Gaia} EDR3. We used combinations of 
a Gaussian mixture model and a supervised machine-learning algorithm to determine cluster membership. This technique is robust, reproducible, and versatile in various cluster environments. We could detect 200--2500 additional members per cluster using our method with respect to previous studies, which helped estimate mean space velocities, distances, number of members and core radii. The UVIT photometric catalogues, including BSSs, main-sequence and red giants, are also provided. The UV--optical catalogues are used to briefly analyse the six clusters using colour-magnitude diagrams.

\textbf{Chapter \ref{ch:UOCS2}} describes the detailed study of open cluster M67. The UV--optical colour-magnitude diagrams suggested the presence of excess UV flux in many members, which could be extrinsic or intrinsic to them. We constructed multi-wavelength spectral energy distributions (SEDs) using photometric data from the UVIT, \textit{Gaia} DR2, 2MASS and \textit{WISE} surveys along with optical photometry. We fitted model SEDs to 7 WDs and found 4 of them have mass $>$ 0.5 \Msun\ and cooling age of less than 200 Myr, thus demanding BSS progenitors. SED fits to 23 stars detected ELM WD companions to WOCS2007, WOCS6006 and WOCS2002, and a low mass WD to WOCS3001, which suggest these to be post-mass-transfer systems. 12 sources with possible WD companions need further confirmation. 9 sources have X-ray and excess UV flux, possibly arising from stellar activity. The increasing detection of post-mass-transfer systems among BSSs and main-sequence stars suggest a strong mass-transfer pathway and stellar interactions in M67. 

\textbf{Chapter \ref{ch:UOCS4}} presents the study of BSSs in open cluster King 2 with an age of $\sim$ 6 Gyr and a distance of $\sim 5700$ pc. The {\it Gaia} EDR3 membership showed the presence of 39 BSS candidates in the cluster. We created multi-wavelength SEDs of all the BSSs. Out of 10 UV detected BSSs, 6 bright ones fitted with double component SEDs and were found to have hotter companions with properties similar to extreme horizontal branch/subdwarf B (sdB) stars, with a range in luminosity and temperature, suggesting a diversity among the hot companions. We suggest that at least 15\% of BSSs in this cluster are formed via the mass transfer pathway.

\textbf{Chapter \ref{ch:BSS_catalogue}} presents the census and a statistical study of BSSs in the Galactic open clusters. We first created a catalogue of BSSs using \textit{Gaia} DR2 data. Among the 670 clusters older than 300 Myr, we identified 868 BSSs in 228 clusters and 500 BSS candidates in 208 clusters. In general, all clusters older than 1 Gyr and having mass greater than 1000 \Mnom\ have BSSs. The average number of BSSs increases with the age and mass of the cluster, and there is a power-law relation between the cluster mass and the maximum number of BSSs in the cluster. We introduced the term fractional mass excess (\Me) for the BSSs. We find that at least 54\% of BSSs have \Me\ $<$ 0.5 (likely to have gained mass through a binary mass transfer), 30\% in the $1.0 <$ \Me\ $< 0.5$ range (likely to have gained mass through a merger) and up to 16\% with \Me\ $>$ 1.0 (likely from multiple mergers/mass transfer). We also find that the percentage of low \Me\ BSSs increases with age, beyond 1--2 Gyr, suggesting an increase in formation through mass transfer in older clusters. The BSSs are radially segregated, and the extent of segregation depends on the dynamical relaxation of the cluster. 

Finally, in \textbf{chapter \ref{ch:summary}}, we present the conclusions and summary of the work. The chapter also contains ideas for future studies, including panchromatic and spectroscopic analysis of noteworthy clusters and simulations to compare the observational properties to the expected properties. 

\end{abstract}

\tableofcontents
\listoffigures
\listoftables
\begin{abbreviations}

\centering
\begin{small}
\begin{tabular}{llll}
    Asymptotic Giant Branch 	& \textbf{	 AGB	}&	    Near Ultra-Violet 	& \textbf{	 NUV	}\\
    Blue Straggler Star 	& \textbf{	 BSS 	}&	    Open Cluster 	& \textbf{	 OC	}\\
    Charge-Coupled Device 	& \textbf{	 CCD	}&	    Proper Motion 	& \textbf{	 PM	}\\
    Colour-Magnitude Diagram 	& \textbf{	 CMD 	}&	Radial Velocity	& \textbf{	RV	}\\
Extreme Horizontal Branch	& \textbf{	EHB	}&	    Red Giant 	& \textbf{	 RG	}\\
    Extremely low-mass White Dwarf 	& \textbf{	 ELM WD	}&	    Red Giant Branch 	& \textbf{	 RGB	}\\
    Far Ultra-Violet 	& \textbf{	 FUV 	}&	    Spectral Energy Distribution 	& \textbf{	 SED 	}\\
    Field Of View 	& \textbf{	 FOV 	}&	    UltraViolet 	& \textbf{	 UV  	}\\
    Full Width Half Maximum 	& \textbf{	 FWHM	}&	    Ultra-Violet Imaging Telescope 	& \textbf{	  UVIT	}\\
    Giant Molecular Cloud 	& \textbf{	 GMC 	}&	    UVIT Open Cluster Study 	& \textbf{	 UOCS	}\\
    Hertzsprung--Russell Diagram 	& \textbf{	 HRD	}&	    Vector Point Diagram 	& \textbf{	 VPD	}\\
    Horizontal Branch 	& \textbf{	 HB	}&	    White Dwarf 	& \textbf{	 WD	}\\
    InfraRed 	& \textbf{	 IR	}&	    WIYN Open Cluster Study 	& \textbf{	 WOCS	}\\
    Main Sequence 	& \textbf{	 MS	}&	    Yellow Straggler Star 	& \textbf{	 YSS 	}\\
    Main Sequence Turn-Off 	& \textbf{	 MSTO	}&	    Zero-Age Horizontal Branch 	& \textbf{	 ZAHB 	}\\
    Mass Transfer 	& \textbf{	 MT	}&	    Zero-Age Main Sequence 	& \textbf{	 ZAMS 	}\\
Membership Probability	& \textbf{	MP	}&		&		\\

\end{tabular}
\end{small}

\end{abbreviations}
\mainmatter
\setcounter{page}{1}

\begin{savequote}[100mm]
An outlier is an observation that differs so much from other observations as to arouse suspicion that it was generated by a different mechanism
\qauthor{D. M. Hawkins}
\end{savequote}

\chapter{Introduction}
\label{ch:intro}
\begin{quote}\small
\end{quote}

The night sky is filled with uncountable stars with assorted brightness and colours. All these stars are born through the gravitational collapse of gaseous nebulae known as \textit{giant molecular clouds} (GMCs). The GMCs are massive ($\sim 10^7$ \Msun) cold dense gas clouds which break into smaller fragments due to self-gravity and inherent turbulence of the diffuse interstellar medium \citep{McKee2007ARA&A..45..565M}. 
The temperature and pressure at the cores of these gravitationally collapsing fragments increase with time, and eventually, they become hot and dense enough to start hydrogen fusion. And a star is born.

\section{Star formation and star clusters}

A star is rarely born in isolation. The GMCs break into multiple fragments of various sizes. The hierarchical fragmentation leads to the observed initial mass function \citep{Salpeter1955ApJ...121..161S, Kroupa2001MNRAS.322..231K}. The protostars accrete the gas from the surroundings and simultaneously ionise the surrounding material with radiation and winds. The formation of an individual star depends on the complex interplay between the GMC properties (temperature, density, pressure, magnetic field) and stellar feedback (winds, supernovae). 
Over time, most of the gas in the GMC is driven out by the stellar feedback or accreted by individual stars resulting in star formation efficiency of 0.2\%--20\% \citep{Shu1987ARA&A..25...23S, Murray2011ApJ...729..133M}. Once the stars are formed, their future is defined by the internal and external dynamical forces. If the stars stay gravitationally bound to each other, they remain as a cluster; otherwise, they will be dispersed in the galaxy \citep{Adamo2020SSRv..216...69A}.

\citet{Krause2020SSRv..216...64K} defined a \textit{cluster} as a gravitationally bound group of at least 12 stars not dominated by dark matter. However, kinematic information is needed to verify the bound-ness of a group of stars, which is not possible in every scenario, especially extragalactic. \citet{Gieles2011MNRAS.410L...6G} provided a kinematic definition for differentiating between clusters and associations.
\begin{equation}
    T_{cr} \equiv 10\left(\frac{R_{eff}^3}{GM} \right)^{1/2} \qquad
    \Pi = Age/T_{cr}
\end{equation}
Where, $T_{cr}$ is stellar crossing time of in the cluster, $R_{eff}$ is the projected half-light radius, $M$ is cluster mass and $Age$ is cluster age. For stellar agglomerates older than 10 Myr, $\Pi>1$ would indicate clusters while $\Pi<1$ would indicate stellar associations.

There are two main classes of clusters in the Milky Way: i) open clusters (OCs) and ii) globular clusters.

\begin{figure}[!ht]
    \centering
    \includegraphics[width=0.98\textwidth]{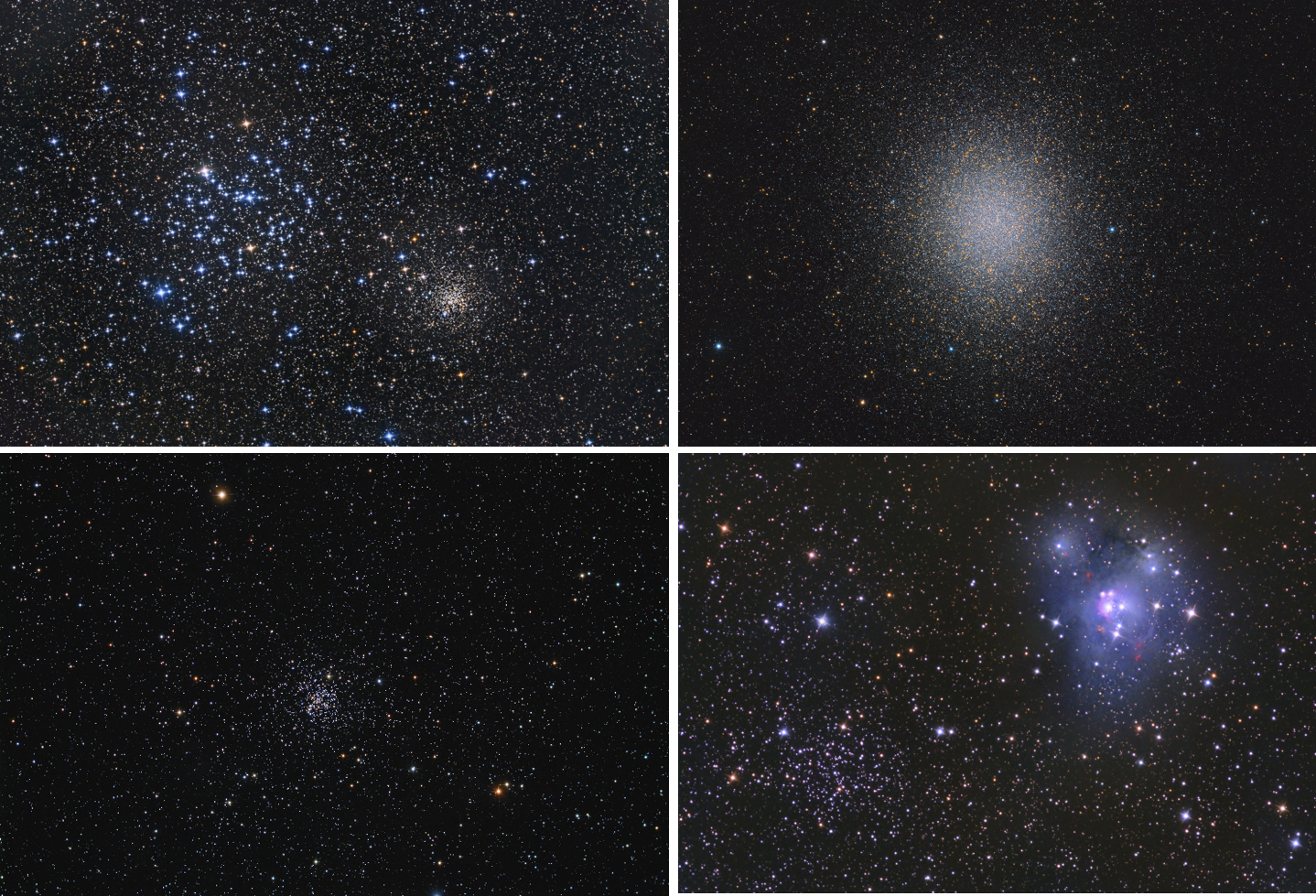}
    \caption{Composite images of open and globular star clusters (Credits: NASA APOD). \textit{Top left:} a pair of OCs, M35 and NGC 2158 (Credits: Dieter Willasch), \textit{top right:} globular cluster Omega Centauri (Credits: Ignacio Diaz Bobillo), \textit{bottom left:} OC NGC 2682 (Credits: Noel Carboni and Greg Parker) and \textit{bottom right:} OC NGC 7142 and nearby reflection nebula NGC 7129. (Credits: Steve Cannistra)}
    \label{fig:1_clusters}
\end{figure}

OCs are smaller clusters found throughout the Milky Way disc. OCs are typically 
\begin{itemize}
    \item young (a few Myr to a few Gyr)
    \item less massive ($\lesssim10^5$ \Msun)
    \item up to 10000 stars
    \item near the Galactic plane
\end{itemize}

The globular clusters are aptly named as such due to their obvious spherical shape. The Galactic globular clusters are typically 
\begin{itemize}
    \item old ($>10$ Gyr)
    \item massive ($>10^4$ \Msun)
    \item more than 10000 stars
    \item present in the Galactic halo
\end{itemize}

Fig.~\ref{fig:1_clusters} shows images of a few OCs and a globular cluster. 
However, this distinction in classification is not absolute. The Magellanic Clouds are known to harbour massive young clusters, and the Galactic bulge contains ancient and massive clusters \citep{Palma2019MNRAS.487.3140P, Ferraro2021NatAs...5..311F}. There are also loosely bound collections of stars present in the spiral arms known as \textit{stellar associations}.
Overall, there are thousands of OCs, more than 150 globular clusters and thousands of stellar associations in the Milky Way.

The star clusters are test-beds for the study of stellar evolution in diverse physical environments because all the stars are formed from the same GMC \citep{Krause2020SSRv..216...64K}. Consequently, their chemical makeup is similar to each other, and their evolution only differs due to the stellar mass and possible interactions. The coeval nature of all members is then used to observe the different phases of stellar evolution across the stellar mass range \citep{Krumholz2019ARA&A..57..227K}. The same can be done across clusters of different ages to get the complete picture of stellar evolution phases across various ages and masses.
Additionally, as all cluster members are situated at the same location, the improved statistics can be used to accurately calculate their distance, project their extinction and estimate their age, and metallicity. 

\subsection{Cluster membership} \label{sec:1_membership}
\begin{figure}
    \centering
    \includegraphics[width=0.98\textwidth]{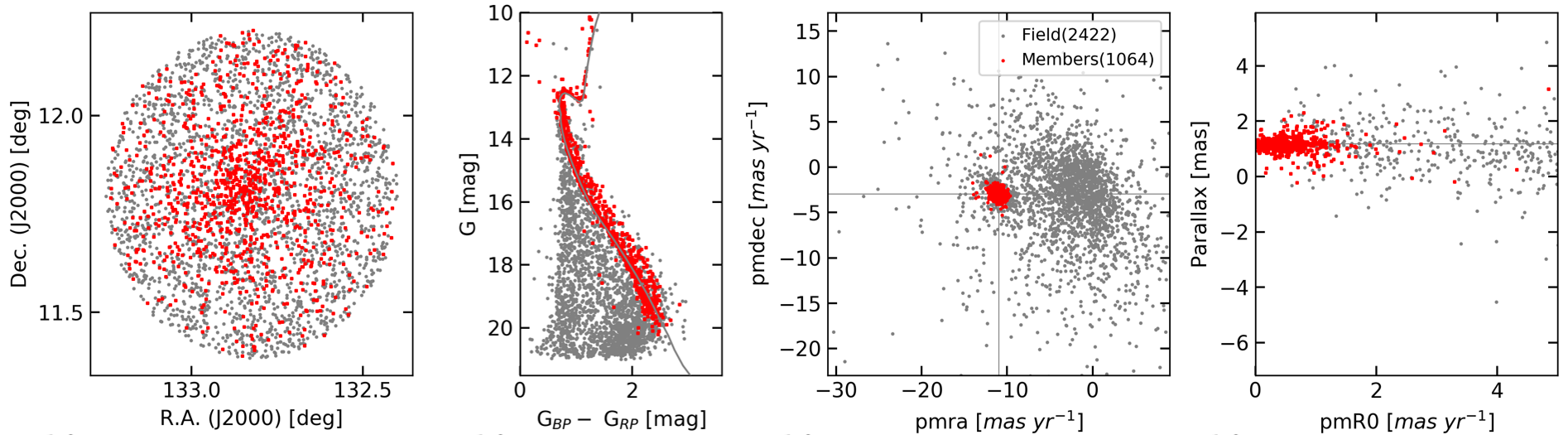}
    \caption{Schematic plots that help in membership determination. Plots show the spatial distribution, CMD, VPD and parallax-proper motion (PM) plots for members (as red) and field stars (as grey) of NGC 2682.
    }
    \label{fig:1_membership}
\end{figure}
    
Before commencing the study of individual stars in a cluster, it is vital to establish their cluster membership. In the early twentieth century, all stars in the vicinity of the cluster were used to study the cluster. \citet{van1942} made significant improvements in the membership determination using the PM of stars in the vicinity of the clusters.
\citet{Vasilevskis1958} and \citet{Sanders1971} pioneered the techniques of membership probability (MP) determination using vector point diagrams (VPDs). As the accuracy of PM measurements improved, membership determination using VPDs were also enhanced (\citealp{Sagar1987, Zhao1990, Balaguer1998, Bellini2009} and references therein). Assuming that the field and cluster stars produce overlapping Gaussian distributions in the VPD, new techniques were developed to separate the two populations \citep{Bovy2011, Vasiliev2019}.

The arrival of \textit{Gaia} was instrumental in study of star clusters \citep{Gaia2016A&A...595A...1G}. Trigonometric parallaxes and accurate PMs from \textit{Gaia} DR2/EDR3 have led to accurate identification of cluster members: using Hertzsprung--Russell diagram (HRD) of star clusters to improve the models of stellar evolution. \citep{Gaia2018a}; \citet{Cantat2018A&A...618A..93C} identified cluster members in 1229 OCs and discovered 60 new clusters using ($\mu_{\alpha}, \mu_{\delta}, \varpi$) clustering. Similarly, \citet{Liu2019, Sim2019, He2020, Castro2020} discovered new OCs by applying visual or machine learning techniques to \textit{Gaia} DR2 and identified cluster members.

Fig.~\ref{fig:1_membership} shows a sample of schematic plots that help in membership determination. The first two panels from the left show that the spatial distribution and CMD location of the field and member stars have significant overlap. Comparatively, VPD shown in the third panel suggests that the cluster members are well separated from the field stars in the velocity plane. The last panel shows the segregation of cluster members in the parallax versus modified PM (see chapter~\ref{ch:UOCS3}) plane. The figure, therefore, demonstrates that a combination of spatial location, CMD, VPD and parallax would be the ideal method to select members in a cluster. Chapter~\ref{ch:UOCS3} gives more details about cluster membership of OCs using the latest \textit{Gaia} EDR3 data \citep{Gaia2021A&A...649A...1G}.

\section{Evolution of low mass stars}

\begin{figure}[!hb]
    \centering
    \includegraphics[width=0.9\textwidth]{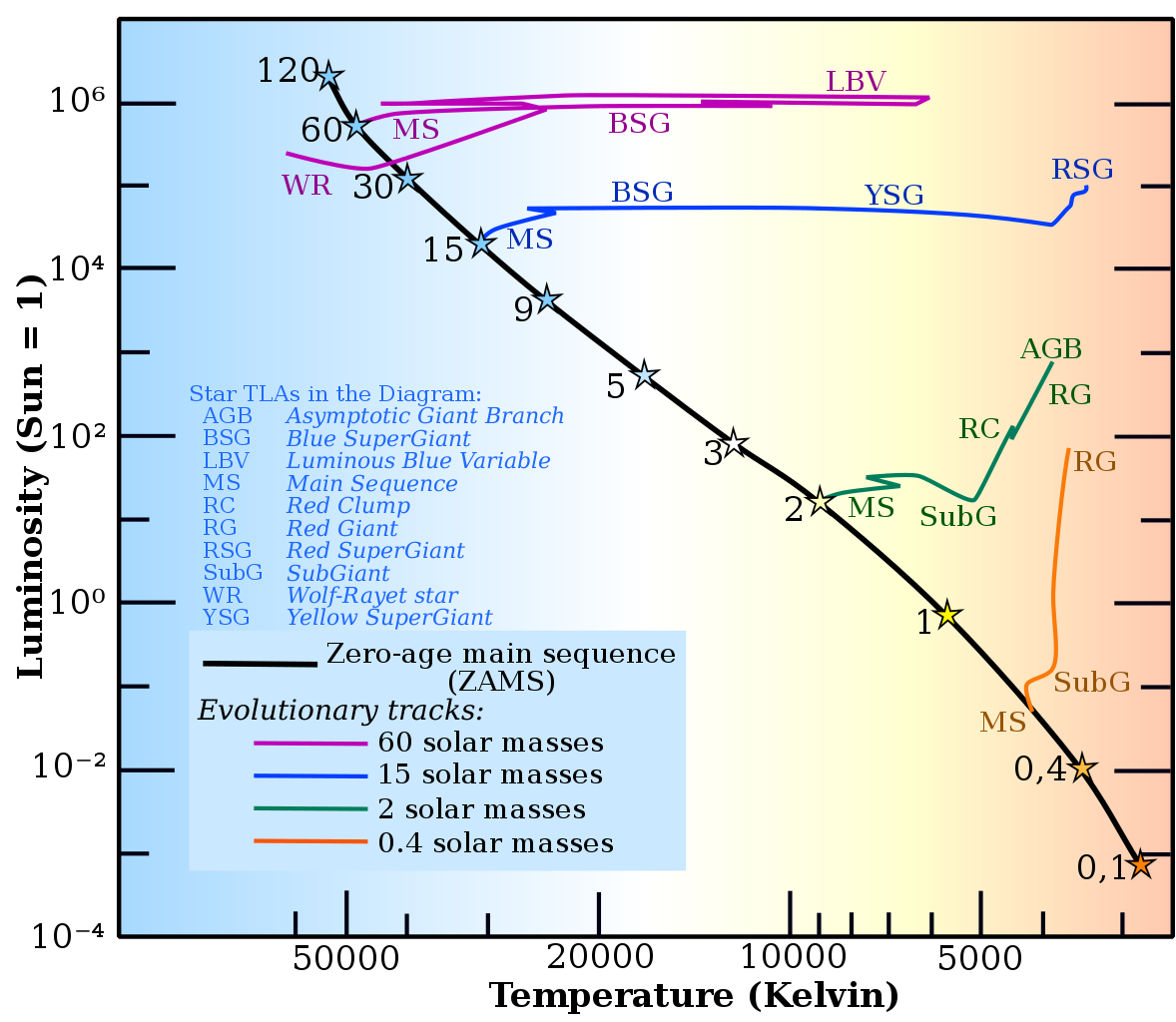}
    \caption{Evolutionary tracks for single stars in mass range 0.5--60 \Mnom\ indicating the various phases each star passes through. (Credits: Wikipedia Commons)}
    \label{fig:1_evo_tracks}
\end{figure}

Fig.~\ref{fig:1_evo_tracks} shows how the stars of different masses move through the HRD. The low mass stars go through the sub-giant, red giant branch (RGB) and asymptotic giant branch (AGB) phases. In comparison, high mass stars go through yellow, blue, and red supergiant phases. This thesis work pertains to less than 8 \Mnom\ stars.

\begin{figure}
    \centering
    \includegraphics[width=0.88\textwidth]{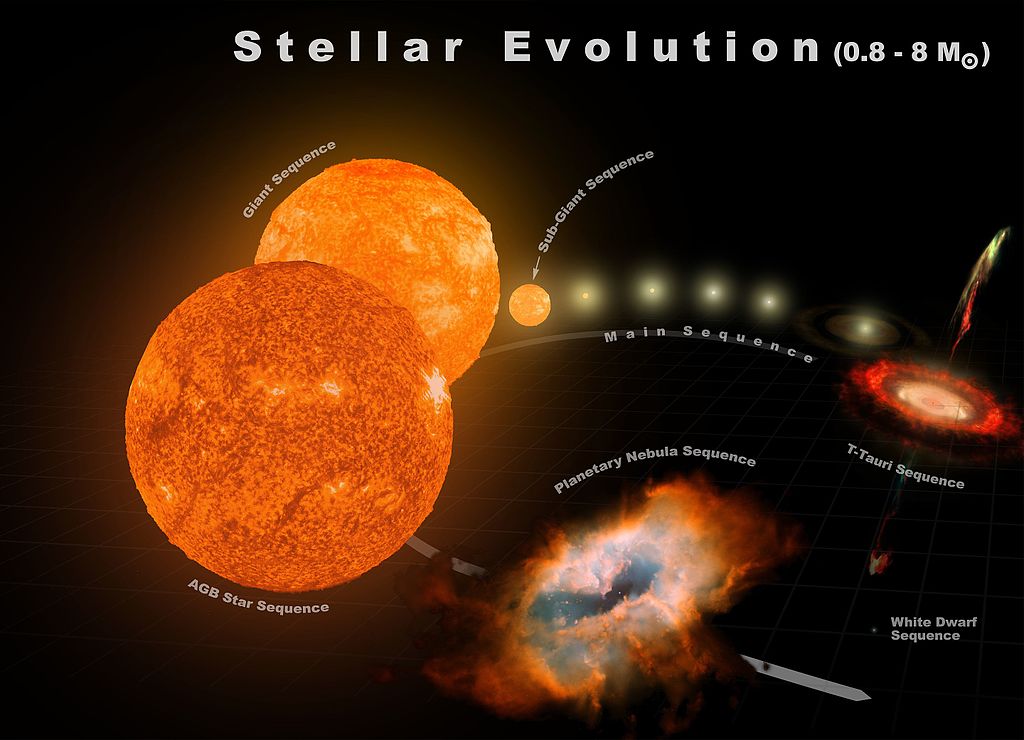}
    \caption{Lifecycle of a low mass star (0.8--8\Mnom). (Credits: Wikimedia Commons)}
    \label{fig:1_lifecycle_low_mass}
\end{figure}

The evolution of single stars primarily depends on its mass. Fig.~\ref{fig:1_lifecycle_low_mass} shows the lifecycle of Sun-like stars. The star starts its life as a T-Tauri star and reaches the main sequence (MS) where it spends most of its time \citep{Kippenhahn1990sse..book.....K}. As the hydrogen in the stellar core depletes, the core begins gravitational collapse. The heat released from this collapse begins H-shell burning, and the H-shell burning gets stronger and stronger as the mass of the core increases due to the He-ash generated from the shell burning. The radiation pressure from the H-shell burning causes the star to expand. The surface temperature drops due to the expansion, and the star becomes a sub-giant. At lower temperatures, the radiation can more freely travel outwards. Hence, the star gets brighter while staying at the same temperature. The continued H-shell forces the star to expand more and become a red giant (RG) \citep{Iben1991ApJS...76...55I, Boehm1992itsa.book.....B}. 

Depending on the mass of the He core, it may or may not ignite. Stars with an initial mass of $<0.5$ \Msun\ will never start He-burning and become He-core white dwarfs (WDs). The cores of 0.5--3 \Msun\ stars contract so much that the electrons become degenerate before the start of He-burning.
The degeneracy allows the temperature to increase ($\sim 10^8$ K) without any increase in pressure, and without increasing pressure, the core cannot expand and cool. This leads to a runaway start to the He-burning, called \textit{He-flash} \citep{Boehm1992itsa.book.....B}. The stellar cores of stars massive than 3 \Msun\ do not become degenerate, and they start the He-burning through the triple-alpha process ($3\ ^4He \longrightarrow ^{12}C$) in a quasi-equilibrium manner. Once the He-burning starts in the core, the energy generation in the He-core and H-shell burning leads to a slight increase in temperature but reduces the luminosity due to the reduction in the radius. Stars in the 0.5--3 \Msun\ range, after He-flash, will have similar core mass, and they form a clump in the HRD called the \textit{red clump}. In older clusters, in the He-core-burning phase, their luminosity is almost constant, but their temperature is determined by the amount of H envelop present around the core. Overall, these stars create a constant luminosity locus near $10^{1.5}$ \Lsun\ collectively called as the \textit{horizontal branch} (HB; \citealt{deBoer2008sse..book.....D}).

The core He burns for roughly 100 Myr, and then the core starts gravitational collapse again. The resulting heat and pressure start the He-shell burning and H-shell burning at the edges of the core. Low mass stars cannot ascend the AGB due to the reduced mass in the envelope, and their cores contract to become a WD. More massive stars ascend the AGB and eventually become a CO core WD \citep{Vassiliadis1994ApJS...92..125V, deBoer2008sse..book.....D}.

\section{Stellar multiplicity}

The evolution scenarios given above are for single stars evolving in isolation. However, the majority of stars in the universe are not single stars. The binary fraction of FGK, B and O type stars is 34\%, 56--58\% and 42--69\% respectively (\citealt{Luo2021arXiv210811120L} and references therein). Most of these stars are formed as bound binary stars during the GMC fragmentation. A minority of binary pairs, typically in dense clusters, can also form by capturing one star by another.

The binary stars can be classified and identified depending on their properties as follows. 
\begin{itemize}
    \item Astrometric binaries:
        These binaries can be resolved through a sufficiently large telescope. These are only identifiable in the solar neighbourhood. However, we need multi-epoch imaging to confirm that it is not a coincidence and that the stars are bound to each other. Recently, resolved binaries are being identified using \textit{Gaia} astrometric data   \citep{ElBadry2021MNRAS.506.2269E}. In addition, \citet{Belokurov2020MNRAS.496.1922B} showed that it is possible to identify nearly unresolved binary stars using the wobble in the \textit{Gaia} astrometry.
        
    \item Spectroscopic binaries: 
        Unresolved binary stars have their own predefined spectra. The orbital motion of the stars shifts the lines in the stellar spectra in a cyclic fashion, which can be used to identify and characterise the binary components. The binary is called \textit{double-lined spectroscopic binary} (SB2) if lines from both components are seen oscillating. This happens when both components have similar flux.
        If one of the component is too faint, only one set of lines oscillates and the binary is called \textit{single-lined spectroscopic binary} (SB1). We can calculate orbital periods, eccentricity, mass function and primary temperature from SB1 binaries. SB2s can provide further information about the mass ratio, minimum masses, and temperature of both stars. 
        However, time-consuming multi-epoch spectroscopic monitoring and significant post-processing are required to reduce the data and determine binary parameters (e.g., \citealt{Mathieu2000ASPC..198..517M, Pourbaix2004A&A...424..727P}). 
    \item Photometric binaries:
        Unresolved binaries with different temperatures can be identified using their position in the colour-magnitude diagram (CMD) and their spectral energy distribution (SED), assuming multiwavelength observations spanning ultraviolet (UV) to infrared (IR) are available. \citet{Thompson2021AJ....161..160T} demonstrated that OC binaries could be identified using optical--IR SEDs. Binaries with different temperatures (especially WD+MS) can be identified using UV--IR SEDs \citep{Jadhav2019ApJ...886...13J, Rebassa2021MNRAS.506.5201R}.
        
    \item Eclipsing binaries:
        If the binary orbital plane lies along the line of sight, the binary components can eclipse one another. Eclipsing binaries provide the most information such as distance, inclination, radii, and masses of components in addition to the typical information from spectroscopic binaries. Detecting such eclipses requires continuous monitoring of the sky with high precision photometric instruments. Recent monitoring by \textit{Kepler} and \textit{TESS} missions has greatly increased the sample of known eclipsing binaries \citep{Kirk2016AJ....151...68K}.
\end{itemize}

\subsection{Evolution of binary systems}

If the binary stars are far apart, both stars evolve independently without affecting the end products. However, components of an interacting binary evolve quite differently from single stars. Close tidal interactions can lead to non-spheroidal shapes and also induce/reduce rotational periods affecting the evolution of one or both components. However, the most significant factor in determining the evolution of a binary system is the mass transfer (MT) between the two components. There are two main ways for MT:

\begin{figure}[!ht]
    \centering
    \includegraphics[width=0.8\textwidth]{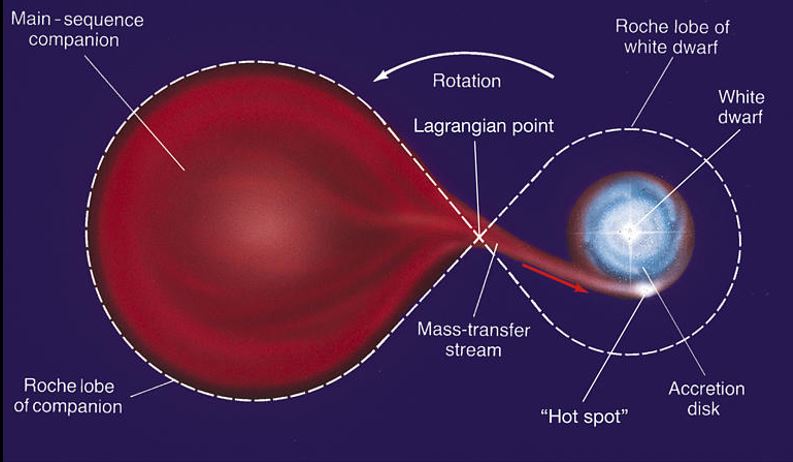}
    \caption{Example of the Roche lobe of an accreting binary system. (Credits: Pearson Prentice Hall, Inc, 2005)}
    \label{fig:1_roche_lobe}
\end{figure}

\begin{itemize}
    \item Roche lobe overflow:
    Roche lobe is the gravitational equipotential surface of a binary system. The gravitational force on the stellar content within one's own Roche lobe is dominated by the same star. In contrast, the material outside the Roche lobe is pulled away from the star by its companion. 
    Fig.~\ref{fig:1_roche_lobe} shows an example of a Roche lobe overflow in a binary system. Here, the left star has evolved to the point where its surface touches the Roche lobe. Hence, MT is happening through the L$_1$ Lagrange point towards its companion.
    \item Wind accretion:
    As a star evolves and increases in radius, the radiation pressure pushes material outside the star in the form of stellar wind. The companion can accrete this gas. The gas typically forms an accretion disc due to the angular momentum in the material. Although most of the stellar wind is lost, simulations have shown that wind accretion can reach an efficiency of up to 45\% for binaries with 2000--10000 day orbits \citep{Abate2013A&A...552A..26A}. 
\end{itemize}

The method of MT primarily depends on the separation of the binaries. Wide binaries do not allow Roche lobe overflow, while wind accretion occurs in both close or wide binaries. Short period binaries can become common envelop binaries due to the small Roche lobe. Both forms of MT lead to angular momentum transfer and changes in orbital period and eccentricity. In general, the MT process leads to loss of angular momentum and circularisation of the orbit. If the binaries are close, they can even become tidally locked.
Binary systems can even undergo multiple instances of MT depending on which component is undergoing Roche lobe overflow. Overall, the ultimate fate of the binary depends on the duration of MT, initial masses of the components, initial/final separation of the binary and instances of MT. This dependency on so many parameters is the reason for the diversity seen in the binary systems and their evolution.

\section{Stellar exotica}

\subsection{Blue straggler stars}
Blue straggler stars (BSSs) are the most massive stars in a cluster. They stand out from other cluster members due to their bluer and brighter position in the CMD. Since their first reporting \citep{Sandage_1953AJ.....58...61S}, multiple mechanisms have been proposed for their formation. All the stars in a cluster are formed almost simultaneously, so there should not be stars brighter than the MS turn-off (MSTO). This apparently longer life is justified by some type of mass accretion by the progenitor of the BSS. The primary formation pathways are as follows:

\begin{enumerate}
    \item \citet{McCrea_1964MNRAS.128..147M} proposed that MT from a binary companion can lead to rejuvenation of the acceptor and formation of a BSS. The MT efficiency depends on the orbital periods: wider orbits have non-conservative MT and leave a remnant behind, while close binaries can have conservative MT and lead to mergers.
    Depending on the type of MT, the binary becomes a BSS and a WD with lower (case A/B MT) or normal (case B/C MT) mass.
    Signatures of hotter/compact companion can be used to detect such systems, using methods such as deconvolving SEDs (e.g., \citealt{Sindhu2019ApJ...882...43S}) and variability in radial velocity (RV).
    MT systems are typically expected to have circular orbits due to the past MT event, but MT in elliptical orbits has also been detected (e.g., \citealt{Boffin2014A&A...564A...1B}). 
    \item Collisions of individual stars or mergers from collisions of binary pairs (2+2) are also linked to the BSS formation \citep{Hills_1976ApL....17...87H,Leonard_1989AJ.....98..217L}. Collisions typically happen in dense environments (such as globular clusters) and do not show chemical peculiarities of an MT event \citep{Sills_1999ApJ...513..428S}. However, \citet{Leigh_2013MNRAS.428..897L} found no correlation between collision rate and number of BSSs, meaning collisions are not the dominant pathway, even in dense globular clusters.
    \item \citet{Naoz_2014ApJ...793..137N} showed that the eccentric Kozai--Lidov mechanism could tighten the inner binary in a hierarchical triple system. \citet{Perets_2009ApJ...697.1048P} suggested that such merger of inner binary has a significant role in BSSs formation in OCs. Presently, such systems will likely be MS+BSS binaries with long eccentric orbits. 
\end{enumerate}

However, different mechanisms are said to dominate in different cluster environments: 
(i) less dense clusters favour binary MT pathway while high-density clusters favour collisional pathway \citep{Davies_2004MNRAS.349..129D}, (ii) binary pathway is dominant in globular clusters of all masses  \citep{Leigh_2007ApJ...661..210L, Knigge_2009Natur.457..288K}, (iii) core collapse of a globular cluster can trigger a burst of BSS formation \citep{Ferraro_2009Natur.462.1028F}, (iv) old, less dense and relaxed clusters favour binary pathway \citep{Mathieu_2015ASSL..413...29M}.

\subsubsection{Evolution of blue stragglers}

\begin{figure}
    \centering
    \includegraphics[width=0.98\textwidth]{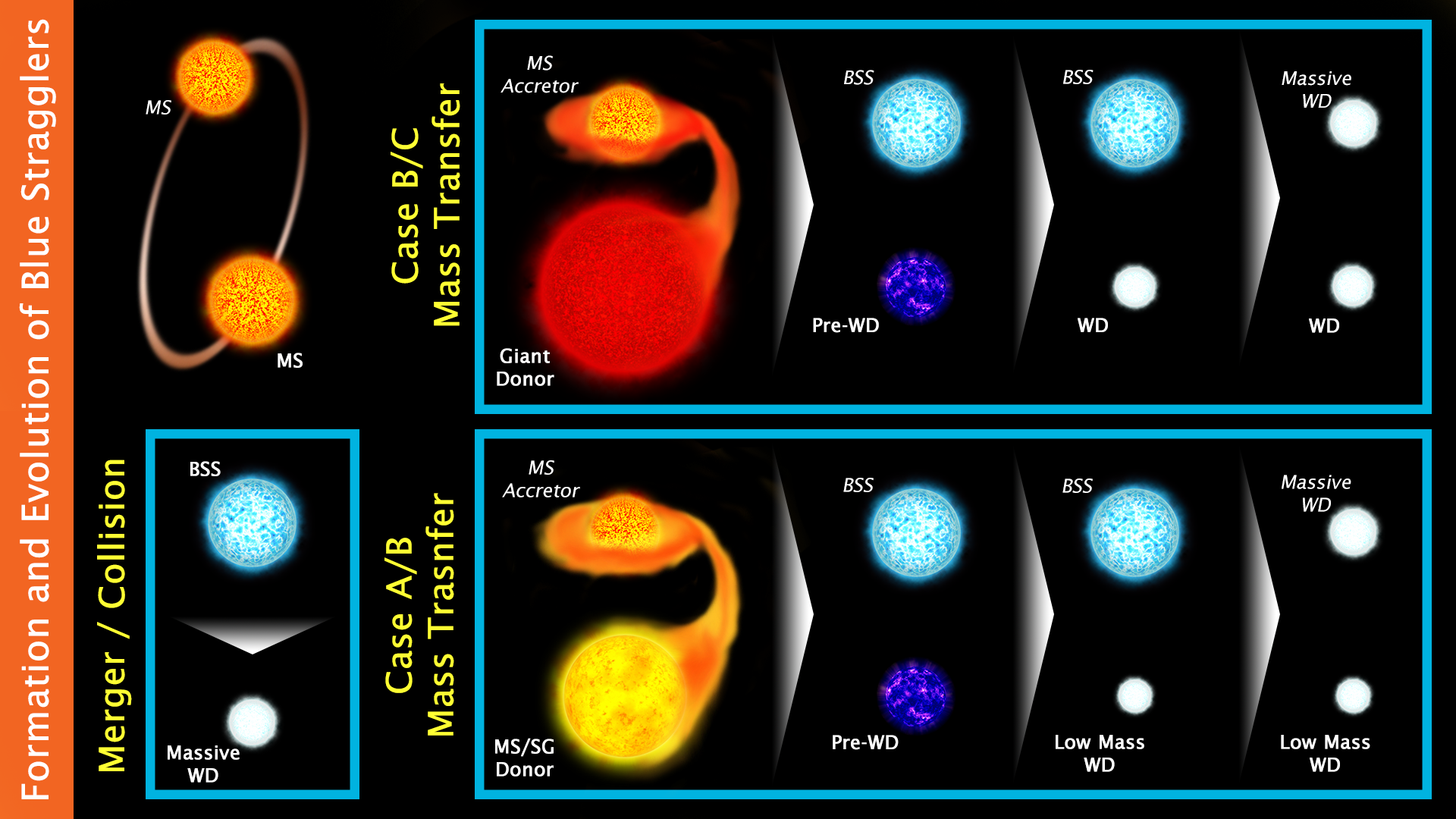}
    \caption{Primary formation pathways of BSSs: collisions/mergers, case A/B MT and case B/C MT.}
    \label{fig:1_BSS_formation_scenarios}
\end{figure}

Fig.~\ref{fig:1_BSS_formation_scenarios} shows a schematic of formation and evolution of BSSs. The collisional/merger products follow a relatively straightforward path of two stars forming a BSS and evolving to a WD. Comparatively, the MT pathway is quite complicated. The MT efficiency depends on the separation of two components, the mass of the two stars and influence from their neighbours. The evolutionary phase of stars at the start of the MT is of utmost importance in terms of the final products. There are three main MT types:
\begin{itemize}
    \item Case A: MT when the donor is in MS
    \item Case B: MT when the donor is in RG phase of evolving towards it
    \item Case C: MT in supergiant phase
\end{itemize}

In the case B and C MT scenario, the donor's core is already formed. Thus, the resultant mass loss does not affect the final evolutionary product of the donor. The top-right panel in Fig.~\ref{fig:1_BSS_formation_scenarios} shows such a scenario where the donor becomes a typical WD while the accretor becomes a BSS. The resultant BSS+WD system can again go through MT if the BSS can fill its Roche lobe. However, the system is unlikely to merge as it has already survived one MT phase. Thus, the final product of such a system will be WD+WD. Here, the donor WD mass will correlate with donor mass, but the accretor WD will have mass corresponding to the BSS mass, which is larger than the original accretor star. 

If the binary stars are close enough for case A or early case B MT and still far enough not to merge, then they can form a BSS+WD system. In this case, the donor's core is not fully formed before the MT begins. Hence, the mass loss will also lead to a lower mass core. Depending on the severity of mass loss, the donor can evolve into a low mass WD. Furthermore, identifying such low mass WD can confirm the case A/B MT pathway as the formation scenario for the BSS.

\subsection{Extremely low-mass white dwarfs}

\begin{figure}[!hb]
    \centering
    \includegraphics[width=0.6\textwidth]{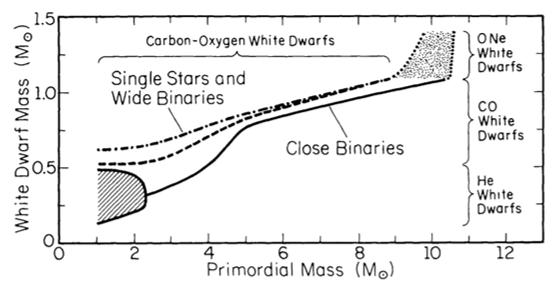}
    \caption{The relation between the WD mass and the progenitor mass in isolated and binary systems. (Credits: \citealt{Iben1991ApJS...76...55I})}
    \label{fig:1_WD_mass_relation}
\end{figure}

90\% of all stars evolve into a WD. However, the mass of the WD depends on the mass of the progenitor. More massive stars lead to massive WDs and vice versa \citep{Iben1991ApJS...76...55I, Cummings2018}. The exact relation depends on the star's metallicity, mass, rotation, and environment. However, Fig.~\ref{fig:1_WD_mass_relation} shows the rough relation between the mass of a progenitor and the resultant WD. 
The most massive stars massive become ONe-core WDs, while lower mass stars become CO-core WDs.
Determining the final WD mass relation and the exact dependencies is an ongoing problem that can only be solved with observations of WDs in clusters, WD binaries and their complete characterisation.

The lower end of the WD mass also has one other limiting factor: the age of the universe. As the low mass stars evolve slower, such stars will not evolve within the $\sim$13 Gyr passed since the big bang. Hence, there is a lower limit of $\sim$0.4 M$_{\odot}$ on the mass of a WD formed through single stellar evolution \citep{Kilic2007ApJ...671..761K}. The oldest stars exhaust their H-fuel and begin the rise in the RGB. At the tip of the RGB, the He-core ignites and begins the conversion of He to C. Hence, the WDs of low mass stars have CO cores. 
However, there are observations of lower mass WDs with He-cores.
WDs of mass of 0.1--0.4 \(M_\odot\) were found to be part of compact binaries \citep{Marsh1995MNRAS.275..828M, Benvenuto2005, Brown2010}. Such extremely low-mass WDs (ELM WDs) are generally found in binary systems where the companions are neutron stars/pulsars \citep{Driebe1998, Lorimer2008}, WDs \citep{Brown2016}, or A/F MS stars in EL CVn-type systems \citep{Maxted2014, Wang2018}. Recently two R CMa-type eclipsing binaries are suggested to have precursors of low mass He WDs \citep{Wang2019}.

The mass loss in the early stage of the evolution can explain the lower mass of the ELM WD. Fig.~\ref{fig:1_WD_mass_relation} shows that WDs from close binaries have lower mass than WD from isolated stars. 
Case-A/B MT can lead to early loss of envelope and result in the He-core WD \citep{Iben1991ApJS...76...55I, Marsh1995MNRAS.275..828M}. 
Thus, MT in close binary systems is a must for the formation of ELM WDs, where companion strips of the envelope of the ELM WD progenitor and the low mass core fails to ignite the core He. 
The MT scenario can also be linked to the acceptor star. While the ELM WD progenitor lost mass, the companion star gained some portion. The companion thus becomes a BSS or a blue lurker (BL; see \S \ref{sec:1_blue_lurker}). The similar formation scenarios mean that detection of an ELM WD alongside a BSS confirms the post-MT nature of both components.

\subsection{Hot subdwarf stars}

\begin{figure}[!ht]
    \centering
    \includegraphics[width=0.7\textwidth]{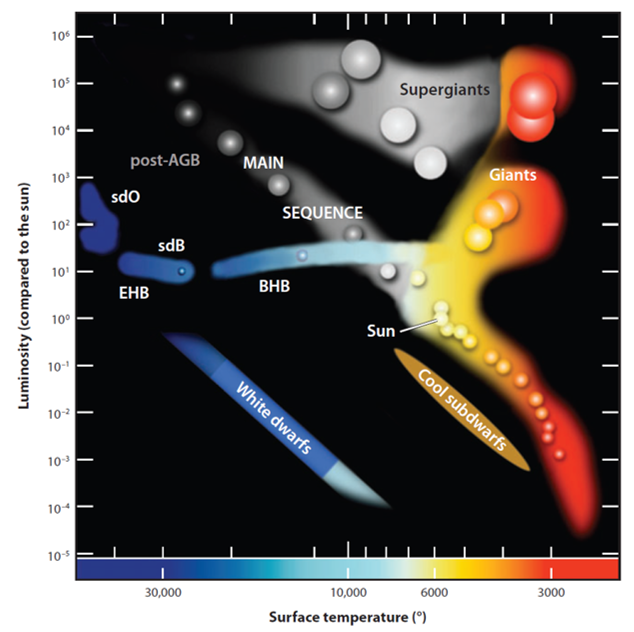}
    \caption{HRD showing position of hot subdwarfs with respect to MS, giants, WDs and cool subdwarf stars. (Credits: \citealt{Heber2009ARA&A..47..211H}).}
    \label{fig:1_CMD_subdwarfs}
\end{figure}

Hot subdwarfs are blue and compact stars that are more luminous than WDs. Depending on their spectral type, they are classified as subdwarf B (sdB) and subdwarf O (sdO). Fig.~\ref{fig:1_CMD_subdwarfs} shows the position of hot subdwarfs in the HRD. Hot subdwarfs have $\sim40$\% binary fraction with a period of fewer than 10 days. 
The review by \citet{Heber2009ARA&A..47..211H} provides detailed observational findings and evolutionary scenarios of hot subdwarfs.
There are two primary scenarios proposed for the formation of hot subdwarfs:

\begin{itemize}
    \item Close binary scenarios: Common envelop ejection(s) in close binaries can lead to formation of sdB+MS star \citep{Han2003MNRAS.341..669H}. A merger of two He-core WDs can also form a rapidly rotating hot subdwarf \citep{Saio2000MNRAS.313..671S}. 
    \item Single star evolution: Hot-flash in the He-core of an RGB star can lead to the star leaving the RGB towards the blue end of the zero-age HB (ZAHB; \citealt{Dcruz1996ApJ...466..359D}). If such a hot flash occurs while the star is on the WD cooling curve, it can become a hot subdwarf and enrich it in He, C, and N.
\end{itemize}
Extreme HB (EHB) stars in globular clusters also occupy similar space as sdB stars in the HRD, and they are proposed to have similar formation mechanisms. However, field hot subdwarfs have more binary fraction than globular cluster EHB stars, likely due to higher merger fractions in globular clusters. Most field sdB stars have WDs or low mass MS stars as companions. However, the exact formalisms for the sdB/sdO origin are still being debated due to the complicated nature of hot flashes and common envelope ejection.

\subsection{Blue lurkers}
\label{sec:1_blue_lurker}

\begin{figure}
    \centering
    \includegraphics[width=0.48\textwidth]{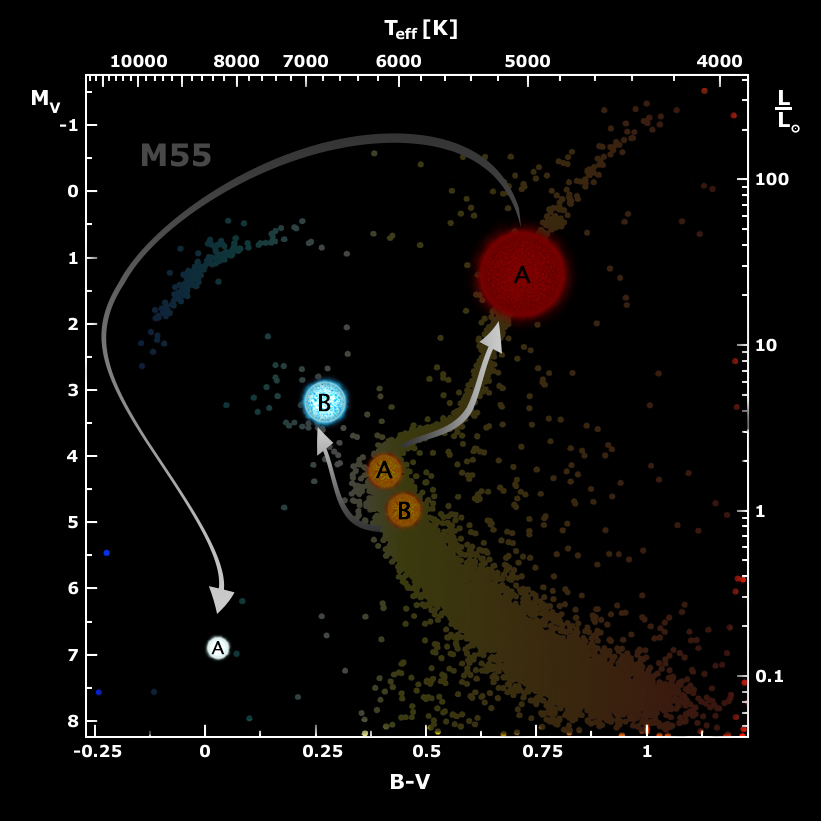}
    \includegraphics[width=0.48\textwidth]{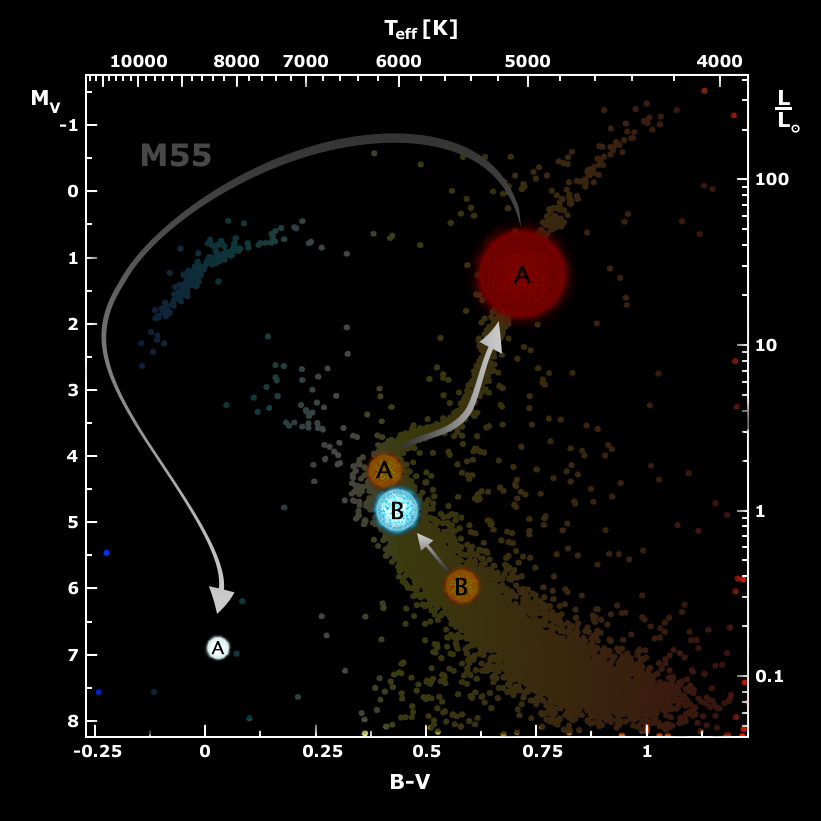}
    \caption{Evolution of binary components in BSSs and BLs in the HRD.
    The left HRD shows the evolution of a binary system, which goes through MT to form a BSS. While the right HRD shows an example of a system which forms a BL.
    Both panels show a donor (A) and accretor (B) moving through the HRD according to their evolution, assuming that there was an MT event which transferred mass from A to B.
    }
    \label{fig:1_BSS_BL_CMD}
\end{figure}

BSSs are the post-MT systems that are brighter and bluer than the MSTO. However, this only happens because the accreted mass is enough to make the progenitor brighter than MSTO. If the amount of accreted gas was less or the accretor star was too small, the jump in the CMD would not be enough to make the accretor brighter than the MSTO. Fig.~\ref{fig:1_BSS_BL_CMD} shows the two scenarios. The left panel gives an example of BSS formation, while the right panel gives an example of an accretor that failed to cross the MSTO. The faint accretor is essentially the same as a BSS, but it cannot be classified as a BSSs simply because it is not bluer and brighter than the MSTO. Hence, these stars were dubbed as \textit{blue lurkers} (\citealt{Leiner2019ApJ...881...47L}). 

Although the figure shows formation through MT, BLs are products of binary (or multiple) evolution which became brighter due to mass accretion similar to the BSSs. Detection of BLs is difficult because they appear as typical stars on the MS. There are a handful of techniques to identify BLs: 
\begin{itemize}
    \item Observation of higher than average rotation ($v$ sin$i$) which is an indication of recent MT
    \item Identification of a companion which can only form through mass donation (e.g., ELM WDs, hot sub-dwarfs)
    \item Presence of chemical peculiarities (r-process, s-process) which are only possible from mass accretion
\end{itemize}

Unfortunately, the abundance studies alone are not convincing evidence of MT without further research into the explicit effects of mergers and MT on stellar abundances. Rotational signatures and hot companions are easily detectable with high-resolution spectra and multiwavelength photometry. However, the uncharacteristic nature of cool compact companions, slowing down of BL after MT, transient nature of atmospheric signatures make the identification of BLs challenging. Similarly, the classification of merger products as BLs is difficult due to the absence of companions and our novice understanding of chemical signatures of mergers.
Chapter~\ref{ch:UOCS2} presents the detection of BLs in NGC 2682 using the characterisation of the hotter companions to MS stars.

\subsection{Yellow stragglers/yellow giants}
\label{sec:1_yss}
Yellow straggler stars (YSSs) or yellow giants are brighter than the subgiant branch but hotter than the RGB. They are thought to be evolving BSSs moving through the subgiant phase. Their cluster membership and stellar mass estimates are required before theorising about the formation mechanisms. \citet{Landsman1997} characterised a YSS+WD (WOCS 2002) pair in NGC 2682 using GHRS and FOS spectra. The WD was found to be an ELM (0.22 \Msun), and thus the most likely formation pathway is the Roche lobe overflow from the WD progenitor leading to the formation of BSS+ELM WD pair which evolved into the YSS+ELM WD. 
\citet{Leiner2016} used \textit{Kepler} K2 observations to study a YSS (WOCS 1015) in NGC 2682 and found that it has a mass twice that of MSTO. The possible formation scenario proposed was one or more binary encounters. However, the formation mechanisms of all YSSs are still not confirmed.
Recently, \citet{Rain_2021arXiv210306004R} found 77 YSS candidates among 408 OCs. A detailed multi-epoch study to determine orbital parameters and spectroscopic/multiwavelength study to characterise the evolutionary states of YSSs and possible companions is necessary to understand the different formation pathways and their frequency clearly.

\section{Dynamical evolution of star clusters}
A star cluster, with its members in motion with respect to the common centre of mass, is located in the gravitational potential of the parent galaxy. The internal movement of the members with a given velocity dispersion in the presence of the galactic potential drives the dynamical evolution of the cluster. Over the life of a cluster, it loses mass due to stellar evolution and dynamical evolution. And the star clusters eventually dissolve into the field of the galaxy.

It takes up to a few million years for the cluster to form and get rid of the residual gas in the parent molecular cloud. The most massive stars evolve through the supernova phase in the next 3--40 Myr. Their stellar remnants (black holes and neutron stars) are also kicked out of the cluster \citep{Faucher2006ApJ...643..332F} leading to a 20\% reduction in mass. In the next 40--100 Myr period, the stellar winds from AGB stars become the dominant mass loss phenomenon. The stellar evolution ceases to be the dominant phenomenon for timescales longer than 100 Myr and the dynamical relaxation becomes the most critical factor in the mass loss of a cluster. 

The stars (and multiple systems) in a cluster exchange kinetic energy through random interactions, increasing the velocity of low mass stars and reducing for low mass stars. The stars with high velocity escape the cluster's gravitational potential, while the massive stars sink to the centre. The timescale over which these interactions substantially affect the motion of each star is known as \textit{dynamical relaxation time} ($T_{relax}$). In a collisional system like a cluster, $T_{relax}$ depends on the number density and mass of a cluster. We calculated the two-body dynamical relaxation time for the cluster as follows \citep{Spitzer1971ApJ...164..399S, Subramaniam1993A&A...273..100S}:
\begin{equation}
\label{eq:1_relaxation}
    \begin{split}
        T_{relax} &= \frac{8.9 \times 10^5 (N_{cluster}\ r^3)^{0.5} }{\langle m \rangle^{0.5} \textrm{log}(0.4N_{cluster})} [yr] \\
        N_{relax} &= Age/T_{relax}
    \end{split}
\end{equation}
\noindent where $r$ is the half-mass radius and $\langle m \rangle$ is the average mass of the stars in the cluster.

The typical $T_{relax}$ is $\sim10^7$ yr for OCs, $\sim10^8$ yr for globular clusters, and $>10^{14}$ yr for galaxies \citep{Binney1987gady.book.....B}. Comparing this to the typical ages of OCs (0.1--10 Gyr), globular clusters ($>10$ Gyr), and galaxies ($>12$ Gyr) indicates that most clusters are relaxed while galaxies are not. Thus, we expect to find mass segregation in OCs. 
In addition to the two body relaxation, interaction with Galactic disc also affects the dynamical evolution of clusters. \citep{Piatti2019MNRAS.489.4367P} showed that the radii of globular cluster decrease with increasing Galactic potential. Similarly, \citep{Piatti2020RNAAS...4..248P} found that the outer globular clusters have smaller BSS segregation compared to the inner globular clusters which have experienced stronger tidal forces by Milky Way. However, most OCs lie within the disc of the Milky Way, hence the tidal forces would strongly depend on the local neighbourhood and their orbit, and the detailed analysis of its effect is beyond the scope of this work.
Within a cluster, BSSs are the most massive stars, hence they are good indicators of mass segregation and dynamical evolution of the cluster \citep{Ferraro2012Natur.492..393F, Alessandrini2016ApJ...833..252A}. Binary stars are also more massive than single stars; hence their segregation can also be used to study the dynamical evolution of clusters \citep{Jadhav2021AJ....162..264J}.

\section{Thesis aim and structure}
Close binary systems can evolve into immensely different exotic systems such as BSSs, YSSs, cataclysmic variables, type Ia supernovae depending on the orbital parameters and how the two components evolve. The formation and evolution scenario for some of these exotic objects are still ambiguous, as they differ significantly from standard single star evolution theory. This thesis is focused on understanding the binary stars and their evolutionary products in OCs.

The structure of this thesis is as follows:
\begin{itemize}
    \item Chapter \ref{ch:intro} contains the introduction of star formation, stellar evolution and stellar exotica such as BSS and BLs.
    \item Chapter \ref{ch:data_and_methods} contains the information about the various telescopes used in this thesis. The chapter also details the methods used for data analysis, such as photometry and SED.
    \item Chapter \ref{ch:UOCS3} shows the method to determine cluster membership using \textit{Gaia} data. We also provide the UV--optical catalogues of the six OCs and their brief analysis.
    \item Chapter \ref{ch:UOCS2} presents the study of BSSs and BLs in OC NGC 2682.
    \item Chapter \ref{ch:UOCS4} presents the study of BSSs in OC King 2.
    \item Chapter \ref{ch:BSS_catalogue} provides the BSS catalogue based on \textit{Gaia} DR2 data of all known OCs.
    \item Chapter \ref{ch:summary} summaries the studies done in this thesis and presents the possible future directions we can take.
\end{itemize}

\begin{savequote}[100mm]
Light brings the news of the universe
\qauthor{William Bragg}
\end{savequote}

\chapter{Data and Methods}
\label{ch:data_and_methods}
\begin{quote}\small
\end{quote}

Stellar emissions in all wavelengths of the electromagnetic spectrum are essential to understand stellar properties. Optical data are the most prevalent due to the possibility of observations from the ground and our advanced understanding of the data. In addition, UV data are vital to study hot and young stellar populations, while IR data is important for colder stars. Thus, multiwavelength observations can allow us to study unresolved binary stars with different temperatures, such as MS+WD systems. In this work, I have used UV data (UVIT and \textit{GALEX}), optical data (\textit{Gaia}, 0.9 m KPNO, 3.5 m CAO, Pan-STARRS, 6.5 m MMT and 2.2 m MPG/ESO) and IR data (2MASS and \textit{WISE}). The details of the telescopes and used data are given in \S~\ref{sec:2_telescopes}. The details about methods developed for analysing the data are provided in \S~\ref{sec:2_methods}

\section{Telescopes} \label{sec:2_telescopes}
\subsection{Ultra-Violet Imaging Telescope}

\begin{figure}
    \centering
    \includegraphics[width=0.5\textwidth]{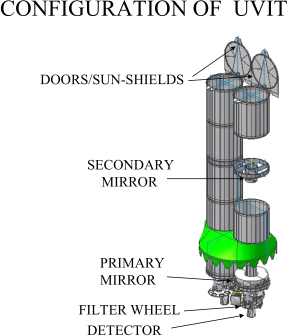}
    \caption{Schematic of the UVIT showing the twin FUV and NUV/VIS telescopes. (Credits: \textit{AstroSat} support cell).}
    \label{fig:2_UVIT_schematic}
\end{figure}

\begin{table}[!hb]
    \centering
    \caption{Details of UVIT filters \citep{Tandon2017AJ....154..128T,Tandon2020AJ....159..158T}.}
    \begin{tabular}{cc ccc}
    \toprule
        Name & Filter & Mean $\lambda$ [\AA]& $\Delta \lambda$ [\AA] & Zero Point\\ \hline
        F148W & CaF2-1 & 1481 & 500 & 18.097$\pm$0.010\\
        F154W & BaF2 & 1541 & 380 & 17.771$\pm$0.010\\
        F169M & Sapphire & 1608 & 290 & 17.410$\pm$0.010\\
        F172M & Silica & 1717 & 125 & 16.274$\pm$0.020\\
        N242W & Silica-1 & 2418 & 758 & 19.763$\pm$0.002\\
        N219M & NUVB15 & 2196 & 270 & 16.654$\pm$0.020\\
        N245M & NUVB13 & 2447 & 280 & 18.452$\pm$0.005\\
        N263M & NUVB4 & 2632 & 275 & 18.146$\pm$0.010\\
        N279N & NUVN2 & 2792 & 90 & 16.416$\pm$0.010\\ 
         \bottomrule
    \end{tabular}
    \label{tab:UVIT_filters}
\end{table}

Ultra-Violet Imaging Telescope (UVIT) onboard \textit{AstroSat} is the first Indian space observatory launched on 2015 September 28. The Ritchey--Chretien UV telescope consists of two 37 cm telescopes. One observing in far UV (FUV; 130--180 nm) and another in near UV (NUV; 200--300 nm) and visible (VIS; 350--550 nm). The VIS channel is only used to correct the drift of the spacecraft. Each channel has multiple narrower filters along with a grating for slit-less spectroscopy. Fig.~\ref{fig:2_UVIT_schematic} shows the configuration of the UVIT telescope. 
The details of effective area curves, UVIT calibrations and instrumentation can be found in \citet{Tandon2017a,Tandon2020AJ....159..158T} and \citet{Kumar2012}.

The NUV and FUV channels work on photon counting mode with CMOS detectors. Each incoming photon is focused on a photocathode. The electron shower from the photocathode is amplified by a factor of $\sim10^7$ using a microchannel plate \citep{Hutchings2007PASP..119.1152H}. This \textit{event} causes the illumination of several pixels on the CMOS. $3\times3$ pixel centroiding is performed onboard for each event and is typically read-out at a rate of 29 Hz. The final resolution of the UVIT images depends on the spread of photoelectrons on the detector ($\sim$1\arcsec) and tracking jitter ($\sim$0\farcsec5).
The VIS channel is used to calculate the telescope's drift due to the large abundance of optically bright sources compared to UV sources. It works in integration mode with exposures of 1/16 s. 16 such exposures are stacked onboard before transmitting the VIS data resulting in effective exposure of 1 s. 

\begin{figure}
    \centering
    \includegraphics[width=0.8\textwidth]{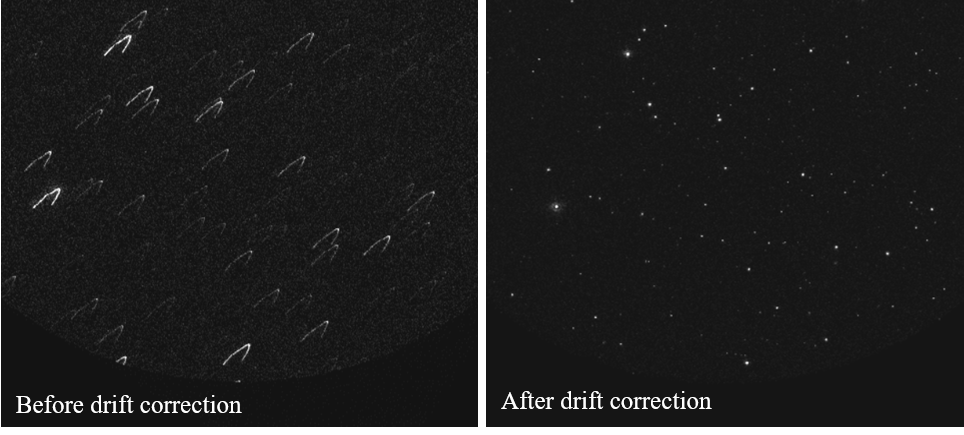}
    \caption{A UVIT NUV image before and after drift correction.}
    \label{fig:2_ccdlab_demo}
\end{figure}

The spacecraft transmits Level 0 (L0) data to ISSDC, ISRO. 
The L0 data is combined with spacecraft metadata and reformatted to Level 1 (L1) data. I used {\sc ccdlab} \citep{Postma2017PASP..129k5002P, Postma2021JApA...42...30P} for processing the L1 UVIT data. The left panel in Fig.~\ref{fig:2_ccdlab_demo} shows an example of an integrated NUV image for one orbit without any corrections. 
{\sc ccdlab} performs a number of functions including removal of duplicated data from L1, performing field distortion correction, correcting for centroiding bias and flat fielding. A critical functionality performed by {\sc ccdlab} is drift correction. As mentioned earlier, the VIS channel is used for correcting the drift. In case of corruption of VIS data, UV data can also be drift corrected by integrating short exposures (2--5 s) and then stacking these exposures. 
The drift corrected images for each orbit are then aligned and merged to create the final image. 
The merged images are still imperfect due to the 1 Hz sampling of VIS data compared to 29 Hz for the UV data. Furthermore, there is a slight difference found in the pointing of the VIS and FUV telescope due to thermal stick-slip \citep{Postma2021JApA...42...30P}. To correct this, {\sc ccdlab} has an \textit{optimising the PSF} (point spread function) functionality which stacks the images of bright sources with $\sim$20 s exposures to obtain a minor drift correction and the best possible PSF. 
The right panel of Fig.~\ref{fig:2_ccdlab_demo} shows an example of a UVIT image after correcting for the satellite drift. 
The circular field of view (FOV) of the UVIT has a radius of 14$\arcmin$, whereas the typical radius of the usable UVIT images is slightly less \citep{Tandon2017a}. For the images used in this thesis, the range for full width half maximum (FWHM) of the PSF was 1\arcsec--2\arcsec\ for various targets and filters. 

After creating science ready images, we used \textit{Gaia} and \textit{GALEX} point source catalogues for doing astrometry. We used astrometric coordinates from \textit{GALEX} images and {\sc ccmap} task of {\sc iraf} to create astrometric solution of NGC 2682 data from April 2017. For all other UVIT images, we have used the coordinate matching algorithm within the updated {\sc ccdlab} \citep{Postma2020PASP..132e4503P}. It first extracts point sources within UV images. Then selects bright UV sources ($\sim50$) and bright \textit{Gaia} G$_{BP}$ sources as reference (this reference catalogue can be changed if needed). A comparison of similar triangles within two catalogues is made using a least-square solution to obtain corresponding triangles within UV and reference catalogues. Further refining is done by adding more bright sources, and the final WCS solution is appended in the fits file header. An error of $\sim 0\farcsec2$ can be expected in the WCS solution. 

\subsection{\textit{Gaia}}

\begin{figure}[!hb]
    \centering
    \includegraphics[width=0.8\textwidth]{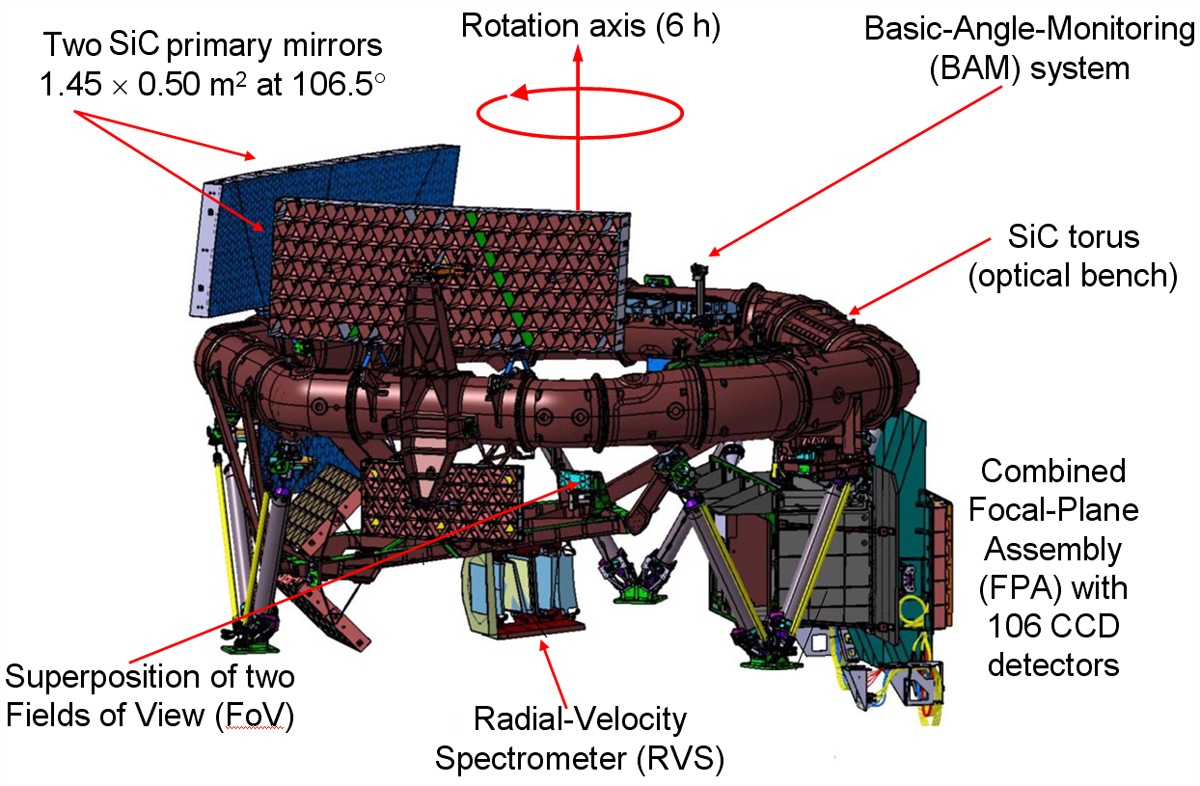}
    \caption{A schematic of the \textit{Gaia} telescope. (Credits: EADS Astrium)}
    \label{fig:2_Gaia_schematic}
\end{figure}

\textit{Gaia} is the revolutionary space telescope launched by European Space Agency (ESA) in 2013 \citep{Gaia2016A&A...595A...1G}. \textit{Gaia} consists of two 1-m class optical telescopes sharing their focal plane, which contains 106 charge-coupled devices (CCDs). Fig.~\ref{fig:2_Gaia_schematic} shows the schematic of \textit{Gaia} telescope indicating the position of primary mirrors, focal plane assembly and spin axis. Fig.~\ref{fig:2_Gaia_focal_plane} shows the CCD arrangement in the focal plane. The primary components of the focal plane are as follows:

\begin{figure}
    \centering
    \includegraphics[width=0.98\textwidth]{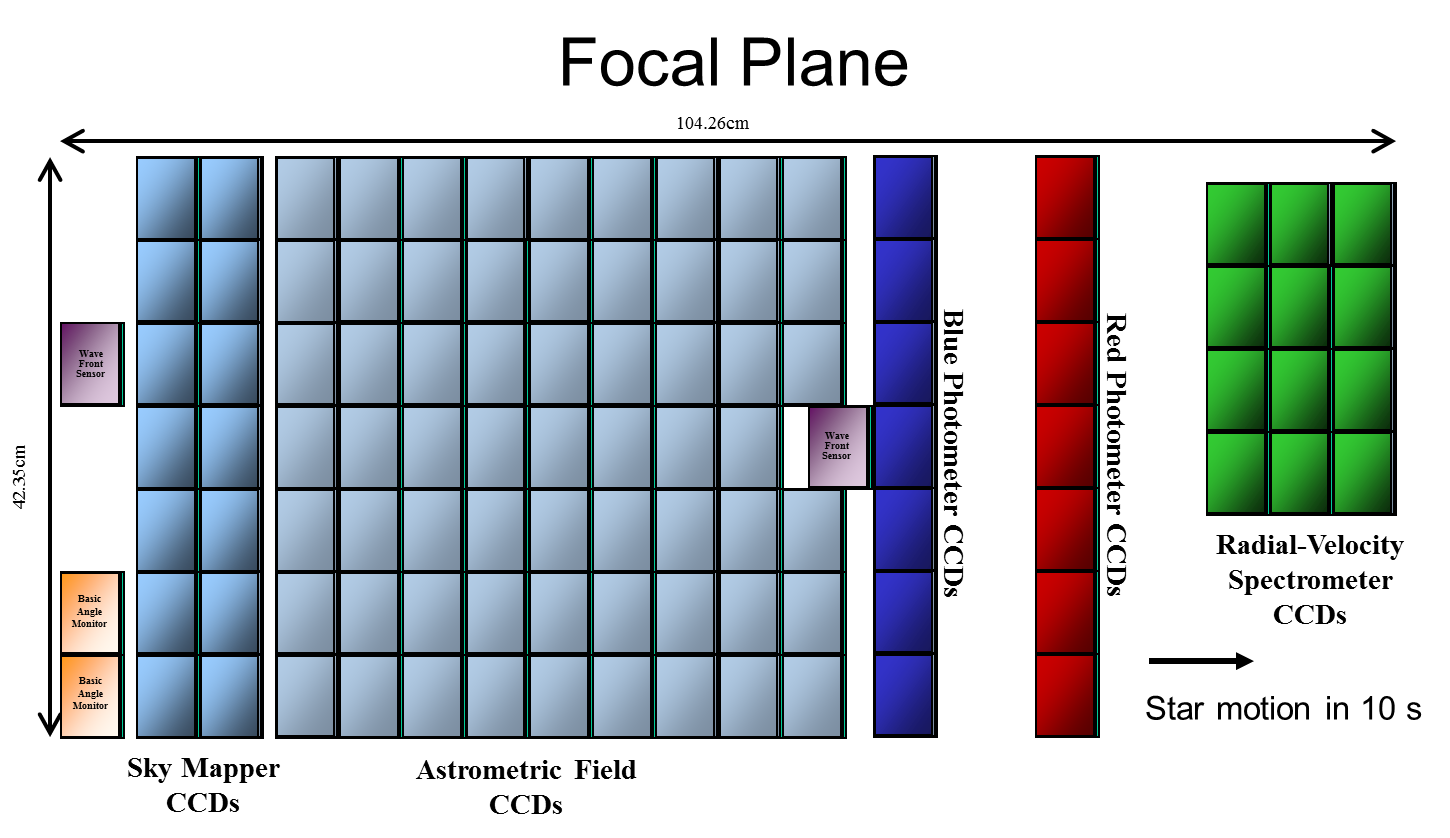}
    \caption{Focal plane of the \textit{Gaia} telescope. (Credits: European Space Agency)}
    \label{fig:2_Gaia_focal_plane}
\end{figure}

\begin{itemize}
    \item Sky mapper CCDs: These 14 CCDs are used to detect sources up to $\sim$20 mag and convey the position of each source to the astrometric field CCDs for tracking purposes.
    \item Astrometric field CCDs: 62 CCDs for astrometric measurements (position, PM, and parallax) and broadband photometric measurements (G-band) up to 21 mag.
    \item Blue and red photometers: 14 CCDs dedicated to spectrophotometric observations of stars in G$_{BP}$ (330--680 nm) and G$_{RP}$ (640--1050 nm) bands.
    \item Radial velocity spectrometer (RVS): 12 CCDs collecting medium resolution spectra ($\lambda / \Delta \lambda \sim 11500$) of the Calcium triplet are used to estimate RV, metallicity, log $g$ and temperature of objects brighter than $\sim15$ mag.
\end{itemize}

The twin telescopes in \textit{Gaia} continuously scan the sky defined by the \textit{Gaia} scanning law (spin rate of 60\arcsec\ s$^{-1}$). The sky mapper CCDs detect the source, and then time-delayed integration is used in the rest of the detectors to measure position, flux (G, G$_{BP}$ and G$_{RP}$) and RVS spectra in a collectively $\sim$80 sec exposure. Around 70--80 transits are expected for each object brighter than 20 mag in the 5-year operation.

The first data release (\textit{Gaia} DR1) was done with 14 months of data; however, PM data were not available for all sources, and spectrophotometric data were omitted entirely. The major data release was \textit{Gaia} DR2 \citep{Gaia2018A&A...616A...1G}, which used 22 months of data. It contained position, PM and 3 band photometric data of more than a billion sources. The most recent data release is \textit{Gaia} EDR3 \citep{Gaia2021A&A...649A...1G}, which contains a similar number of stars as DR2, but has more precise parameters due to observations spanning 34 months. 
The typical performance of \textit{Gaia} EDR3 is as follows: 
\begin{itemize}
    \item Completeness: Essentially complete for 12--17 G-mag, but actual completeness depends on the crowding and the scanning law pattern.
    \item Astrometry: Positional accuracy of better than 1 mas, PM accuracy of better than 1 mas yr$^{-1}$, parallax accuracy of better than 1.3 mas at 21 G-mag.
    \item Photometry: Accuracy of 6 mmag in G-band, 108 mmag in G$_{BP}$ and 52 mmag in G$_{RP}$ at 20 G-mag.
\end{itemize}

In this thesis, I have used both \textit{Gaia} DR2 (chapter \ref{ch:UOCS2}, \ref{ch:BSS_catalogue}) and EDR3 (chapter \ref{ch:UOCS3}, \ref{ch:UOCS4}) data.

\subsection{\textit{GALEX}}

\begin{figure}[!hb]
    \centering
    \includegraphics[width=0.75\textwidth]{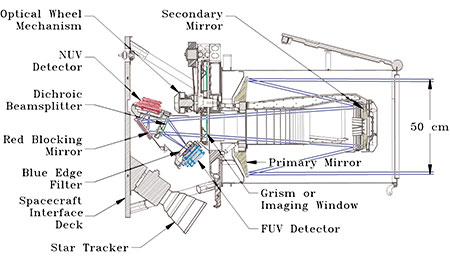}
    \caption{Instruments schematic of \textit{GALEX} telescope.}
    \label{fig:2_galex}
\end{figure}

The Galaxy Evolution Explorer, \textit{GALEX}, is a 50 cm Ritchey--Chretien space telescope launched in 2003 by the National Aeronautics and Space Administration (NASA). It simultaneously observes in two filters, FUV (1344--1786 \AA) and NUV (1771--2831 \AA). It has a large FOV of 1\farcdeg2 with an effective area of 36.8 and 67.7 cm$^2$ in FUV and NUV, respectively. The plate scale of \textit{GALEX} detectors is 1\farcsec5 pix$^{-1}$ with FWHM of 4\farcsec2 and 5\farcsec3 in FUV and NUV channels, respectively. Fig.~\ref{fig:2_galex} shows the cross-section of \textit{GALEX} along with the light path and focal plane instruments. 
\textit{GALEX} has done two main sky surveys: 
\begin{itemize}
    \item All-sky imaging survey (AIS): Typical exposure of 100 s with FUV and NUV depth of 20 and 21 ABmag, respectively.
    \item Medium-depth imaging survey (MIS): Typical exposure of 1500 s with FUV and NUV depth of 22.7 ABmag in both filters.
\end{itemize}
In this thesis, I have used the photometric data from the revised catalogue of \textit{GALEX} UV sources \citep{Bianchi2017ApJS..230...24B}.

\subsection{0.9 m Kitt Peak National Observatory}
We have used archival photometric data of NGC 2682 \citep{Montgomery1993} from the 0.9 m telescope at Kitt Peak National Observatory (KPNO). It is a Cassegrain telescope with a FOV of 6\farcmin6.
\citet{Montgomery1993} observed NGC 2682 in \textit{UBVRI} filters\footnote{\url{https://www.noao.edu/kpno/mosaic/filters/filters.html}} using 25 overlapping field to cover the cluster. Typical exposures were 4$\times$60 s (\textit{VI}), 4$\times$120 s (\textit{B}) and 900 s (\textit{U}). More details about the observations can be found in \citet{Montgomery1993}.

\subsection{3.5 m Calar Alto Observatory}
We have used archival data from the 3.5 m telescope at Calar Alto, Spain for OC King 2. \citet{Aparicio1990A&A...240..262A} observed King 2 in 1988 with \textit{UBVR} filters and exposure time of 2500, 590, 230 and 110 s respectively. The FWHMs of the images were $\leq$ 1\arcsec.

\subsection{1.8 m Panoramic Survey Telescope and Rapid Response System}
Panoramic Survey Telescope and Rapid Response System (Pan-STARRS1) is the first data release that used the 1.8 m Ritchey--Chretien telescope in Haleakala, Hawaii \citep{Chambers2016arXiv161205560C}. It has 3$^{\circ}$ FOV with a 1.4 Gigapixel camera. The survey was carried out in \textit{grizy$_{P1}$} filters with sensitivity of 23.3, 23.2, 23.1, 22.3 and 21.4 respectively (S/N = 5). Typical exposure times were 30--60 s in each filter.

\subsection{6.5 m MMT}
MMT is a classical Cassegrain telescope with a 6.5 m primary mirror with a large FOV of up to 1$^{\circ}$ depending on the secondary mirror. \citet{Williams2018} used the wide field CCD imager Megacam with FOV of $25\arcmin \times 25\arcmin$ to observe OC NGC 2682 in three filters: \textit{u} (35$\times$600 s), \textit{g} (10$\times$240 s) and \textit{r} (10$\times$180 s). The FWHM in the \textit{ugr} filters were 0\farcsec6, 0\farcsec8 and 1\farcsec3 respectively.  

\subsection{2.2 m MPG/ESO telescope}
MPG/ESO 2.2-metre telescope is a Ritchey--Chretien reflector based at La Silla, Chile. \citet{Yadav2008} used the NGC 2682 images taken by Wide Field Imager (WFI) with FOV of $34\arcmin \times 33\arcmin$ in 2000 and 2004 to calculate PMs. The WFI camera has a pixel scale of 238 mas, which results in positional precision of $\sim$7 mas for a bright star. The PM accuracy from these observations was 3 mas yr$^{-1}$ at 18 V-mag and 6 mas yr$^{-1}$ at 20 V-mag.

\subsection{Two Micron All-Sky Survey}
Two Micron All-Sky Survey (2MASS) is an all IR sky survey carried out at two 1.3 m telescopes at Mount Hopkins, Arizona, and Cerro Tololo, Chile between June 1997 and February 2001 \citep{Skrutskie2006AJ....131.1163S}. 2MASS covered 99.998\% of the sky simultaneously in three near-IR bands (\textit{JHK}). 2MASS used 6$\times$1.3 s exposures at each sky location with FOV of 8\farcmin5$\times$8\farcmin5. The sensitivity of the final catalogue is 15.8, 15.1 and 14.3 mag in \textit{JHK} respectively (S/N = 10). The typical FWHM of 2MASS images is 2.5--3\arcsec with photometric uncertainties of 0.02--0.03 mag till 13 mag. 

\subsection{\textit{Wide-field Infrared Survey Explorer}}
\textit{Wide-field Infrared Survey Explorer} (\textit{WISE}) is an IR space telescope launched by NASA in 2009 \citep{Wright2010AJ....140.1868W}. The telescope has a 40 cm mirror with a FOV of 47\arcmin. The FOV was split into 4 detectors using dichroic beam splitters. \textit{WISE} operated in continuous scanning mode using a secondary mirror to freeze the sky onto the detector for the exposure time of 8.8 s. Each source was observed at least 8 times with sensitivity of 17.11, 15.66, 11.40 and 7.97 mag for W1, W2 W3 and W4 filters, respectively (S/N = 5). Typical FWHM for the \textit{WISE} filters are 6\farcsec1, 6\farcsec4, 6\farcsec5 and 12\farcsec0 respectively.

\section{Research methods and models} \label{sec:2_methods}

\subsection{Photometry}
In this thesis, I have used the {\sc daophot} package of Image Reduction and Analysis Facility ({\sc iraf}; \citealt{Tody1993}) to do photometry on the UVIT images. I have followed the photometry manual by W. E. Harris\footnote{\url{https://physics.mcmaster.ca/~harris/daophot_irafmanual.txt}} for doing the photometry. The summary of the tasks used is given below.
\begin{itemize}
    \item \textit{imexam:} Estimation of background counts and FWHM of the image.
    \item \textit{daofind:} Identification of sources above a certain threshold (typically 3--10 times background).
    \item \textit{phot:} Doing aperture photometry on the stars selected by \textit{daofind}.
    \item \textit{pstsel and psf:} Selection of isolated stars to create a PSF model.
    \item \textit{allstar:} Estimating PSF magnitudes of all stars simultaneously through multiple iterations of PSF fitting to groups of stars.
    \item \textit{wcsctran:} Applying world coordinate solution (WCS) using manually cross-correlated sources between UVIT and \textit{GALEX/Gaia} reference sources.
\end{itemize}

The PSF magnitudes from \textit{allstar} are then updated using (i) aperture correction: effect of the curve of growth (ii) PSF correction: systematic difference between \textit{allstar} magnitudes and \textit{phot} magnitudes. As the UVIT FUV \& NUV detectors work in photon counting mode, further saturation correction was performed using the steps given in \citet{Tandon2017AJ....154..128T}. 

\subsection{Spectral energy distribution}

\begin{figure}[!hb]
    \centering
    \includegraphics[width=0.98\textwidth]{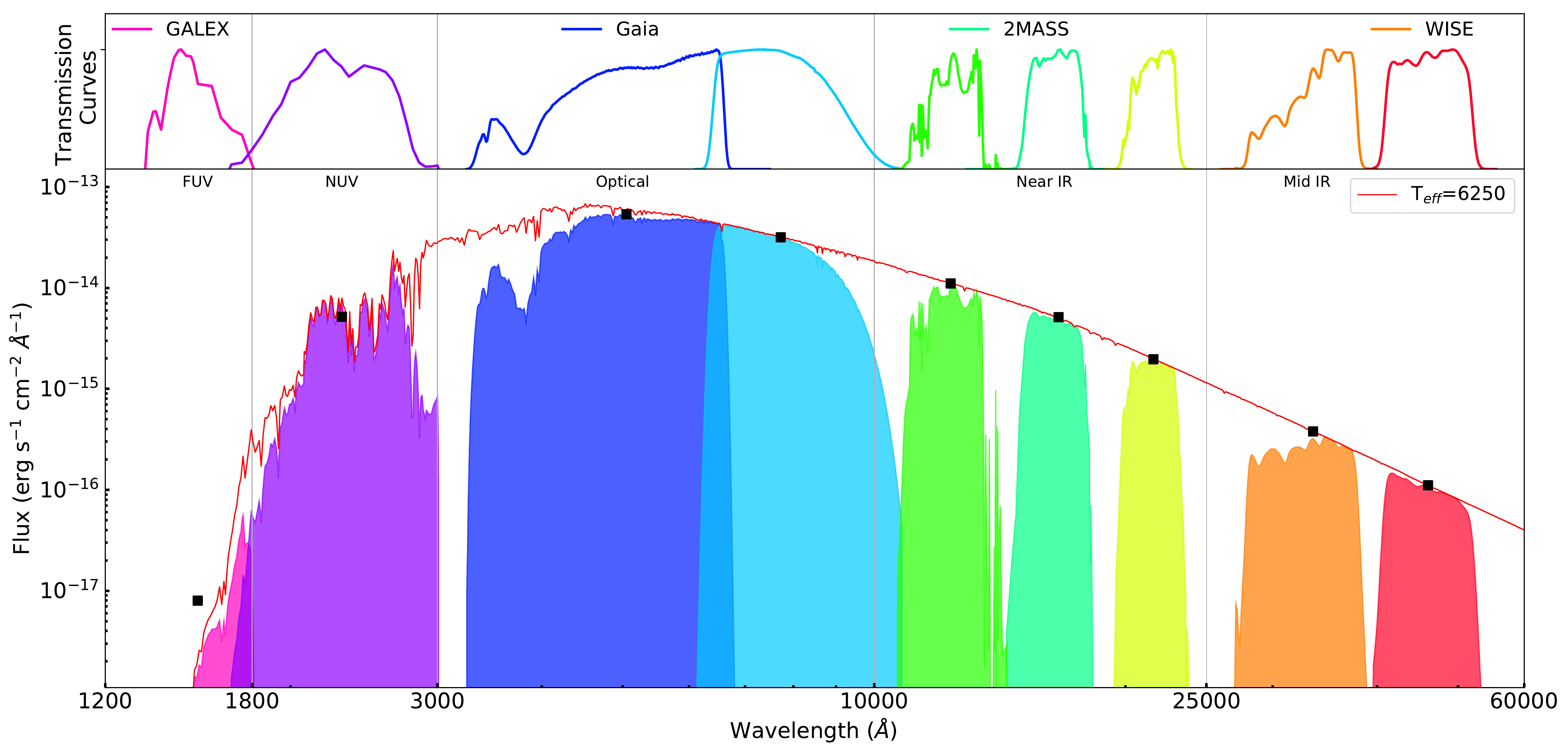}
    \caption{Schematic example SED of a model star with a temperature of 6250 K. The Kurucz spectrum is shown in red. Selective filters profiles from \textit{GALEX}, \textit{Gaia}, 2MASS and \textit{WISE} are shown in the top panel. The effective flux sampled by each filter is shown by the coloured regions in the bottom panel. The black square markers show the integrated flux detected by each filter. The integrated flux plotted as a function of effective wavelengths forms the SED.}
    \label{fig:2_SED_example}
\end{figure}

Ideally, the spectrum of a star is used to estimate the temperature and log $g$ of a star. However, obtaining spectra of a large number of stars is time-consuming and prohibitively expensive. The cheaper alternative (observation time wise, instrument complexity wise and monetarily) is an SED. An SED can be created by combining imaging done across the electromagnetic spectrum. Fig.~\ref{fig:2_SED_example} shows an example SED of a model star assuming it was observed with \textit{GALEX}, \textit{Gaia}, 2MASS and \textit{WISE}. The SED (shown with black points) contains information about the luminosity and temperature of the star. Additionally, log $g$ and/or metallicity can also be derived from SEDs depending on the sensitivity of the models. The archival photometry of stars is available from multiple wide-field surveys and through dedicated observations. Hence, SEDs are an effective tool to characterise a large number of stars.

We used the virtual observatory tool, VO SED Analyzer ({\sc vosa}\footnote{\url{http://svo2.cab.inta-csic.es/theory/vosa/index.php}}; \citealt{Bayo2008}), for SED analysis. {\sc vosa} includes spectral libraries and virtual observatory facilities. The conjoined SVO Filter Profile Service\footnote{\url{http://svo2.cab.inta-csic.es/theory/fps/}} provides the filter profiles and characteristics (such as effective wavelengths and zero points). One can create an SED from spatial location alone using the virtual observatory services included in {\sc vosa}. 

For the work done in this thesis, I uploaded the UVIT photometry to {\sc vosa} and used the integrated virtual observatory service to get archival photometry from \textit{GALEX}, \textit{Gaia}, Pan-STARRS1, 2MASS and \textit{WISE}. Depending on the availability of targeted observations, I also uploaded photometry from MMT, MPG/ESO, Calar Alto and KPNO to increase the data points in the SEDs. The SEDs are then corrected for extinction before fitting the models. The corrections for distance are applied later to estimate the luminosity and radius.

After getting the flux of the source in all filters, {\sc vosa} calculates synthetic photometry, for a selected theoretical model, using filter transmission curves. 
It performs a $\chi^{2}$ minimisation test by comparing the synthetic photometry with observed data to get the best-fit parameters of the SED. The reduced $\chi^{2}$, $\chi^{2}_{r}$, value is given by
\begin{equation}
\small
     \chi^{2}_{r} =\frac{1}{N-N_{f}} \chi^{2}= \frac{1}{N-N_{f}} \sum_{i=1}^{N}\Big\{\frac{(F_{o,i}-M_{d}F_{m,i})^{2}}{\sigma_{o,i}^{2}}\Big\}
\label{chi2}
\end{equation}

where N is the number of photometric data points, N$_{f}$ is the number of free parameters in the model, $F_{o,i}$ is the observed flux, $M_{d}F_{m,i}$ is the model flux of the star, $\displaystyle{M_{d}=\bigg(\frac{R}{D}\bigg)^{2}}$ is the scaling factor corresponding to the star (where R is the radius of the star and D is the distance to the star) and $\sigma_{o,i}$ is the error in the observed flux. In this thesis, the value of $N$ for stellar sources changes from 10--16 depending on the number of detections in all available filters. The $N_{f}$ is 2 for single fits (temperature and scaling factor). 

The radius and luminosity of the source can be calculated from the scaling factor and temperature as follows:
\begin{equation}
    \begin{split}
        R &= \sqrt{D^2M_d} \\ 
        L &= 4 \pi \sigma_{SB} R^2 T^4
    \end{split}   
\end{equation}

Typically, $\chi^2_r \sim 1$ fits are considered as good fits while $\chi^2_r < 1$ is considered overfitting and $\chi^2_r > 1$ is considered bad fit. However, this is only true assuming the model is correct and is linear \citep{Andrae2010arXiv1009.2755A, Andrae2010arXiv1012.3754A}. The models in SEDs are not linear, and the $\chi^{2}_r$ value is highly dependent on errors in the photometry. It can be as high as 100 for data with very small errors. Hence, keeping a lower limit to the fractional errors is recommended so that each data point has enough weightage during the fitting process. For example, \citet{Lodieu2019A&A...628A..61L} limited the smallest fractional errors to half of the averaged fractional error by increasing the smaller errors. We have also used a similar technique to increase errors in Chapter \ref{ch:UOCS4}. 

\subsection{Binary SED fitting} \label{sec:2_binary_SED_fitting}
Upon investigation of SED fits obtained via {\sc vosa}, we found some stars show poor fits, especially in the UV region. A UV excess flux can be caused by lower than expected metallicity, chromospheric activity or hot companion. The hot companion can be characterised by using double component SEDs.
SEDs have previously been used to distinguish between single and binary stars. Recently, \citet{Thompson2021AJ....161..160T} used optical--IR (\textit{griJH$K_s$}[3.6][4.5]) SEDs to study binary population in OCs: NGC 1960, NGC 2099, NGC 2420, and NGC2682. They presumed that the members should follow the mass-luminosity-radius relations given by modified PARSEC isochrones. However, this study was designed to identify MS+MS type binary stars with mass ratio $>0.3$. This approach does not work for MS+hot companion systems as hot companions are wide variety, and they do not follow a simple mass-luminosity-radius.

\begin{figure}[!ht]
    \centering
    \includegraphics[width=0.98\textwidth]{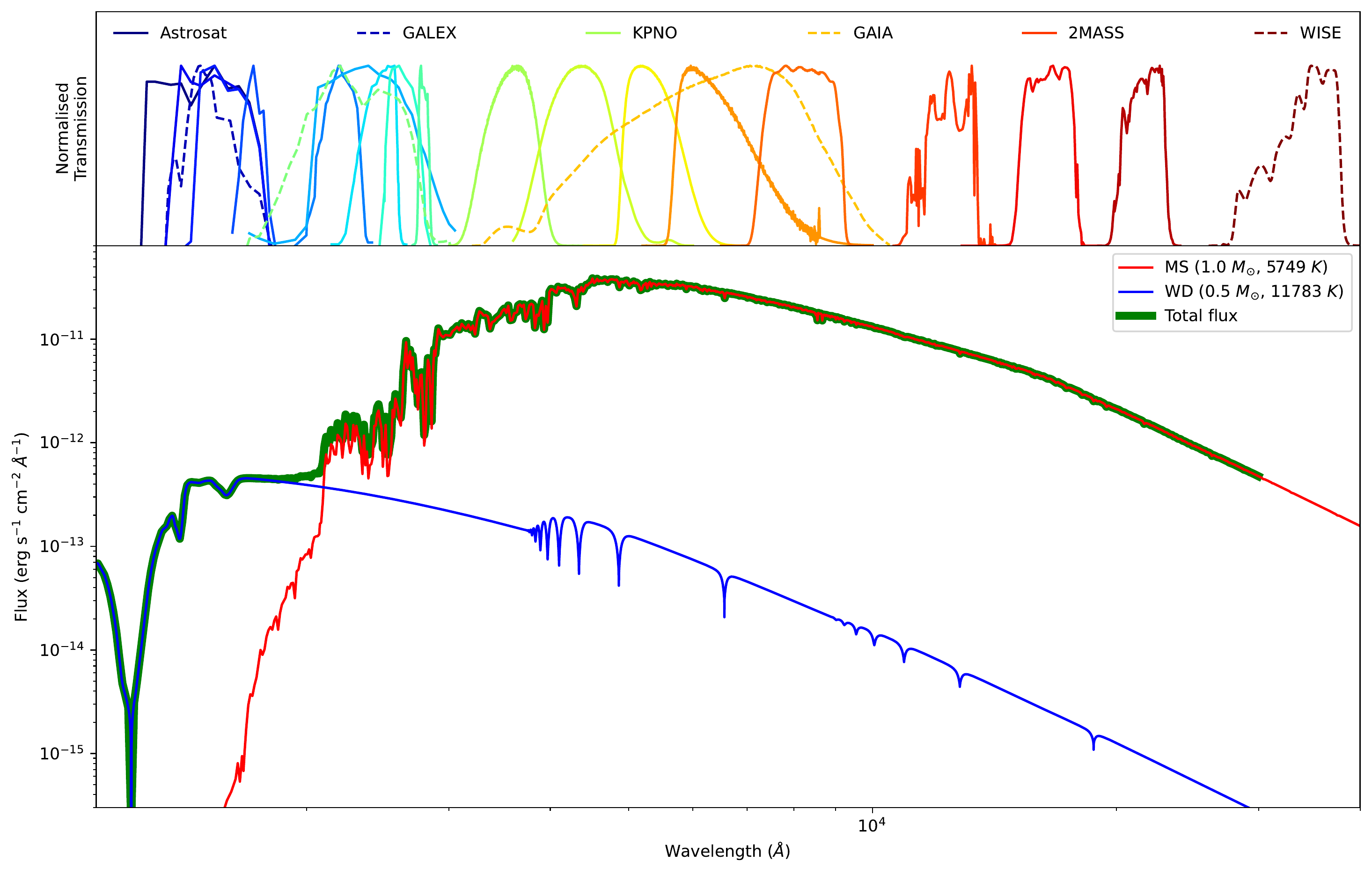}
    \caption{Schematic of a composite spectrum of MS-WD binary star (green spectrum). The red and blue spectra correspond to a typical MS (1 \Msun) and WD (0.5 \Msun) respectively. The top panel shows scaled transmission curves of various missions used for NGC 2682.}
    \label{fig:2_binary_SED}
\end{figure}

Fig~\ref{fig:2_binary_SED} shows an example spectrum of an MS+WD system. The optical--IR emission is dominated by the MS star, while FUV emission is dominated by the WD with a mix of two in the NUV. The exact transition region depends on the relative temperature difference between the two stars, but usually, the previous statement applies to most MS+WD systems.

I have developed a python code\footnote{\url{https://github.com/jikrant3/Binary_SED_Fitting}} to fit two-component SEDs and give the component parameters. 
In this thesis, I have searched for optically sub-luminous hotter companions. Hence, I will refer to the cooler MS type star as primary and the hotter compact star as secondary.
The steps for double fitting are as follows:
\begin{itemize}
    \item Selection of sources with no neighbours within 5\arcsec. This step makes sure that all the photometric points (e.g., \textit{GALEX}, 2MASS, \textit{WISE}) have accurate flux. 
    \item Fitting of single Kurucz model \citep{Castelli1997A&A...318..841C} SEDs to all these isolated stars using UV--IR data points using {\sc vosa}.
    \item Identification of sources with unsatisfactory SED fits (e.g., unusually large $\chi^2$, significant fractional residual).
    \item Refitting of these sources using only optical--IR data ($> 3000$\AA) in {\sc vosa}. The fits should have minimal residual in the optical--IR region.
    \item Selection of models for primary and secondary components. In this work, all cooler components were fitted with Kurucz models \citep{Castelli1997A&A...318..841C}, while the hotter components were fitted using either Koester models \citep{Koester2010} for WDs (in chapter~\ref{ch:UOCS2}) or Kurucz models for hot sub-dwarfs and MS type stars(in chapter~\ref{ch:UOCS2} and \ref{ch:UOCS4}).
    \item The residual flux after primary fitting is then fitted with a hotter SED. This step gives the temperature and radius of the secondary source.
    \item The errors in temperature and radii are estimated from a combination of the steps sizes in the models, errors in distance and errors in photometry.
\end{itemize}
After fitting a satisfactory double fits, we checked whether these temperatures and radii are physically possible and what type of objects these could be. Chromospherically active sources, X-ray sources and binaries with ongoing MT can give off excess UV flux. Hence, such sources were not confirmed to have a hotter component even after successful double component fitting.

\subsubsection{Error estimation}
Estimating the errors in $\chi^{2}$ fitting is not a straightforward process due to the non-linear nature of SED fitting. One way to estimate errors is using the grid size in the temperature to estimate the errors. In this case, the errors in temperature, scaling factor, radius, and luminosity are estimated as follows:

\begin{equation} \label{eq:2_error_step}
\begin{split}
    {\Delta T} &= \frac{Temperature\ step\ size}{2}\\
    \frac{\Delta R}{R} &= \frac{\Delta D}{D}\\
    \frac{\Delta L}{L} &= 2\frac{\Delta R}{R} + 4\frac{\Delta T}{T}
\end{split}
\end{equation}

Alternatively, we can use bootstrapping to get statistical errors in temperature and other parameters. We first add a Gaussian noise proportional to the error in the observed flux. We fit this noisy observed flux with a model SED and get a new set of $T$, $R$ and $L$. We use 100 random noisy SED fits to get a distribution of fitting parameters. The 32nd, 50th and 68th percentiles (e.g., $T_{32},\ T_{50}$ and $T_{68}$) of distributions are used to get the asymmetric 1-$\sigma$ errors in $T$, $R$ and $L$. For example,
\begin{equation} \label{eq:2_error_boot_T}
\begin{split}
    T    &= T_{50}\\
    left\_error  &= T_{50}-T_{32}\\
    right\_error &= T_{68}-T_{50}
\end{split}  
\end{equation}
To get the total errors in $R$ and $L$, we combine the statistical errors and distance errors as follows:
\begin{equation} \label{eq:2_error_boot_R}
    \begin{split}
        R    &= R_{50}\\
        left\_error  &= \sqrt{(R_{50}-R_{32})^2 + \left( \frac{R\Delta D}{D} \right)^2}\\
        right\_error &= \sqrt{(R_{68}-R_{50})^2 + \left( \frac{R\Delta D}{D} \right)^2}        
    \end{split}
\end{equation}

\begin{equation} \label{eq:2_error_boot_L}
    \begin{split}
        L    &= L_{50}\\
        left\_error  &= \sqrt{(L_{50}-L_{32})^2 + \left( \frac{2L\Delta D}{D} \right)^2}\\
        right\_error &= \sqrt{(L_{68}-L_{50})^2 + \left( \frac{2L\Delta D}{D} \right)^2}
    \end{split}
\end{equation}

The methods and techniques mentioned above evolving since 2018 and are under active development. Hence, there will be differences between the exact method and codes used in Chapter \ref{ch:UOCS2}, Chapter \ref{ch:UOCS4}, \citet{Vaidya2022arXiv220108773V} and the latest version available in GitHub. Chapter \ref{ch:UOCS2} used eq.~\ref{eq:2_error_step} to estimate errors while Chapter \ref{ch:UOCS4} used eq.~\ref{eq:2_error_boot_T}--\ref{eq:2_error_boot_L} for estimating errors.

\subsection{Isochrones and evolutionary tracks}

An isochrone is the locus of coeval stars with different masses in the HRD/CMD. The stars lie in various locations according to their evolutionary phase. As stars in a cluster are formed at approximately the same time, a CMD of a cluster looks like an isochrone except for a few deviations: (i) varying number density according to the initial mass function and (ii) small shifts in the position due to unresolved stars (binaries, multiple systems and coincidental spatial overlap).

\begin{figure}[!ht]
    \centering
    \includegraphics[width=0.98\textwidth]{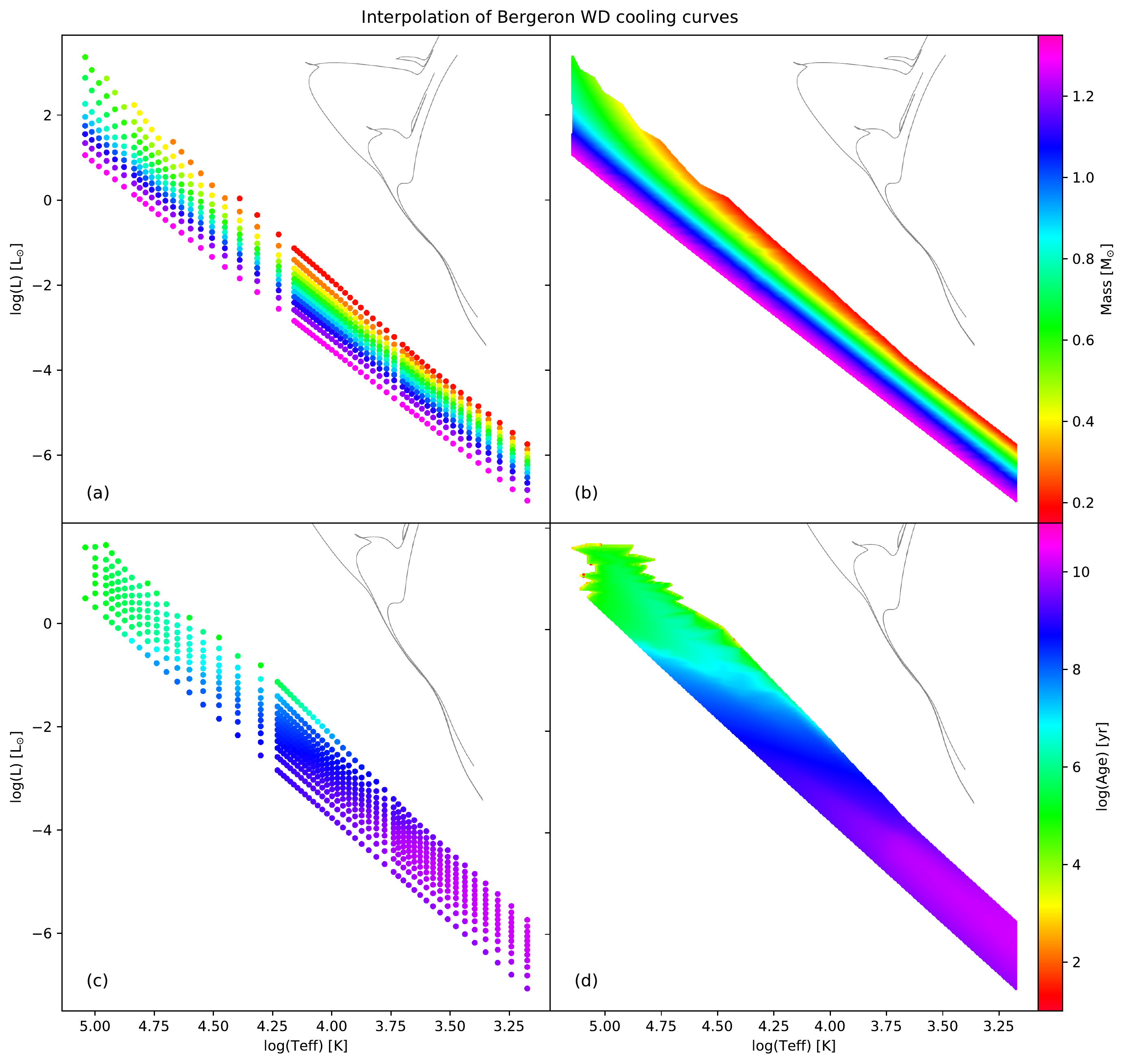}
    \caption{Interpolated Bergeron WD cooling curves. Left panels show the publicly provided cooling curves for 0.2--1.3 \Msun\ WDs. The right panels show the interpolated models. The WD cooling curves are coloured according to their mass and age in the top and bottom panels. Solar metallicity isochrones of log(age) 8.0, 9.0 and 10.0 are shown in grey for reference.}
    \label{fig:2_interpolated_WD_curve}
\end{figure}

I have primarily used {\sc parsec} isochrones \footnote{\url{http://stev.oapd.inaf.it/cgi-bin/cmd}} \citep{Bressan2012MNRAS.427..127B} generated with known cluster metallicity and age. The isochrones used \citet{Kroupa2001MNRAS.322..231K} initial mass function and the extinction correction was done using \citet{Cardelli1989} and \citet{Odonnell1994}. Examples of isochrones are given as grey curves in Fig.~\ref{fig:2_interpolated_WD_curve}.

The WD cooling curves give the luminosity, temperature, and log $g$ of WDs with different masses and ages. We have used the Bergeron WD models\footnote{\url{http://www.astro.umontreal.ca/~bergeron/CoolingModels/}} \citep{Fontaine2001, Tremblay2011} for hydrogen atmosphere (DA type). The models span cooling curves of masses 0.2--1.3 \Msun. Fig.~\ref{fig:2_interpolated_WD_curve} (a) shows the HRD of the WD models. The models are convolved with commonly used filters to obtain the model magnitudes of WDs in different telescopes. The convolutions with UVIT filters were obtained via personal communication with P. Bergeron. 
To estimate the mass and age of detected WDs, we needed finer data points than provided. Hence, we interpolated the models into a finer regular grid of 500 points spanning the log(L) and log(T) plane. Fig.~\ref{fig:2_interpolated_WD_curve} (b) and (d) show the interpolated models in the HRD. 
In Chapter \ref{ch:UOCS2}, we use these models to estimate the mass and age of the WDs. The WD temperature and luminosity errors are compared with the interpolated models, and corresponding errors in age and mass are derived.

For lower mass WDs, I have used He-core WD models \footnote{ \url{http://fcaglp.fcaglp.unlp.edu.ar/~panei/models.html}} by \citet{Panei2007MNRAS.382..779P}. These models span a mass range of 0.1869--0.4481 M$_{\odot}$. Due to the unavailability of convolved magnitudes, I have only used these models in the luminosity--temperature plane.

\begin{savequote}[100mm]
The skies are painted with unnumber'd sparks
\qauthor{William Shakespeare}
\end{savequote}

\chapter[Cluster Membership and UV Catalogues]{Cluster Membership and UV Catalogues \\ \large{\textcolor{gray}{Jadhav et al., 2021, MNRAS, 503, 236}}}
\label{ch:UOCS3}
\begin{quote}\small
\end{quote}

\section{Introduction} \label{sec:3_intro}

Multi-wavelength studies of stars in clusters help to reveal the possible formation mechanism of non-standard stellar populations \citep{Thomson2012, Jadhav2019ApJ...886...13J}. OCs in the Milky Way span a wide range in ages, distances and chemical compositions \citep{Dias2002A&A...389..871D, Kharchenko2013, Netopil2016, Cantat2020}. The relatively low stellar density in the OCs is also an essential factor that helps in understanding the properties of binary systems in a tidally non-disruptive environment. 

\begin{table}
    \centering
    \caption{Cluster coordinates, ages, distances, mean PMs and radii are taken from \citet{Cantat2018A&A...618A..93C,Cantat2020}. The metallicity of Berkeley 67 is from \citet{Lata2004} and other metallicities are from \citet{Dias2002A&A...389..871D}.}
    \label{tab:3_cluster_details}
    \resizebox{0.98\textwidth}{!}{
    \begin{tabular}{lcccc ccccc r}
    \toprule
Name	&	$\alpha_c$ (J2015.5)	&	$\delta_c$ (J2015.5)	&	l	&	b	&	$D$	&	Age	&	[M/H]	&	$\mu_{\alpha, c}cos\delta$	&	$\mu_{\delta, c}$	&	r50	\\
	&	($^{\circ}$)	&	($^{\circ}$)	&	($^{\circ}$)	&	($^{\circ}$)	&	(pc)	&	(Gyr)	&		&	(mas yr$^{-1}$)	&	(mas yr$^{-1}$)	&	(')	\\ \hline
Berkeley 67	&	69.472	&	50.755	&	154.85	&	2.48	&	2216	&	1.3	&	+0.02	&	2.3	&	-1.4	&	4.9	\\
King 2	&	12.741	&	58.188	&	122.87	&	-4.68	&	6760	&	4.1	&	-0.41	&	-1.4	&	-0.8	&	3.1	\\
NGC 2420       	&	114.602	&	21.575	&	198.11	&	19.64	&	2587	&	1.7	&	-0.38	&	-1.2	&	-2.1	&	3.2	\\
NGC 2477	&	118.046	&	-38.537	&	253.57	&	-5.84	&	1442	&	1.1	&	+0.07	&	-2.4	&	0.9	&	9.0	\\
NGC 2682       	&	132.846	&	11.814	&	215.69	&	31.92	&	889	&	4.3	&	+0.03	&	-11.0	&	-3.0	&	10.0	\\
NGC 6940	&	308.626	&	28.278	&	69.87	&	-7.16	&	1101	&	1.3	&	+0.01	&	-2.0	&	-9.4	&	15.0	\\
   \toprule
    \end{tabular}
    }
\end{table} 

The OCs of our Galaxy are located at various distances from us. Thus, stars detected in any observation will be a mixture of cluster members as well as both foreground and background field stars. The identification of cluster members using a reliable method is therefore extremely important.
Earlier, this was accomplished using the spatial location of stars in the cluster region, as well as their location on different phases of single stellar evolution, i.e., the MS, sub-giant branch and RGB in the CMDs of star clusters \citep{Shapley1916}. 
However, many intriguing and astrophysically significant stars such as BSSs, sub-sub-giants were not considered members due to their \textit{peculiar} locations in the OC CMDs. In this chapter, we refer to locations other than MS, sub-giant branch and RGB, which are part of the single star evolution, as \textit{peculiar}.

\begin{figure}
    \centering
    \includegraphics[width=0.9\textwidth]{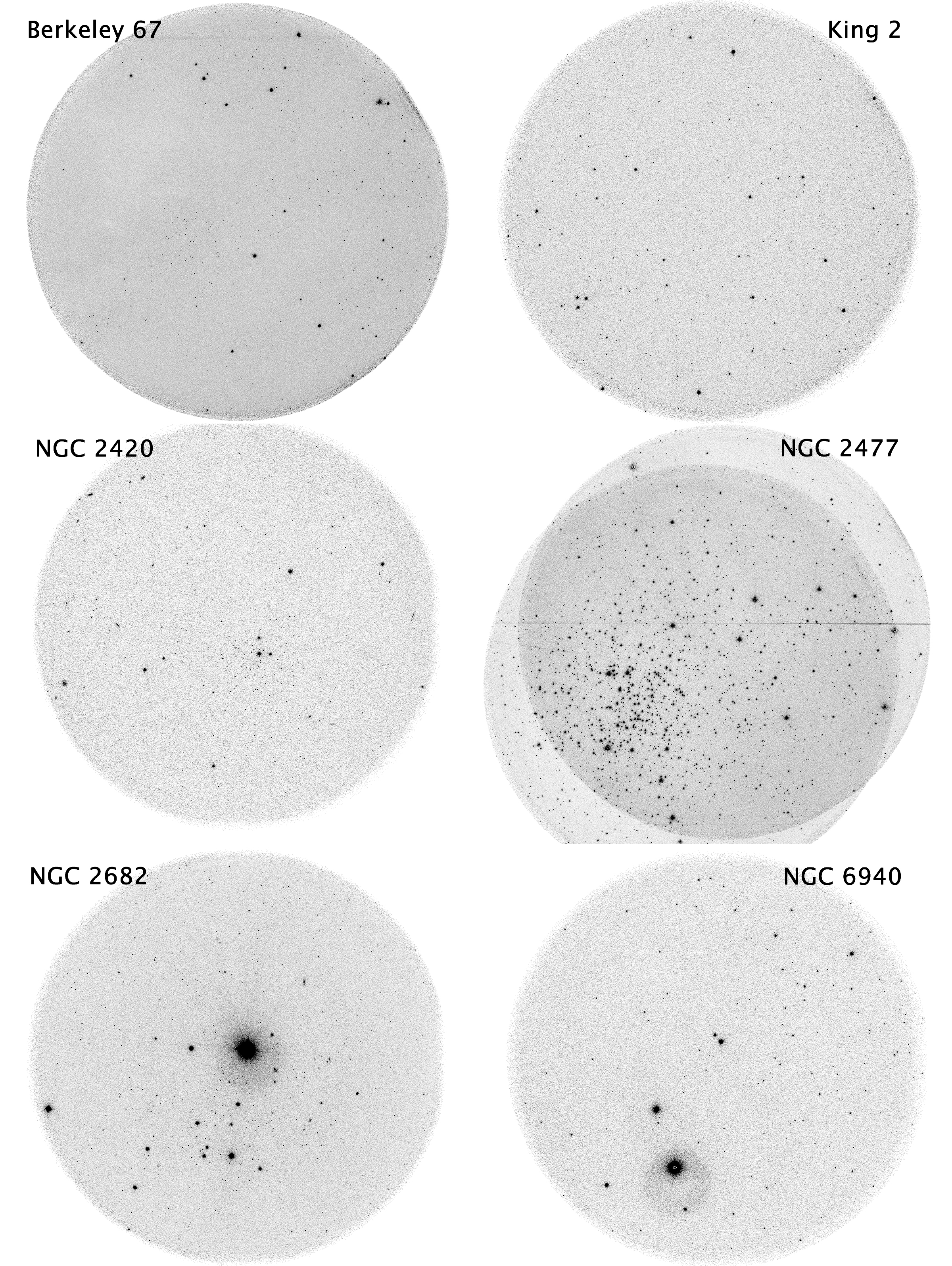}
    \caption{UVIT images of the six clusters: Berkeley 67 in N245W, King 2 in N219M, NGC 2420 in F148W, NGC 2477 in N263M, NGC 2682 in F148W and NGC 6940 in F169M.}
    \label{fig:3_images}
\end{figure}

To study exotic stellar populations in OCs, 
we selected clusters that are safe to be observed using UVIT (those at the high galactic latitude and without bright stars in the UVIT FOV) and have a high probability of detecting UV stars. 
The OCs were selected such that enough bright members will be detected with specific focus on UV bright population such as BSSs and WDs. Some clusters are (and will be) looked into individually.
This work focuses on OCs Berkeley 67, King 2, NGC 2420, NGC 2477, NGC 2682 and NGC 6940. Fig.~\ref{fig:3_images} shows the UVIT images of the six clusters. They span a range of age (0.7--6 Gyr) and distance (0.8--5.8 kpc). Table~\ref{tab:3_cluster_details} lists the parameters such as location in the sky, distance, age, mean PM and radius of the OCs under study. The relevant literature surveys are included in \S \ref{sec:3_intro_1}. 

UVIT study of NGC 2682 is presented in \citet{Sindhu2019ApJ...882...43S, Subramaniam2020JApA...41...45S, Pandey2021MNRAS.507.2373P}, and chapter \ref{ch:UOCS2} \citep{Jadhav2019ApJ...886...13J}. 
However, this chapter includes more recent and deeper photometry for NGC 2682 compared to chapter \ref{ch:UOCS2}.
It is also one of the most studied OCs with well-established CMD; hence we compare the behaviour of other OCs with NGC 2682 to interpret the optical and UV CMDs in further sections. We also use it to validate our membership determination method against previous efforts.
This chapter aims to analyse the UV--optical CMDs and the overall UV characteristics of these clusters.

Here, we have used multi-modal astrometric and photometric data from the latest \textit{Gaia} EDR3 \citep{Gaia2021A&A...649A...1G} for cluster membership. The membership determination of OC stars, in particular the UV bright population of BSSs, binaries and WDs, requires careful incorporation of data quality indicators from \textit{Gaia} EDR3. 
PMs of field and cluster stars can be approximated by Gaussian distributions \citep{Sanders1971} which can be separated analytically, and individual MP can be estimated from the distance of a star from field and cluster centre in the VPD. However, this method does not distinguish between field stars with the same PM as cluster members. 
Therefore, parallax and CMD position could be used to remove such field stars.
Also, parallax, colour, and magnitudes have non-Gaussian distributions. To optimally use all the \textit{Gaia} parameters,
we chose supervised machine learning to segregate the cluster members. The use of machine learning techniques is increasing in astronomy to automate classification tasks, including cluster membership \citep{Gao2018a, Gao2018b, Gao2018c, Gao2018d, Zhang2020, Castro2020}.
However, as most machine learning techniques do not include errors in the data, we used probabilistic random forest ({\sc prf}, \citealt{Reis2019}), which incorporates errors in the data. To train the {\sc prf}, we first selected the cluster members by deconvolving the PM Gaussian distributions using a Gaussian Mixture Model ({\sc gmm}, \citealt{Vasiliev2019}). The overall method also provides the much-needed MPs necessary for stars with non-standard evolution.

This chapter is arranged as follows: \S~\ref{sec:3_intro_1} contains literature surveys of the six clusters.
\S~\ref{sec:3_obs} has the details of UVIT observations, \textit{Gaia} data and isochrone models. 
The membership determination technique is explained in \S~\ref{sec:3_method}.
The membership results and UV--optical photometry are presented in \S~\ref{sec:3_results} and discussed in \S~\ref{sec:3_discussion}.
The full versions of \textit{Gaia} EDR3 membership catalogue (Table~\ref{tab:3_cat_Gaia}) and UV photometric catalogues of the six OCs (Table~\ref{tab:3_cat_UV}) are available online.

\subsection{Literature information of the OCs} \label{sec:3_intro_1}

\textbf{Berkeley 67} is a $\sim$1 Gyr old OC located at a distance of $\sim$2.45 kpc. It is a low-density cluster with an angular diameter of $\sim$14\arcmin. \citet{Lata2004} carried out deep Johnson UBV and Cousins RI CCD photometry of this cluster while \citet{Maciejewski2007} obtained BV CCD data as part of a survey of 42 open star clusters. Both studies are based on the optical CMD of the cluster.

\textbf{King 2} is a $\sim$5 Gyr old OC located at a distance of $\sim$6 kpc towards the Galactic anti-centre direction. It is a faint but rich cluster situated in a dense stellar field. It lags behind the local disc population by 60--100 km s$^{-1}$ and could be part of the Monoceros tidal stream \citep{Warren2009MNRAS.393..272W}. \citet{Kaluzny1989AcA....39...13K} obtained BV CCD photometric data for the cluster. A deep Johnson--Cousins UBVR CCD photometric study of the cluster was carried out by \citet{Aparicio1990A&A...240..262A}. They estimated E(B $-$ V) = 0.31 mag in the cluster's direction and indicated the presence of $>25\%$ of binary stars, based on the observed scatter in the CMD of the cluster.

The OC \textbf{NGC 2420} is $\sim$1 Gyr old and located at a distance of $\sim$3 kpc. \citet{Cannon1970} obtained relative PMs and also determined BV photographic magnitudes. The broadband optical CCD photometric study was carried out by \citet{Sharma2006}. The ubyCaH intermediate-band CCD photometry of this star cluster was performed by \citet{Anthony2006}. All these studies indicate that NGC 2420 is older than 1 Gyr.

The intermediate-age ($\sim$0.9 Gyr) southern rich OC \textbf{NGC 2477} is located at a distance of $\sim$1.4 kpc \citep{Hartwick1974, Smith1983, Kassis1997, Eigenbrod2004, Jeffery2011}. This cluster has a metallicity near Solar ([Fe/H] $\sim$ $-$0.17--0.07 dex; \citealt{Friel2002AJ....124.2693F, Bragaglia2008}) and a high binary frequency ($\sim$36\%) for the RGs \citep{Eigenbrod2004}. Presence of significant differential reddening (E(B $-$ V) = 0.2--0.4 mag) across the cluster was indicated \citep{Hartwick1972, Smith1983, Eigenbrod2004}. Using \textit{Gaia} DR2 data down to $\sim$21 mag, \citet{Gao2018b} identified more than 2000 cluster members. A deep \textit{HST} photometric study of the NGC 2477 was carried out by \citet{Jeffery2011} to identify WD candidates and estimate their age.

\textbf{NGC 2682} (M67) is a nearby OC with an age of $\sim$3--4 Gyr \citep{Montgomery1993, Bonatto2015} and located at a distance of $\sim$800--900 pc \citep{Stello2016}. It is a well-studied cluster from X-rays to IR \citep{Mathieu1986, Belloni1998, Bertelli2018, Sindhu2018}. There are various studies on the membership determination of NGC 2682 \citep{Sanders1977, Yadav2008, Geller2015, Gao2018c}. It contains stars in various stellar evolutionary phases such as MS, RGs, BSSs, and WDs. NGC 2682 contains 38\% photometric binaries \citep{Montgomery1993} and 23\% spectroscopic binaries \citep{Geller2015}. Recently \citet{Sindhu2019ApJ...882...43S} and \citet{Jadhav2019ApJ...886...13J} detected massive and ELM WDs with UVIT observations. The presence of 24 BSSs, four YSSs, two sub-subgiants, massive WDs and ELM WDs indicates that constant stellar interactions occur in NGC 2682.  

\textbf{NGC 6940} is a well-known intermediate-age ($\sim$1 Gyr) OC located at a distance of about 0.8 kpc. The membership of the cluster was investigated by \citet{Vasilevskis1957} and \citet{Sanders1972}; while photometric studies were carried out by \citet{Walker1958}, \citet{Johnson1961}, \citet{Larsson1964} and \citet{Jennens1975}. \citet{Baratella2018} presented medium resolution (R $\sim$13000), high signal-to-noise (S/N $\sim$100), spectroscopic observations of seven RG members.
\section{Data and models} \label{sec:3_obs}

\subsection{UVIT data} \label{sec:3_uvit_data}

\begin{table}[!ht]
    \centering
    \caption{The log of UVIT observations in different filters are given along with exposure time and FWHM. The number of detected stars in each filter as well as the number of cluster \texttt{members/candidates} determined according to the Eq.~\ref{eq:3_classif} are also listed.}
    \label{tab:3_cluster_obs}
    \resizebox{0.98\textwidth}{!}{
    \begin{tabular}{lccc cccr} 
    \toprule
Cluster	&	Filter	&	Observation Date	&	Exposure Time	&	Detected Stars	&	\texttt{Members}	&	\texttt{Candidates}	&	FWHM	\\	
	&		&	(yyyy-mm-dd)	&	(s)	&		&		&		&	(\arcsec)	\\	\toprule
Berkeley 67	&	N242W	&	2016-12-21	&	2700	&	469	&	64	&	5	&	1.15	\\	
	&	N245M	&	2016-12-21	&	2722	&	258	&	19	&	3	&	1.09	\\	\hline
King 2	&	F148W	&	2016-12-17	&	2666	&	150	&	5	&	1	&	1.33	\\	
	&	N219M	&	2016-12-17	&	2714	&	303	&	3	&	1	&	1.35	\\	\hline
NGC 2420	&	F148W	&	2018-04-30	&	2136	&	177	&	57	&	2	&	1.70	\\	\hline
NGC 2477	&	F148W	&	2017-12-18	&	2278	&	301	&	92	&	16	&	1.56	\\	
	&	N263M	&	2017-12-18	&	1881	&	1637	&	576	&	53	&	1.34	\\	\hline
NGC 2682	&	F148W	&	2018-12-19	&	6575	&	918	&	84	&	18	&	1.76	\\	
	&	F154W	&	2017-04-23	&	2428	&	267	&	31	&	7	&	1.47	\\	
	&	F169M	&	2018-12-19	&	6596	&	259	&	58	&	15	&	2.01	\\	\hline
NGC 6940	&	F169M	&	2018-06-13	&	1875	&	151	&	43	&	12	&	1.73	\\	
\toprule
    \end{tabular}
    }
\end{table} 

\begin{figure}[!hb]
\centering
\includegraphics[width=\textwidth]{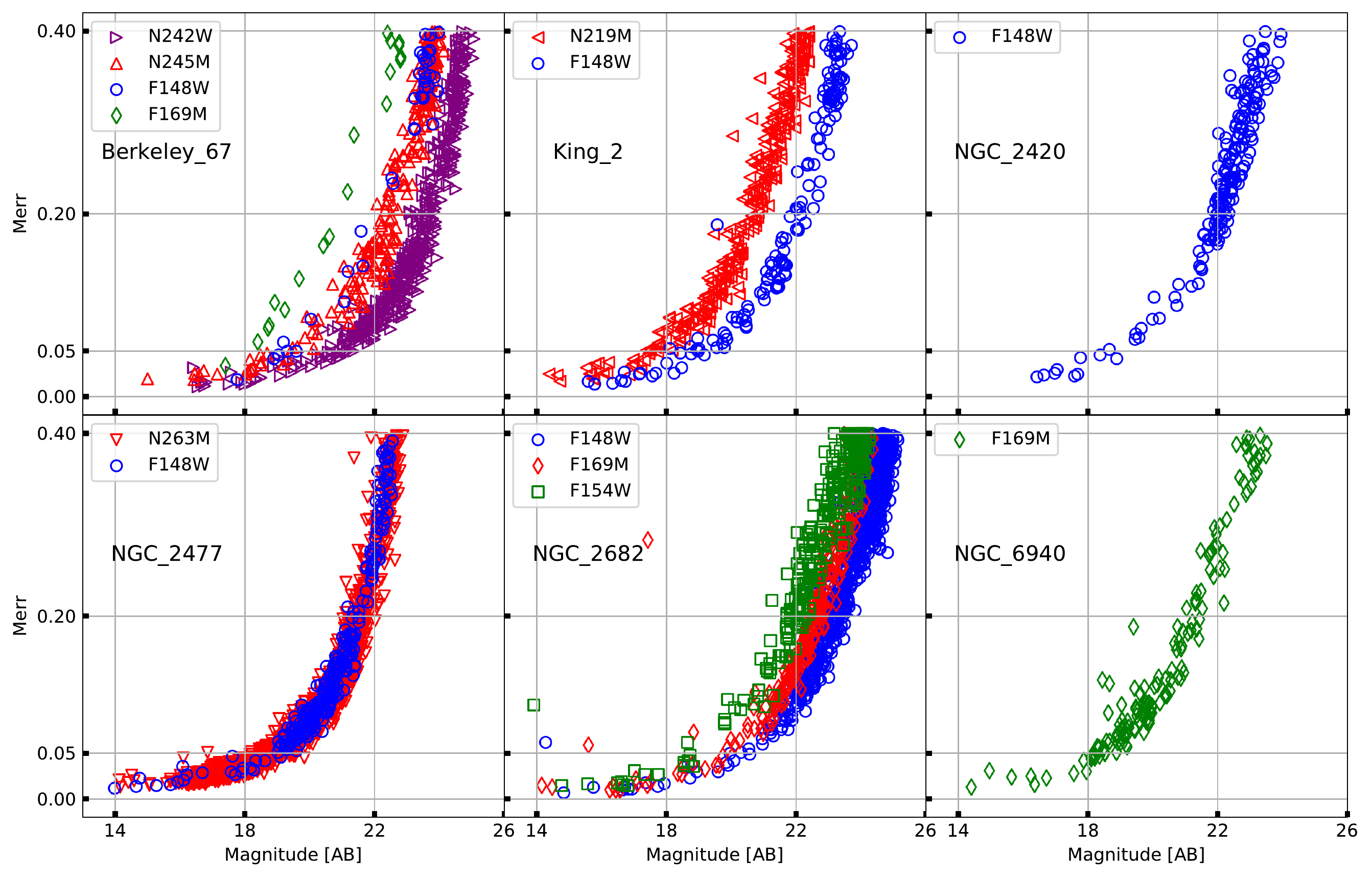}
    \caption{The photometric error in the magnitudes. Each subplot shows the magnitude-error plots for a cluster in all available filters.}
    \label{fig:3_mag_err}
\end{figure} 

The observations were carried out from December 2016 to December 2018 using different UV filters of UVIT (\textit{AstroSat} proposal IDs: A02\_170, A04\_075, G07\_007 and A05\_068). The log of UVIT observations is presented in Table~\ref{tab:3_cluster_obs}, along with total exposure time in each filter. We planned observations in at least one FUV and one NUV broadband filter to get wavelength coverage across the UV regime for detailed study.
Due to payload related issues, NUV observations were done only for early observations such as Berkeley 67, King 2 and NGC 2477.
Unfortunately, no cluster members could be detected in FUV observations of Berkeley 67 despite observing it in two FUV filters due to the lack of FUV bright stars.
The remaining three OCs (NGC 2420, NGC 2682 and NGC 6940) were observed in FUV filters alone.
Exposure time for different filters ranges from 1875--6596 sec with a typical value of $\sim$2000 sec. 

We performed PSF photometry on all UVIT images using {\sc daophot} package of {\sc iraf}. We used 5-$\sigma$ detection and limited the catalogue to the detections with magnitude errors $<$0.4 mag. The magnitude vs PSF error plots for all the images are shown in Fig.~\ref{fig:3_mag_err}. The magnitudes were corrected for saturation following \citet{Tandon2017a}. We removed artefacts arising from saturated/bright stars and false detection at the edge to create the final list of UVIT detected sources for each observed filter. 
We included the saturated stars in the catalogue, however, their magnitudes represent the upper limit (they are brighter than these values), and their astrometry may be incorrect by a few arcseconds.
In this way, over 100 stars were detected in each image and the details of this are listed in Table~\ref{tab:3_cluster_obs}.

\subsection{\textit{Gaia} data} \label{dec:gaia}

The \textit{Gaia} EDR3 data for all clusters were compiled by constraining spatial and parallax measurements. We used the r50 (radius containing half the members) mentioned in \citet{Cantat2020} to get the majority of the members with minimal contamination in the VPD. This region was used to calculate MPs in the {\sc gmm} model. We tripled the radius for running the {\sc prf} algorithm to detect more members in the outer region.
The definition and formulae of independent/derived \textit{Gaia} parameters used in this work are shown in Table~\ref{tab:3_Gaia parameters}.
The errors of \textsc{ra, dec, pmra, pmdec} and \textsc{parallax} are taken from \textit{Gaia} EDR3. Upper limits of the photometric errors in \textsc{g} are calculated using the `mag\_error' formula in Table~\ref{tab:3_Gaia parameters} (similar for errors in \textsc{gbp, grp, bp\_rp bp\_g} and {\sc g\_rp}). Errors in {\sc ruwe} and {\sc aen} are assumed to be zero.

The parallax\_cluster, ra\_cluster, dec\_cluster and radius\_cluster (as mentioned in Table~\ref{tab:3_cluster_details}) are used to select sources near the cluster using the following ADQL query:

\noindent\rule[0.5ex]{\linewidth}{0.5pt}
{\footnotesize {\fontfamily{pcr}\selectfont
\noindent 
select *
\\\textcolor{magenta}{from} gaiaedr3.gaia\_source
\\\textcolor{magenta}{where} 
\\pmra \textcolor{magenta}{is not null and} parallax \textcolor{magenta}{is not null and}
\\ABS(parallax-\textit{\textbf{cluster\_parallax}})$<$3* parallax\_error \textcolor{magenta}{and}
\\\textcolor{magenta}{contains}(\textcolor{magenta}{point}('icrs', gaiaedr3.gaia\_source.ra, gaiaedr3.gaia\_source.dec), \\\textcolor{magenta}{circle}('icrs', \textbf{\textit{cluster\_ra, cluster\_dec, cluster\_radius}})) = 1
}}

\noindent\rule[0.5ex]{\linewidth}{0.5pt}

\begin{table}[!ht]
\centering
\caption{Definition and formulae for \textit{Gaia} parameters and derived parameters used in this study.}
\label{tab:3_Gaia parameters}
\begin{tabular}{C{15cm}} 	\toprule					
\textbf{\textit{Gaia} EDR3 Parameters $^{\dagger}$} 	\\ \midrule					
\textsc{ra, dec, pmra, pmdec, parallax, ra\_error, dec\_error, pmra\_error, pmdec\_error, parallax\_error,} \\ 
{\sc phot\_g\_mean\_mag (\text{AS} g), phot\_bp\_mean\_mag (\text{AS} gbp), phot\_rp\_mean\_mag (\text{AS} grp), bp\_rp, bp\_g, g\_rp, phot\_g\_mean\_flux, phot\_bp\_mean\_flux, phot\_rp\_mean\_flux, phot\_g\_mean\_flux\_error, phot\_bp\_mean\_flux\_error, phot\_rp\_mean\_flux\_error, g\_zero\_point\_error, bp\_zero\_point\_error, rp\_zero\_point\_error, } \\ 
\textsc{astrometric\_excess\_noise (\text{AS} aen), astrometric\_excess\_noise\_sig (\text{AS} aen\_sig), ruwe}	\\ \toprule	
\textbf{Cluster parameters}  (taken from Table~\ref{tab:3_cluster_details})
	\\ \midrule					
\textit{ra\_cluster, dec\_cluster, parallax\_cluster, radius\_cluster}	\\ \toprule					\textbf{Derived parameters}	\\ \midrule	
\end{tabular}			
\begin{footnotesize}
\begin{tabular}{L{3.5cm}L{8.5cm}R{2.5cm}}						
mag\_error	&	$\sqrt{\left(1.086 \times \frac{flux\_error}{flux}\right)^2 + zero\_point\_error^2}$	&	[mag]	\\	
\textsc{pmR0}	&	$\sqrt{(\textsc{pmra}-cluster\_pmra)^2+(\textsc{pmdec}-cluster\_pmdec)^2}$	&	[mas yr$^{-1}$]	\\	
\multirow{2}{*}{$\textsc{qf}\  \text{ (Quality filter)} \quad \bigg\{ $}	&	0, $ \text{if } \textsc{(ruwe > 1.4) \ or \ (aen > 1 and aen\_sig > 2)}$	&		\\	
	&	$1, \qquad \text{otherwise}$	&		\\ \toprule	
\end{tabular}
\end{footnotesize}
\\$^{\dagger}$ {\footnotesize \url{gea.esac.esa.int/archive/documentation/GEDR3/Gaia_archive/chap_datamodel/sec_dm_main_tables/ssec_dm_gaia_source.html}}
\end{table}  

\subsection{Isochrones and evolutionary tracks} \label{dec:iso_tracks}

We used {\sc parsec} isochrones \footnote{http://stev.oapd.inaf.it/cgi-bin/cmd} \citep{Bressan2012MNRAS.427..127B} generated for cluster metallicity and age, adopted from \citet{Dias2002A&A...389..871D} and WEBDA\footnote{https://webda.physics.muni.cz/}. 
As the UV images would detect WDs, we included WD (hydrogen-rich atmosphere, type DA) cooling curves in the CMDs. 
As the turn-off masses of the OCs under study range from 1.1--2.3 M$_{\odot}$, we included WD cooling curves of mass 0.5--0.7 M$_{\odot}$ \citep{Fontaine2001,Tremblay2011,Cummings2018}.
We used reddened isochrones and WD cooling curves in this chapter.

\section{Membership determination} \label{sec:3_method}

\subsection{Gaussian Mixture Model} \label{sec:3_gmm}
The distribution of stars in the PM space is assumed to be an overlap of two Gaussian distributions. The sum of which can be written as,
\begin{equation}
    f(\mu|\overline{\mu_j},\Sigma_j)=\sum_{j=1}^{2} w_j \frac{exp\left[-1/2(\mu-\overline{\mu_j})^T\Sigma^{-1}_{j}(\mu-\overline{\mu_j})\right]}{2\pi\sqrt{det \Sigma_j}}
\end{equation}
\begin{equation}
    w_j>0,\qquad \sum_{j=1}^{2} w_j=1
\end{equation}
where $\mu$ is individual PM vector, $\overline{\mu_j}$ are field and cluster mean PMs, $\Sigma$ is the symmetric covariance matrix and $w_j$ are weights for the two Gaussian distributions.
The generalised formalism for the n-D case and details of fitting the Gaussian distributions to \textit{Gaia} data are available in the appendix of \citet{Vasiliev2019}.

We selected stars within r50 of cluster centre and removed sources with following quality filters \citep{Lindegren2018, Riello2020} to keep stars with good astrometric solutions:
\begin{equation} \label{eq:3_quality_GMM}
    \begin{split}
    \textsc{ruwe} > 1.4 \\
    \textsc{aen > 1.0}\quad \textsc{and} \quad  \textsc{aen\_sig > 2.0} \\
    |\textsc{parallax} - parallax\_cluster| > 3 \times \textsc{parallax\_error}
\end{split}
\end{equation}

For such sources, a {\sc gmm} is created using {\sc pmra} and {\sc pmdec}, as only these parameters have distinct Gaussian distribution for the cluster members. Two isotropic Gaussian distributions are assumed for the field and member stars. These were initialised with previously known values of cluster PM and internal velocity dispersion.
We used {\sc GaiaTools}\footnote{https://github.com/GalacticDynamics-Oxford/GaiaTools} to maximise the likelihood of the {\sc gmm} and get the mean and standard deviation of the two Gaussian distributions. Simultaneously, the MPs of all stars in the field are calculated.

{\sc gmm} cannot use the other parameters provided by \textit{Gaia} EDR3 catalogue ({\sc parallax, ra, dec, g, bp\_rp}, etc.) due to their non-Gaussian distributions. {\sc gmm} does not organically account for systematic parameters leading to loss of interesting stellar systems with variability, binarity and atypical spectra. However, {\sc gmm} can convincingly give the average CMD and VPD distribution of stars in a cluster. This can be further enhanced with the inclusion of photometric and systematic information.
Hence, we used a supervised machine learning method to improve membership determination and utilise the non-Gaussian parameters.

\subsection{Probabilistic Random Forest} \label{sec:3_prf}

\begin{figure}[!ht]
    \centering
    \includegraphics[width=0.8\textwidth]{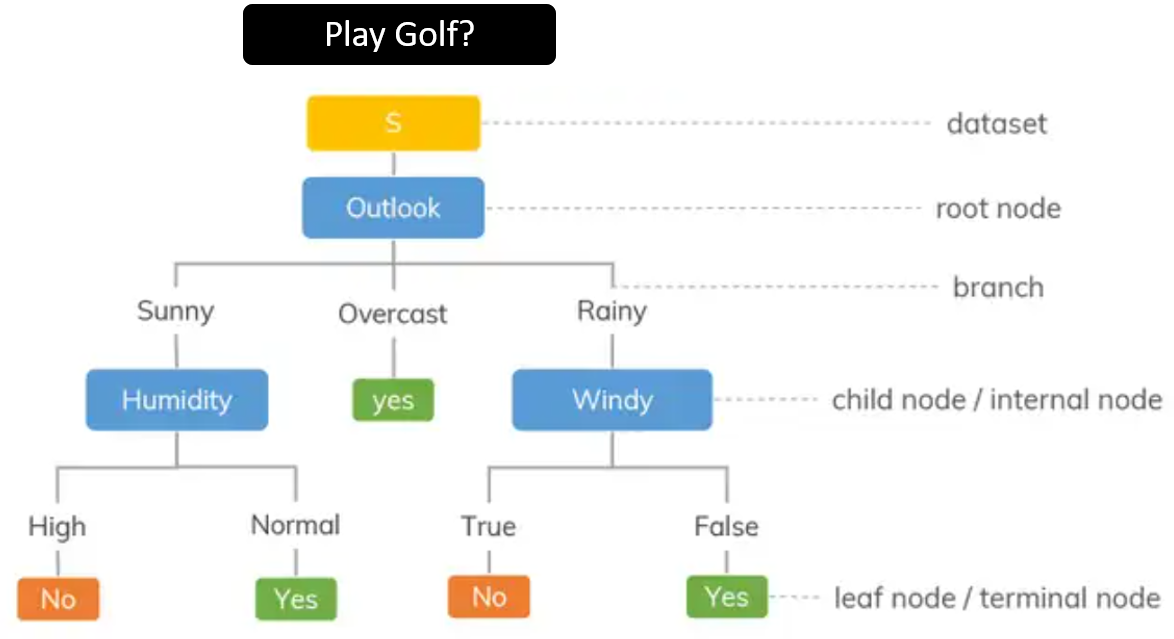}
    \caption{Example of a decision tree.}
    \label{fig:3_decision_tree}
\end{figure}

A random forest consists of multiple decision trees. Fig.~\ref{fig:3_decision_tree} shows an example of a decision tree for deciding whether to play golf. There are many factors to be accounted for, such as outlook, humidity, and wind. These factors are called as \textit{features}. The data-set, \textit{S}, contains the information about the features. Then each node (root and internal node) uses one feature and one threshold to make a decision. The next node uses one of the remaining features to make its decision. The last node, \textit{leaf node}, gives the final decision to play or not to play. 
Overall, the layered nodes make a decision tree. 
The random forest is constructed using multiple such trees. Each tree starts with a randomised subset of the data-set to avoid overfitting and uses random order for nodes with specific features. The average classification of each tree (leaf nodes) gives the decision of the random forest.

The random forest needs to be trained in case of a supervised process. We use a labelled data-set with values of the features and the expected decision in each case. The random forest takes the labelled data-set and modifies the order of nodes, threshold values and the number of layers to replicate the expected decisions closely. This optimising process is known as \textit{training}. Now the random forest can make decisions based on new input features.

The major drawback of traditional random forest is that no uncertainties (errors) are assumed during these calculations. The probabilistic random forest ({\sc prf}) algorithm \footnote{\url{https://github.com/ireis/PRF}}, developed by \citet{Reis2019}, takes care of errors in the data, which is essential for any astronomical data-set.
It assumes all features and labels as probability distribution functions and out-performs traditional random forest algorithms in the case of noisy data sets.

In this study, we used the photometric, astrometric and systematic parameters from \textit{Gaia} as features.
To create a training set for the {\sc prf} algorithm, we first calculated the MPs using the {\sc gmm} method. We used stars within r50 radius \citep{Cantat2020} from the cluster centre to reduce field star contamination.
The training set was created for each cluster by labelling $P\_GMM>0.5$ as members and others as non-members. The training and testing sets were created by randomly splitting the data set into 3:1 ratios.
The {\sc prf} requires the features, their errors and the known class ({\sc gmm} membership labels) as inputs for training. The output contains fractional MP for each star and the feature-importance.
After training the {\sc prf} on stars within radius r50, we applied the algorithm on the stars within 3$\times$r50 of the cluster centre to increase the sample size.

We assessed the performance of the following parameters as features that can impact the membership determination:
{\sc ra, dec, pmra, pmdec, parallax, g, g\_rp, ruwe, aen, pmR0} and many others. 
The meaning of the features used is mentioned in Table~\ref{tab:3_Gaia parameters}. RVs are limited to stars with {\sc g} $<15$ mag; hence they were not used as a feature. 

We tried more than 22 feature-combinations to optimise the membership determination. We judged the different combinations by:

\begin{enumerate}[leftmargin=*]
    \item their ability to recreate MP similar to {\sc gmm} using \textit{accuracy score} in testing phase. The accuracy score is defined as: 
        \begin{equation}    
        accuracy\  score = \frac{correctly\  predicted\  class}{total\  testing\  class}\times 100
        \end{equation}
     Although the accuracy score itself is not enough to select the final feature-combination, one can weed out poorly performing combinations.
    \item the distribution of members in VPDs (Cluster should occupy a compact circular region in the VPD {e.g., Fig.~\ref{fig:3_CV_combined}}).
    \item the distribution of members in CMDs (Minimal contamination to the CMD, although it is a subjective judgement).  
    \item the distribution of members in PM--parallax plot (the cluster should occupy small range in both PM and distance). 
\end{enumerate}

Based on their individual merits, the notable feature-combinations are listed in Table~\ref{tab:3_feature_selection}.

\subsection{Selection of features and membership criteria} \label{sec:3_feature_selection_0}

\begin{table}[!ht]
\centering
\caption{Feature-combinations used in {\sc prf} algorithm to calculate MP.}
\label{tab:3_feature_selection}
\begin{tabular}{L{1.5cm}L{8cm}L{2.5cm}}
\toprule
Name	&	Features	& 	Information	\\ \midrule
F6	& 	{\sc ra, dec, pmra, pmdec, parallax, pmR0} 	&	Astrometry	\\ \hline
F8	&	{\sc ra, dec, pmra, pmdec, parallax, pmR0, g, g\_rp} 	&	Astrometry+ Photometry	\\ \hline
F10	&	{\sc ra, dec, pmra, pmdec, parallax, pmR0, g, g\_rp, ruwe, aen} 	&	Astrometry+ Photometry+ Systematics	\\

 \bottomrule
 \end{tabular}
 \end{table} 
 
We trained the {\sc prf} using 1--1000 trees and saw a plateau in accuracy score after 150--200 trees. As \citet{oshiro_how_2012} suggested, the optimum number of trees lies between 64--128; hence we chose 200 trees for further analysis. 
Almost all feature-combinations had an accuracy score of 92--98\%, as all were designed to select the cluster members. Hence, choosing the best combination was not trivial. 

As expected {\sc pmra, pmdec} and {\sc parallax} are important features for membership. The cluster's distribution in VPD is a 2D Gaussian, and the random forest does not completely replicate this quadratic relation between {\sc pmra} and {\sc pmdec}. Hence, we created a new parameter called {\sc pmR0}, which is the separation of the source from the cluster centre in the VPD. The PM cluster centre was obtained from the {\sc gmm} results. {\sc pmR0} helped constrain the cluster distribution to a circular shape.
To test the importance of individual features, we introduced a column with random numbers as a feature. Among the \textit{Gaia} features, {\sc ra} and {\sc dec} showed very comparable feature-importance as the random column. However, upon further inspection, we found that inclusion of {\sc ra/dec} does not harm the {\sc prf} while improving the membership determination is some cases (King 2 is an example, which is the farthest cluster in our set and has the smallest sky footprint). Due to known overestimation of {\sc gbp} flux in fainter and redder stars \citep{Riello2020}, we have used {\sc g} and {\sc g\_rp} as features.

\begin{figure}[!ht]
    \centering
    \includegraphics[width=0.8\textwidth]{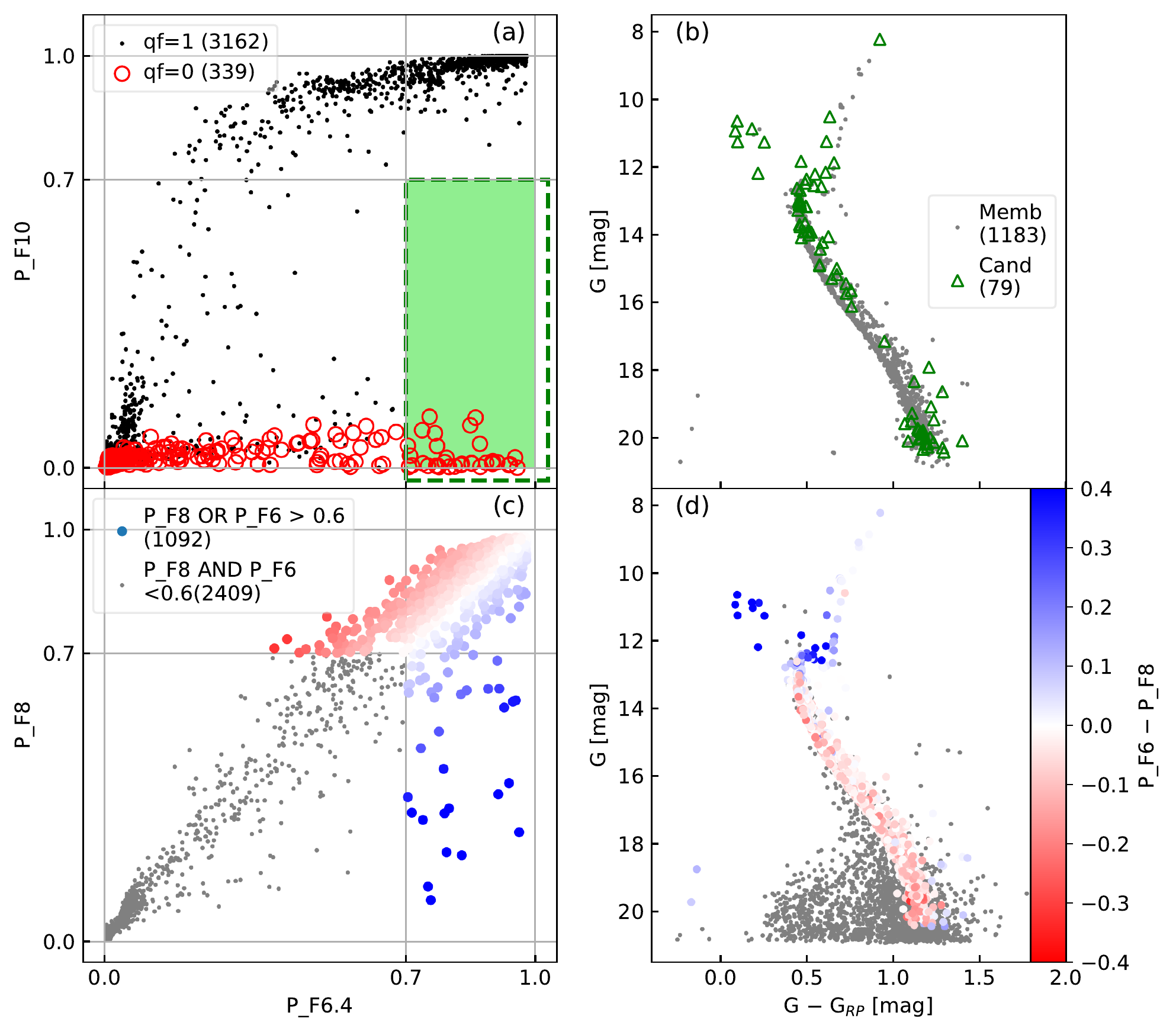}
    \caption{Comparison of different MPs from {\sc prf} feature-combinations for NGC 2682. The numbers in brackets represent the number of stars in that particular category. 
    {(a)} Comparison of F6 and F10, where black dots are good quality sources ({\sc qf} = 1) and red circles are poor quality sources ({\sc qf} = 0). The green dashed box represents sources with \(P\_F10 < 0.7 \leq P\_F6 \). 
    {(b)} CMD of NGC 2682 according to membership criteria in Eq.~\ref{eq:3_classif}, the grey dots are \texttt{members} (\(P\_F10 > 0.7\)), while the green triangles are the \texttt{candidates} (\(P\_F10 < 0.7 \leq P\_F6\)).
    {(c)} Comparison of F6 and F8 to demonstrate the MP dependence on CMD location. The colour is according to (P\_F6 $-$ P\_F8), as shown in the right most panel.
    {(d)} CMD of stars coloured according to (P\_F6 $-$ P\_F8). The stars with \textit{peculiar} CMD position are bluer.
    } \label{fig:3_P_comparison_internal}
\end{figure} 

We added {\sc ruwe} and {\sc aen} as features to include the quality checks in {\sc prf}. This nullifies the need for manually filtering the data. The sources with large {\sc ruwe/aen} are typically binary stars, variables, extended sources or stars with atypical SEDs \citep{Lindegren2020arXiv201203380L, Riello2020, Gaia2021A&A...649A...1G, Fabricius2020}. As binaries and atypical SEDs are intriguing sources, we devised a method to keep such poor quality sources as candidates. Hereafter, we will refer to sources with {\sc qf = 1} as \textit{good quality sources} and sources with {\sc qf = 0} as \textit{bad quality sources}.
The F6 feature-combination uses only astrometric data (see Table~\ref{tab:3_feature_selection}) for the membership determination; hence it can give the MPs for poor quality sources. F10 uses astrometric, photometric and systematic parameters as features; hence it can give membership of good quality sources. Fig.~\ref{fig:3_P_comparison_internal} (a) shows the comparison of MPs from F6 and F10. As seen from the CMD in Fig.~\ref{fig:3_P_comparison_internal} (b), the bad quality sources in the green region (P\_F10 $<$ 0.7 $<$ P\_F6) are likely to be cluster members. For further analysis, we define the \texttt{members, candidates} and \texttt{field}, as follows:
\begin{equation} \label{eq:3_classif}
    \begin{split}
    \texttt{Members} \Rightarrow {}& P\_F10 > cutoff\\
    \texttt{Candidates} \Rightarrow {}& P\_F10 < cutoff \leq P\_F6.4 \\
    \texttt{Field} \Rightarrow {}& P\_F10\  \textsc{and}\  P\_F6 < cutoff
    \end{split}
\end{equation}
\noindent In an ideal scenario without systematic errors, we would use only F6 for the membership. We recommend using the \texttt{candidate} classification for {\sc G $<$ 19} (large intrinsic errors at fainter magnitudes create spread in bottom MS). After looking at the CMDs and residual VPDs with various cutoffs, we recommend a cutoff of 0.7. However, we note that the ideal cutoff varies from cluster to cluster and strongly depends on the separation of cluster--field in the VPD and ratio of field stars to cluster members.

While analysing the different feature combinations, we found that adding photometric information (F8) to astrometric information (F6) leads to lessening the MP of stars in \textit{peculiar} CMD locations. This is demonstrated in Fig.~\ref{fig:3_P_comparison_internal} (c) and (d). Most stars have the same MP from F8 and F6; however, the BSSs in NGC 2682 have larger P\_F6~$-$~P\_F8 due to the absence of many stars in the same location in the training set. For further discussion, we will refer to P\_F6~$-$~P\_F8 as \textit{peculiarity}. As the other clusters do not have many BSSs, the \textit{peculiarity} can be used to distinguish between stars on the MS and sub-giant/giant branch.

\subsection{Comparison with literature}

\begin{figure}[!ht]
    \centering
    \includegraphics[width = 0.98\textwidth]{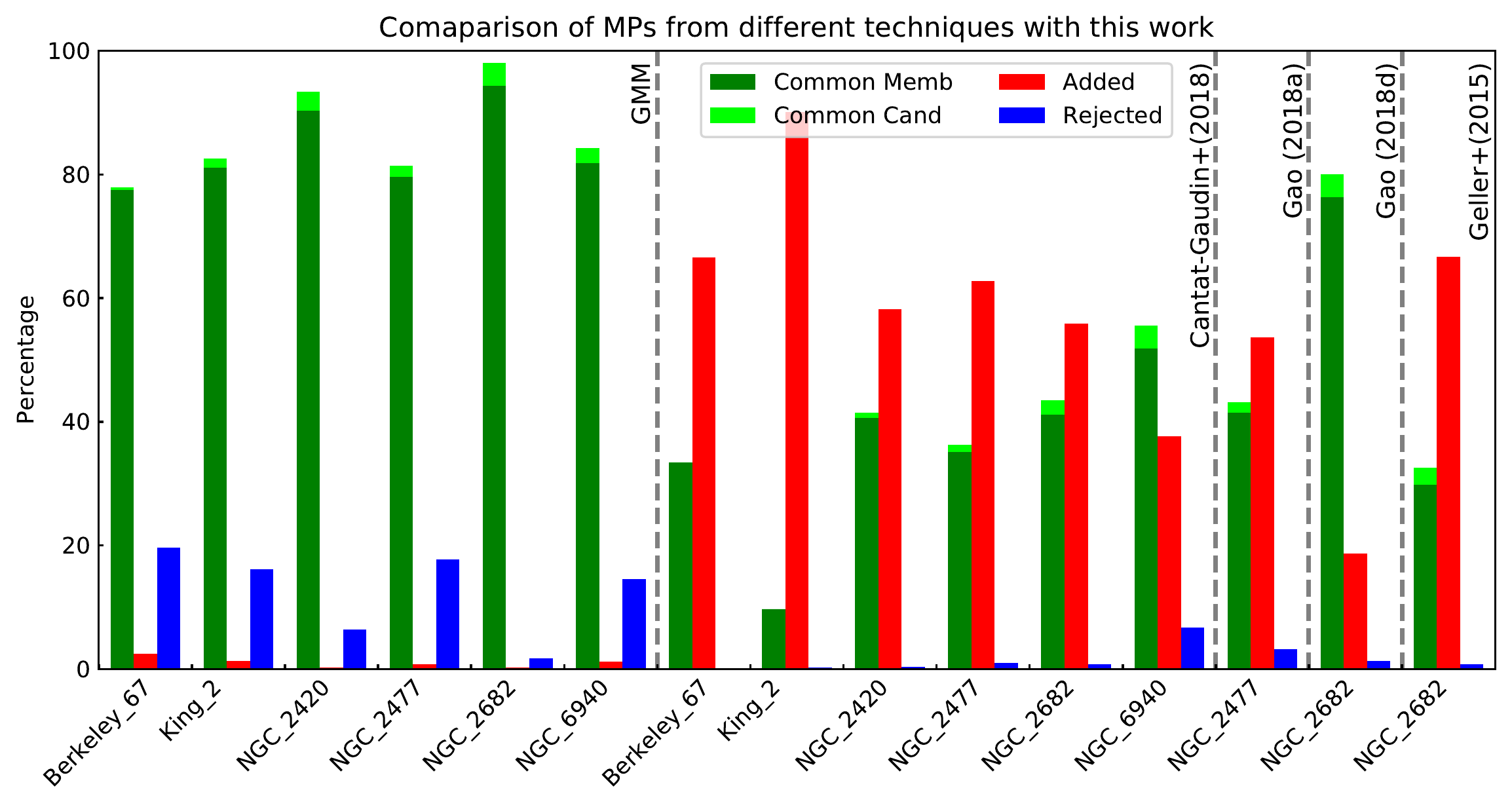}
    \caption{Grouped histogram for common \textit{candidates}, common \textit{members}, added \textit{members} and rejected stars. The totals of four classes are normalised to 100 for easy visualisation and the numbers are tabulated in Table~\ref{tab:3_lit_comparison}. The likely members by {\sc prf} and other techniques are represented by Common Memb (dark green) and Common Cand (lime). Red bars are stars added as \texttt{members} by {\sc prf} (classified as field by other techniques or missing from literature catalogues). Blue bars are stars rejected (classified as \texttt{field}) by {\sc prf}, but these are members in other techniques. Most of the rejected stars are faint and with larger {\sc pmR0}. The dotted lines separate the comparisons with different methods and papers viz. {\sc gmm}, \citet{Cantat2018A&A...618A..93C}, \citet{Gao2018b}, \citet{Gao2018c} and \citet{Geller2015}. All comparisons are done for the same FOV.}
    \label{fig:3_lit_comparison}
\end{figure}

\begin{table}[!ht]
    \centering
    \caption{Comparison of membership classification by {\sc prf} with {\sc gmm}, \citet{Cantat2018A&A...618A..93C}, \citet{Gao2018b}, \citet{Gao2018c} and \citet{Geller2015}.}
    \begin{tabular}{lcc cc}
    \toprule
	&	Common	&	Common	&	Added	&	Rejected	\\
	&	\texttt{Memb}	&	\texttt{Cand}	&	\texttt{Memb}	&	Stars	\\
	&	[1]	&	[2]	&	[3]	&	[4]	\\ \midrule
\multicolumn{5}{c}{Comparison with {\sc gmm}}									\\
Berkeley 67	&	158	&	1	&	5	&	40	\\
King 2	&	506	&	9	&	8	&	101	\\
NGC 2420	&	354	&	12	&	1	&	25	\\
NGC 2477	&	1416	&	33	&	14	&	316	\\
NGC 2682	&	436	&	17	&	1	&	8	\\
NGC 6940	&	338	&	10	&	5	&	60	\\ \midrule
\multicolumn{5}{c}{Comparison with {\citet{Cantat2018A&A...618A..93C}}}									\\
Berkeley 67	&	131	&	0	&	261	&	0	\\
King 2	&	104	&	0	&	968	&	3	\\
NGC 2420	&	357	&	7	&	511	&	3	\\
NGC 2477	&	1396	&	46	&	2492	&	39	\\
NGC 2682	&	502	&	28	&	681	&	9	\\
NGC 6940	&	399	&	29	&	290	&	52	\\ \midrule
\multicolumn{5}{c}{Comparison with \citet{Gao2018b}}									\\
NGC 2477	&	1695	&	67	&	2193	&	133	\\ \midrule
\multicolumn{5}{c}{Comparison with \citet{Gao2018c}}									\\
NGC 2682	&	950	&	46	&	233	&	16	\\ \midrule
\multicolumn{5}{c}{Comparison with \citet{Geller2015}}									\\
NGC 2682	&	365	&	34	&	817	&	10	\\ \bottomrule
\multicolumn{5}{c}{\footnotesize [1] Classified as \texttt{members} by both {\sc prf} and other techniques}									\\
\multicolumn{5}{c}{\footnotesize [2] Members of other techniques classified as \texttt{Candidates} by {\sc prf}}									\\
\multicolumn{5}{c}{\footnotesize [3] Added \texttt{members} by {\sc prf}, which are not members in other catalogues}									\\
\multicolumn{5}{c}{\footnotesize [4] Members from other techniques classified as \texttt{field} by {\sc prf}}									\\
\end{tabular}
    \label{tab:3_lit_comparison}
\end{table} 

Fig.~\ref{fig:3_lit_comparison} shows the comparison of {\sc prf} with {\sc gmm}, \citet{Cantat2018A&A...618A..93C}, \citet{Gao2018b} and \citet{Gao2018c}. The actual numbers of different types of stars are listed in Table~\ref{tab:3_lit_comparison}. Although intuitive, the meaning of \textit{added}, \textit{rejected} etc. is given in the footnote of Table~\ref{tab:3_lit_comparison}. Although \citet{Lindegren2020arXiv201203380L} warns against direct crossmatch between DR2 and EDR3 due to changes in epochs, the astrometric shift for cluster members is $<$10 $\mu$as. All comparisons were done over the same FOV.

{\sc gmm} and {\sc prf} used different set of parameters. As seen Fig.~\ref{fig:3_lit_comparison}, the classification by {\sc prf} was similar to {\sc gmm}. The accuracy score (reproducibility) of {\sc prf} was between 90--99\% for the six clusters. However, there are some differences, which are expected and embraced. The major difference was seen in the rejection of {\sc gmm} members (2--25\%), most of which were in {\sc G} $>$ 19 mag region. Among stars brighter than 19 mag, the percentage of rejected stars drops to 0--4\%, almost all having poor astrometric solutions ({\sc ruwe $>1.2$}). 

\citet{Cantat2018A&A...618A..93C} used clustering in the ($\mu_{\alpha},\mu_{\delta}, \pi$) space to identify the members using \textit{Gaia} DR2 data. They selected the stars with parallax within 0.3 mas and PM within 2 mas yr$^{-1}$ of the cluster mean. The probabilities were calculated using {\sc upmask}, an unsupervised clustering algorithm. {\sc prf} has identified significantly more (260--2500) new members compared to \citeauthor{Cantat2018A&A...618A..93C}. All the added stars have acceptable CMDs, VPDs and PM--parallax distributions. As the magnitude limit of the \citeauthor{Cantat2018A&A...618A..93C} catalogue was 18, many new members were added in the fainter end of the MS. The rejected members (0--52) typically have either high {ruwe}/low {\sc pmR0} or low {\sc ruwe}/high {\sc pmR0}. This is a good optimisation as the added stars far outnumber the rejected stars. Fig.~\ref{fig:3_Cant18_comp_0} shows the comparison between MPs, VPDs, CMDs, parallax--PM plots and \textsc{ruwe}--\textsc{aen} plots for star common and newly added stars. 

\citet{Gao2018b} used \textit{Gaia} DR2 to determine membership of NGC 2477 (and three other clusters) using a {\sc gmm}. Although there are 1695 common members, {\sc prf} has added 2193 stars and rejected 133 stars. Majority of added stars are near the cluster parallax and {\sc pmR0} $<$ 1.5 mas yr$^{-1}$. The rejected stars are again typically results of {\sc ruwe/aen} and {\sc pmR0} trade-off. The top panels of Fig.~\ref{fig:3_gao_comp} show the comparison between the \citet{Gao2018b} and {\sc prf} membership.

\citet{Gao2018c} utilised a random forest of 11 \textit{Gaia} parameters ({\sc ra, dec, parallax, pmra, pmdec, g, gbp, grp, bp\_rp, bp\_g and g\_rp }) to calculate the MPs of NGC 2682. \citeauthor{Gao2018c} did not remove stars with high systematic errors, and the random forest algorithm did not incorporate uncertainties in the astrometric data. The use of EDR3 data, inclusion of errors and using the F6 and F8 feature-combinations has led to 233 more \texttt{members}. 
\citet{Geller2015} calculated the MPs in NGC 2682 using a combination of RV measurements (up to 40 yr baseline) and previous PM data \citep{Yadav2008}. Their catalogue is magnitude limited due to its spectroscopic nature. Among the crossmatched \citeauthor{Geller2015} members, we classify 3\% stars as \texttt{field}, due to larger {\sc ruwe} or different PM/{\sc parallax}.
The middle and lower panel of Fig.~\ref{fig:3_gao_comp} show the comparison of the {\sc prf} membership with \citet{Gao2018c} and \citet{Geller2015}, respectively.

Comparison with previous literature confirms that F10 membership is adequate for membership, and we will use F10 as primary membership criteria.
Due to the limitations (systematic and statistical errors) in \textit{Gaia} EDR3, we included the \texttt{candidate} classification to account for the likely cluster members (using F6).
We used both \texttt{members} (selected from F10)  and \texttt{candidates} (selected from F10 and F6) for further analysis. 

Importantly, our method was able to add a significantly large number of stars in all the clusters, ranging from 200--2500 stars per cluster (Table~\ref{tab:3_lit_comparison}). Therefore, this is a significant improvement over the previous studies using \textit{Gaia} DR2, mainly in the faint MS. This will undoubtedly help in the detailed analysis of the clusters and locate interesting candidates that are bright in the UV.

\begin{landscape}

\begin{figure*}
  \centering
    \includegraphics[height=0.83\textwidth]{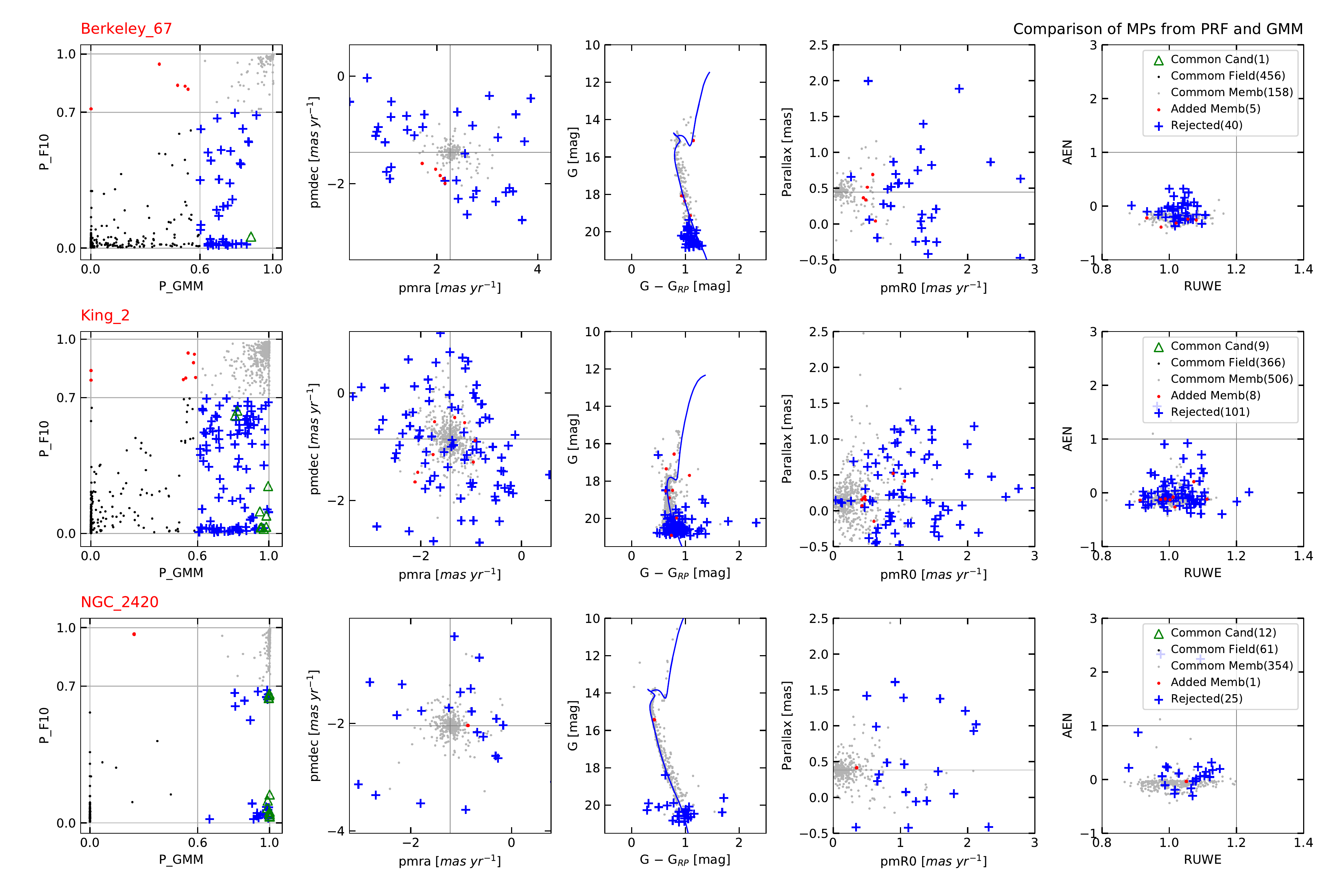}
    \caption{Comparison of MPs from {\sc prf} and {\sc gmm}. Common members (P\_F10.3 $>$ cutoff {\sc and} P\_GMM) $>$ 0.6) are shown as grey dots, common candidates (P\_F10.3 $<$ cutoff {\sc and} P\_F6.4 $>$ cutoff {\sc and} P\_GMM) $>$ 0.6) are shown as green triangles, common field stars (P\_F10.3 $<$ cutoff {\sc and} P\_F6.4 $<$ cutoff {\sc and} P\_GMM) $<$ 0.6) are shown as black dots, added members (P\_F10.3 $>$ cutoff {\sc and} P\_GMM) $<$ 0.6) are shown as red dots and rejected stars (P\_F10.3 $<$ cutoff {\sc and} P\_GMM) $>$ 0.6).}
    \label{fig:3_GMM_comp_0}
\end{figure*}

\begin{figure}
  \ContinuedFloat
  \centering
    \includegraphics[height=0.9\textwidth]{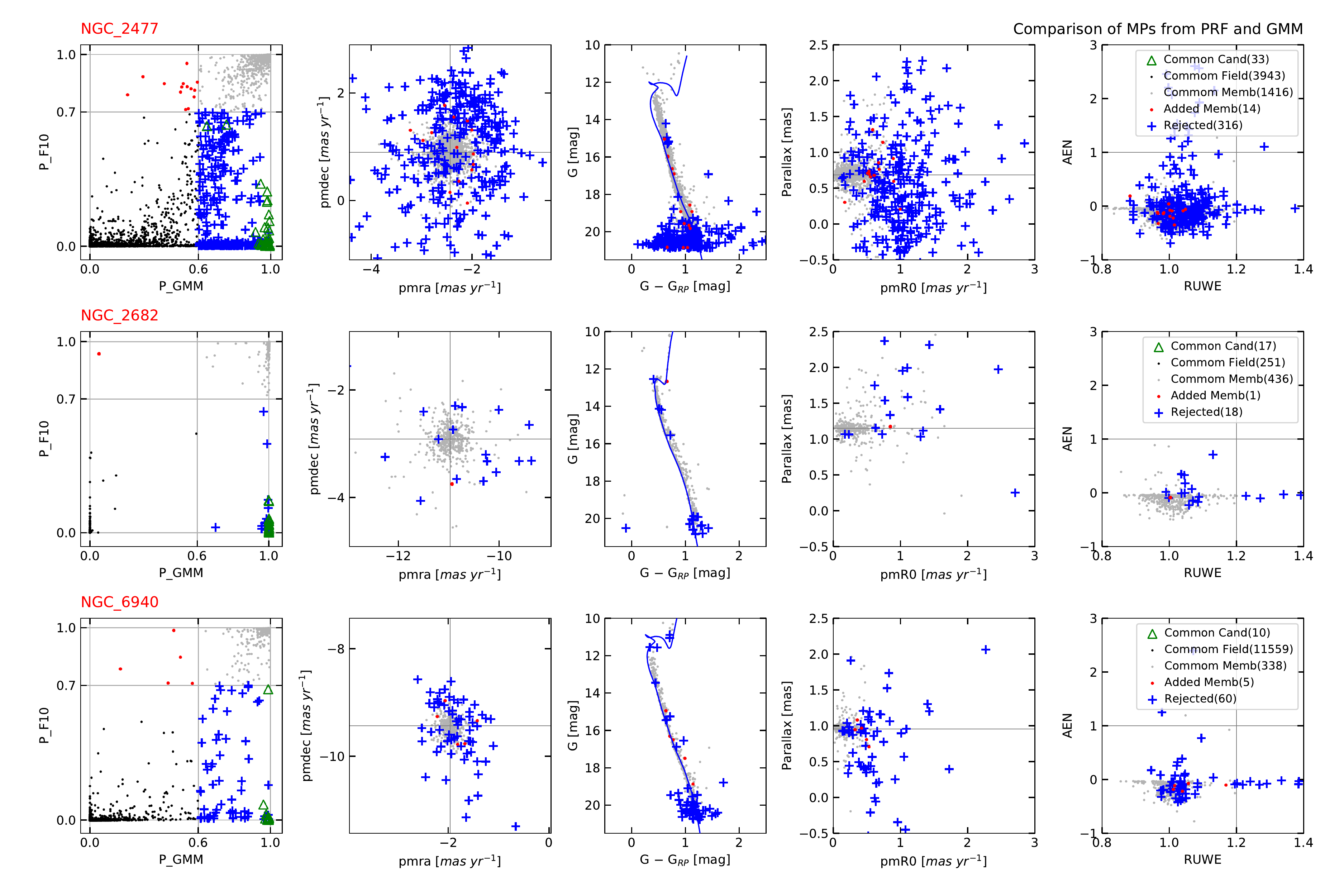}
    \caption[]{(continued...)}
\end{figure}

\begin{figure}
  \centering
    \includegraphics[height=0.9\textwidth]{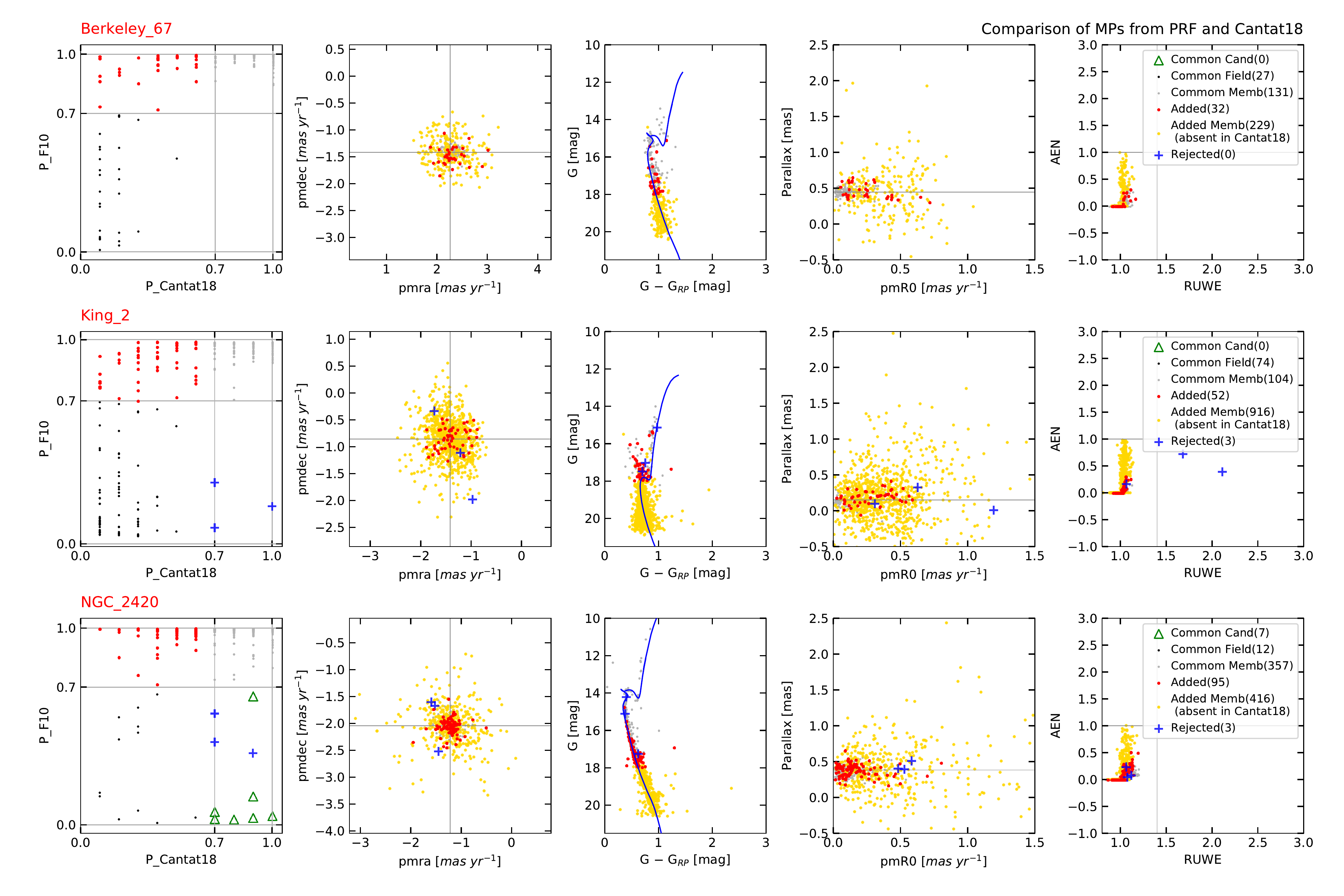}
    \caption{Comparison of MPs from {\sc prf} and \citet{Cantat2018A&A...618A..93C}. The markers for different types of stars and the individual panels are similar to Fig.~\ref{fig:3_GMM_comp_0}.}
    \label{fig:3_Cant18_comp_0}
\end{figure}
\begin{figure}
  \ContinuedFloat
  \centering
    \includegraphics[height=0.9\textwidth]{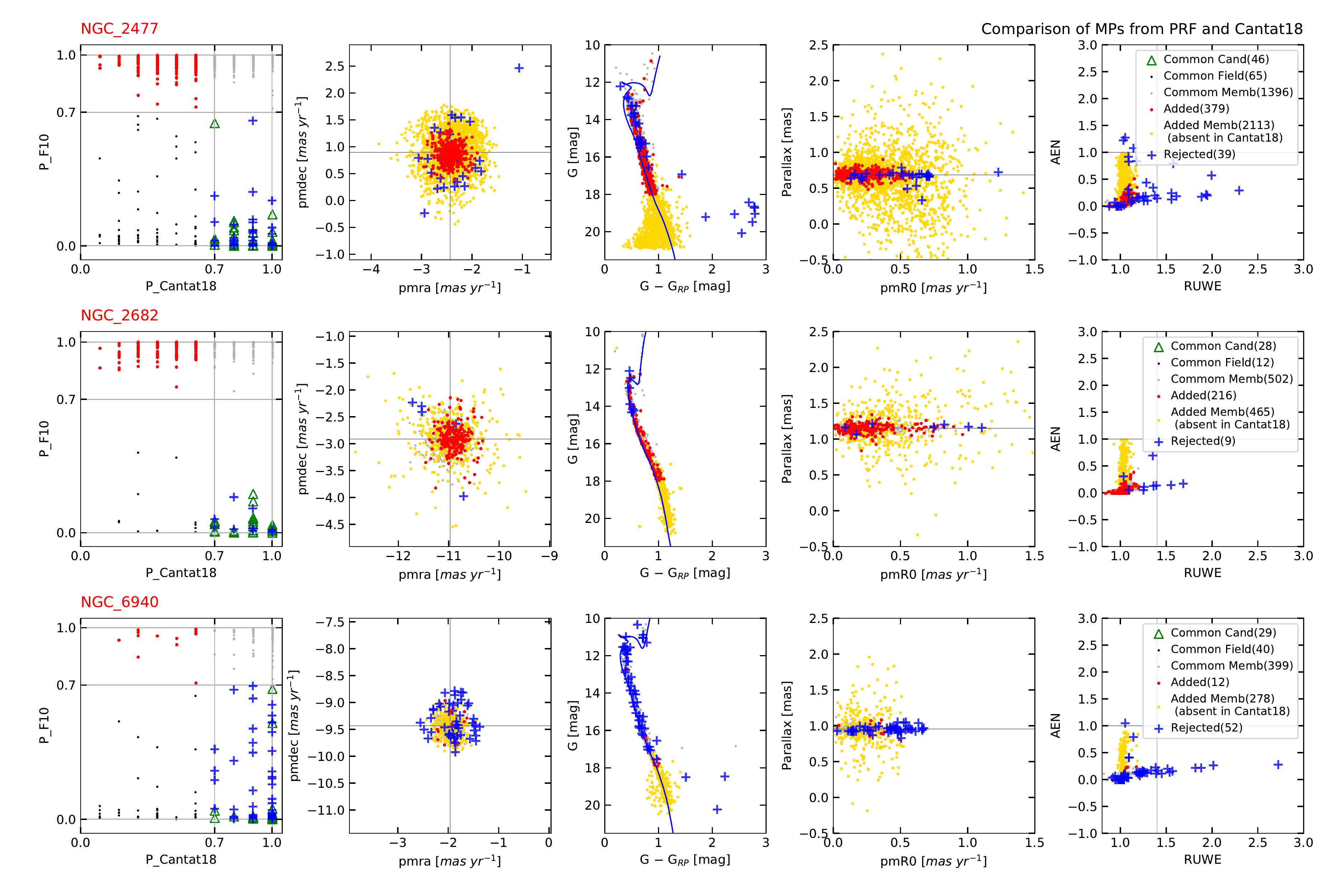}
    \caption[]{(continued...)}
\end{figure}

\begin{figure}
  \centering
    \includegraphics[height=0.9\textwidth]{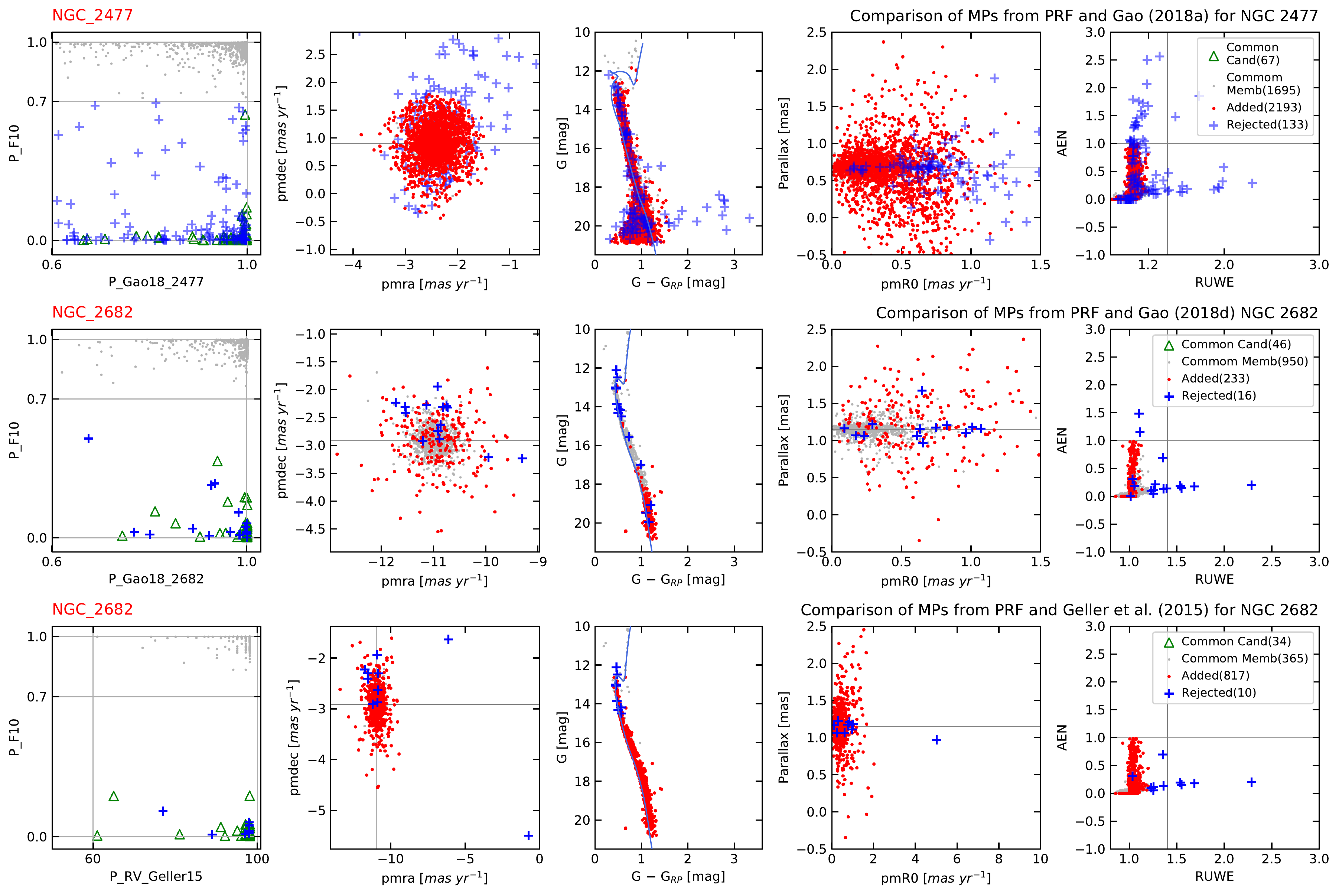}
    \caption{Analysis of membership from {\sc prf} and \citet{Gao2018b,Gao2018c} and \citet{Geller2015}. The markers for different types of stars and the individual panels are similar to Fig.~\ref{fig:3_GMM_comp_0}.}
    \label{fig:3_gao_comp}
\end{figure}

\end{landscape}

\section{Results}\label{sec:3_results}

\subsection{The catalogues} \label{sec:3_catalogue}

\begin{table}[!ht]
\centering
\caption{Example of \textit{Gaia} EDR3 membership catalogue 
with MP using {\sc gmm} and machine learning. The spatial coordinates, {\sc g} and {\sc g\_rp} along with MP obtained by {\sc gmm}, F6, F8 and F10 feature-combinations are included. The `class' column shows the classification according to Eq.~\ref{eq:3_classif} (M: \texttt{member}, C: \texttt{candidate} and F: \texttt{Field}).}
\label{tab:3_cat_Gaia}
\resizebox{0.98\textwidth}{!}{
\begin{tabular}{ccccc ccccc cc}
\toprule
source\_id &         RAdeg &        DEdeg &  g\_mag &  g\_rp & qf &   P\_F6 &   P\_F8 &  P\_F10 &  P\_GMM & class &  cluster \\
\midrule
 260364731415812736 &  69.623768 &  50.538089 &  19.93 &  1.27 &  0 &  0.431 &  0.345 &  0.038 &     --- &     F &  Berkeley\_67 \\
 260364804431166080 &  69.683631 &  50.556330 &  20.43 &  1.06 &  0 &  0.342 &  0.192 &  0.039 &     --- &     F &  Berkeley\_67 \\
 260364804433635840 &  69.687125 &  50.552245 &  19.76 &  1.10 &  1 &  0.081 &  0.092 &  0.104 &     --- &     F &  Berkeley\_67 \\
 260364834495034880 &  69.653163 &  50.556060 &  19.95 &  1.06 &  1 &  0.370 &  0.358 &  0.532 &     --- &     F &  Berkeley\_67 \\
 260364838790438784 &  69.650741 &  50.557053 &  20.07 &  1.04 &  1 &  0.118 &  0.133 &  0.136 &     --- &     F &  Berkeley\_67 \\
\bottomrule
\end{tabular}
}
\end{table} 

\begin{table}[!h]
\centering
\caption{Example of photometric catalogue of all the detected stars in the UVIT images of NGC 6940. The catalogue includes UV magnitudes and errors along with the membership classification. Similar tables for each cluster are available online. The magnitudes of saturated stars are listed as `F169M\_sat'. The last column shows the classification according to Eq.~\ref{eq:3_classif} (M: \texttt{member}, C: \texttt{candidate} and F: \texttt{Field}).}
\label{tab:3_cat_UV}
\begin{tabular}{ccc ccc cc}
\toprule
    RAdeg &     DEdeg &  F169M &  F169M\_sat &  e\_F169M &  P\_F10 &   P\_F6 & class \\
\midrule
 308.6433 &  28.25829 &  19.78 &        --- &     0.08 &  0.003 &  0.005 &     F \\
 308.7407 &  28.22939 &  19.88 &        --- &     0.09 &    --- &    --- &   --- \\
 308.6312 &  28.23331 &  19.70 &        --- &     0.10 &  0.960 &  0.992 &     M \\
 308.9526 &  28.28288 &  17.86 &        --- &     0.04 &  0.732 &  0.006 &     C \\
 308.8039 &  28.35747 &  20.92 &        --- &     0.18 &  0.972 &  0.994 &     M \\
\bottomrule
\end{tabular}

\end{table} 

The results of this study are presented in the form of seven catalogues, a membership catalogue and six catalogues of UVIT photometry. 
The membership catalogue (for sources with $P\_F6\ \textsc{or}\ P\_F10 > 0.1$) contains \textit{Gaia} EDR3 astrometry and photometry (\textsc{ra, dec, g, g\_rp}), MPs (P\_GMM, P\_F6, P\_F8 and P\_F10), quality filter (\textsc{qf}) and membership classification (M: \texttt{member}, C: \texttt{candidate} and F: \texttt{Field}). The example of the catalogue is given in Table~\ref{tab:3_cat_Gaia}.
Table~\ref{tab:3_cat_UV} shows the example of the UV catalogue of NGC 6940, which is observed in F169M filter\footnote{The membership catalogue and UV photometric catalogues of six clusters are available at CDS via anonymous ftp to \url{https://cdsarc.cds.unistra.fr/viz-bin/cat/J/MNRAS/503/236}}.
The catalogues contain R.A.(J2016), Dec.(J2016), UV magnitudes, magnitude errors, MPs (P\_F10 and P\_F6) and membership classification (M: \texttt{member}, C: \texttt{candidate} and F: \texttt{Field}). We have included saturated stars in the catalogue, whose magnitudes give the upper bound to the actual numerical value.

We cross-matched the \textit{Gaia} and UVIT catalogues with a radius of 1$^{\prime\prime}$, to get merged catalogues using \textsc{topcat}\footnote{http://www.star.bris.ac.uk/~mbt/topcat/}. We checked for crowding and issues during the cross-matching process (e.g., duplicity). However, both \textit{Gaia} and UVIT catalogue showed an insignificant number of stars within 1$^{\prime\prime}$ of each other (2 for all UVIT detections and $<$0.4\% in \textit{Gaia} detections). These merged catalogues were used for further analysis.
 
\subsection{Cluster properties} \label{sec:3_cluster_properties}

We derived the following mean cluster properties by fitting Gaussian distribution to the \texttt{members}: R.A., Dec., parallax, PM and RV. Additionally, we included distances calculated by isochrone fitting.
We removed a few outliers while calculating the mean parallax and RV.
We fitted King's surface density profile to cluster surface density. 
\begin{equation}
    \rho(R) = F_{bg}+ \frac{F_{0}}{1+(R/R_{core})^2}
\end{equation}
\noindent where $F_{bg}$ is background counts, $F_{0}$ is count in the bin, R is the RMS of each bin limits (in degree) and $R_{core}$ is the core radius. 
We binned the \texttt{members} such that the bin area was constant for each bin in the spatial plane. This method decreased the thickness of the bin as we moved outwards from the cluster centre. The smallest bin width was kept equal to the mean separation between nearby \texttt{members}, and the rest of the bins were scaled accordingly. The $F_{bg}$ was assumed to be nil for the profile fitting. 

\begin{table}[!ht]
\centering
\caption{The cluster parameters as derived from the \texttt{members} (see Eq.~\ref{eq:3_classif}) of the six clusters.
} 
\label{tab:3_cluster_parameters}
\resizebox{0.98\textwidth}{!}{
\begin{tabular}{lcc ccc cc} 
\toprule
Cluster			&	Berkeley 67	&	King 2	&	NGC 2420	&	NGC 2477	&	NGC 2682	&	NGC 6940	\\ \midrule
Total stars in 3$\times$r50			&	4962	&	4343	&	1604	&	37649	&	3501	&	99769	\\
\texttt{Members}			&	392	&	1072	&	868	&	3888	&	1183	&	689	\\
\texttt{Candidates}			&	33	&	46	&	47	&	174	&	79	&	43	\\ \midrule
ra\_mean		[degree] $^{\alpha}$	&	69.471	&	12.727	&	114.603	&	118.048	&	132.844	&	308.632	\\
dec\_mean		[degree] $^{\alpha}$	&	50.743	&	58.186	&	21.577	&	$-$38.534	&	11.827	&	28.300	\\
pmra\_mean		[mas yr$^{-1}$] $^{\alpha}$	&	2.28$\pm$0.28	&	$-$1.43$\pm$0.27	&	$-$1.22$\pm$0.30	&	$-$2.43$\pm$0.26	&	$-$10.96$\pm$0.33	&	$-$1.96$\pm$0.15	\\
pmdec\_mean		[mas yr$^{-1}$] $^{\alpha}$	&	$-$1.42$\pm$0.22	&	$-$0.85$\pm$0.35	&	$-$2.05$\pm$0.26	&	0.90$\pm$0.27	&	$-$2.91$\pm$0.29	&	$-$9.44$\pm$0.16	\\
Stars with RV			&	2	&	1	&	6	&	16	&	39	&	15	\\
RV\_mean		[km s$^{-1}$] $^{\alpha}$	&	$\sim$ $-$1	&	$\sim$ $-$41	&	73$\pm$2	&	8$\pm$4	&	34$\pm$4	&	8$\pm$4	\\ \midrule
parallax\_mean		[mas] $^{\alpha}$	&	0.43$\pm$0.29	&	0.15$\pm$0.39	&	0.38$\pm$0.26	&	0.64$\pm$0.33	&	1.16$\pm$0.28	&	0.95$\pm$0.19	\\
distance from isochrone	[pc] $^{\beta}$	&	2023	&	5749	&	2512	&	1514	&	848	&	1000	\\
R\_core		[\arcmin] $^{\gamma}$	&	1.3	&	0.5	&	1.2	&	6.4	&	6.3	&	2.2	\\
R\_core 		[pc]	&	0.76	&	0.84	&	0.88	&	2.82	&	1.55	&	0.64	\\
\toprule															
\multicolumn{7}{c}{ $^{\alpha}$ The means and errors are mean and standard deviations of Gaussian fit to the \texttt{member} parameters.}	\\
\multicolumn{7}{c}{ $^{\beta}$ The errors in isochrone distance could be $\sim$10\% due to the degeneracy in extinction, mettalicity and age.} \\ 
\multicolumn{7}{c}{ $^{\gamma}$ Projected R\_core is calculated using distance from isochrone fits}
\end{tabular}
}
\end{table} 

\begin{landscape}
\begin{figure}
   \centering
    \includegraphics[height=0.8\textwidth]{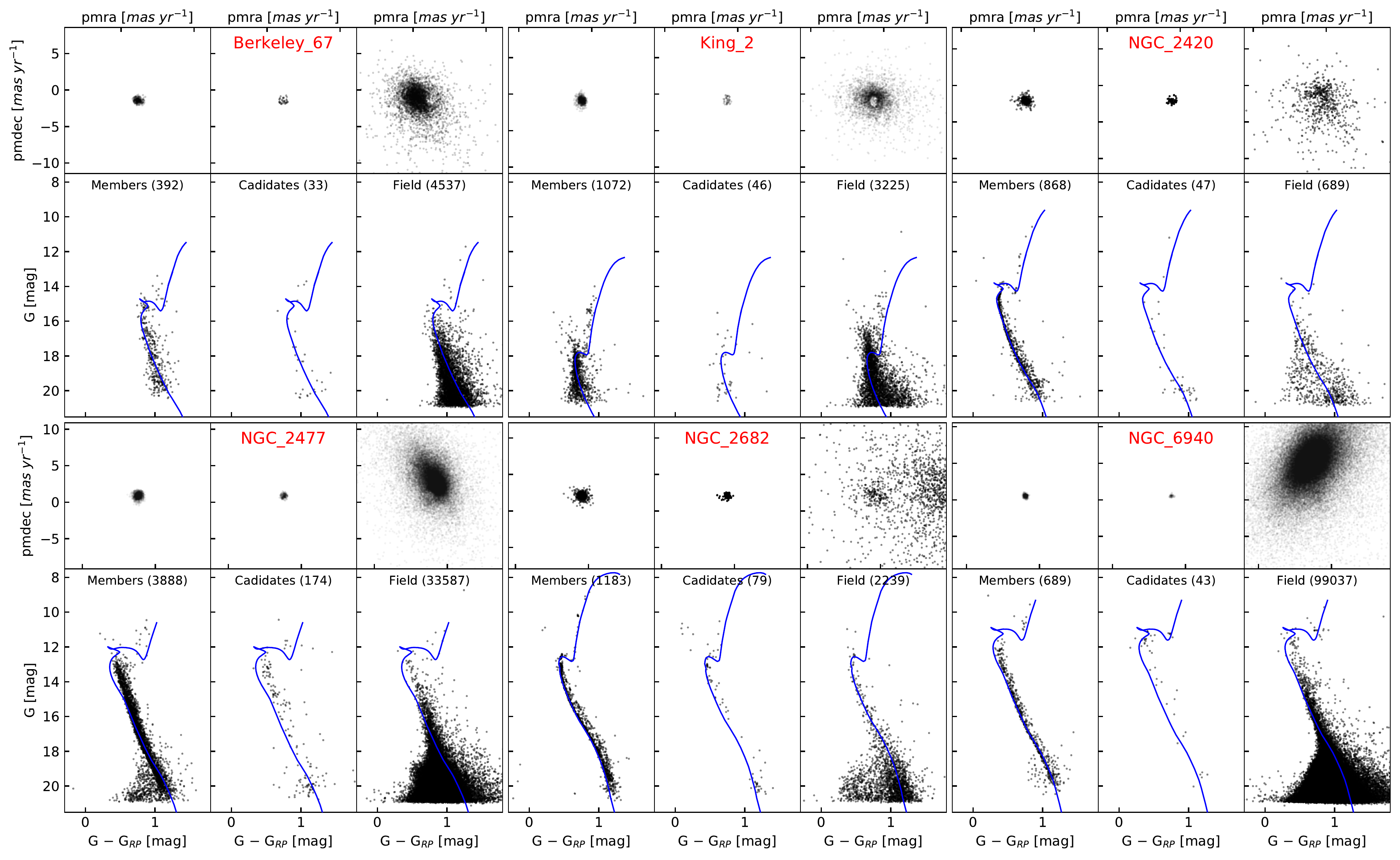}
    \caption{VPDs and CMDs of the six clusters. First and third rows are VPDs of \texttt{members, candidates} and \texttt{field} for respective clusters. 
    Second and fourth rows are CMDs of \texttt{members, candidates} and \texttt{field} for respective clusters.
    Isochrones are plotted as a blue line for reference.}
    \label{fig:3_CV_combined}
\end{figure}
\end{landscape}

\begin{figure}[!ht]
   \centering
    \includegraphics[width=0.87\textwidth]{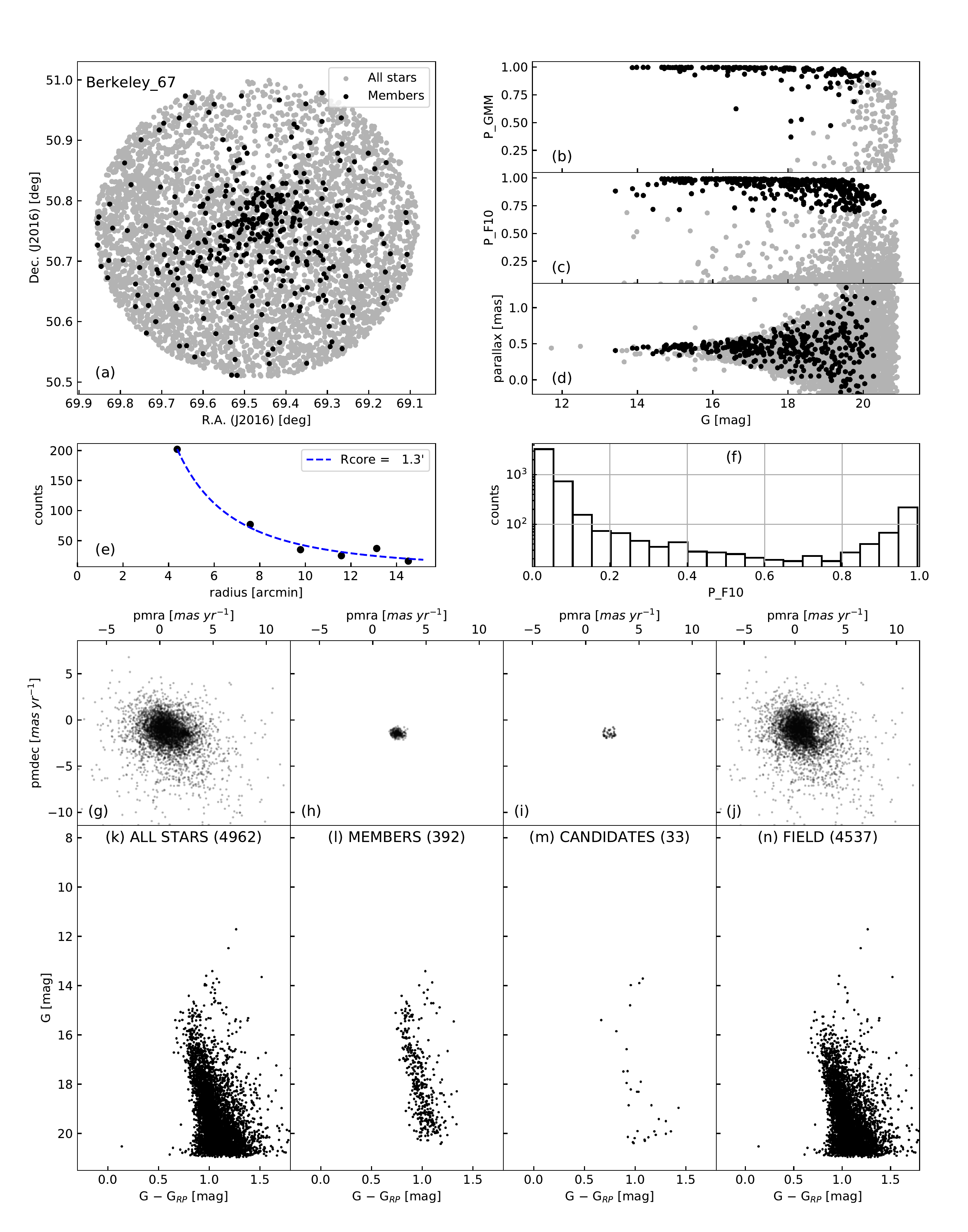}
    \caption{(a) Spatial distribution for Berkeley 67. All \textit{Gaia} stars are denoted by grey dots and cluster \texttt{members} are denoted by black points. (b)--(d) Distribution of {\sc g} magnitude with P\_GMM, P\_F10.3 and parallax. (e) Calculation of core radius by fitting radial density with King's surface density profile. (f) Histogram of P\_10.3. (g)--(j) and (k)--(n) show the VPD and CMD of all stars, \texttt{members, candidates} and \texttt{field} respectively. (k)--(n) show the number of stars plotted in the brackets along with the Padova isochrone as a blue curve.}
    \label{fig:3_rVC_be67}
\end{figure}

\begin{figure}[!ht]
   \centering
    \includegraphics[width=0.97\textwidth]{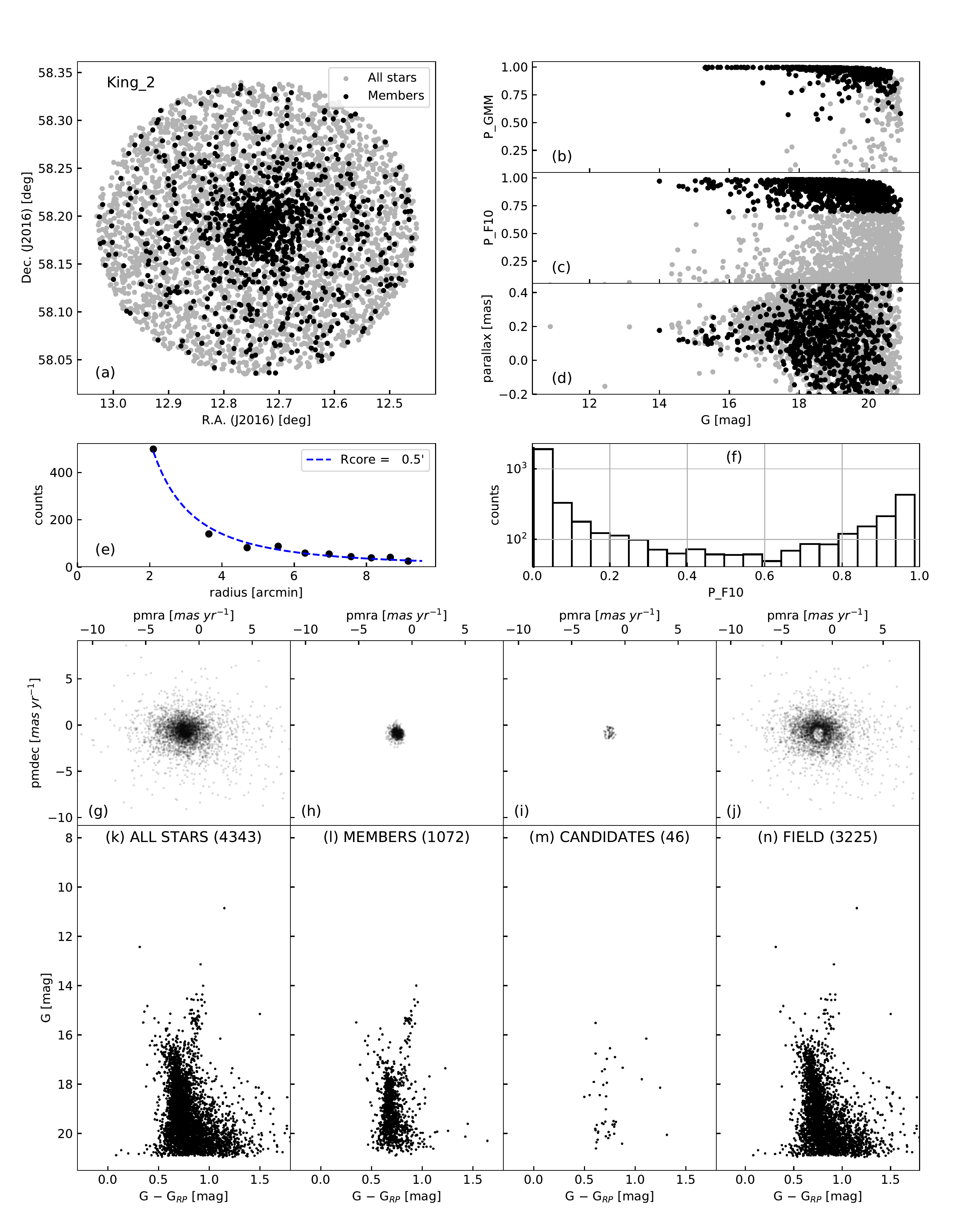}
    \caption{Spatial distribution, VPD and CMDs of King 2. The subplot descriptions are the same as Fig.~\ref{fig:3_rVC_be67}.}
    \label{fig:3_rVC_K2}
\end{figure}

\begin{figure}[!ht]
   \centering
    \includegraphics[width=0.97\textwidth]{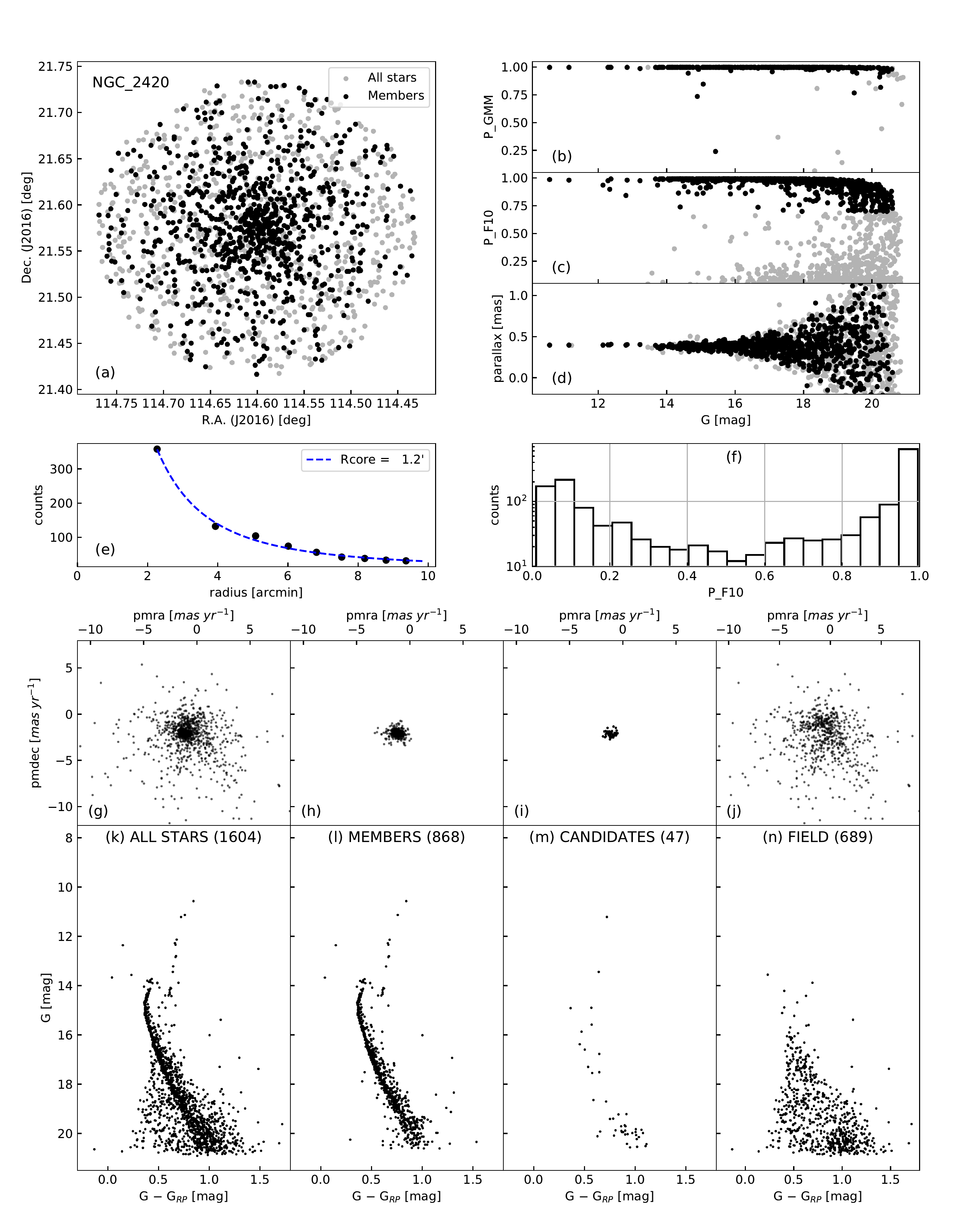}
    \caption{Spatial distribution, VPD and CMDs of NGC 2420. The subplot descriptions are the same as Fig.~\ref{fig:3_rVC_be67}.}
    \label{fig:3_rVC_2420}
\end{figure}

\begin{figure}[!ht]
   \centering
    \includegraphics[width=0.97\textwidth]{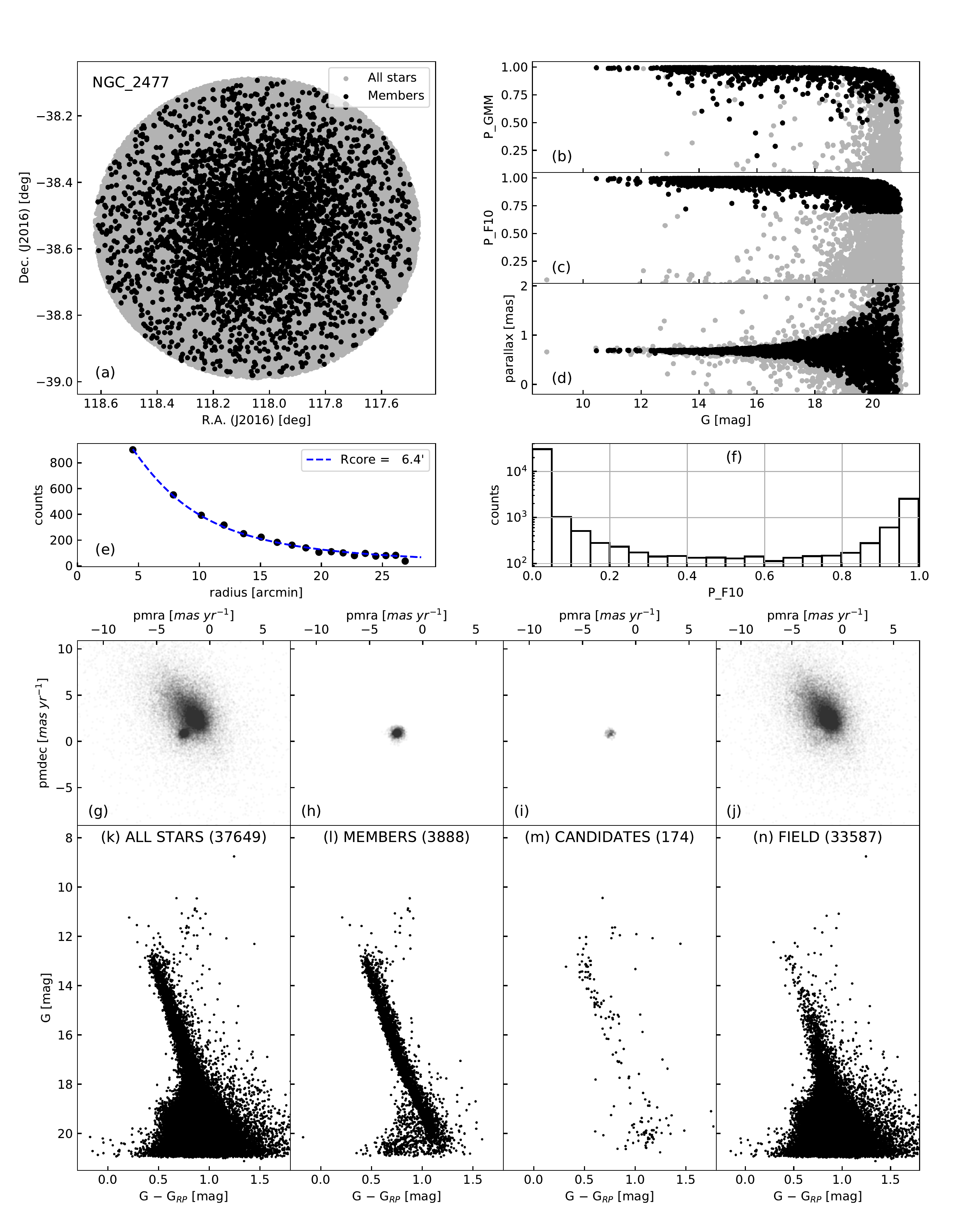}
    \caption{Spatial distribution, VPD and CMDs of NGC 2477. The subplot descriptions are the same as Fig.~\ref{fig:3_rVC_be67}.}
    \label{fig:3_rVC_2477}
\end{figure}

\begin{figure}[!ht]
   \centering
    \includegraphics[width=0.97\textwidth]{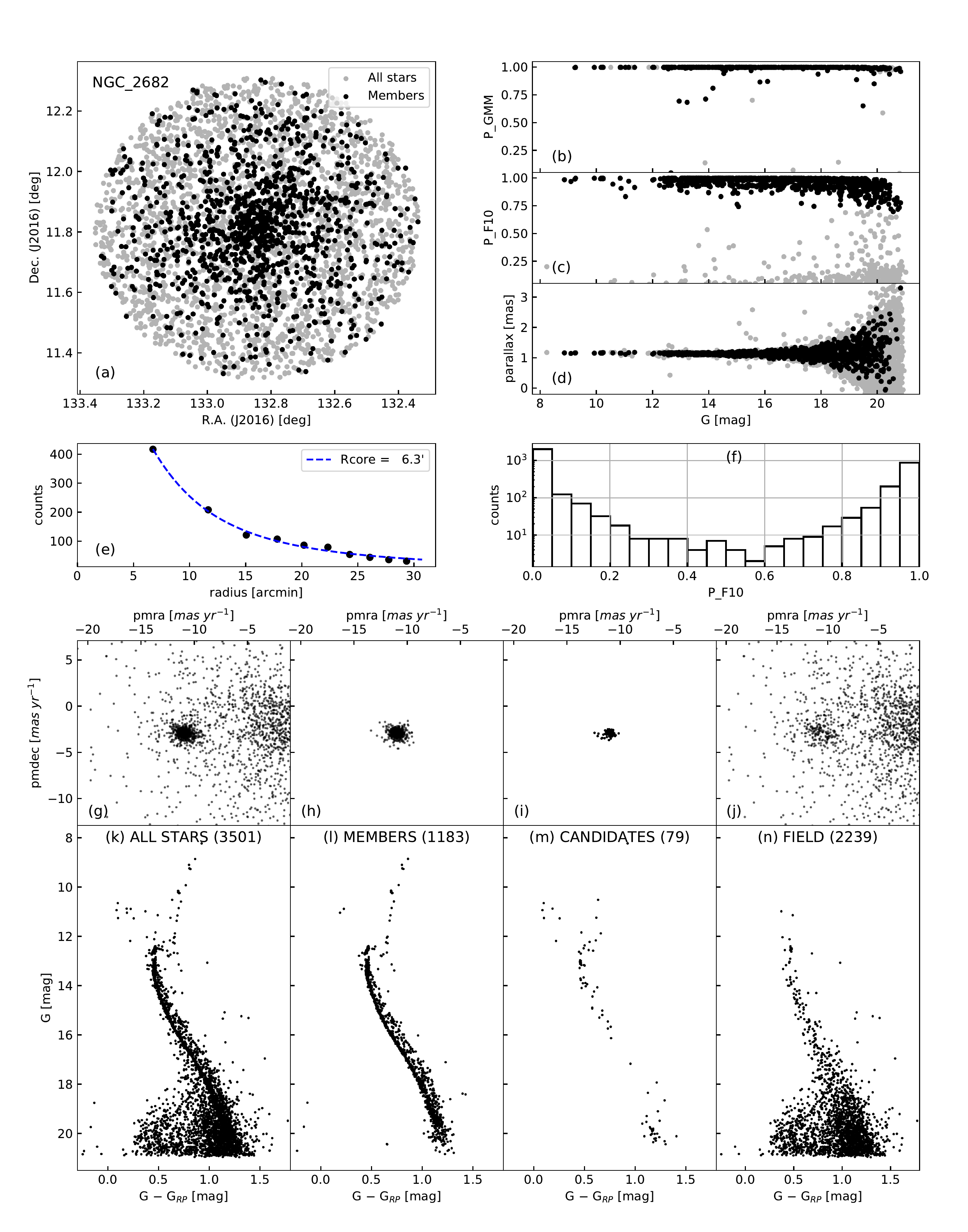}
    \caption{Spatial distribution, VPD and CMDs of NGC 2682. The subplot descriptions are the same as Fig.~\ref{fig:3_rVC_be67}.}
    \label{fig:3_rVC_2682}
\end{figure}

\begin{figure}[!ht]
   \centering
    \includegraphics[width=0.97\textwidth]{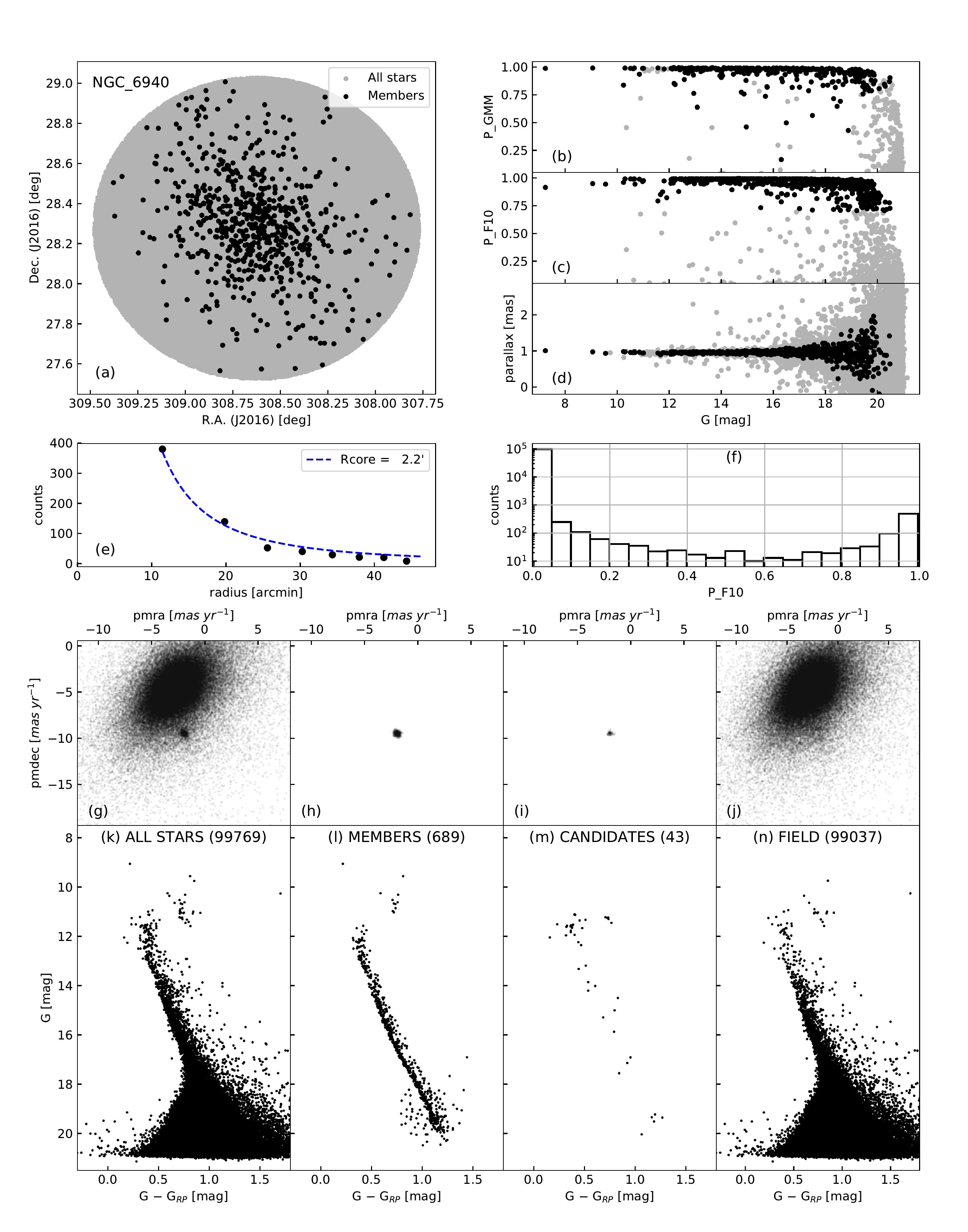}
    \caption{Spatial distribution, VPD and CMDs of NGC 6940. The subplot descriptions are the same as Fig.~\ref{fig:3_rVC_be67}.}
    \label{fig:3_rVC_6940}
\end{figure}

All the parameters are tabulated in Table \ref{tab:3_cluster_parameters}. \textit{Gaia} EDR3 sources near each cluster are divided into three subsets: \texttt{members, candidates} and \texttt{field}. The VPDs and CMDs of all clusters for these individual subsets are shown in Fig.~\ref{fig:3_CV_combined}.
The spatial distribution, VPD, CMD, {\sc g} vs MP etc. for clusters is shown in Fig.~\ref{fig:3_rVC_be67}--\ref{fig:3_rVC_6940}. 
For each cluster, Fig. (a) shows the spatial distributions of \texttt{members} and non-\texttt{member} population. 
Fig. (b) and (c) show the distribution of GMM and F10.3 probabilities as a function of \textsc{g}. We expected clear separation between \texttt{members} for bright stars, which is seen in all the clusters up to 16--18 mags. 
Fig. (d) shows the distribution parallax as a function of \textsc{g}. All clusters, except King 2, show a peak in parallax for the \texttt{member} stars. 
Fig. (e) shows the King's surface density profile fitted to \texttt{members}' surface density.
Fig. (f) shows the histogram of F10 MP.
Fig. (g) and (k) show the VPD and CMD for all stars.
Fig. (h) and (l) show the VPD and CMD for \texttt{members}.
Fig. (i) and (m) show the VPD and CMD for \texttt{candidates}. 
Fig. (j) and (n) show the VPD and CMD for \texttt{field} stars.


\begin{figure}
    \centering
    \begin{tabular}{c}
    \includegraphics[width=0.97\textwidth]{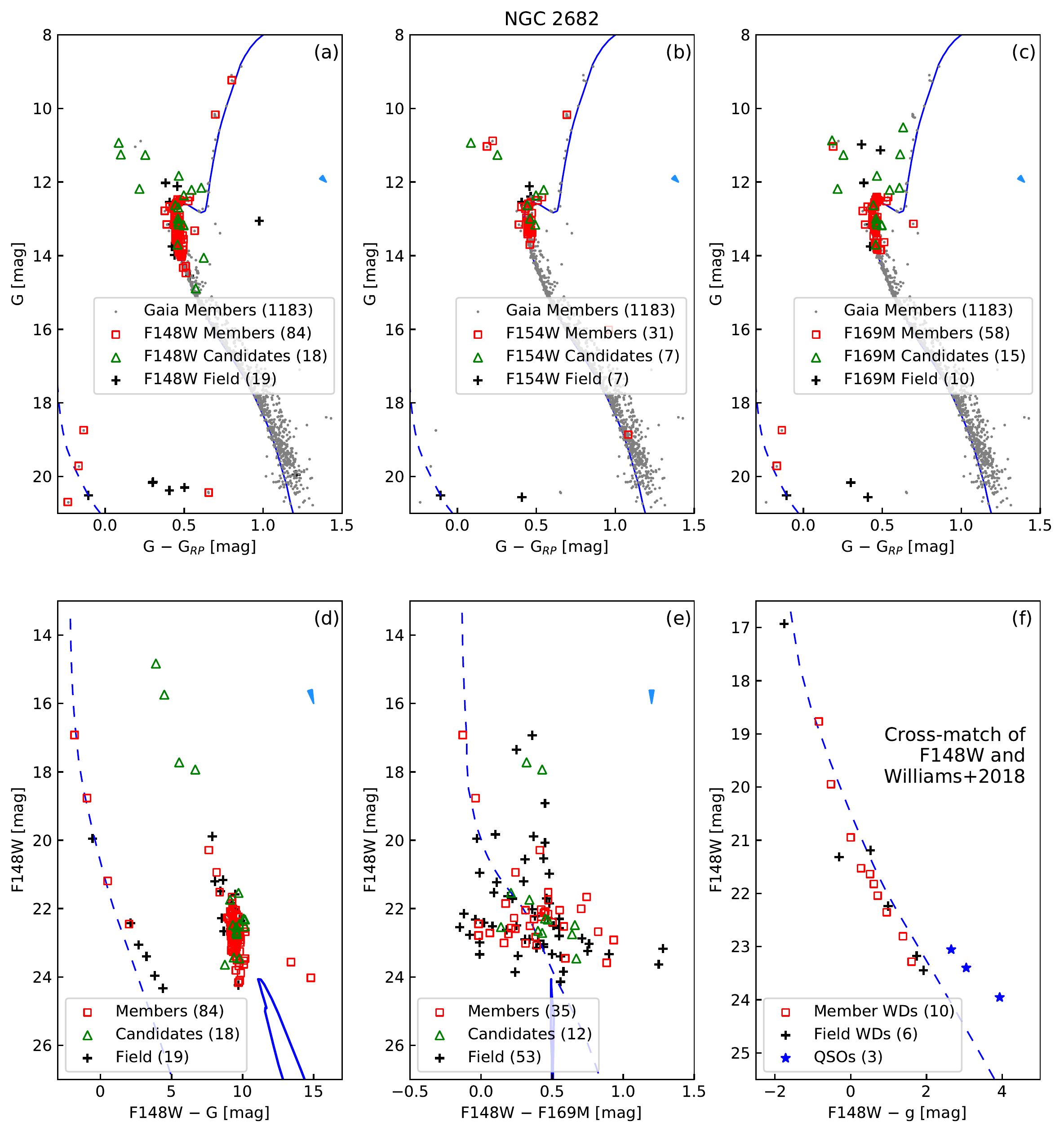}
     \end{tabular}
     \caption{The CMDs of NGC 2682 with UVIT and \textit{Gaia} photometry. Gray dots are \textit{Gaia} EDR3 \texttt{members} according to Eq.~\ref{eq:3_classif}. Red squares, green triangles and black crosses are \texttt{members}, \texttt{candidates} and \texttt{field} stars detected in particular filters.
     The blue line is an isochrone with $Log(Age)=9.6, Distance Modulus=9.64, E(B-V)=0.05\ mag, [M/H]=0\ dex$; the dashed blue line is a WD cooling curve for a 0.5 M$_{\odot}$ WD; the light blue arrows represent the reddening vectors with direction and magnitude in each CMD.
    {(a), (b) and (c)} The optical CMDs with stars detected in F148W, F154W and F169M filters respectively.
    {(d)} The UV--optical CMDs with F148W and {\sc g} filters. 
    {(e)} The UV CMD with F148W $-$ F169M colour.
    {(f)} UV--optical CMD of sources cross-matched with \citet{Williams2018} catalogue of possible WDs. The CMD also shows three quasars (blue stars) detected by F148W filter.
    }
     \label{fig:3_CMD_2682}
\end{figure} 

\begin{figure}
    \centering
    \begin{tabular}{c}
    \includegraphics[width=0.9\textwidth]{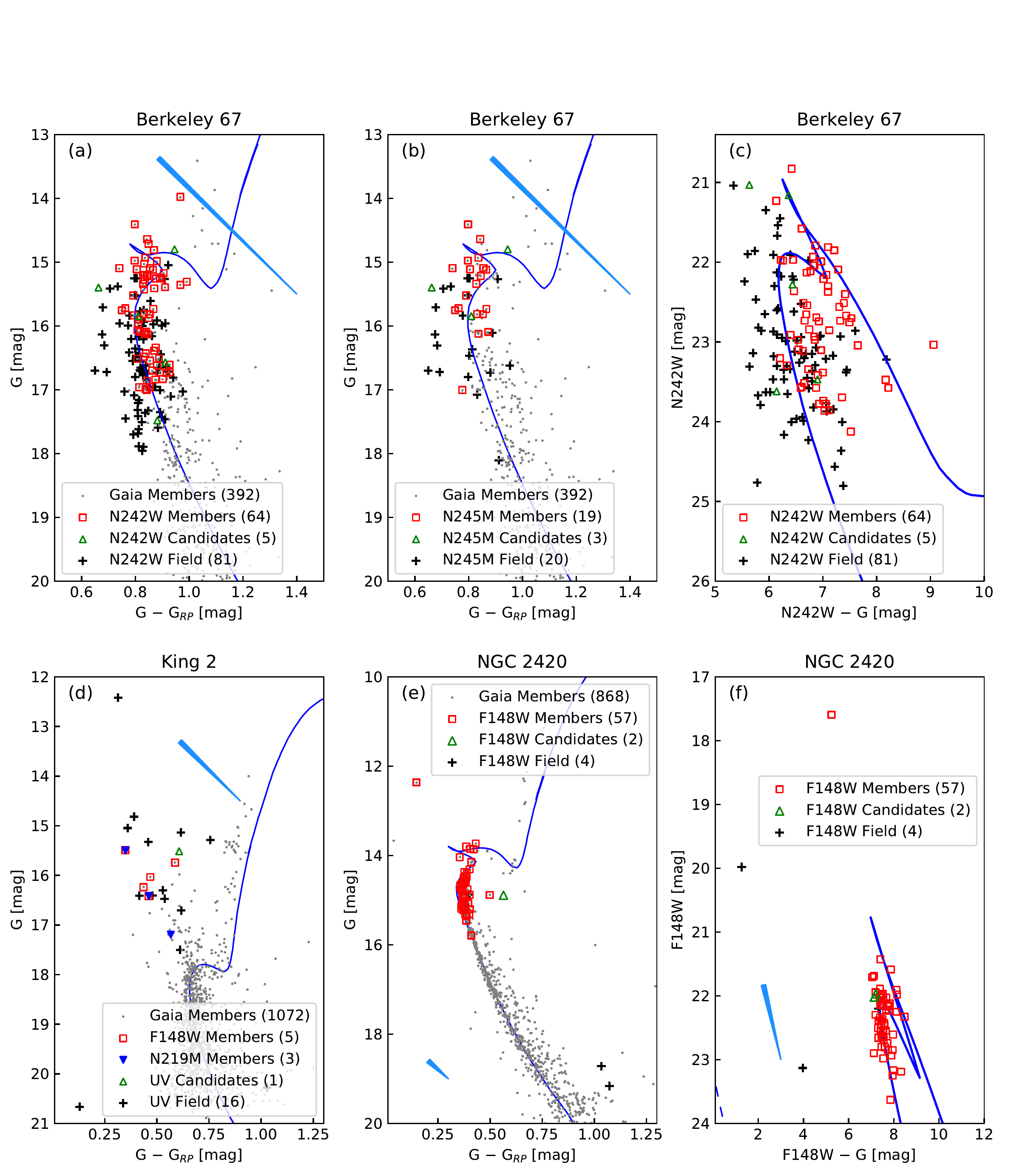} \\
    \end{tabular}
    \caption{{(a), (b) and (c)} The optical and optical--UV CMDs of Berkeley 67.
    The markers are the same as Fig.~\ref{fig:3_CMD_2682}. The isochrone for Berkeley 67 has $Log(Age)=9.2, Distance Modulus=11.53, E(B-V)=0.8\ mag, [M/H]=0\ dex$.
    {(d)} Optical CMD of King 2. with F148W detected \texttt{members} denoted by red squares and N219M detected \texttt{members} denoted by blue filled triangles. 
    The isochrone for King 2 has $Log(Age)=9.7, Distance Modulus=13.8, E(B-V)=0.45\ mag, [M/H]=-0.4\ dex$.
    {(e) and (f)} The optical and UV--optical CMDs of NGC 2420. 
    The markers are the same as Fig.~\ref{fig:3_CMD_2682}. The isochrone for NGC 2420 has $Log(Age)=9.3, Distance Modulus=12.0, E(B-V)=0.15\ mag, [M/H]=-0.4\ dex$; WD cooling curve is for 0.6 M$_{\odot}$ WD.}
    \label{fig:3_CMD_Be67}
\end{figure} 

\begin{figure}
    \centering
    \begin{tabular}{c}
    \includegraphics[width=0.97\textwidth]{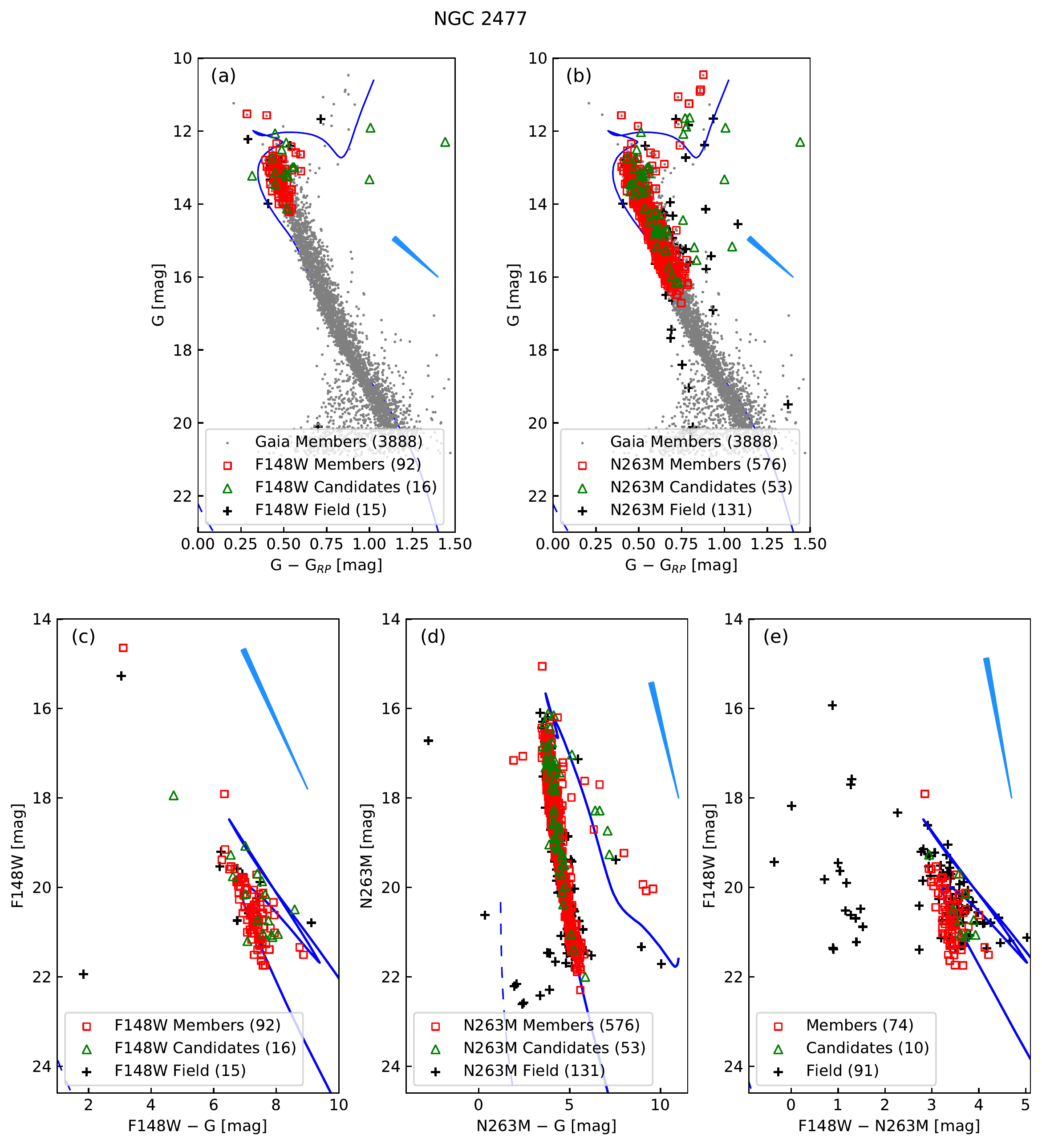} \\
    \end{tabular}
    \caption{{(a) and (b)} The optical CMDs of NGC 2477. 
    {(c), (d) and (e)} The UV--optical and UV CMDs of NGC 2477. 
    The markers are the same as Fig.~\ref{fig:3_CMD_2682}. The isochrone is for $Log(Age)=8.9, Distance Modulus=10.9, E(B-V)=0.4\ mag, [M/H]=0\ dex$; WD cooling curve is for a 0.7 M$_{\odot}$ WD.}
    \label{fig:3_CMD_2477}
\end{figure} 

\begin{figure}[!ht]
    \centering
    \begin{tabular}{c}
    \includegraphics[width=0.97\textwidth]{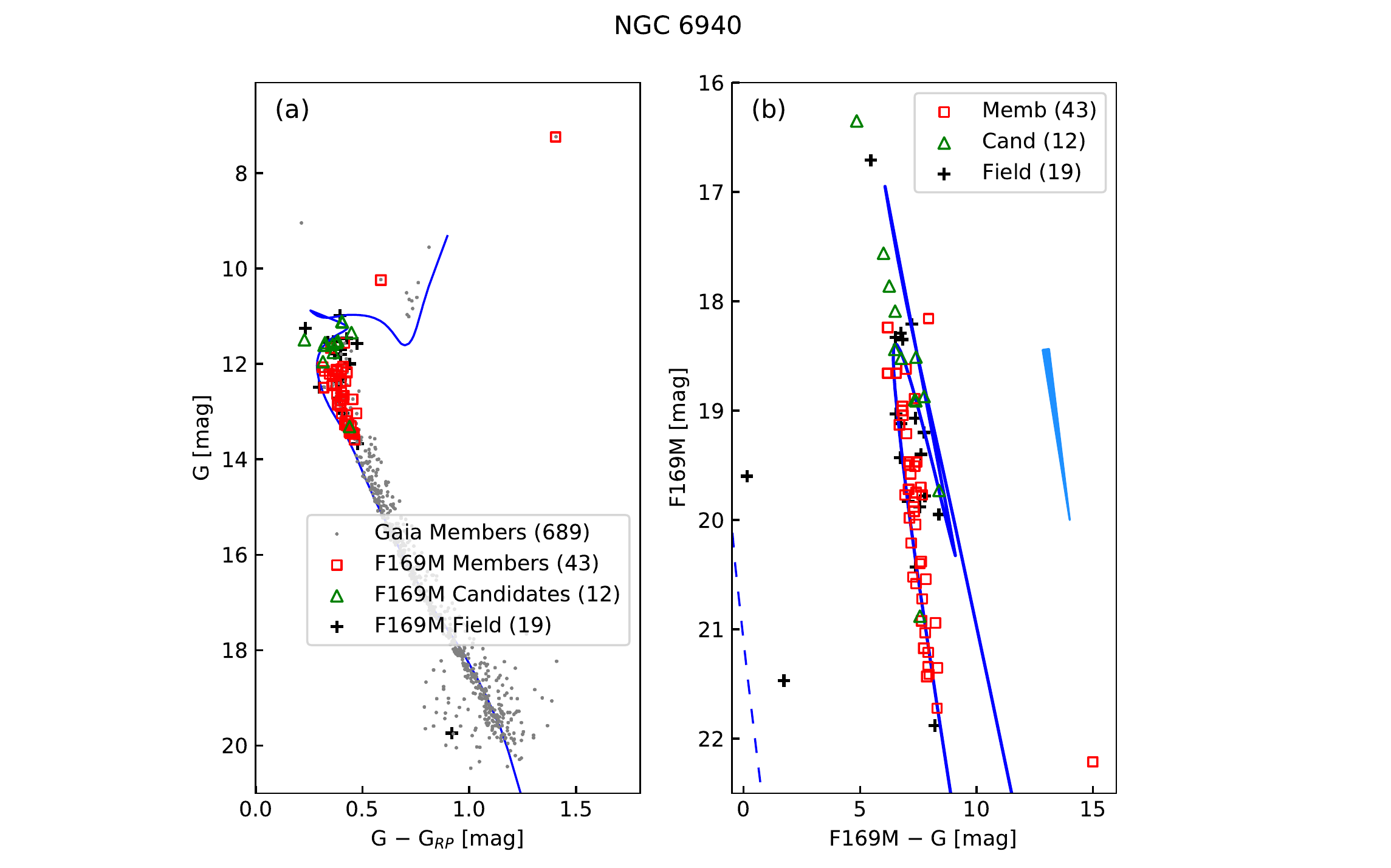} \\
    \end{tabular}
    \caption{{(a) and (b)} The optical and UV--optical CMDs of NGC 6940. 
     The markers are the same as Fig.~\ref{fig:3_CMD_2682}. The isochrone is for $Log(Age)=9.0, Distance Modulus=10.0, E(B-V)=0.2\  mag, [M/H]=0.0\ dex$; WD cooling curve is for a 0.6 M$_{\odot}$ WD.}
    \label{fig:3_CMD_6940}
\end{figure} 


\subsection{Individual clusters} \label{sec:3_2682}

\textbf{NGC 2682} has a prominent binary sequence, RGB and BSS population (Fig.~\ref{fig:3_CV_combined}). We detected many \texttt{candidates} as BSSs, MS stars and a few RG stars. The MS \texttt{candidates} typically have $\textsc{qf}=0$. 
The optical CMDs of stars detected in the FUV filters are shown in Fig.~\ref{fig:3_CMD_2682} (a), (b) and (c). The CMDs contain all \textit{Gaia} \texttt{members}, UVIT detected sources, isochrone and WD cooling curve of 0.5 M$_{\odot}$.
Overall, we detected 84, 31 and 58 \texttt{members} in F148W, F154W and F169M respectively. 
Fig.~\ref{fig:3_CMD_2682} (d) shows the UV--optical CMD of sources cross-matched between \textit{Gaia} EDR3 and F148W. The turn-off of the isochrone lies at 24 mag, which is the limiting magnitude of F148W observations. 
All the stars on optical MS are located above the turn-off in UV--optical CMD.
Hence, as previously seen in \citet{Jadhav2019ApJ...886...13J}, almost all MS stars detected in optical CMD have FUV excess. 
The photometry presented here is two magnitudes fainter than \citet{Jadhav2019ApJ...886...13J} in F148W, and we have detected a good number of MS stars in the 22--24 magnitude range, with FUV excess.
Fig.~\ref{fig:3_CMD_2682} (e) shows the F148W, (F148W $-$ F169M) CMD (the MS/turn-off is the vertical line at $\sim$0.5 colour).
The sources show a large spread in FUV colour, ranging from 0.0 to 1.3 mag.

FUV images can be used to detect WDs. However, as they are faint in the optical wavelengths, \textit{Gaia} is not suitable to detect them. To effectively identify the WDs, we cross-matched the F148W detected sources with the WD catalogue of \citet{Williams2018}. Fig.~\ref{fig:3_CMD_2682} (f) shows the CMD of all cross-matched sources and the WD cooling curve. The membership information and g-band photometry (for this subplot alone) are taken from \citet{Williams2018}. F148W has detected ten member WDs, six field WDs and three quasars. All these sources follow the WD cooling curve.

\textbf{Berkeley 67}'s VPD (Fig.~\ref{fig:3_CV_combined}) shows that the mean cluster motion and mean field motion are within a few mas yr$^{-1}$ of each other.
The cluster has only $\sim$400 \texttt{members}; hence there is not much over-density in the VPD. Thus, it is particularly challenging to determine the membership for Berkeley 67. The CMD shows a large spread in the MS. We suspect the large spread is the result of the differential reddening in the cluster region, whose effect is enhanced by the high extinction towards the cluster. 
UVIT images of Berkeley 67 in F148W (exp. time = 2683 s) and F169M (exp. time = 1317s) filters detected no \texttt{member} stars. Therefore, the cluster does not have any FUV bright members.
N242W and N245M images detected 64 and 19 \texttt{members} respectively. There are no BSSs in Berkeley 67.
Fig.~\ref{fig:3_CMD_Be67} (a) and (b) show the \textit{Gaia} CMDs of these NUV detected \texttt{members}. We observe turn-off stars in both NUV filters, with the wider N242W filter going till 17 mag.
N242W filter detects two RGs which are rarely detected in the UV regime.
Fig.~\ref{fig:3_CMD_Be67} (c) shows the N242W, (N242W $-$ {\sc g}) CMD for Berkeley 67, which again confirms the large scatter.

\textbf{King 2} is the farthest cluster included in this work. This is evident from the small apparent core radius (0.$^\prime$5) and high feature-importance for {\sc ra} and {\sc dec}. 
The parallax measurements are unreliable at the distance of $\sim$5 kpc. The cluster and field centres in the VPD are very close, and hence, there may be contamination from field stars among the members. 
Nevertheless, many BSSs and red clump stars are present in the cluster (Fig.~\ref{fig:3_CV_combined}).
We detected 5 and 3 \texttt{members} each in F148W and N219M filter respectively (Fig.~\ref{fig:3_CMD_Be67} d). This 5 Gyr old cluster located at a large distance has the MSTO at 18 mag (in {\sc g}-band), and hence we detected only the brightest of the BSSs in UV. Overall, there are 5 \texttt{member} BSSs and 1 \texttt{candidate} BSS in UV images. Two of the BSSs have both FUV and NUV detections. We detect one blue and faint ({\sc g}-band) object, which is likely a foreground WDs.

\textbf{NGC 2420} lies in a relatively less dense region with the field to member stars ratio of $\sim$1 (Fig.~\ref{fig:3_CV_combined}). It has a clearly defined binary sequence and an RGB, with a few BSSs.
The UVIT image has stars up to two magnitudes below the MSTO, including a BSS. The UV CMD shows that the \texttt{candidates} are located close to the turn-off, though they are much fainter in the optical CMD, suggesting a brightening in the UV. However, the excess UV flux is not as prominent as NGC 2682. Similarly, a few \texttt{members} also show UV excess flux, but the change in the magnitude is not as large as in NGC 2682.

\textbf{NGC 2477} is a very dense cluster in a high stellar density region (38000 stars and 3900 \texttt{members} in 18 {\arcmin} radius). The \textit{Gaia} CMD shows a well-defined binary sequence, RC stars and $\sim$5 BSSs. The turn-off has a large spread in {\sc g}-band. There is spread in the \texttt{members} below 18 mag, indicating that the probability cutoff should be slightly higher than 0.7.
We detected 92 and 576 \texttt{members} in F148W and F263M respectively (Fig.~\ref{fig:3_CMD_2477} a and b). Fig.~\ref{fig:3_CMD_2477} (c) shows the F148W, (F148W $-$ {\sc g}) CMD with 2 BSSs and MSTO stars.
Fig.~\ref{fig:3_CMD_2477} (d) shows the N263M, (N263 $-$ {\sc g}) CMD with a large range in NUV magnitude consisting of red clump stars and MS stars.
Fig.~\ref{fig:3_CMD_2477} (e) shows the F148W, (F148W $-$ N263M) CMD, here we see the turn-off stars and many \texttt{field} stars with bluer UV colour. 

\textbf{NGC 6940} is situated in a very dense stellar environment (stellar density is 13 times that of NGC 2682 neighbourhood). The cluster is well separated from the central field in the VPD and has a significant parallax ($\sim$1 mas); hence it is easy to extract (Fig.~\ref{fig:3_CV_combined}). The CMD shows a clear binary sequence and red clump stars. NGC 6940 also has a relatively broad MS with a spread near its turn-off, although less prominent than NGC 2477.
We detected \texttt{members} up to 2 magnitudes below turn-off (Fig.~\ref{fig:3_CMD_6940} a) including a giant. We did not detect any BSS in this 1 Gyr old cluster.


\section{Discussion} \label{sec:3_discussion}
\subsection{Membership determination} \label{sec:3_membership_discussion}

There are multiple ways of determining memberships. 
The aim of the study mainly dictates the choice of method. 
Any simple method such as VPDs for membership estimation is adequate for studies requiring the computation of cluster parameters such as mean PM, age and distance. 
Here, our objective is to identify UV bright member stars in OCs, that could be in non-standard evolutionary stages. This requires the implementation of rigorous methods to assign membership to such stars, as discussed below:

\begin{figure}[!ht] 
\centering
\begin{tabular}{c}
\includegraphics[width=0.95\textwidth]{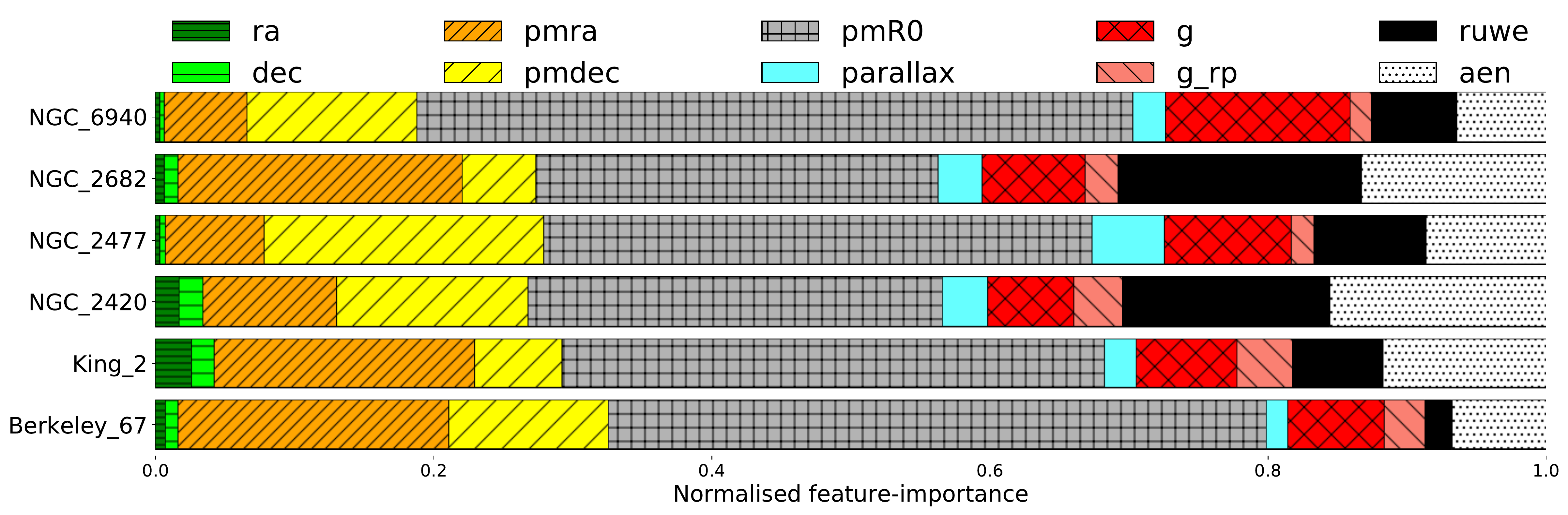}
    \end{tabular}
    \caption{Normalised feature-importance for F10 for all six OCs.}
    \label{fig:3_feature_importance}
\end{figure}

\textbf{Feature-importance:} 
The {\sc prf} method gives the importance of each feature as one of the results during the training phase. The normalised feature-importance is shown in  Fig.~\ref{fig:3_feature_importance}.
The most important features are {\sc pmR0, ruwe, aen, pmra, pmdec} and {\sc g}.
The distance of the stars from the cluster centre in the VPD, {\sc pmR0}, is an important feature, as expected.
The high importance of {\sc ruwe/aen} is because they are cutoffs established during the training of the algorithm.
The {\sc g} importance is aided by the dependence of all errors on the magnitude. 
The {\sc ra/dec} importance is more for King 2 and NGC 2420 when compared to other clusters. They have core radii of 0.$^\prime$5--1.$^\prime$2, while other clusters are typically larger. The smaller spatial distribution of members is causing an increase in the importance of {\sc ra/dec}. As expected, the importance of {\sc parallax} increases for nearby clusters (NGC 2420, NGC 2477, NGC 2682 and NGC 6940). 

\textbf{Efficacy in various environments:}
The {\sc prf} technique works for OCs with diverse cluster-members to field-stars ratio (0.01--1.3), thereby helping in efficient detection of members. 
The presence of systematic errors and including CMD locations through magnitude and colour tends to remove poor quality as well as \textit{peculiar} stars. 
Therefore, we introduced the \texttt{candidate} classification to list such stars.
For these six clusters, we found the \texttt{candidates} to \texttt{members} ratio to be 0.04--0.08. 

\textbf{Versatility of the technique:}
The algorithm is adaptable, and one can choose a particular feature-combination depending on the requirements. For example, for a statistical study of clusters, a feature combination with \textsc{ra, dec} and \textsc{parallax} would be enough. To find \textit{peculiar} stars in the CMD, one could measure the difference between F6 ({\sc ra/dec, pmra/pmdec, parallax} and {\sc pmR0}) and F8 (F6 + {\sc g} and {\sc g\_rp}).
\textit{Peculiar} stars typically have lower P\_F8.
The {\sc prf} technique can also be applied to any data-set besides \textit{Gaia} EDR3. Moreover, the inclusion of {\sc ruwe/aen} as features indicates any systematic terms, if present in any other data-set, can also be incorporated in the algorithm.

\textbf{Classification of BSSs:}
In the field of NGC 2682, there were 10 potential BSSs (bluer and brighter than the turn-off). {\sc prf} classified two as \texttt{members}, six as \texttt{candidates} and two as \texttt{field}. Many of these stars are photometric variables or binaries \citep{Geller2015}, which can lead to high $\textsc{ruwe}$ and hence classification as \texttt{candidates}. The two field stars have cluster parallax and RV (not considered as a membership criterion in {\sc prf}). However, they have larger {\sc pmR0} leading to their rejection as members. For such stars with large PM deviation from the cluster mean, deeper RV measurements and accurate parallax will be useful in constraining membership.
High \textit{peculiarity} in combination with high {\sc ruwe} of the BSSs is the reason for these stars to be categorised as \texttt{candidates}. 
Hence, the technique (Eq~\ref{eq:3_classif}) is capable of selecting BSSs (albeit as \texttt{candidates}).

\textbf{Existence of the \texttt{Candidate} class:}
All the cluster \texttt{candidates} lie near the cluster centre in the VPDs. Their number increases as they get fainter; this mirrors the fact that the systematic errors in \textit{Gaia} EDR3 are larger for fainter stars. The CMD of NGC 6940 \texttt{candidates} (Fig.~\ref{fig:3_CV_combined}) shows that the majority of them lie on the binary sequence. A similar but lesser effect is seen in NGC 2477 and NGC 2682. Binary systems are known to produce high {\sc ruwe} values due to variability or unsymmetrical PSF \citep{Deacon2020}, hence they can have low P\_F10 and get classified as \texttt{candidates}. 

\textbf{Detection of \textit{peculiar} stars using multiple feature-combinations:}
In Fig.~\ref{fig:3_P_comparison_internal} (e), we compared the MPs with and without \textsc{g} and \textsc{g\_rp} as features (F8 and F6 respectively).
F6 has no knowledge of CMD positions, so it uses only spatial location and velocity to classify stars. However, F8 selects stars with common CMD positions and rejects stars with uncommon CMD positions.
This effect is demonstrated by the positive values of $P\_F6-P\_F8$ for BSSs in NGC 2682 (Fig.~\ref{fig:3_P_comparison_internal} (d)). 
We refer to large $P\_F6-P\_F8$ as \textit{peculiarity}.
Such \textit{peculiarity} can be seen for the BSSs in NGC 2682, NGC 2477 and King 2. However, for King 2 majority of stars bluer than {\sc bp\_rp} $<$ 1.1 mag have similar \textit{peculiarity} regardless of their magnitude. Other clusters in this study do not have many BSSs, and they only showed large $P\_F6 - P\_F8$ near limiting magnitude.

\subsection{Individual clusters} \label{sec:3_cluster_discussion}
\textbf{NGC 2682:}
The detection of stars on the MS in the UV CMDs, suggests that many MS stars have excess UV flux. \citet{Jadhav2019ApJ...886...13J} presented the reasons for UV excess, such as the presence of hot WD components, chromospheric activity and hot spots on contact binaries. Such UV excess detected among the MS stars is unique to NGC 2682; as for all other clusters, only a few stars on the MS show UV excess.
In Fig.~\ref{fig:3_CMD_2682} (a) and (d), a few WD \texttt{members} and many \texttt{field} stars are found near the WD cooling curve. The CMD location suggests these can be WDs.
Their MPs are low due to astrometric and photometric errors.
As NGC 2682 is a well studied OC, we used a deep photometric catalogue of \citet{Williams2018}. They identified hot and faint stars in u,(u $-$ g) plane and carried out spectroscopic observations to confirm the WDs from their atmospheric signatures.
We cross-matched all F148W detections with the \citet{Williams2018} catalogue and found ten member WDs and six field WDs, along with the three quasars. Therefore, UVIT observations are well suited to detect WDs in NGC 2682.

\begin{figure} 
\centering
\begin{tabular}{c}
\includegraphics[width=0.6\textwidth]{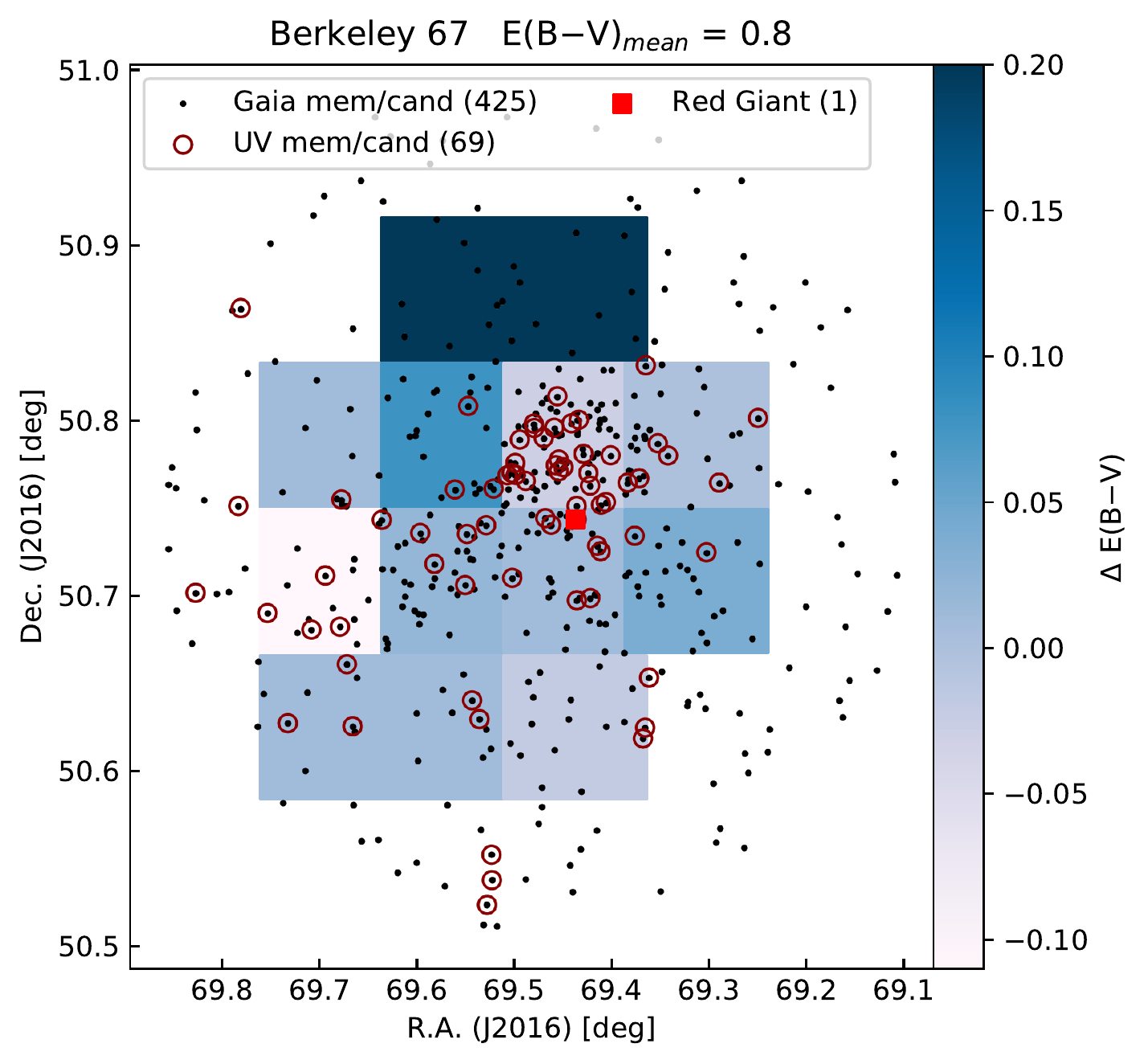}
    \end{tabular}
    \caption{Spatial location of RGs in Berkeley 67 over-plotted on the reddening map. The plot contains all \textit{Gaia} \texttt{members/candidates} (as black dots), N242W \texttt{members/candidates} (as red circles) and the RG detected in N242W filter (as red square).}
    \label{fig:3_be67_red}
\end{figure}

\textbf{Berkeley 67:}
The cluster has very high reddening, E(B $-$ V) = 0.8 mag. Thus, small relative changes in reddening have a substantial impact on magnitude/colour and cause a broadening of MS in the CMD.  
We tried binning the \textit{Gaia} \texttt{members} spatially and analysed the distributions in the CMD plane. Initial estimates suggested that the E(B $-$ V) values range from 0.7--1.0 mag. The reddening map is shown in Fig.~\ref{fig:3_be67_red}. We detected one RG \texttt{member} in NUV, which lies in the low reddening region. Further investigation is needed to determine the exact cause of UV brightening of the RG.
Fig.~\ref{fig:3_CMD_Be67} (c) shows \texttt{members} distributed in MS and sub-giant branch.
The UV CMD indicates that the spread at the optical turn-off can be due to subgiant stars or differential reddening. As the extinction vector is parallel to the subgiant branch, any differential reddening will increase the spread in the same direction.

\textbf{King 2:}
As the oldest and farthest cluster in this work, only bright BSSs are detected by UVIT. Chapter \ref{ch:UOCS4} presents the detailed analysis of the detected BSS population, including detection of EHB/sdB type stars as companions to BSSs.

\textbf{NGC 2420:}
We detected the BSS present in the cluster in the UV.
We found $\sim$6 stars (out of 59), located on the MS, to show signs of excess UV flux. Some stars are found at the turn-off, and three are \texttt{candidates}. 
One of the \texttt{candidate} has {\sc ruwe} = 3.9 while the other has no \textit{Gaia} colours. High {\sc ruwe} is known to be caused by variability and/or binarity \citep{Deacon2020}. The missing colour/high {\sc ruwe} and excess UV flux points towards a hotter companion or variability. Multi-wavelength analysis and X-ray observations of these stars can shed light on their evolutionary status.

\textbf{NGC 2477:}
We have detected a large number of stars in FUV (108) and NUV (629); hence it is ideal for studying the UV properties from the MS up to the red clump. This massive cluster is also ideal for exploring the UV properties of stars in the broad MSTO present in this cluster.
Overall the UV CMDs are aligned with the UV isochrones, not indicating a collective UV brightening like NGC 2682 among the MS stars. However, a couple of stars show considerable UV excess, which require a multi-wavelength study.

\textbf{NGC 6940:}
The UV CMD has some field stars near the WD cooling curve. These can be members or runaway WDs, which are quite faint in {\sc g}.
Some turn-off stars are brighter than the turn-off in the UV CMD, suggesting excess FUV flux.
We detected a giant with \textsc{g = 7.3} mag and \textsc{bp\_rp = 3.7} mag in FUV at the limiting magnitude. It is a variable star of spectral type M5II-III D \citep{Wallerstein1962}. The isochrone suggests that this is likely to be a post-AGB star. Due to the low temperature (T$_{eff}^{Gaia}$ = 3355 K), the stellar continuum cannot emit detectable FUV flux. Further study is needed to characterise the emission mechanism.

\subsection{General discussion for all clusters} \label{OC_with_Gaia_UVIT}
\textbf{WD detections:}
Cross-match of \citet{Williams2018} catalogue and \textit{Gaia} EDR3 resulted in only 4 WDs in NGC 2682. While the cross-match with F148W resulted in the detection of 16 WDs. The F148W image has detected stars up to 21.7 mag in g-band and 21.8 mag in u-band. This indicates that UV images are more suitable to detect hot WDs as compared to \textit{Gaia}.
CMDs of a few other clusters imply presence of photometric WDs: King 2 (Fig.~\ref{fig:3_CMD_Be67} d), NGC 2477 (Fig.~\ref{fig:3_CMD_2477} d), NGC 6940 (Fig.~\ref{fig:3_CMD_6940} b).
However, a comparison with deeper catalogues is required to detect WDs in these clusters. The membership determination of WDs is challenging due to the deficiency of long-baseline deep observations in other OCs.
Although all the WDs have significant PM errors (\textit{Gaia} EDR3 and \citealt{Yadav2008}), the spread in PM is clearly visible. It is interesting to note that three of the \textit{Gaia} detected WDs in NGC 2682 lie at/just outside the edge of the cluster in the VPD.  

\textbf{Comparison of \textit{Gaia} DR2 and EDR3:}
The membership analysis was done for both \textit{Gaia} DR2 and EDR3. As EDR3 has halved the errors in PM, there were some changes in the members. EDR3 data has led to the addition of sources in the fainter end, which have PM similar to the cluster. We could probe the membership of all the stars in the cluster without any magnitude cutoff due to better accuracy in PM and inclusion of errors in the MP determination. The total percentage of \texttt{candidates} has dropped from 11\%--5\%, and the VPD distribution of NGC 2682 members was elliptical in DR2, which is now circular in EDR3 data reflecting better handling of systematics in EDR3.

\textbf{Future Improvements:}
There are scopes to improve the membership determination process in future based on the following points:
\begin{enumerate}[leftmargin=*]
    \item The use of RV can constrain the spatial motion of members; however, deeper RV data is needed to get cluster membership for fainter stars.
    \item The use of distance from the fiducial isochrone in the CMD could constrain the spread visible in the fainter region of the CMDs. 
    \item We miss some stars with slightly different space velocity, as the primary selection criterion is PM (e.g., 2 PM and RV members from \citealt{Geller2015}). Such stars are essential to understanding the kinematics of the cluster. Increasing the weightage and accuracy of parallax and CMD location can help identify such stars.
\end{enumerate}

The method developed here is generic and can be applied to non \textit{Gaia} data as well. We recommend a feature-combination similar to F8 (\textsc{ra, dec, pmra, pmdec, pmR0, parallax}/distance, photometric information) to constrain the spread in CMD and VPD. Additionally, comparison with equivalent F6 (Only astrometric information) will be helpful to identify \textit{peculiar} stars in the CMD.

\section{Conclusions and summary} \label{sec:3_conclusions}
\begin{enumerate}[leftmargin=*]
    \item We developed an machine learning-based method to determine the individual stellar membership within OCs using \textit{Gaia} EDR3. We have tried more than 22 different feature-combinations to calculate the MPs. The stars are classified as \texttt{members, candidates} and \texttt{field} using a combination of two {\sc prf} methods. Our primary method (F10) identifies stars that have properties similar to the mean cluster properties and have small systematic errors as \texttt{members}. To incorporate \textit{peculiar} stars (uncommon CMD locations) and stars with large systematic errors, we utilised another method (F6) which only uses spatial location and velocity coordinates. We compared and validated the performance of our methods with past membership studies. Additionally, we created a technique to identify stars with \textit{peculiar} CMD position and demonstrated that it could identify BSSs.
    \item We demonstrated that the {\sc prf} algorithm could be used to determine the MPs in a variety of clusters. It is found to be robust, reproducible, versatile, and efficient in various environments (such as variation in stellar density, reddening, age). We have identified 200--2500 more cluster \texttt{members}, primarily in the fainter MS, compared to previous studies (which used \textit{Gaia} DR2 data). The algorithm presented here is generic and could be changed to suit other data sets or scientific problems. It is editable by selecting different features or creating new features, as required.
    \item We present a catalogue (\textit{Gaia} EDR3 based) of six clusters which provides spatial location, MPs, and classification in Table~\ref{tab:3_cat_Gaia}. The presence of \texttt{candidate} stars suggests a need for better astrometry and photometry, which will be available in future \textit{Gaia} releases and other large scale surveys. We used the \textit{Gaia} catalogue to identify cluster members in UVIT images. We present the UVIT catalogue of six OCs in one or more filters along with its membership information in Table~\ref{tab:3_cat_UV} (full catalogues are available online). We estimated cluster properties such as mean PM, distance, mean RV and core radii from the identified \texttt{member} population. 
    \item We detected 3--700 \texttt{member} stars in various UVIT images of six clusters, apart from $\sim$13\% \texttt{candidates}. We detected BSSs in King 2, NGC 2477, NGC 2420 and NGC 2682. FUV photometry presented here will be used to understand the formation pathways of BSSs. We also detected giant \texttt{members} in FUV (NGC 2682, NGC 6940) and NUV (Berkeley 67, NGC 2477). While most of the NUV detections are expected due to their luminosity and temperature, their FUV detections are unusual. We detected 10 WD members in FUV images of NGC 2682. UV CMDs indicate a few possible WDs in NGC 2477, NGC 2682 and NGC 6940.
    \item As seen in earlier studies, NGC 2682 has unusually high UV bright MS members. We detect no such systematic UV brightening among MS stars in other clusters. Some individual stars do show excess UV flux (RGs, a post-AGB star and a few MS stars). These are good contenders for detailed individual studies. The VPD of NGC 2682 is also notable due to its elliptical shape.
    \item The massive cluster NGC 2477 has 92/576 \texttt{members} detected in FUV/NUV, which will be helpful to study the UV properties of stars in the extended turn-off and various evolutionary stages from MS to red clump.
\end{enumerate}

\begin{savequote}[100mm]
Somewhere, something incredible is waiting to be known
\qauthor{Carl Sagan}
\end{savequote}

\chapter[Extremely Low Mass White Dwarfs and Post--Mass Transfer Binaries in NGC 2682
]{Extremely Low Mass White Dwarfs and Post--Mass Transfer Binaries in NGC 2682 \\ \large{\textcolor{gray}{Jadhav et al., 2019, ApJ, 886, 13}}}
\label{ch:UOCS2}
\begin{quote}\small
\end{quote}

\section{Introduction} \label{sec:4_intro}
The evolution of stars in close binaries and that of multiple stellar systems within star clusters will be different owing to the interactions with their neighbours. The high stellar density in globular clusters causes collisions leading to mergers, creation, and disruption of binary systems. OCs, on the other hand, provide ample examples of binaries that have more chance of remaining relatively undisturbed due to the low stellar densities.

The evolution of a binary star occurs in multiple pathways in OCs, as it depends on their orbital parameters. Very long period binaries are likely to evolve independent of each other, while closer binaries may merge or undergo MT \citep{Perets2015}. Among contact binaries, W UMa type binaries evolve into a contact configuration from initially detached systems by angular momentum loss via magnetic torques from a stellar wind in which the spin angular momentum and the orbital angular momentum are coupled through tides \citep{Vilhu1982, Guinan1988, Eggen1989}. Estimates based on the level of chromospheric and coronal activity exhibited by components of short-period MS binaries suggest that systems with initial orbital periods of a few days may evolve into a contact configuration on a timescale of a few giga years. W UMa systems ultimately coalesce into single stars \citep{Webbink1985}, which provide a natural pathway for the formation of BSSs.  

The stars formed from the binary evolution in OCs are generally detected in UV and X-rays \citep{Eaton1980, Geske2005}. On the other hand, young WDs emit in the UV region owing to their high temperature \citep{Sindhu2018}. Hot spots in contact and semi-detached binaries also show enhanced UV flux. \citet{Kouzuma2019} gave examples of stellar hot spots in contact binaries showing hot spots with $T_{eff}$ of 4500--11000 K where the hotter hot spots can give significant UV flux. Single and binary stars that show magnetic activity contribute to the total UV flux emitted by intermediate-age star clusters. Chromospheric activity on the stellar surface can reach temperatures of 7000--8000 K \citep{Linsky2017, Hall2008}, which could also produce UV flux. Flares on the stars are also sources of transient UV radiation. Many of these systems also contribute to the X-ray flux. Coronal emissions at very high temperatures, capable of producing X-rays, can emit in the UV region. It is important to note that the hot spots, flares, coronal activity, and very hot WDs also produce X-rays, as well as a significant flux in the UV \citep{Dempsey1993, Mitra2005}. Some active stars like the RS CVn type stars have spots resulting in excess emission in UV and X-rays \citep{Walter1981}. Among contact binaries, W Uma type stars are the most common and are found in intermediate-age OCs like NGC 2682 and NGC 188 \citep{Geller2015, Chen2016}. These systems are known to have excess UV flux, along with detectable X-ray flux. Semicontact binaries may also develop hot spots, resulting in excess UV flux \citep{Polubek2003}. Therefore, it is essential to identify the source of UV flux in known binary systems, as it could be due to the intrinsic property of the star or due to the presence of a hot companion. This is particularly important in the case of single-lined spectroscopic binaries (SB1s), where a sub-luminous companion is expected.

\begin{figure}[!ht]
    \centering
    \includegraphics{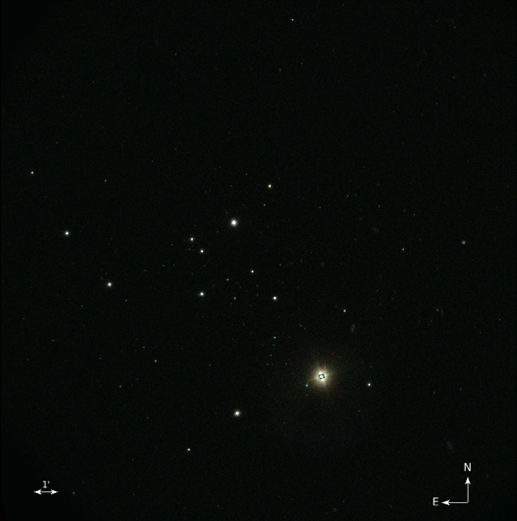}
    \caption{UVIT F148W image of NGC 2682.}
    \label{fig:4_UVIT_image_M67}
\end{figure}

Two intermediate-age OCs with well-identified member stars, along with well-studied binary properties, through PM and RV studies, are NGC 188 and NGC 2682. These clusters are well known to have a significant fraction of various types of binaries, including contact binaries. The NGC 2682 star cluster is well studied through photometry in several wavelength bands, covering from the X-rays to the IR regions \citep{Mathieu1986, Belloni1993, Landsman1997, Belloni1998, Van2004, Sarajedini2009}, and through spectroscopy \citep{Mathieu1990, Shetrone2000, Bertelli2018}. The cluster has a rich population of exotic stellar types, which do not follow the standard single stellar evolutionary theory. 

Old OCs are also ideal sites to study the properties of WDs \citep{Kalirai2010}.
In general, WDs detected in OCs are the end products of single stellar evolution. Hence, typically the mass of a WD that is recently formed in OCs is from a progenitor with the MSTO mass of the cluster. \citet{Williams2018} detected $\sim$50 WD candidates in NGC 2682 and estimated their mass and spectral type, where many WDs required a progenitor more massive than a single star at the MSTO of NGC 2682. Therefore, they concluded that these high-mass WDs are likely to be evolved from BSSs. Similarly, \citet{Sindhu2019ApJ...882...43S} detected WDs with mass $<$0.3 \Msun\ as a companion to a BSS in NGC 2682. They suggest that the formation pathway MT in a binary produces a BSS with an initially hot companion, such as a WD. As single-star evolution takes more time than the age of the universe to form such ELM WDs, they must have undergone significant mass loss during their evolution in close binary systems \citep{Brown2010} and have never ignited helium in their cores.

NGC 2682 has been studied in UV by \citet{Landsman1998}, \citet{Siegel2014}, and \citet{Sindhu2018,Sindhu2019ApJ...882...43S}. \citet{Landsman1998} studied 11 BSSs, 7 WDs and 1 YSS using UVIT with 1210 s exposure in a single FUV filter. They found that BSSs dominated the integrated UV spectrum, and some stars indicate hot subluminous companions. \citet{Siegel2014} used the Ultraviolet Optical Telescope (UVOT) aboard the \textit{Swift Gamma-Ray Burst Mission} to study the NUV CMD of NGC 2682.

We have recently started a program to understand the UV properties of binary and single stars in intermediate and old OCs. The first paper in this series was a study of the UV properties of NGC 2682 stars by \citet{Sindhu2018}, which identified several UV-bright stars using \textit{GALEX}. Some were found to be bright in the FUV, whereas a larger number were found to be bright in the NUV. They found two RGs to be bright in the FUV, which was explained by the presence of chromospheric activity in them, as traced by the Mg\textsc{ii} line emission in the \textit{International Ultraviolet Explorer (IUE)} spectra. The authors also speculated that many UV-bright stars located near the MSTO could be chromospherically active.

To understand the properties of the FUV-bright stars detected by \citet{Sindhu2018}, we carried out an imaging study of NGC 2682 in FUV, using the UVIT. UVIT has a superior spatial resolution (1\farcsec5) when compared to \textit{GALEX} ($>$4$^{\prime\prime}$; \citealt{Martin05}). Hence, UVIT will obtain relatively uncontaminated photometry of individual members of NGC 2682 that have detectable flux in the FUV. We observed NGC 2682 using three FUV filters to comprehensively study the UV-bright population.
\citet{Sindhu2019ApJ...882...43S} detected an ELM WD companion to one of the BSSs in NGC 2682 using these observations. We further this study by analysing other UV sources in NGC 2682 and ascertaining the connection between the stellar type and UV/X-ray emission.
We created CMDs and SEDs to estimate the fundamental parameters such as luminosity, temperature, and radius. Here, we analyse 30 members of NGC 2682 individually to assess the source of the UV flux.

This chapter is arranged as follows: observations and archival data are provided in \S~\ref{sec:4_obs}. Membership, CMDs, SED fitting and mass estimation are discussed in \S~\ref{sec:4_Analysis}. Results and discussion are given in \S~\ref{sec:4_results} and \S~\ref{sec:4_Discussion} respectively.

\section{Observations and analysis}
\label{sec:4_obs}
\subsection{UVIT data}
\label{sec:4_UVIT_Data} 

\begin{figure}[!hb]
\centering
\includegraphics[width=0.45\textwidth]{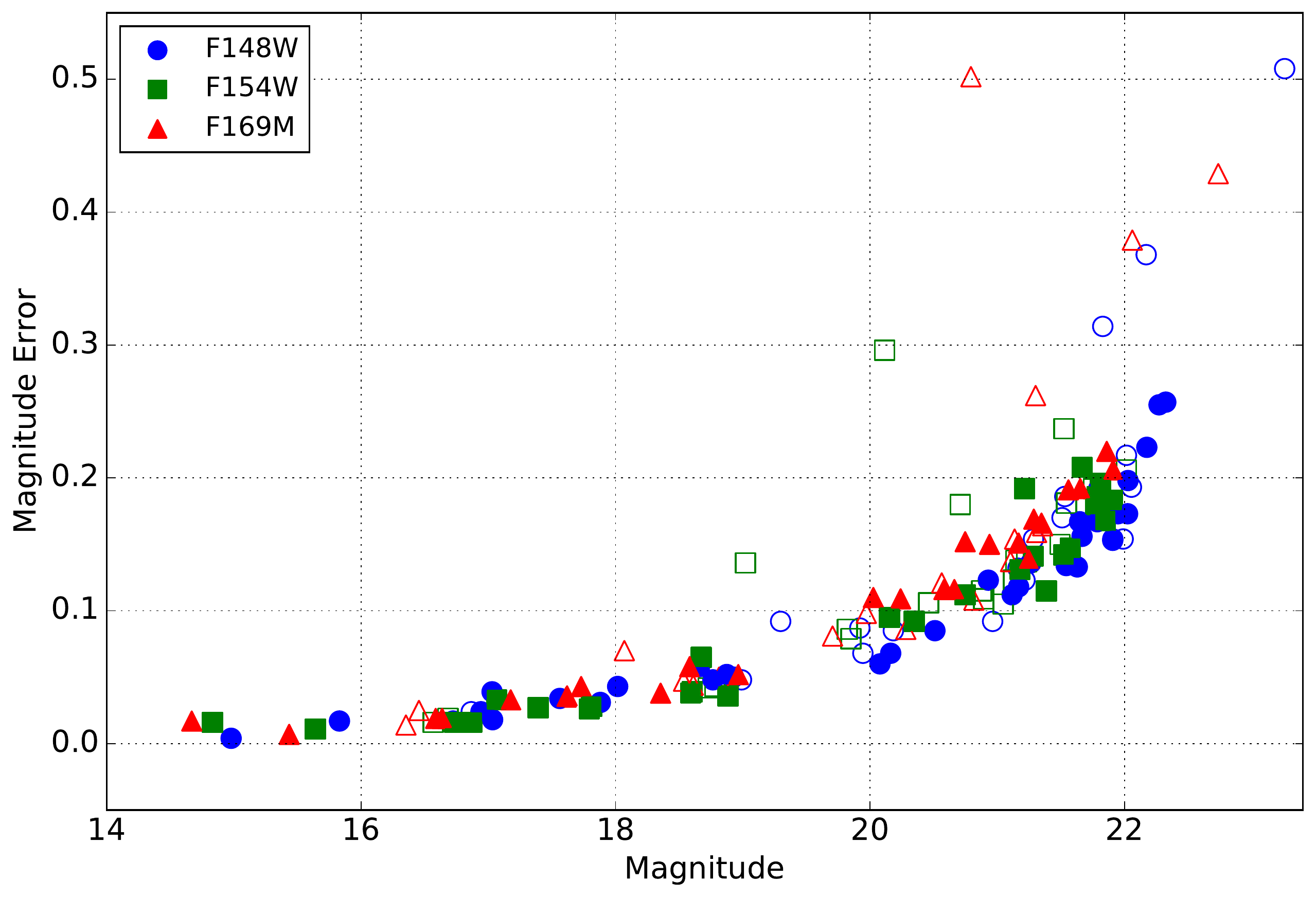}
    \caption{Photometric errors in magnitudes for all three filters. Filled points are NGC 2682 members, while hollow points are other detected stars.}
    \label{fig:4_Err_V_mag}
\end{figure}

\begin{table}
\footnotesize
\centering
\caption{The measured magnitudes (not corrected for extinction) of WD and other stellar sources of NGC 2682 from three UVIT filters. Probability of PM membership (PPM) and probability of RV membership (PRV)  are obtained from $^a$ \citet{Yadav2008}, $^b$ \citet{Zhao1993}, $^c$ \citet{Girard1989}, $^d$ \citet{Geller2015}, $^e$ \citet{Williams2018}, $^f$ Photometric member \citep{Williams2018}.}
\label{tab:4_photometry}
\resizebox{0.98\textwidth}{!}{
\begin{tabular}{lcccc cccr} 
\toprule
\textbf{Name}	&	\textbf{RA} 	&	\textbf{Dec}	&	\textbf{F148W} 	&	\textbf{F154W}	&	\textbf{F169M} &	\textbf{PPM}	&	\textbf{PRV$^d$}	\\ 
 & (deg) & (deg) & (AB mag) & (AB mag) & (AB mag) &  & \\
\toprule
WOCS1001	&	132.84560	&	11.81378	&	21.89$\pm$0.18	&	21.67$\pm$0.21	&	21.65$\pm$0.19	&	99$^a$	&	98	\\
WOCS1006	&	132.86270	&	11.86466	&	14.98$\pm$0.00	&	14.83$\pm$0.02	&	14.67$\pm$0.02	&	99$^c$	&	--	\\
WOCS1007	&	132.89310	&	11.85297	&	17.03$\pm$0.02	&	16.87$\pm$0.02	&	16.64$\pm$0.02	&	99$^c$	&	90	\\
WOCS11005	&	132.83390	&	11.77832	&	21.91$\pm$0.15	&	--	&	--	&	100$^b$	&	98	\\
WOCS11011	&	132.77015	&	11.76581	&	22.02$\pm$0.17	&	--	&	--	&	98$^c$	&	--	\\
WOCS2002	&	132.84918	&	11.83040	&	18.66$\pm$0.06	&	18.67$\pm$0.07	&	18.58$\pm$0.06	&	99$^c$	&	98	\\
WOCS2003	&	132.82937	&	11.83498	&	--	&	21.85$\pm$0.17	&	--	&	100$^a$	&	98	\\
WOCS2007	&	132.83850	&	11.76469	&	21.54$\pm$0.13	&	21.52$\pm$0.14	&	21.56$\pm$0.19	&	99$^c$	&	89	\\
WOCS2008	&	132.84077	&	11.87721	&	20.93$\pm$0.12	&	20.75$\pm$0.11	&	20.58$\pm$0.12	&	99$^c$	&	97	\\
WOCS2009	&	132.83674	&	11.89064	&	17.56$\pm$0.03	&	17.39$\pm$0.03	&	17.73$\pm$0.04	&	99$^c$	&	98	\\
WOCS2011	&	132.86013	&	11.73081	&	15.83$\pm$0.02	&	15.64$\pm$0.01	&	15.43$\pm$0.01	&	99$^c$	&	97	\\
WOCS2012	&	132.76361	&	11.76317	&	22.32$\pm$0.26	&	--	&	--	&	100$^a$	&	98	\\
WOCS2015	&	132.73204	&	11.87076	&	21.63$\pm$0.13	&	--	&	--	&	99$^c$	&	98	\\
WOCS3001	&	132.84580	&	11.82036	&	21.67$\pm$0.16	&	21.39$\pm$0.12	&	21.29$\pm$0.17	&	100$^a$	&	98	\\
WOCS3005	&	132.88590	&	11.81452	&	16.94$\pm$0.02	&	16.74$\pm$0.02	&	16.59$\pm$0.02	&	99$^c$	&	94	\\
WOCS3009	&	132.91350	&	11.83443	&	21.80$\pm$0.19	&	21.57$\pm$0.15	&	21.17$\pm$0.15	&	99$^c$	&	98	\\
WOCS3010	&	132.80980	&	11.75018	&	18.76$\pm$0.05	&	18.59$\pm$0.04	&	18.35$\pm$0.04	&	90$^d$	&	98	\\
WOCS3012	&	132.78017	&	11.88390	&	21.65$\pm$0.17	&	21.28$\pm$0.14	&	20.66$\pm$0.12	&	100$^a$	&	98	\\
WOCS3013	&	132.76464	&	11.75078	&	18.02$\pm$0.04	&	17.79$\pm$0.03	&	17.62$\pm$0.04	&	99$^c$	&	61	\\
WOCS4003	&	132.86730	&	11.82434	&	20.51$\pm$0.09	&	20.35$\pm$0.09	&	20.03$\pm$0.11	&	100$^a$	&	--	\\
WOCS4006	&	132.88590	&	11.84466	&	17.88$\pm$0.03	&	17.81$\pm$0.03	&	17.61$\pm$0.04	&	99$^c$	&	--	\\
WOCS4015	&	132.97231	&	11.80585	&	--	&	21.81$\pm$0.20	&	--	&	99$^c$	&	98	\\
WOCS5005	&	132.83309	&	11.78349	&	20.08$\pm$0.06	&	--	&	--	&	99$^c$	&	98	\\
WOCS5007	&	132.86807	&	11.87159	&	--	&	21.77$\pm$0.18	&	21.35$\pm$0.17	&	100$^a$	&	98	\\
WOCS5013	&	132.93230	&	11.75419	&	22.03$\pm$0.20	&	--	&	--	&	100$^a$	&	93	\\
WOCS6006	&	132.89300	&	11.82889	&	21.79$\pm$0.17	&	21.79$\pm$0.19	&	21.25$\pm$0.14	&	100$^a$	&	98	\\
WOCS7005	&	132.88410	&	11.83439	&	--	&	--	&	21.86$\pm$0.22	&	99$^a$	&	91	\\
WOCS7009	&	132.90787	&	11.84922	&	--	&	21.90$\pm$0.18	&	--	&	97$^c$	&	--	\\
WOCS7010	&	132.86439	&	11.89071	&	--	&	21.81$\pm$0.18	&	--	&	100$^a$	&	98	\\
WOCS8005	&	132.88843	&	11.81432	&	22.27$\pm$0.26	&	--	&	--	&	100$^a$	&	98	\\
WOCS8006	&	132.83588	&	11.77128	&	21.26$\pm$0.14	&	--	&	20.75$\pm$0.15	&	100$^a$	&	98	\\
WOCS8010	&	132.81020	&	11.75099	&	--	&	21.21$\pm$0.19	&	--	&	--	&	91	\\
WOCS9005	&	132.81447	&	11.79214	&	--	&	21.18$\pm$0.13	&	--	&	99$^c$	&	98	\\
WOCS9028	&	132.70291	&	12.00243	&	--	&	--	&	21.91$\pm$0.21	&	99$^c$	&	94	\\ \hline
Y1168	&	132.83310	&	11.81147	&	17.03$\pm$0.04	&	17.07$\pm$0.03	&	17.18$\pm$0.03	&	Y$^d$	&	--	\\
Y563	&	132.89530	&	11.70523	&	18.87$\pm$0.05	&	18.88$\pm$0.04	&	18.96$\pm$0.05	&	Y$^d$	&	--	\\
Y886	&	132.91900	&	11.76718	&	20.16$\pm$0.07	&	20.15$\pm$0.10	&	20.24$\pm$0.11	&	Y$^d$	&	--	\\
Y1157$^f$	&	132.94060	&	11.80987	&	21.12$\pm$0.11	&	--	&	--	&	--	&	--	\\
Y701	&	132.77230	&	11.73277	&	21.95$\pm$0.17	&	--	&	--	&	Y$^d$	&	--	\\
Y856	&	132.78730	&	11.76274	&	22.17$\pm$0.22	&	--	&	--	&	Y$^d$	&	--	\\
Y1487	&	132.62320	&	11.88052	&	21.17$\pm$0.12	&	--	&	20.94$\pm$0.15	&	Y$^d$	&	--	\\ \hline
\bottomrule
\end{tabular}
}
\end{table}

The near-simultaneous observations of NGC 2682 were carried out by the UVIT on 2017 April 23.
We used three filters in the FUV region viz. F148W (1481$\pm$250 \AA), F154W (1541$\pm$190 \AA), F169M (1608$\pm$145 \AA). 
The band-passes are shown in the top panel of Fig.~\ref{fig:4_SED_3009} (c).
NUV data could not be obtained due to some instrument-related issues. The science ready images have the following exposure times: F148W = 2290 s, F154W = 2428 s, and F169M = 2428 s. 
Fig.~\ref{fig:4_UVIT_image_M67} shows the FUV image of the NGC 2682.
We detected a total of 133, 114, and 92 stars in F148W, F154W, and F169M filters, respectively. The error variation with magnitudes in all three filters is shown in Fig.~\ref{fig:4_Err_V_mag}. It shows that we have detected objects up to 22 mag with a maximum error of 0.25 mag for the faintest members.

\subsection{Archival data}
\label{sec:4_Archival_Data}

We combined the UVIT data with the data in the longer wavelengths to identify and characterise the detected stars. All cross-matches were done with a maximum separation of 3$^{\prime\prime}$.

Flux measurements from UV to IR bands were obtained as follows: FUV (1542$\pm$200 \AA) and NUV (2274$\pm$530 \AA) from \textit{GALEX} \citep{Bianchi2017ApJS..230...24B}; U (3630$\pm$296 \AA), B (4358$\pm$502 \AA), V (5366$\pm$470 \AA), R (6454$\pm$776 \AA) and I (8100$\pm$912 \AA) from Kitt Peak National Observatory (KPNO, \citealt{Montgomery1993}); Gbp (5050$\pm$1172 \AA), G (6230$\pm$2092 \AA) and Grp (7730$\pm$1378 \AA) from {\it Gaia} DR2 \citep{Gaia2018a}; J (12350$\pm$812 \AA), H (16620$\pm$1254 \AA) and Ks (21590$\pm$1309 \AA) from Two Micron All-Sky Survey (2MASS, \citealt{Skrutskie2006AJ....131.1163S}, \citealt{Ochsenbein2000}); W1 (33526$\pm$3313 \AA), W2 (46028$\pm$5211 \AA) and W3 (115608$\pm$27528 \AA) from Wide-Field Infrared Survey Explorer (WISE, \citealt{Ochsenbein2000}, \citealt{Wright2010AJ....140.1868W}). 

For WDs, we included photometry from two NGC 2682 catalogues: B (4525$\pm$590 \AA), V (5340$\pm$520 \AA) and I (9509$\pm$2000 \AA) from LaSilla \citep{Yadav2008}; G (4581$\pm$830 \AA), U (3550$\pm$350 \AA) and R (6248$\pm$800 \AA) from MMT \citep{Williams2018}. IR photometry is not available for the WDs.

\section{Analysis}
\label{sec:4_Analysis}

\subsection{Membership}
\label{sec:4_Membership}

\begin{figure}[!ht]
\centering
\includegraphics[width=0.6\textwidth]{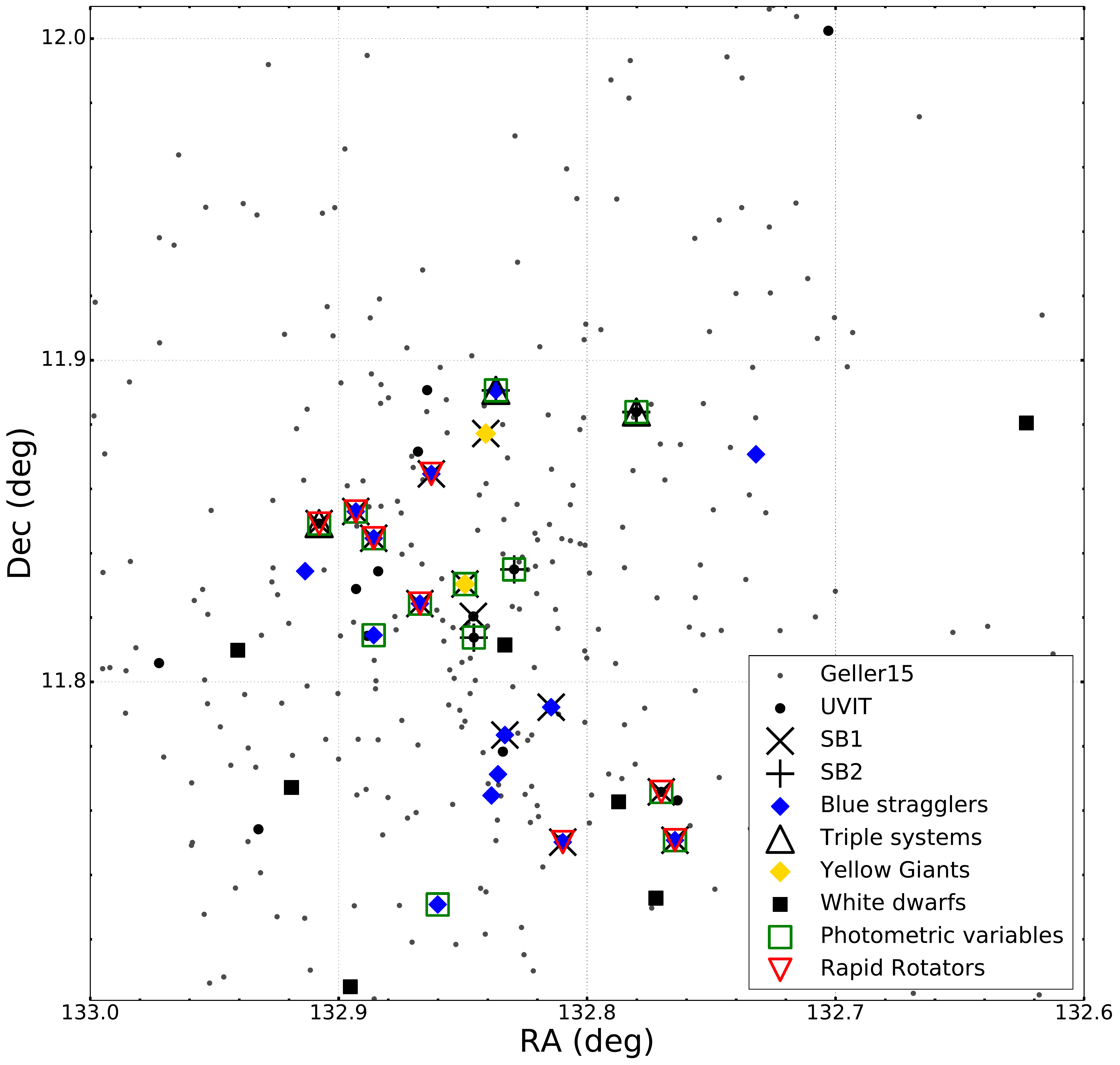}
    \caption{Spatial distribution of NGC 2682 members as observed by UVIT along with the members from the \citet{Geller2015} catalogue. The NGC 2682 members according to \citet{Geller2015} are shown in grey, while the members detected by UVIT are stylised according to their known classification.}
    \label{fig:4_Members_UVIT_and_Geller}
\end{figure}

Among the detected stars, we identified 34 members by cross-matching with the \citet{Geller2015} catalogue with PM MP \citep{Yadav2008} or RV MP over 90\%. Among these members, 16 stars are catalogued by \citealt{Geller2015} as BSSs. However, it is to be noted that not all sources labelled as BSSs in their catalogue are confirmed BSSs; some stars are blue straggler candidates. We also identify 2 YSSs and 2 triple systems (WOCS2009, WOCS3012) (WOCS: WIYN Open Cluster Study) as classified by \citet{Geller2015}. These stars are further categorised as 13 SB1s and 4 SB2s. We used the \citet{Yadav2008} and \citet{Williams2018} catalogues to identify 6 WDs with PM membership and 1 WD with photometric membership.

The spatial distribution of all member stars identified by \citet{Geller2015}, along with 41 stars detected by UVIT is shown in Fig.~\ref{fig:4_Members_UVIT_and_Geller}. The photometry in the three UVIT FUV filters, their probabilities of PM and their RV membership are tabulated in Table~\ref{tab:4_photometry}.

\subsection{Colour-magnitude diagrams}
\label{sec:4_CMD}
CMDs are very useful to detect stars in various evolutionary phases. As we have three filters in the FUV, we can use UV CMDs, and UV--optical CMDs to identify UV-bright stars. 
\citet{Sindhu2018} demonstrated that the UV and UV--optical CMDs, along with the optical CMDs, are good tools to identify the UV bright stars. 
We have overlaid the isochrones generated from the flexible stellar population synthesis (\textsc{fsps}) code \citep{Conroy2009, Conroy2010} on the CMDs. The \textsc{fsps} code can generate modified isochrone models of BaSTI and Padova to include multiple phases of the stellar evolutionary track such as HB, AGB, BSS, WD, etc. We have used the \textsc{fsps} code to generate both optical and UV isochrones of the BaSTI model (\citealp{Pietrinferni2004}, \citealp{Cordier2007}) by providing the input parameters of the cluster viz. distance modulus $V-M_{v}= 9.57 \pm 0.03$ mag \citep{Stello2016}, solar metallicity, reddening of $E(B-V)= 0.05 \pm 0.01$  mag \citep{Montgomery1993}, and age of $\sim$4 Gyr.
The isochrones are corrected for reddening and extinction.
The \textsc{fsps}-generated locus of BSSs, assuming them to be MS stars with masses more than the turn-off mass, which uniformly populates 0.5--2.5 magnitudes brighter than the MSTO, is also shown. 

\begin{figure}[!ht]
  \centering
  \begin{tabular}{c c}
    \includegraphics[width=.45\textwidth]{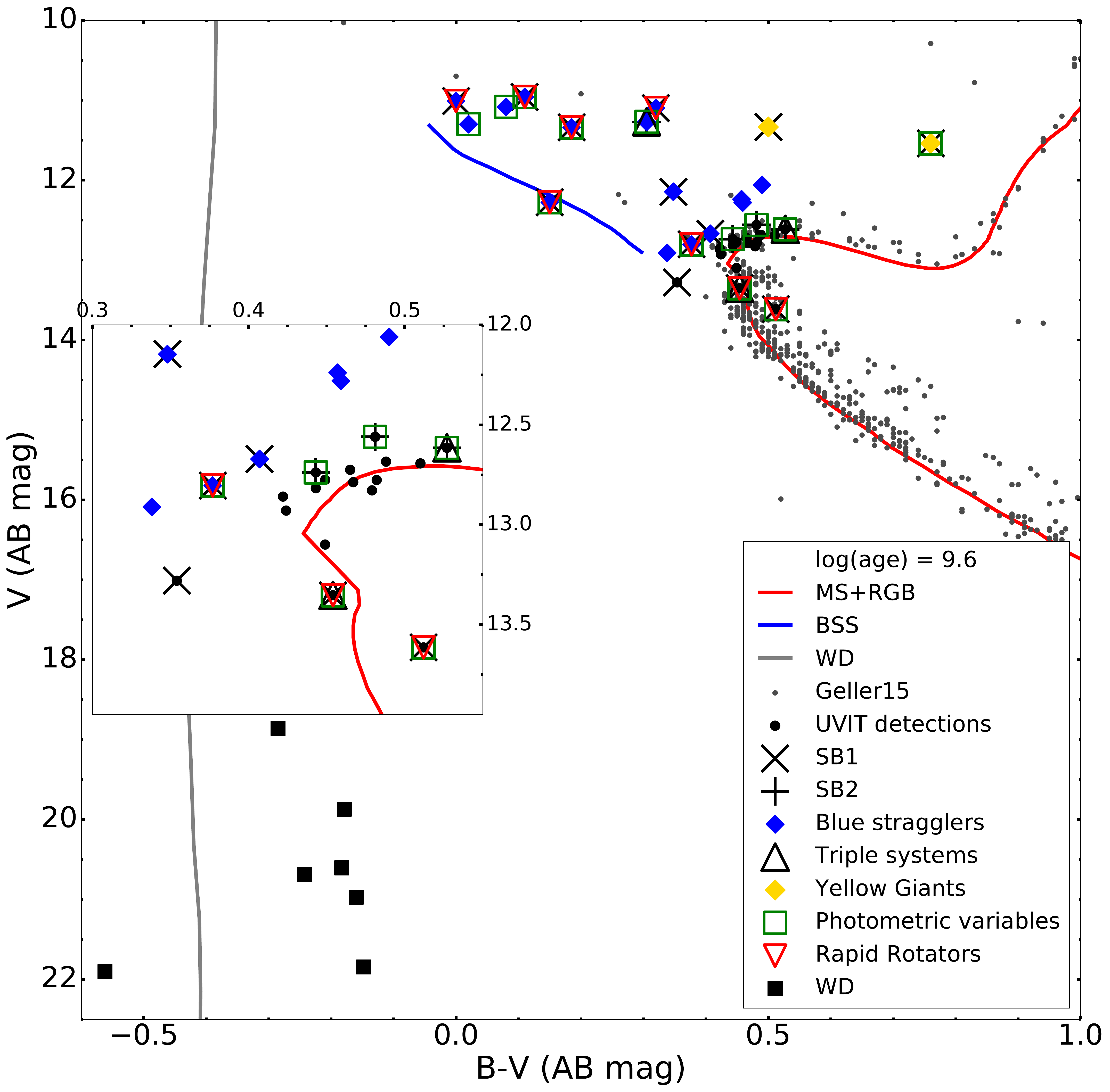} &
    \includegraphics[width=.45\textwidth]{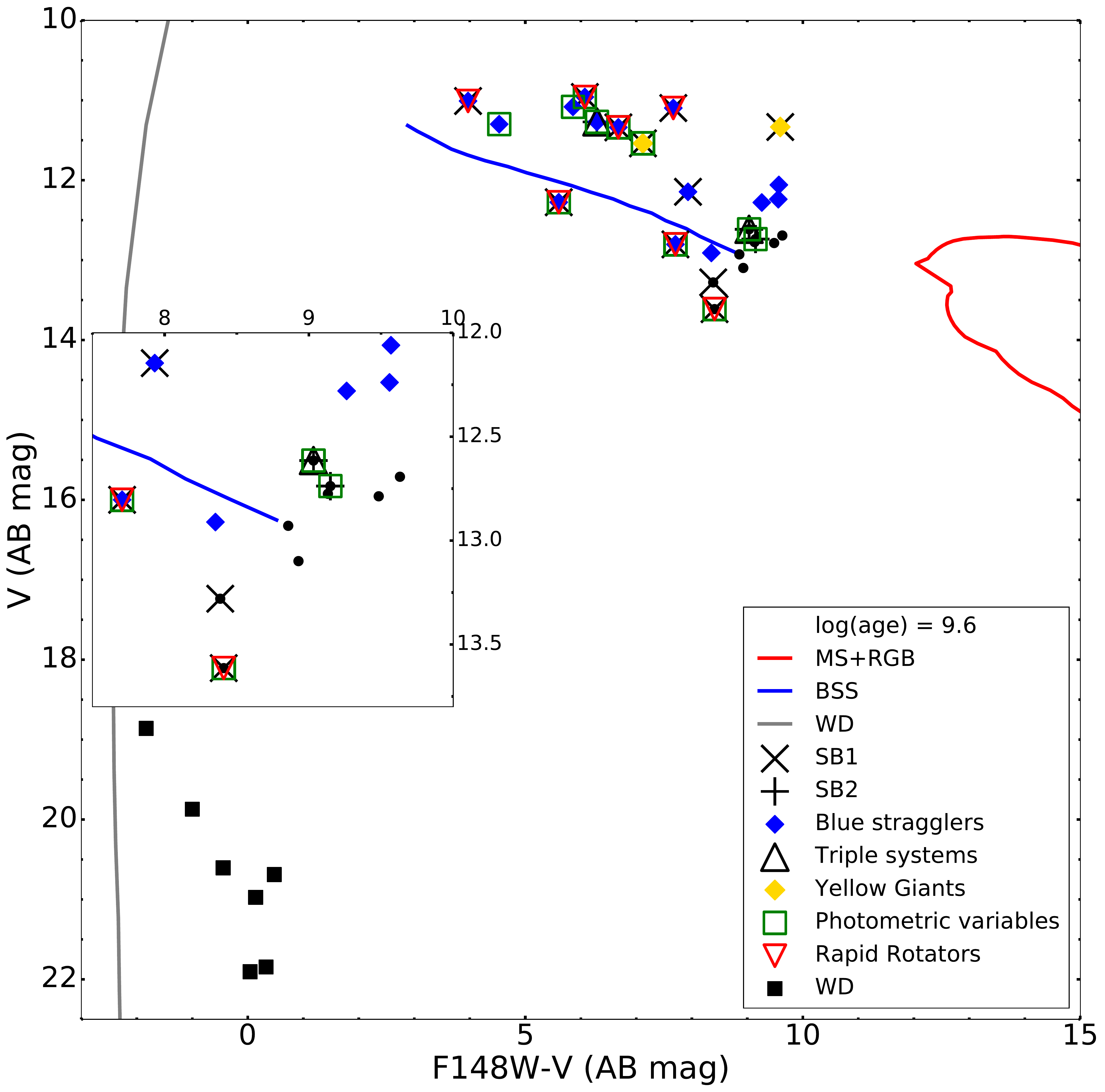} \\
    (a) & (b) \\
    \includegraphics[width=.45\textwidth]{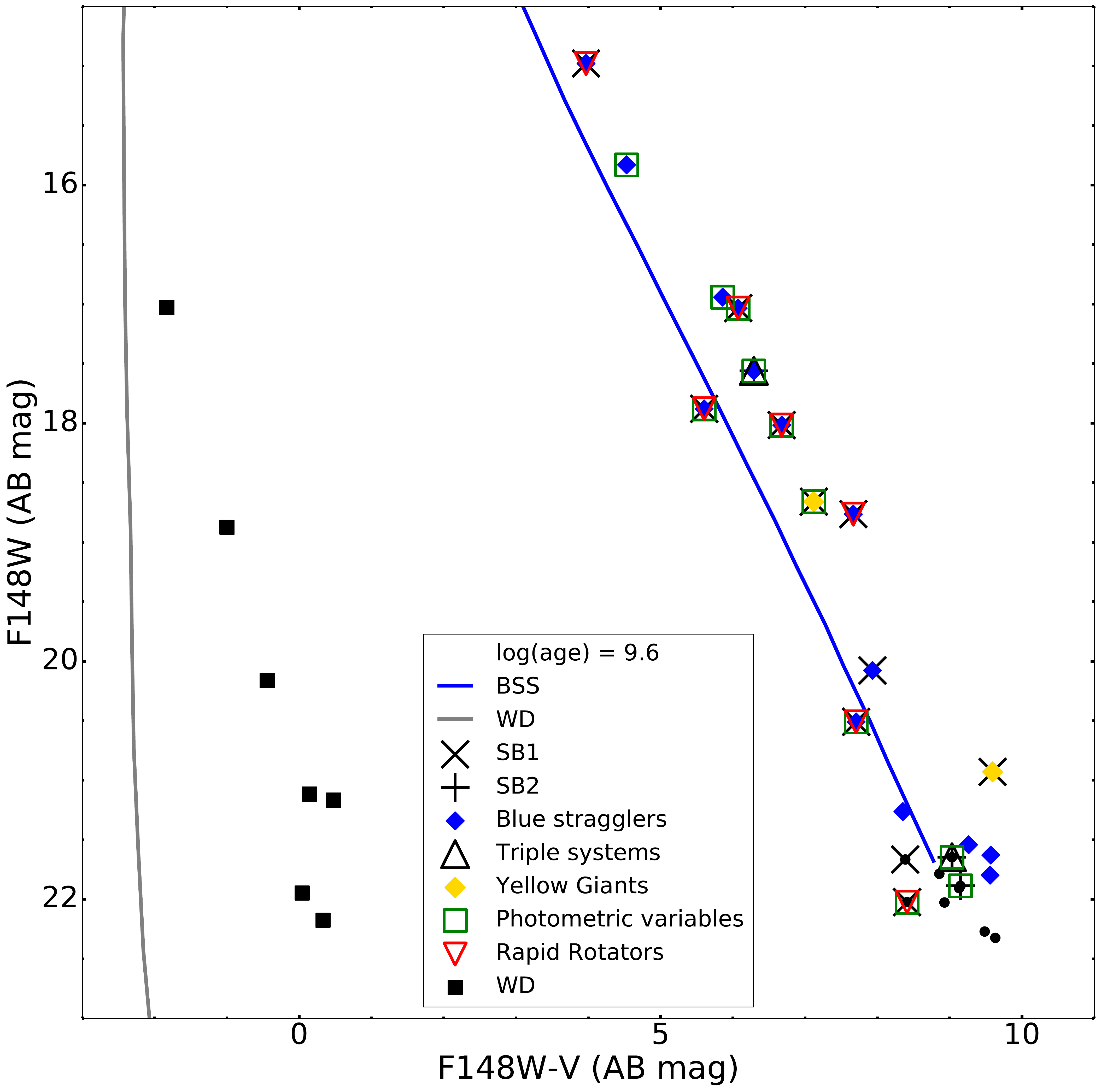} &
    \includegraphics[width=.45\textwidth]{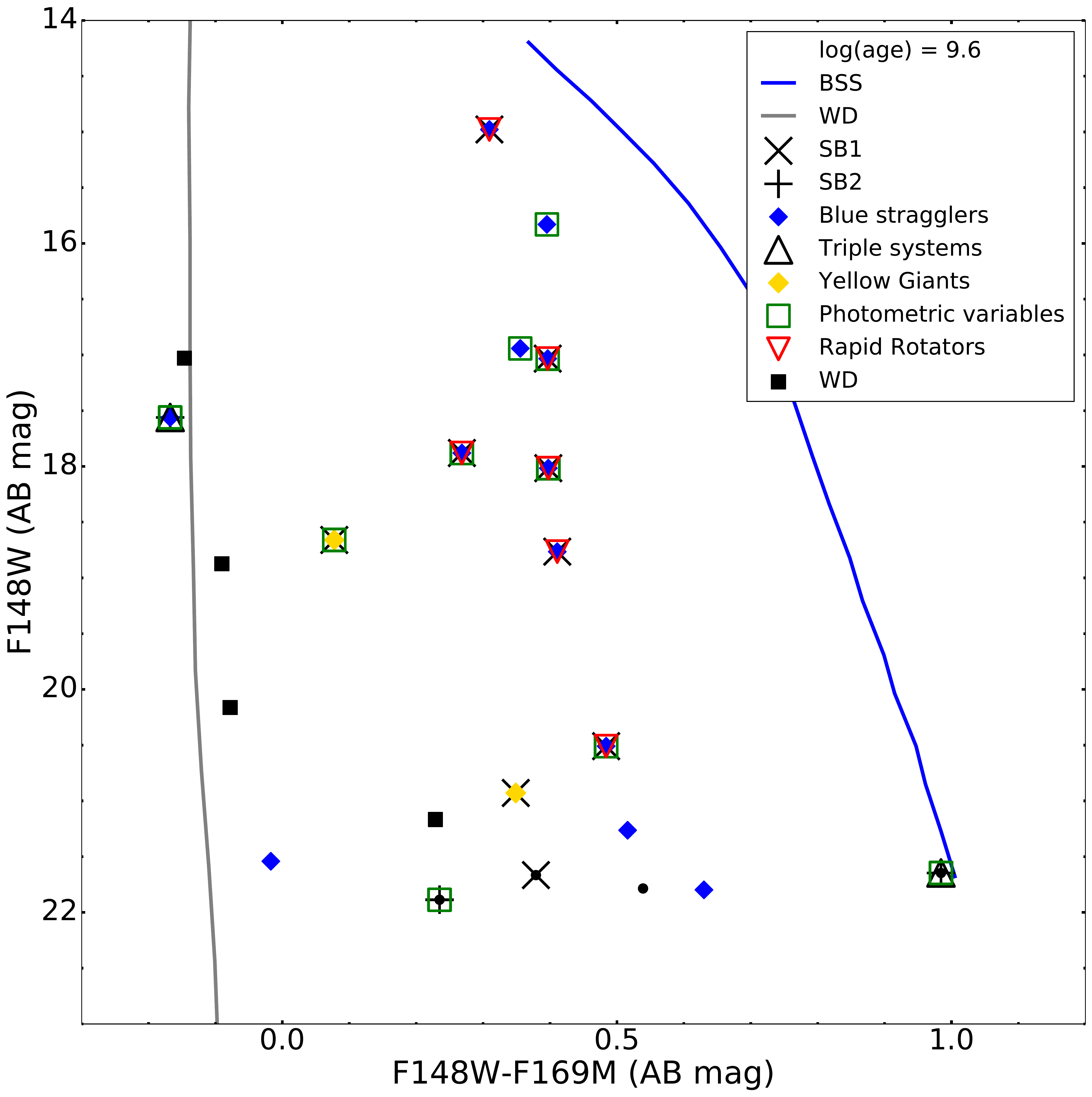} \\
    (c) & (d) \\
  \end{tabular}
  \caption{(a) Optical CMD of NGC 2682. The members listed by \citet{Geller2015} are shown in grey dots, while the unclassified members detected by UVIT are shown as black dots. The isochrone (age = 3.98 Gyr) is generated with the \textsc{fsps} code using the BaSTI model. The MS/subgiant/RG phase is shown in red; expected locations of BSSs and WDs are shown in blue and grey, respectively.
  The inset shows the expanded view of the turn-off region. (b) UV--optical CMD (V, F148W$-$V) of NGC 2682. The inset shows the expanded view of the fainter end of the BS sequence. (c) UV--optical CMD (F148W, F148$-$V) of NGC 2682. The inset shows the expanded view of the turn-off region. (d) UV CMD (F148W, F148W$-$F169M) of NGC 2682.}
  \label{fig:4_CMD}
\end{figure}

We present an optical CMD, two UV--optical CMDs and a UV CMD in Fig.~\ref{fig:4_CMD}. The optical CMD shows all 41 members detected by UVIT according to their respective known categories along with the members identified in \citet{Geller2015} shown as grey dots in Fig.~\ref{fig:4_CMD} (a). The optical photometries for non-WD and WD sources are adopted from \citet{Montgomery1993} and \citet{Yadav2008} respectively. As we have marked the UVIT-detected sources, it can be seen that we have detected only stars near the MSTO and hotter stars, including BSSs and WDs. We do not have UVIT detections for most of the MS, as these stars are relatively cooler and much fainter in the FUV. The blown-up view near the MSTO is shown in the inset.

In Fig.~\ref{fig:4_CMD} (b), we have shown the V, (F148W$-$V) CMD of the detected members (hereafter referred to as UV--optical CMD). This figure has the same y-axis as panel (a), but the x-axis uses the F148W flux. We also note that the  colour spread of BSSs increases from 0.5 in ($B-V$) as seen in Fig.~\ref{fig:4_CMD} (a) to 6 magnitudes in (F148W$-$V) as seen in Fig.~\ref{fig:4_CMD} (b). We can see that the BSSs follow the model BSS line, whereas the stars located near the MSTO in the optical CMD get bluer and are located close to the red end of the BSS model line. We can also notice that the MSTO of the isochrone is at $(F148W-V)$ $\sim$ 12 mag, much redder than the detections, which have a colour range of 4--10 mag. The inset shows the blown-up view of the red end of the BSS model line. It is clear that stars near the MSTO in the optical CMD have an excess of at least 2 mag in the (F148W$-$V) colour with respect to their expected location in this CMD. 
This indicates the possibility of a significant amount of excess flux in FUV for some UVIT-detected stars.

In Fig.~\ref{fig:4_CMD} (c), we have plotted the F148W, (F148W$-$V) CMD. Excluding the WDs, we find that all detected members are close to the BSS model line. The limiting magnitude of our observations is $\sim$22 mag, and the tip of the MS is found to be at $\sim$25 mag in F148W. This demonstrates that our observations are not suited to detect the MS stars in the F148W filters, as they are at least 3 mag fainter than the limiting magnitude. Therefore, the UVIT observations cannot detect any normal MS star due to the detection limit. Noticeably, a few stars on the MSTO in the optical are detected in UVIT filters, suggesting that these stars have excess flux in the F148W filter. Similar brightening of stars in the FUV was found by \citet{Sindhu2018} in the UV--optical CMDs constructed using the {\it GALEX}  data.

To compare the flux of the detected members in the F169M filter with respect to the F148W filter, we created the F148W, (F148W $-$ F169M) CMD. In Fig.~\ref{fig:4_CMD} (d), we have shown the UV CMD for stars detected in both the UVIT filters. The y-axes for panels (c) and (d) are the same, but the colour axis in panel (d) is a UV colour. The UVIT-detected stars belong to various classes, and they are identified in the CMDs, including BSSs, YSSs, photometric variables, rapid rotators, MSTO stars, WDs. We detected 7 WDs, which are located close to the WD model line. The BSS model line shows a slope in the UV colour, suggesting a range in temperature. We note that the stars that were located along the model BSS line in panel (c) no longer fall on the BSS model line. The stars have a range in (F148W$-$F169M) colour, suggesting a difference of up to 1.0 mag between the F148W and F169M magnitudes. Many stars have colour around $\sim$0.4 mag, suggesting that these are likely to have similar $T_{eff}$. 

In panel (d), we note that one triple system and a YSS are located close to the WD region, appearing as hot as the WDs.
In the CMDs presented here, the excess UV flux for the detected stars could be due to either intrinsic or extrinsic factors. To investigate this further, we estimate the properties of these stars using SEDs in the next section.

Note that not all 41 members are detected in all the 3 UVIT filters; thus, CMDs in Fig.~\ref{fig:4_CMD} (b)--(d) will not have all 41 members. The number of stars present in each CMD will depend on whether they were detected in the respective filters.

\subsection{Spectral energy distribution}
\label{sec:4_SED}

\begin{table}[!ht]
\centering
\caption{The $\chi^2_{red}$ comparison between single fits and double fits. Single fits are done by fitting single Kurucz model SED to all available points ($\chi^2_{S1}$) or excluding UV points ($\chi^2_{S2}$). Double fits are the combination of one Kurucz SED ($T_{A}$) and one WD SED ($T_B$ ) at log $g=$ 7 ($\chi^2_{Dob}$).}
\label{tab:4_chi}
\begin{tabular}{lccccr}
\toprule
        \multirow{2}[3]{*}{\textbf{WOCS}}  & \multicolumn{2}{c}{\textbf{Single Fit}} & \multicolumn{3}{c}{\textbf{Double Fit}}\\ 
        \cmidrule(lr){2-3}         \cmidrule(lr){4-6} 
&  $T_{S1}$ (K) & $\chi^{2}_{S1} (\chi^{2}_{S2})$ & $T_{A}$ (K) &  $T_B$ (K) & $\chi^2_{Dob}$ \\ 
        \toprule
1001	&	6750	&	2.4(1.4)	&	6250	&	11500	&	0.31	\\
11005	&	6500	&	4.3(1.1)	&	6250	&	11500	&	0.41	\\
11011	&	6000	&	3.8(3.3)	&	6000	&	11500	&	0.65	\\
2002	&	5250	&	78(4.3)	&	5250	&	14750	&	1.6	\\
2003	&	6500	&	2.5(4.8)	&	6250	&	9250	&	0.34	\\
2007	&	6500	&	11(12)	&	6000	&	11500	&	6.3	\\
2008	&	6500	&	7.6(1.1)	&	6000	&	11500	&	0.25	\\
2012	&	6000	&	4.4(1.1)	&	6000	&	11500	&	0.6	\\
2015	&	6500	&	5.4(14)	&	6250	&	9750	&	2.2	\\
3001	&	7000	&	2(2.9)	&	6750	&	12500	&	0.65	\\
3009	&	6750	&	2.6(1.3)	&	6250	&	10000	&	0.28	\\
4003	&	7250	&	8.2(3)	&	6500	&	10250	&	1.2	\\
4015	&	6500	&	1.9(2.7)	&	6250	&	11500	&	0.59	\\
5007	&	6750	&	2.9(1.5)	&	6250	&	9750	&	0.21	\\
5013	&	6500	&	2.4(12)	&	6250	&	11500	&	3.2	\\
6006	&	6750	&	1.8(3.2)	&	6250	&	10250	&	0.57	\\
7005	&	6500	&	1.5(6.4)	&	6000	&	11500	&	0.71	\\
7010	&	6500	&	3(3)	&	6250	&	11500	&	0.062	\\
8005	&	5750	&	17(23)	&	6000	&	10750	&	3.9	\\
8006	&	7000	&	9.9(13)	&	6750	&	11500	&	1.1	\\
9005	&	6750	&	1.3(5.6)	&	6500	&	11500	&	0.43	\\
\bottomrule
\end{tabular}
\end{table}

\begin{figure}
  \centering
  \begin{scriptsize}
  \begin{tabular}{c c}
    \includegraphics[width=.45\textwidth]{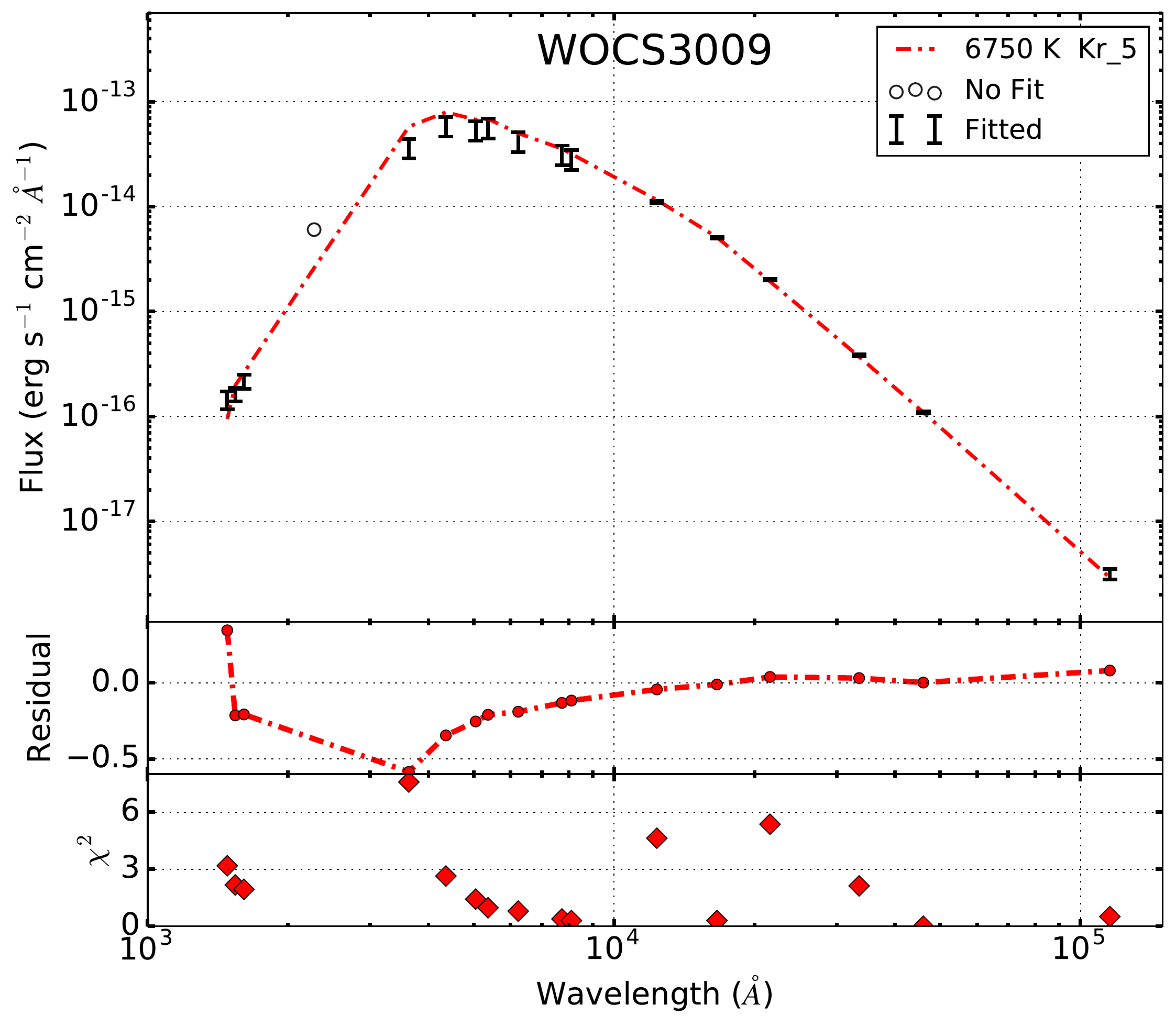} &
    \includegraphics[width=.45\textwidth]{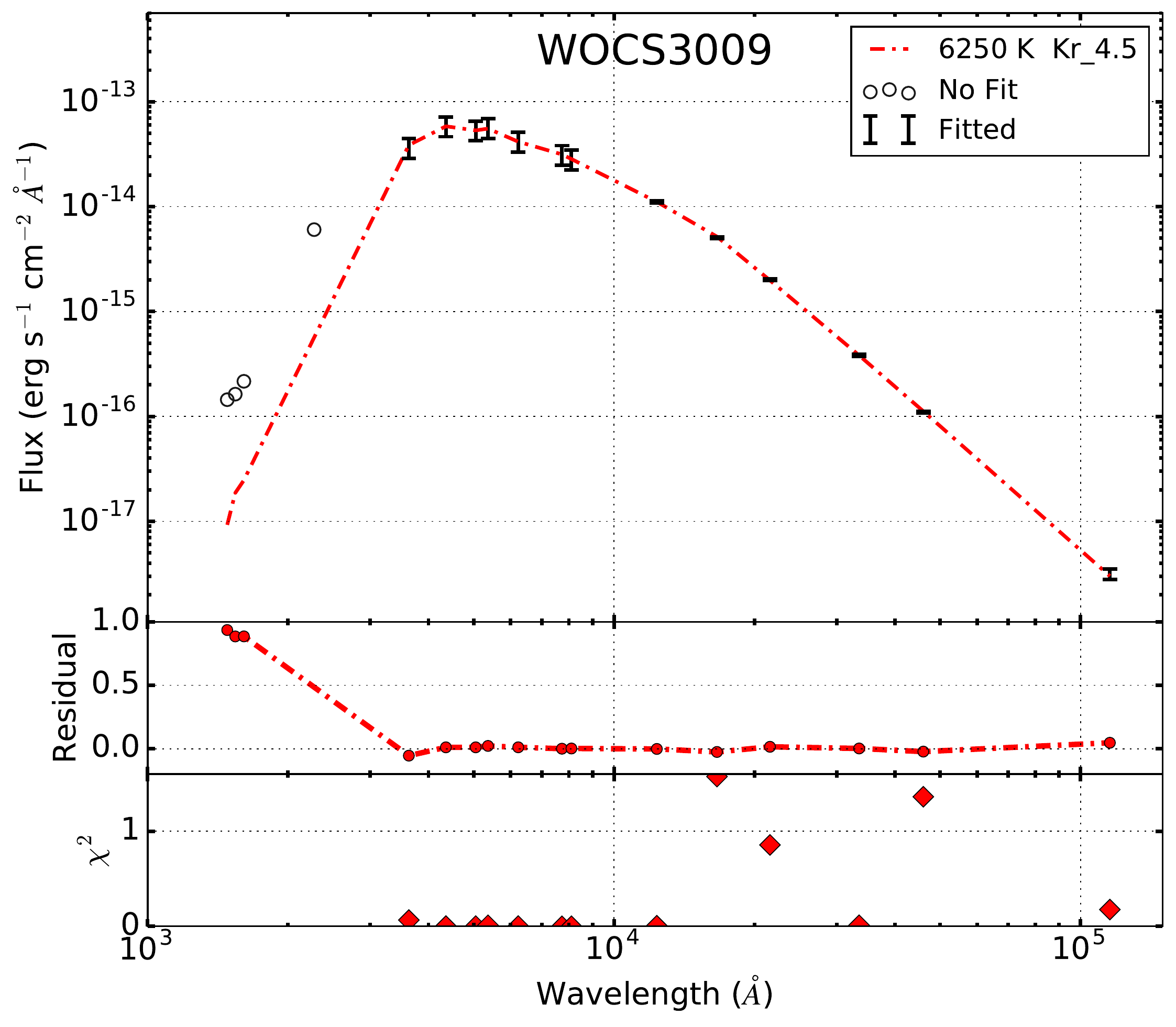}\\
    (a) & (b) \\
    \multicolumn{2}{c}{\includegraphics[width=.95\textwidth]{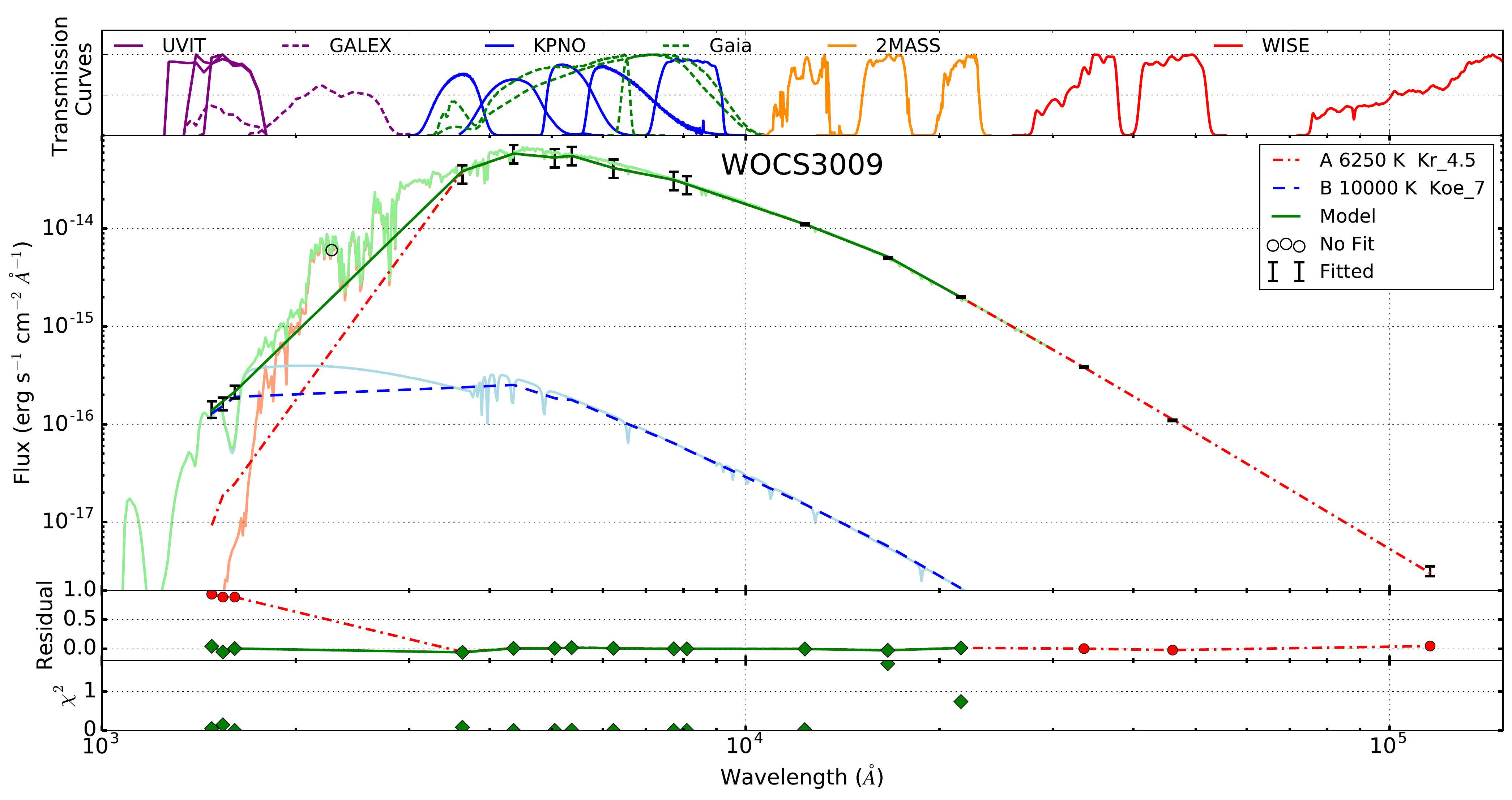}} \\
    \multicolumn{2}{c}{(c)} \\
  \end{tabular}
    \end{scriptsize}
  \caption{Example of the method used to fit SEDs using WOCS3009. (a) The top panel shows the least $\chi^2$ fit for a single SED over UV--IR data points. The legend notes the $T_{eff}$ of the fit, the model used (Kr: Kurucz; Koe: Koester WD models), and the log $g$. The fitted points are shown with error bars, while not fitted points are shown as circles. The middle panel shows the residual for the fit. The bottom panel shows the individual $\chi^2$ values calculated at each point.
  (b) Same as panel (a) but fit is done over the optical--IR region.
  (c) The top bar shows the transmission curves (not to scale) for the filters of respective telescopes. 
  The second panel shows the composite double fit SED. 
  The figure shows the `A component' (Kurucz model; red dot--dashed), `B component' (Koester WD model; blue dashed), `Model' i.e., total flux of 2 components (green solid) and observed flux (black, as only error bars to simplify the graph) and unfitted points (hollow circle). The light-coloured solid lines in blue, red, and green show the higher-resolution spectra corresponding to Kurucz, Koester (WD), and composite models, respectively. The third and fourth panels are similar to the residual and $\chi^2$ panels in panel (a).}
  \label{fig:4_SED_3009}
\end{figure}

In order to characterise the excess UV flux as suggested by the optical and UV CMDs, we performed a detailed study using their SEDs. We compiled the fluxes of 30 sources (23 stars and 7 WDs) from UV to IR. The multi-wavelength SEDs were created and compared with model SEDs to determine their characteristics. The analysis presented here is similar to that presented in \citet{Subramaniam2016ApJ...833L..27S} and \citet{Sindhu2019ApJ...882...43S}.

We use the theoretical stellar models, that span the UV--IR wavelength coverage, as our SEDs cover from 130--16,000 nm. We use updated Kurucz stellar atmospheric models \citep{Castelli1997A&A...318..841C} for the stellar (non-WD) sources that cover the same wavelength range. The theoretical spectra for WDs of spectral type DA with pure hydrogen atmospheres were obtained from \citet{Koester2010}, which are mentioned as WD models in the rest of the chapter. The spectra were converted to synthetic photometry for the required filters using {\sc vosa} \citep{Bayo2008}, according to the individual filter profiles \citep{Rodrigo2012}.
Extinction of $A_V = 0.1736$ and distance 831$\pm$11 pc were used to normalise the SEDs. We used \citet{Fitzpatrick1999PASP..111...63F} extinction curves to calculate extinction coefficients in all other bands.

Out of the detected stars, the study of 9 bright bona fide BSSs (WOCS numbers: 1006, 1007, 2011, 2013, 3005, 3010, 3013, 4006, 5005) is presented in \citet{Sindhu2019ApJ...882...43S, Sindhu2019b} and \citet{Pandey2021MNRAS.507.2373P}.
We could not successfully cross-match WOCS8010 with archival data to create a useful SED due to its closeness to WOCS3010. WOCS9028 lay near the edge of our FOV and was only detected in F169M. We could not analyse WOCS3012 as a result of it being a triple system with no known individual parameters.
After removing these stars from consideration, we fitted the observed flux distribution of 30 stars with SED models, of which the 7 WDs were fitted with WD model SEDs with a range of $T_{eff}$ = 5000--80000 K. \citet{kepler2015} showed that most WDs have surface gravity near log $g \sim$ 8. We have thus used two surface gravity values (log $g$ = 7 and 9) for the WD models. The results of SED fittings of WDs are discussed in \S~\ref{sec:4_WD}. 
The stellar SEDs were fitted with Kurucz model SEDs with solar metallicity and limited the fits to log $g$ = 3--5 dex, and T = 3500--50000 K. Each fit provides us with the $T_{eff}$ and radius corresponding to the star. The {\it GALEX} DR6 magnitudes are not available for all stars. We also observe variations in {\it GALEX} and UVIT magnitudes in the FUV region for some stars; the reasons may be the non-simultaneous nature of observations or UVIT's superior resolution of 1\farcsec5 in FUV when compared to 4\farcsec5 of \textit{GALEX}. Thus, we did not use the {\it GALEX} photometry for fitting SEDs in the case of stellar sources. We show the {\it GALEX} flux in SED only for comparison.

The $\chi^2_{red}$ values for the stellar SEDs fits are listed in Table~\ref{tab:4_chi} as $\chi^2_{S1}$. Almost all stars show large $\chi^2$ values. As we suspected excess flux in UV as suggested from UV--optical and UV CMDs, we tried to fit the SEDs again by ignoring the flux below 1800 \AA. The $\chi^2_{red}$ values of modified fits are given in the brackets as $\chi^2_{S2}$. These values are relatively less when compared to $\chi^2_{S1}$ values. 
As the flux in the UV region is ignored, the residual flux, which consists mainly of the excess UV flux, was then fitted with a WD model, such that the flux due to two models is added up to fit the full range of observed flux.
Details of the double fits are also shown in the table, where the $\chi^2_{red}$ for the double model fit is denoted as $\chi^2_{Dob}$.
Note that these S2 fits also have fewer data points when compared to S1 fits and do not cover the full wavelength range, and hence it is better to compare $\chi^2_{S1}$ with $\chi^2_{Dob}$ directly. In most of the SED fits, it can be seen that $\chi^2_{Dob}$ is significantly less than $\chi^2_{S1}$. It is important to note that we are able to fit most of the SEDs satisfactorily using a double fit, as suggested by the low $\chi^2_{Dob}$ values. The interpretation of the changes in $\chi^2$ for each source can be found in \S~\ref{sec:4_results}.

Fig.~\ref{fig:4_SED_3009} demonstrates the above procedure using WOCS3009 as an example. Panel (a) shows the SED when we fit all available data points with a single Kurucz spectrum ($\chi^2_{red}$ = 2.6), panel (b) shows the SED when we fit only optical and IR data points ($\chi^2_{red}$ = 1.3). A good SED fit in the optical--IR region and a lower value of $\chi^2_{red}$ support the presence of a UV excess in WOCS3009. Panel (c) shows the result of fitting a two component SED (a Kurucz SED of 6250 K and log $g$ = 4.5 and a WD SED of 11,000 K and log $g$ = 9) with $\chi^2_{Dob}$ = 0.2. The residuals in panel (c) and in all SED figures are calculated as
\begin{equation}
\small
     Residual=(Flux_{Obs}-Flux_{Model})/Flux_{Obs}
\label{residual}
\end{equation}
where $Flux_{Model}$ is the flux for the model (single/composite double) SED. 

\begin{figure}
  \centering
   \includegraphics[width=0.95\textwidth]{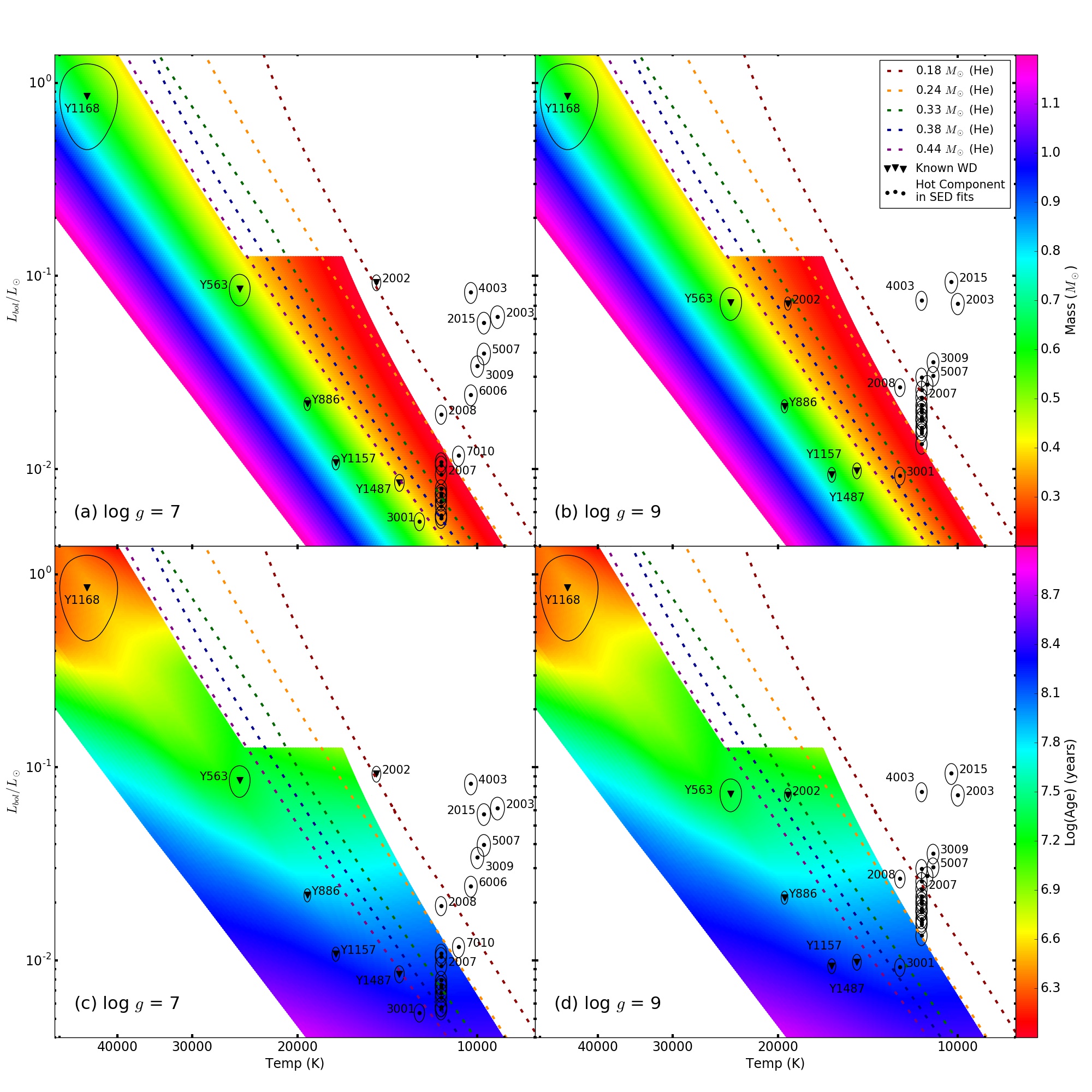}\\
   \caption{HRD of WDs and hotter components plotted over interpolated DA type \citep{Tremblay2009ApJ...696.1755T} (as solid band) and He-core \citep{Panei2007MNRAS.382..779P} (as dashed lines) WD cooling curves. The errors are plotted as ellipses. Legends for all figures are same as in (b).
   (a) The gradient corresponds to the mass of the DA model WDs. The single WDs and the hotter components' parameters from 'log $g$ = 7' fits are over-plotted to estimate their mass.
   (b) Same as 'a' for the SED fits with 'log $g$ = 9'.
   (c) The gradient corresponds to the cooling age of the DA model WDs. The single WDs and the hotter components' parameters from 'log $g$ = 7' fits are over-plotted to estimate their age.
   (d) Same as 'c' for the SED fits with 'log $g$ = 9'.
  }
  \label{fig:4_WD_mass_radius}
\end{figure}

\subsection{Mass and age estimation}
\label{sec:4_mass}
The SED fits provided the temperature and radius of all the components. The bolometric luminosity of the components is calculated using luminosity relation
\begin{equation}
\small
     L=4 \pi R^2 \sigma T_{eff}^4
\label{luminosity}
\end{equation}

We use DA (pure hydrogen) WD models \citep{Tremblay2009ApJ...696.1755T} to estimate the age and mass of the WDs. The model cooling curves were available for 0.2--1.2 \Msun\ in increments of 0.2 \Msun. We assumed log $T_{eff} \propto$ log $M$,  log $L \propto$ log $M$, log $Age \propto$ log $M$ \citealt{Myakutin1995} 
and linearly interpolated log $L$, log $M$, log $Age$ and log $T_{eff}$. The interpolation was done producing 100 steps in mass range (0.2--1.2 \Msun) and 500 steps in luminosity range ($5\times10^{-6}$--1.4 \Lsun), with each point having corresponding age and temperature. 

Fig.~\ref{fig:4_WD_mass_radius} (a) and (b) show the HRD of the known WDs and hotter components in double fits with log $g$ = 7 and 9, respectively, plotted over the interpolated DA WD model with mass as the auxiliary colour. Similarly, panels (c) and (d) are the same data points plotted over the DA WD model with age as the auxiliary colour. We also include cooling curves from the He-core low-mass WD model \citep{Panei2007MNRAS.382..779P} for the sake of completeness, as many of the hotter components lie near the lower mass range.
We used the intrinsic errors in the SED fits to estimate the errors in the mass and age. These errors are plotted as ellipses in the figure.
For the points outside the interpolated DA model, the mass is stated in Table~\ref{tab:4_All_para} as $<$0.2 \Msun\ and the upper limit of age is calculated from the vertical intercept to the DA model cooling curve at 0.2 \Msun.

\section{Results}
\label{sec:4_results}

\begin{table}
\begin{tiny}
\centering
\caption{The best-fit parameters of all sources estimated using $\chi^2$ fits. First column has identification from
\newline$^a$ WOCS: \citet{Geller2015}, S: \citet{Sanders1977}, Y: \citet{Yadav2008}, WD: \citet{Williams2018}. }
\label{tab:4_All_para}
\resizebox{0.98\textwidth}{!}{
\begin{tabular}{lcccc ccccr} 
\toprule
\textbf{Name$^a$}	&	\textbf{Comp}	&	\textbf{$T_{eff}$}	&	\textbf{log $g$}	&	\textbf{R}	&	\textbf{L}	&	\textbf{$M_{WD}$}	&	\textbf{$Age_{WD}$}	&	\textbf{Comments}	&	\textbf{Remark}	\\	
&	&	(K)	&	(cm s$^{-2}$)	&	(R$_{\odot}$)	&	(\Lsun)	&	(\Msun)	&	(Myr)	&	&	\\	\toprule
WOCS1001	&	A	&	6250	$\pm$	125	&	5	&	1.99$\pm$0.03	&	5.1	&	&	&	BM, SB2,PV,	&	WD/Ch	\\	
(S1024)	&	B	&	11500	$\pm$	250	&	9	&	0.0344$\pm$0.0005	&	0.0190	&	$<$0.2	&	$<$120	&	CX111, X46	&	\\	
&	B	&	11500	$\pm$	250	&	7	&	0.0218$\pm$0.0003	&	0.0075	&	0.27$\pm$0.02	&	178$\pm$13	&	&	\\	\midrule
WOCS11005	&	A	&	6250	$\pm$	125	&	4	&	1.94$\pm$0.03	&	4.8	&	&	&	SM	&	WD?	\\	
(S995)	&	B	&	11500	$\pm$	250	&	9	&	0.0369$\pm$0.0005	&	0.0210	&	$<$0.2	&	$<$120	&	&	\\	
&	B	&	11500	$\pm$	250	&	7	&	0.0224$\pm$0.0003	&	0.0079	&	0.26$\pm$0.02	&	164$\pm$5	&	&	\\	\midrule
WOCS11011	&	A	&	6000	$\pm$	125	&	3.5	&	1.46$\pm$0.02	&	2.3	&	&	&	BLM, SB1, HS Cnc,	&	WD/Ch	\\	
(S757)	&	B	&	11500	$\pm$	250	&	9	&	0.036$\pm$0.0005	&	0.0200	&	$<$0.2	&	$<$120	&	RR, W Uma, PV,	&	\\	
&	B	&	11500	$\pm$	250	&	7	&	0.0218$\pm$0.0003	&	0.0075	&	0.27$\pm$0.02	&	178$\pm$13	&	CX23, NX21	&	\\	\midrule
WOCS2002	&	A	&	5250	$\pm$	125	&	5	&	5.34$\pm$0.07	&	18	&	&	&	BM, SB1, PV,	&	WD+Ch	\\	
(S1040)	&	B	&	19250	$\pm$	250	&	9	&	0.0242$\pm$0.0003	&	0.0720	&	0.31$\pm$0.01	&	25$\pm$1	&	YSS, WD,	&	\\	
&	B	&	14750	$\pm$	250	&	7	&	0.0467$\pm$0.0006	&	0.0930	&	$<$0.2	&	$<$110	&	CX6, X10, NX5	&	\\	\midrule
WOCS2003	&	A	&	6250	$\pm$	125	&	4	&	2.17$\pm$0.03	&	6.4	&	&	&	BM, SB2, PV	&	Ch/Sp	\\	
(S1045)	&	B	&	10000	$\pm$	250	&	9	&	0.0896$\pm$0.001	&	0.0720	&	$<$0.2	&	$<$200	&	CX88, X41	&	\\	
&	B	&	9250	$\pm$	250	&	7	&	0.0969$\pm$0.001	&	0.0620	&	$<$0.2	&	$<$240	&	&	\\	\midrule
WOCS2007	&	A	&	6000	$\pm$	125	&	4	&	2.65$\pm$0.04	&	8.2	&	&	&	SM, BSS	&	WD	\\	
(S984)	&	B	&	11500	$\pm$	250	&	9	&	0.0387$\pm$0.0005	&	0.0240	&	$<$0.2	&	$<$120	&	&	\\	
&	B	&	11500	$\pm$	250	&	7	&	0.0245$\pm$0.0003	&	0.0094	&	0.24$\pm$0.02	&	150$\pm$10	&	&	\\	\midrule
WOCS2008	&	A	&	6500	$\pm$	125	&	4.5	&	4.18$\pm$0.06	&	19	&	&	&	BM, BSS, SB1, YSS	&	WD/Ch	\\	
(S1072)	&	B	&	12500	$\pm$	250	&	9	&	0.0348$\pm$0.0005	&	0.0270	&	$<$0.2	&	$<$120	&	CX24, X37, NX16	&	\\	
&	B	&	11500	$\pm$	250	&	7	&	0.035$\pm$0.0005	&	0.0190	&	$<$0.2	&	$<$120	&	&	\\	\midrule
WOCS2009	&	Aa	&	7250	$\pm$	250	&	4	&	2.04$\pm$0.1	&	10	&	&	&	BM, BSS, SB2, PV	&	Ch/Sp	\\	
(S1082)	&	Ab	&	6000	$\pm$	250	&	4	&	2.15$\pm$0.1	&	5.4	&	&	&	ES Cnc, Triple, &	\\	
&	B	&	6000	$\pm$	250	&	4.5	&	2.02$\pm$0.03	&	4.8	&	&	&	CX3, X4, NX37	&	\\	\midrule
WOCS2012	&	A	&	6000	$\pm$	125	&	3	&	2.2$\pm$0.03	&	5.3	&	&	&	SM	&	WD?	\\	
(S756)	&	B	&	11500	$\pm$	250	&	9	&	0.0313$\pm$0.0004	&	0.0150	&	$<$0.2	&	$<$120	&	&	\\	
&	B	&	11500	$\pm$	250	&	7	&	0.0188$\pm$0.0002	&	0.0056	&	0.34$\pm$0.03	&	219$\pm$14	&	&	\\	\midrule
WOCS2015	&	A	&	6250	$\pm$	125	&	3.5	&	2.81$\pm$0.04	&	9.7	&	&	&	SM, BSS	&	Ch	\\	
(S792)	&	B	&	10250	$\pm$	250	&	9	&	0.0969$\pm$0.001	&	0.0930	&	$<$0.2	&	$<$200	&	&	\\	
&	B	&	9750	$\pm$	250	&	7	&	0.0842$\pm$0.001	&	0.0570	&	$<$0.2	&	$<$240	&	&	\\	\midrule
WOCS3001	&	A	&	6750	$\pm$	125	&	4.5	&	1.32$\pm$0.02	&	3.1	&	&	&	BM, SB1	&	WD	\\	
(S1031)	&	B	&	12500	$\pm$	250	&	9	&	0.0205$\pm$0.0003	&	0.0093	&	0.30$\pm$0.02	&	149$\pm$9	&	&	\\	
&	B	&	12500	$\pm$	250	&	7	&	0.0157$\pm$0.0002	&	0.0054	&	0.45$\pm$0.04	&	235$\pm$22	&	&	\\	\midrule
WOCS3009	&	A	&	6250	$\pm$	125	&	4.5	&	2.46$\pm$0.03	&	7.8	&	&	&	SM, BSS	&	WD?	\\	
(S1273)	&	B	&	11000	$\pm$	250	&	9	&	0.0522$\pm$0.0007	&	0.0360	&	$<$0.2	&	$<$150	&	&	\\	
&	B	&	10000	$\pm$	250	&	7	&	0.0618$\pm$0.0008	&	0.0340	&	$<$0.2	&	$<$200	&	&	\\	\midrule
WOCS4003	&	A	&	6500	$\pm$	125	&	4.5	&	1.78$\pm$0.02	&	4.6	&	&	&	BM, BSS, SB1, RR,	&	Ch/Sp	\\	
(S1036)	&	B	&	11500	$\pm$	250	&	9	&	0.069$\pm$0.0009	&	0.0750	&	$<$0.2	&	$<$120	&	EV Cnc, PV	&	\\	
&	B	&	10250	$\pm$	250	&	7	&	0.0912$\pm$0.001	&	0.0820	&	$<$0.2	&	$<$200	&	CX19, X45, NX20	&	\\	\midrule
WOCS4015	&	A	&	6250	$\pm$	125	&	4.5	&	1.99$\pm$0.03	&	5.1	&	&	&	SM	&	WD?	\\	
(S1456)	&	B	&	11500	$\pm$	250	&	9	&	0.0338$\pm$0.0004	&	0.0180	&	$<$0.2	&	$<$120	&	&	\\	
&	B	&	11500	$\pm$	250	&	7	&	0.0209$\pm$0.0003	&	0.0068	&	0.29$\pm$0.02	&	191$\pm$7	&	&	\\	\midrule
WOCS5007	&	A	&	6250	$\pm$	125	&	5	&	1.92$\pm$0.03	&	4.5	&	&	&	SM	&	WD?	\\	
(S1071)	&	B	&	11000	$\pm$	250	&	9	&	0.0481$\pm$0.0006	&	0.0300	&	$<$0.2	&	$<$150	&	&	\\	
&	B	&	9750	$\pm$	250	&	7	&	0.07$\pm$0.0009	&	0.0400	&	$<$0.2	&	$<$240	&	&	\\	\midrule
\end{tabular}
}
\end{tiny}
\end{table}
\begin{table}
\ContinuedFloat
\begin{tiny}
\centering
\caption[]{\textit{Continued...}}
\resizebox{0.98\textwidth}{!}{
\begin{tabular}{lcccc ccccr} 
\toprule
\textbf{Name$^a$}	&	\textbf{Comp}	&	\textbf{$T_{eff}$}	&	\textbf{log $g$}	&	\textbf{R}	&	\textbf{L}	&	\textbf{$M_{WD}$}	&	\textbf{$Age_{WD}$}	&	\textbf{Comments}	&	\textbf{Remark}	\\	
&	&	(K)	&	(cm s$^{-2}$)	&	(R$_{\odot}$)	&	(\Lsun)	&	(\Msun)	&	(Myr)	&	&	\\	\toprule
WOCS5013	&	A	&	6250	$\pm$	125	&	5	&	1.68$\pm$0.02	&	3.6	&	&	&	SM	&	WD?	\\	
(S1230)	&	B	&	11500	$\pm$	250	&	9	&	0.0354$\pm$0.0005	&	0.0200	&	$<$0.2	&	$<$120	&	&	\\	
&	B	&	11500	$\pm$	250	&	7	&	0.0215$\pm$0.0003	&	0.0073	&	0.28$\pm$0.02	&	181$\pm$13	&	&	\\	\midrule
WOCS6006	&	A	&	6250	$\pm$	125	&	4.5	&	1.82$\pm$0.02	&	4.3	&	&	&	SM	&	WD	\\	
(S1271)	&	B	&	11250	$\pm$	250	&	9	&	0.0437$\pm$0.0006	&	0.0270	&	$<$0.2	&	$<$120	&	&	\\	
&	B	&	10250	$\pm$	250	&	7	&	0.0495$\pm$0.0007	&	0.0240	&	$<$0.2	&	$<$200	&	&	\\	\midrule
WOCS7005	&	A	&	6000	$\pm$	125	&	4	&	2.19$\pm$0.03	&	5.3	&	&	&	SM	&	WD?	\\	
(S1274)	&	B	&	11500	$\pm$	250	&	9	&	0.0293$\pm$0.0004	&	0.0130	&	$<$0.2	&	$<$120	&	&	\\	
&	B	&	11500	$\pm$	250	&	7	&	0.0192$\pm$0.0003	&	0.0058	&	0.33$\pm$0.03	&	209$\pm$11	&	&	\\	\midrule
WOCS7009	&	Aa	&	6250	$\pm$	250	&	4.5	&	1.3$\pm$0.1	&	2.8	&	&	&	BLM, SB1, RR,	&	Ch+WD?	\\	
(S1282)	&	Ab	&	6250	$\pm$	250	&	4.5	&	0.68$\pm$0.05	&	0.64	&	&	&	AH Cnc, W Uma,	&	\\	
&	B	&	11500	$\pm$	250	&	9	&	0.0324$\pm$0.0004	&	0.0160	&	$<$0.2	&	$<$120	&	CX16, X40, NX10	&	\\	
&	B	&	11500	$\pm$	250	&	7	&	0.0202$\pm$0.0003	&	0.0064	&	0.306$\pm$0.03	&	198$\pm$14	&	&	\\	\midrule
WOCS7010	&	A	&	6750	$\pm$	125	&	3.5	&	1.97$\pm$0.03	&	4.8	&	&	&	SM	&	WD?	\\	
(S1083)	&	B	&	11500	$\pm$	250	&	9	&	0.0338$\pm$0.0004	&	0.0180	&	$<$0.2	&	$<$120	&	&	\\	
&	B	&	10750	$\pm$	250	&	7	&	0.0313$\pm$0.0004	&	0.0120	&	$<$0.2	&	$<$170	&	&	\\	\midrule
WOCS8005	&	A	&	6000	$\pm$	125	&	4.5	&	2.18$\pm$0.03	&	5.1	&	&	&	SM	&	WD?	\\	
(M5951)	&	B	&	11500	$\pm$	250	&	9	&	0.0317$\pm$0.0004	&	0.0160	&	$<$0.2	&	$<$120	&	&	\\	
&	B	&	11500	$\pm$	250	&	7	&	0.0192$\pm$0.0003	&	0.0058	&	0.33$\pm$0.03	&	209$\pm$11	&	&	\\	\midrule
WOCS8006	&	A	&	6750	$\pm$	125	&	4	&	1.56$\pm$0.02	&	4.1	&	&	&	SM, BSS	&	WD?	\\	
(S2204)	&	B	&	11500	$\pm$	250	&	9	&	0.0404$\pm$0.0005	&	0.0260	&	$<$0.2	&	$<$120	&	&	\\	
&	B	&	11500	$\pm$	250	&	7	&	0.0258$\pm$0.0003	&	0.0100	&	0.22$\pm$0.02	&	139$\pm$10	&	&	\\	\midrule
WOCS9005	&	A	&	6500	$\pm$	125	&	4.5	&	1.86$\pm$0.02	&	5.6	&	&	&	BM, BSS, SB1	&	WD?	\\	
(S1005)	&	B	&	11500	$\pm$	250	&	9	&	0.0437$\pm$0.0006	&	0.0300	&	$<$0.2	&	$<$120	&	&	\\	
&	B	&	11500	$\pm$	250	&	7	&	0.0264$\pm$0.0003	&	0.0110	&	0.22$\pm$0.02	&	136$\pm$10	&	&	\\	\midrule
Y1157	&	&	16250	$\pm$	250	&	9	&	0.0122$\pm$0.0002	&	0.0094	&	0.66$\pm$0.04	&	180$\pm$17	&	&	\\	
(WD30)	&	&	17250	$\pm$	250	&	7	&	0.0117$\pm$0.0002	&	0.011	&	0.70$\pm$0.04	&	163$\pm$15	&	DB	&	\\	\midrule
Y1168,  	&	&	45000	$\pm$	5000	&	9	&	0.0152$\pm$0.0002	&	0.85	&	0.66$\pm$0.20	&	2.6$\pm$0.9	&	&	\\	
(WD15)	&	&	45000	$\pm$	5000	&	7	&	0.0152$\pm$0.0002	&	0.85	&	0.66$\pm$0.20	&	2.6$\pm$0.9	&	DA	&	\\	\midrule
Y1487	&	&	14750	$\pm$	250	&	9	&	0.0152$\pm$0.0002	&	0.0098	&	0.50$\pm$0.03	&	151$\pm$12	&	&	\\	
(WD1)	&	&	13500	$\pm$	250	&	7	&	0.0169$\pm$0.0002	&	0.0085	&	0.41$\pm$0.03	&	157$\pm$12	&	DB	&	\\	\midrule
Y563	&	&	24000	$\pm$	1000	&	9	&	0.0157$\pm$0.0002	&	0.073	&	0.52$\pm$0.08	&	21$\pm$2	&	&	\\	
(WD2)	&	&	25000	$\pm$	1000	&	7	&	0.0157$\pm$0.0002	&	0.086	&	0.52$\pm$0.07	&	18$\pm$2	&	DA	&	\\	\midrule
Y701	&	&	15750	$\pm$	250	&	9	&	0.00941$\pm$0.0001	&	0.0049	&	&	&	&	\\	
(WD9)	&	&	16250	$\pm$	250	&	7	&	0.00934$\pm$0.0001	&	0.0055	&	&	&	DA	&	Unrel	\\	\midrule
Y856	&	&	14500	$\pm$	250	&	9	&	0.0113$\pm$0.0001	&	0.0051	&	&	&	&	\\	
(WD10)	&	&	12750	$\pm$	250	&	7	&	0.0128$\pm$0.0002	&	0.0039	&	&	&	DA, DD &	Unrel	\\	\midrule
Y886	&	&	19500	$\pm$	250	&	9	&	0.0128$\pm$0.0002	&	0.021	&	0.65$\pm$0.02	&	85$\pm$4	&	&	\\	
(WD25)	&	&	19250	$\pm$	250	&	7	&	0.0133$\pm$0.0002	&	0.022	&	0.61$\pm$0.03	&	77$\pm$7	&	DA	&	\\	\bottomrule
\multicolumn{10}{c}{\textit{WD:Candidate WD, WD?: possible WD, ch: chromospheric, Sp: hot-spots, Unrel:Unreliable}} \\
\multicolumn{10}{c}{\textit{PV:Pulsating Variable, RR:Rapid Rotator, DD: Double Degenerate WD, DA/DB: WD spectral types}} \\
\multicolumn{10}{c}{\textit{BM:Binary Member, BLM:Binary Likely Member, SM:Single Member}} 
\end{tabular}
}
\end{tiny}
\end{table}

The parameters estimated from the best SED fits are listed in Table~\ref{tab:4_All_para}. All 'A' components (including Aa and Ab) are fitted with Kurucz model SEDs, while all 'B' components (except WOCS2009, a triple system, where the B component is also fitted with a Kurucz SED) are fitted with WD model SEDs suitable for hotter companions. The temperature $T_{eff}$, log $g$ and radii are obtained from SED fit parameters.
The `Comments' column in Table~\ref{tab:4_All_para} includes the comments by \citet{Geller2015} and X-ray detection identifiers from \textit{ROSAT} (X; \citealt{Belloni1998}), \textit{Chandra} (CX; \citealt{Van2004}) and \textit{XMM-Newton} observations (NX; \citealt{Mooley2015}).
In the case of WDs, the SEDs with fewer data points are noted as unreliable (Unrel).

\begin{figure}[!ht]
  \centering
  \begin{tabular}{cc}
    \includegraphics[width=0.45\textwidth]{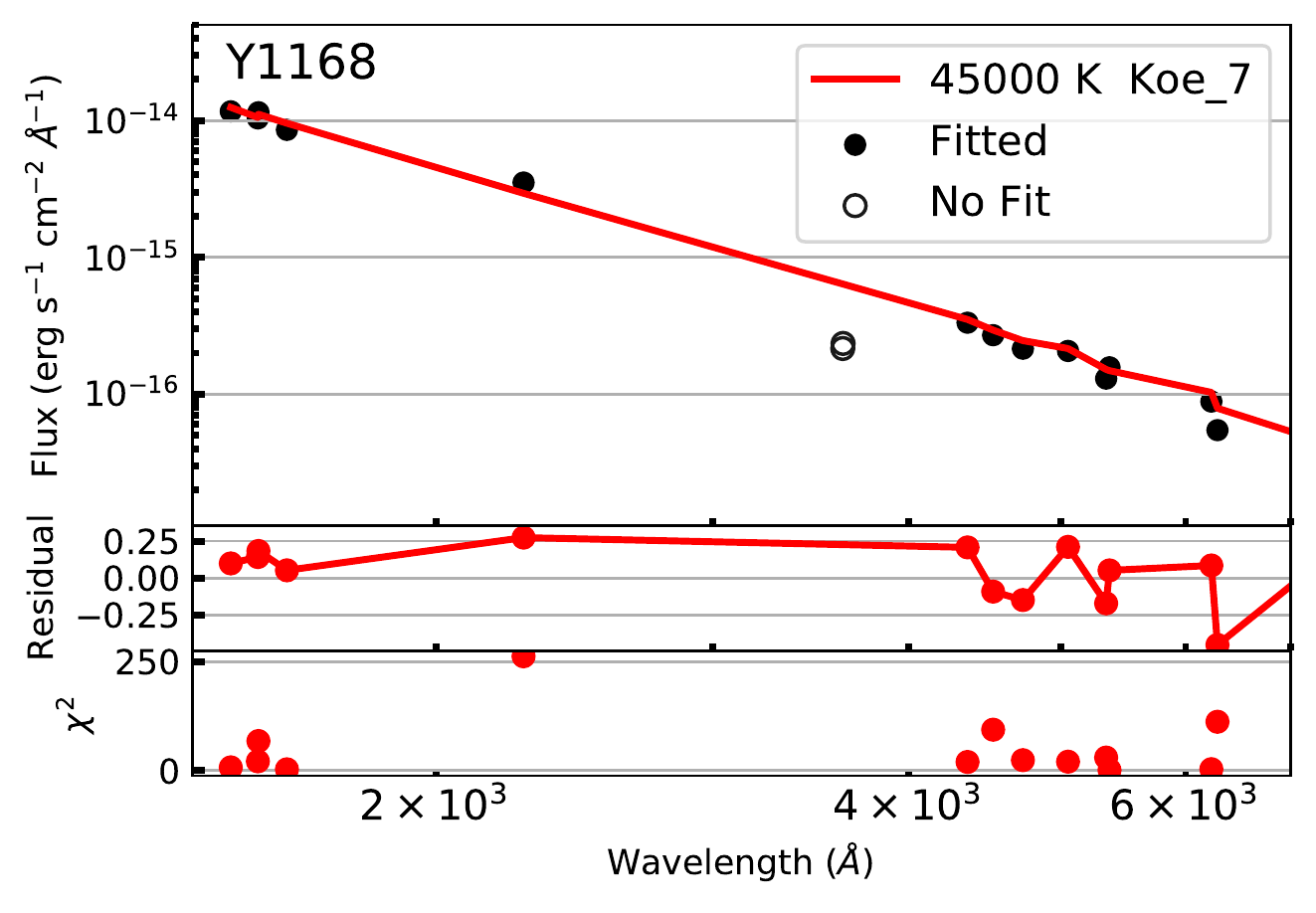}&
    \includegraphics[width=0.45\textwidth]{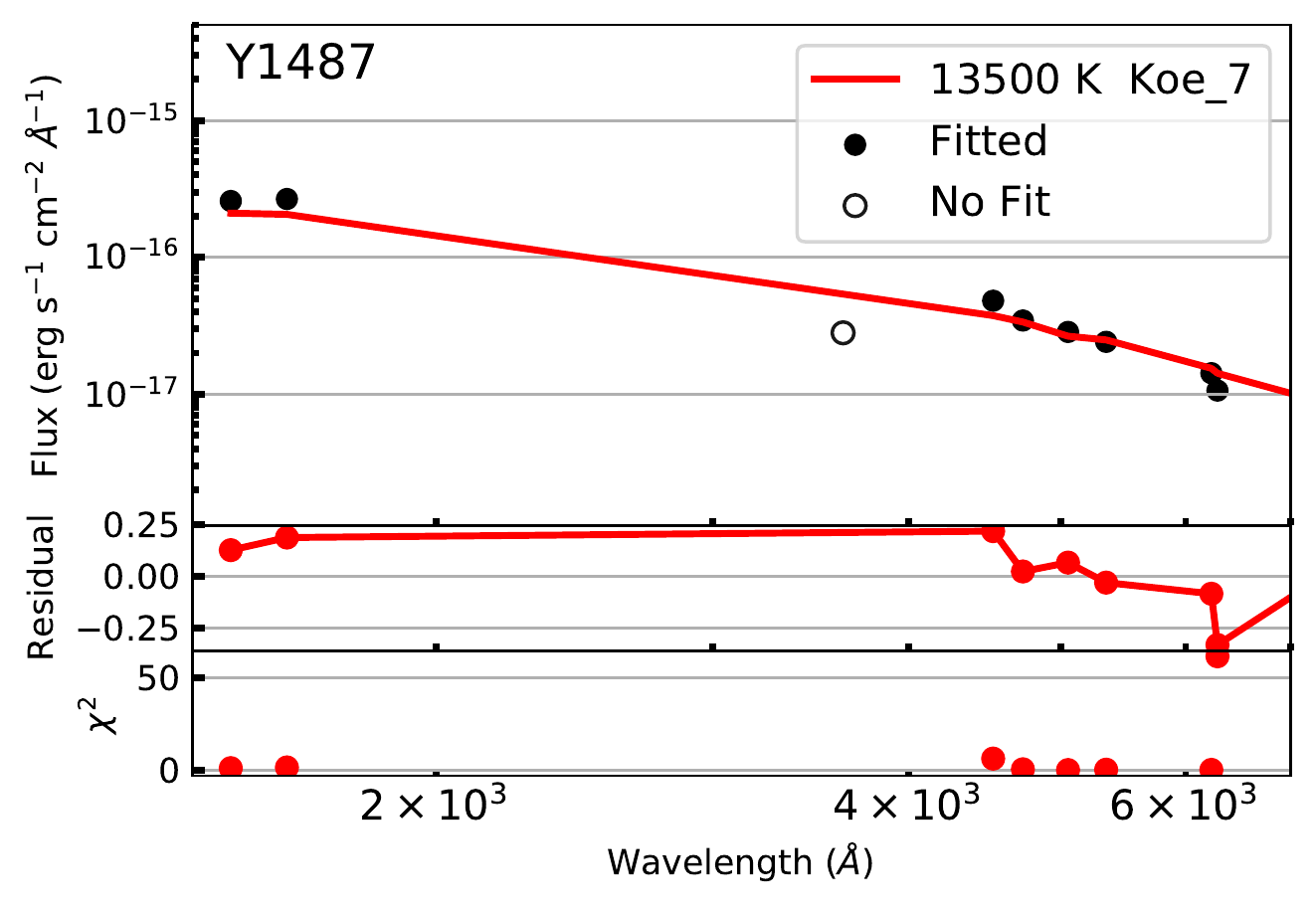} \\
    \includegraphics[width=0.45\textwidth]{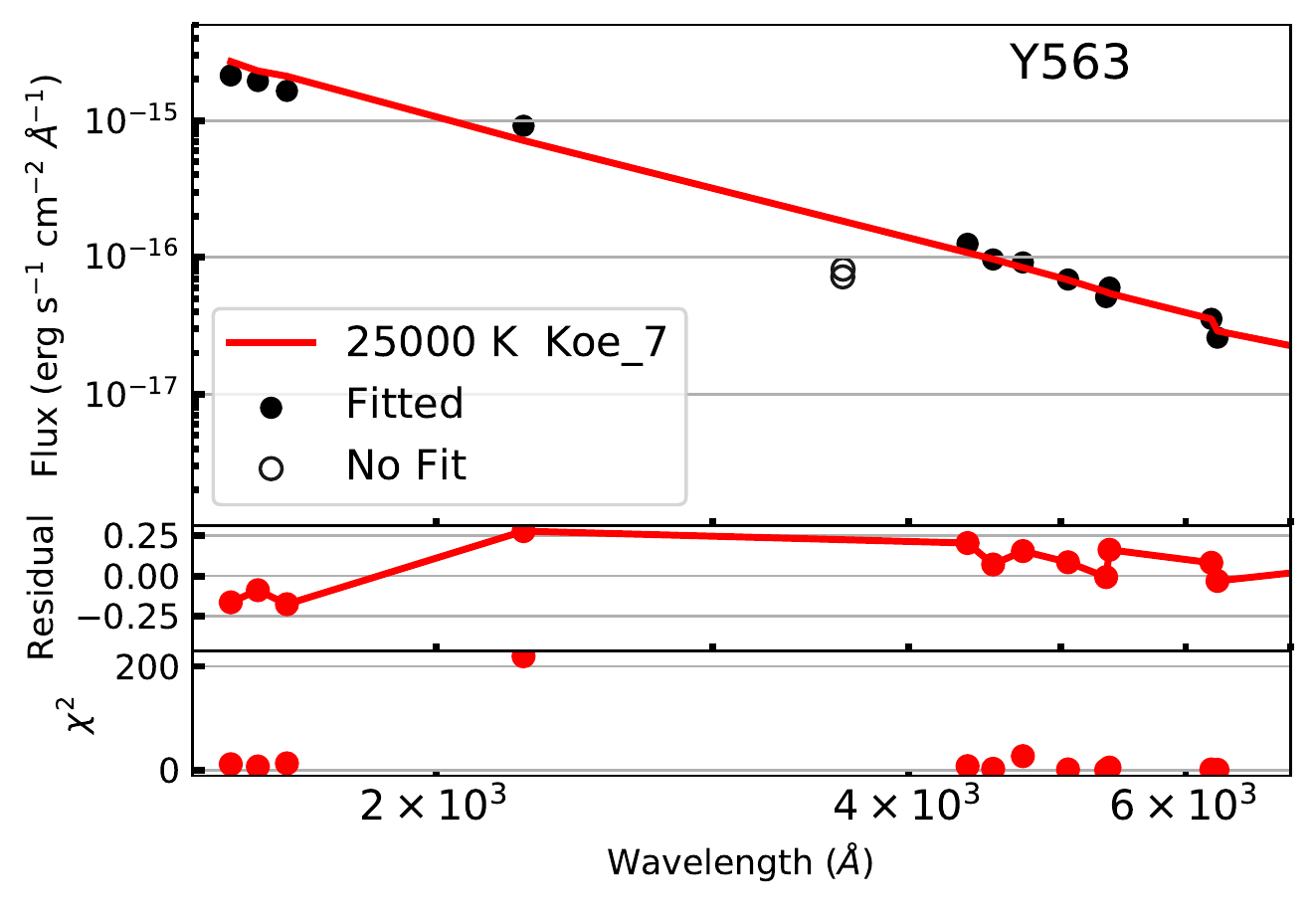} &
    \includegraphics[width=0.45\textwidth]{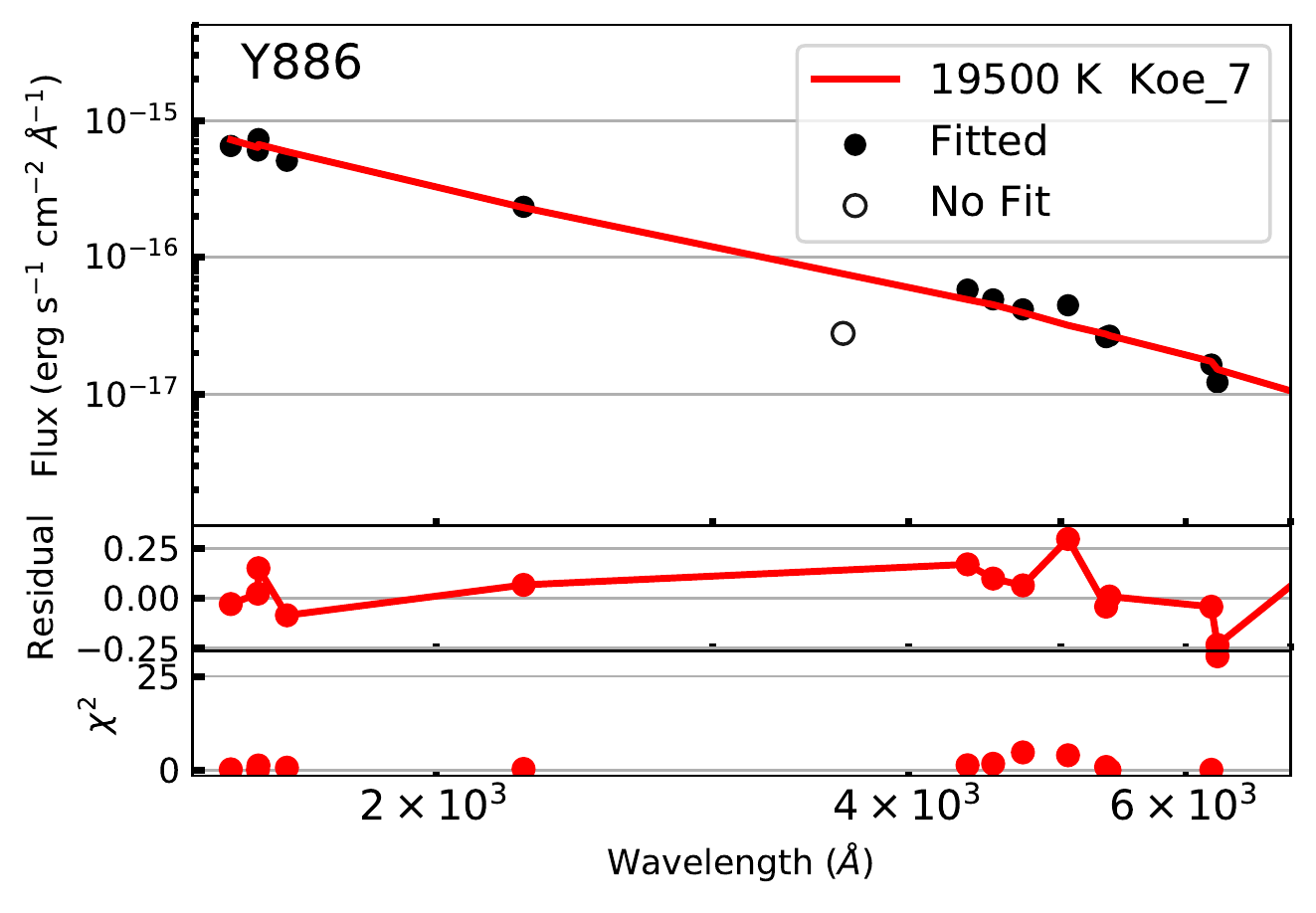} \\
  \end{tabular}
  \begin{tabular}{ccc}
    \includegraphics[width=0.3\textwidth]{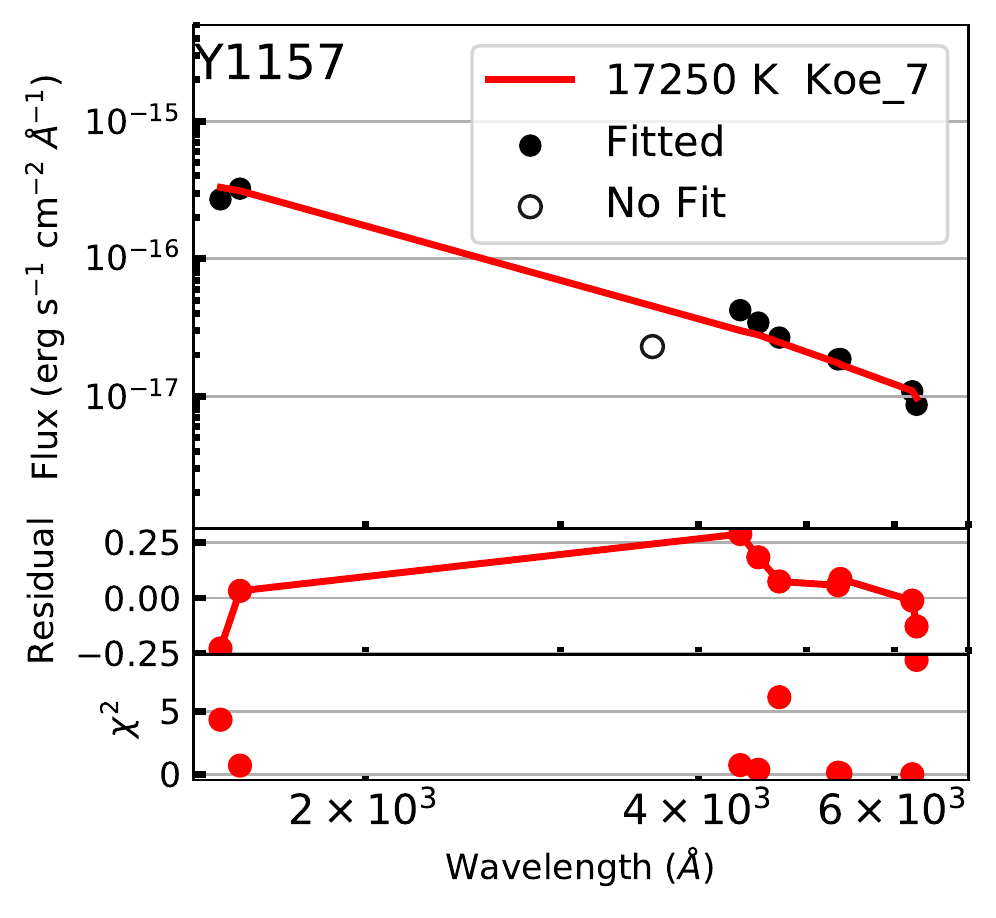} &
    \includegraphics[width=0.3\textwidth]{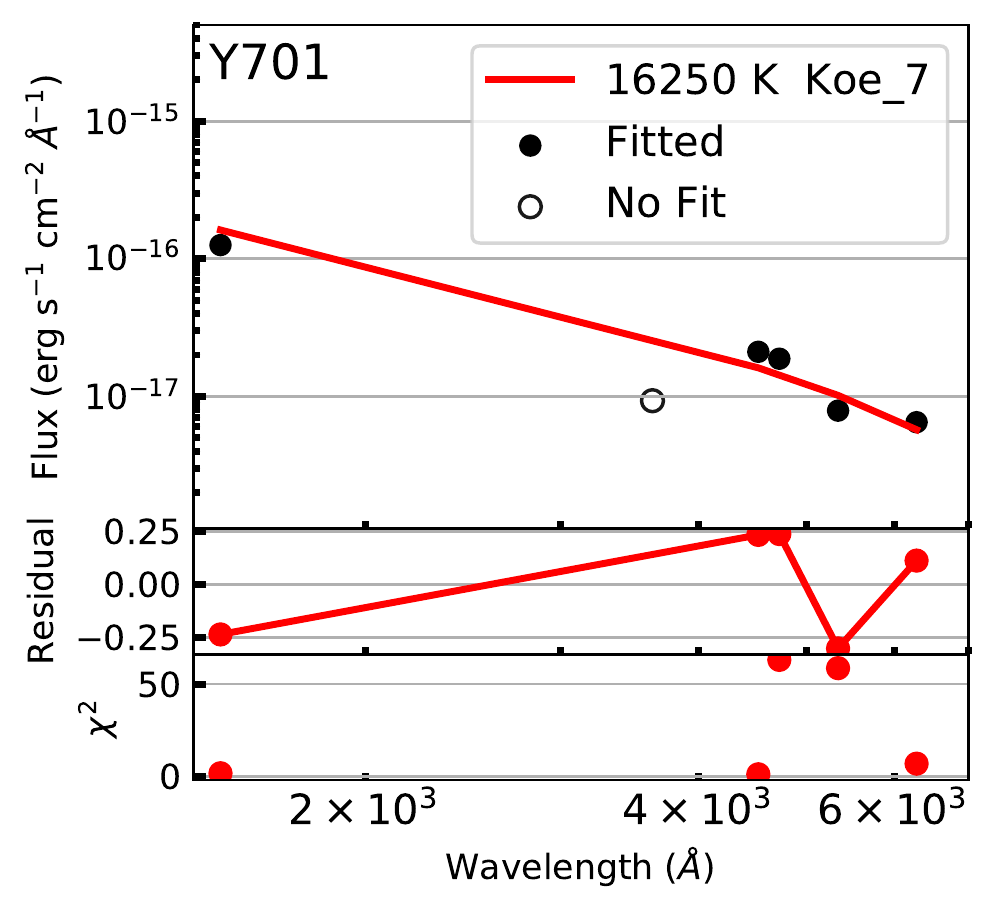} &
    \includegraphics[width=0.3\textwidth]{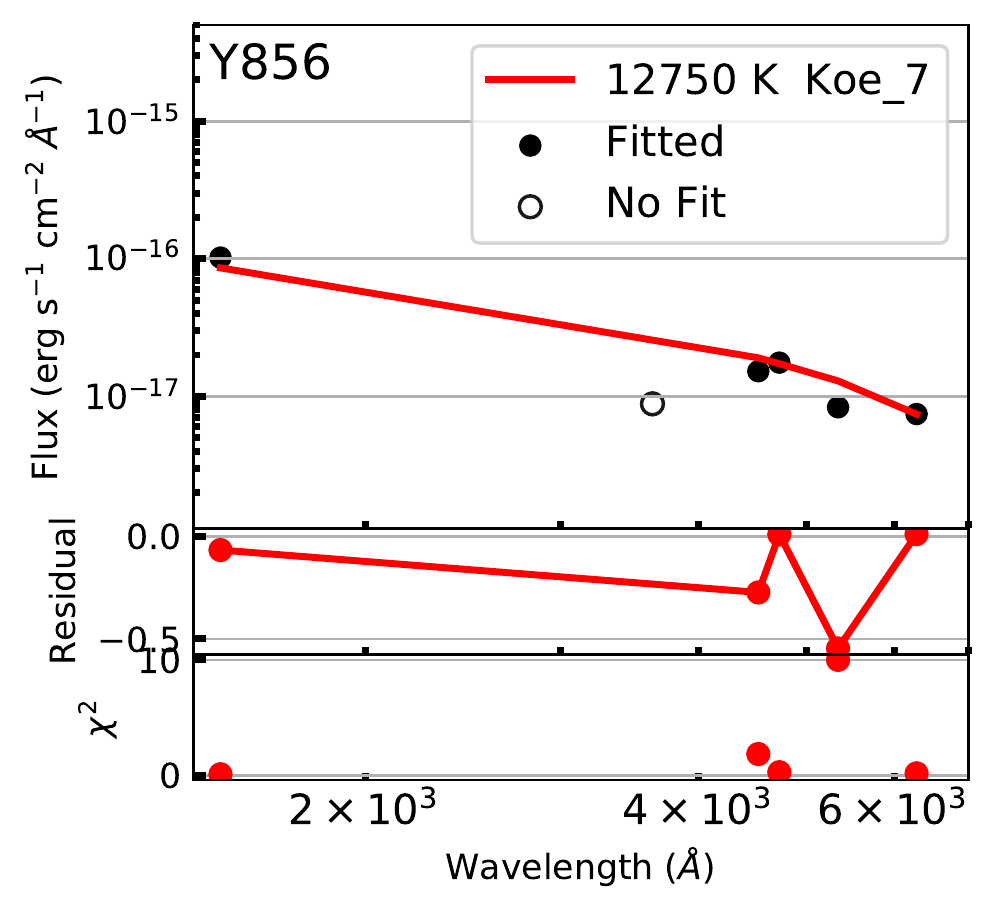} \\
  \end{tabular}
   \caption{SED fits of all isolated WDs with Koester WD model SEDs of log $g$ = 7.}
  \label{fig:4_SED_WD}
\end{figure}

The single WD fits are shown in Fig.~\ref{fig:4_SED_WD}. The stellar SEDs consisting of double fits are shown in Fig.~\ref{fig:4_SED_double_2002} and  Fig.~\ref{fig:4_SED_double_1001}--\ref{fig:4_SED_double_7010} and triple systems in Fig.~\ref{fig:4_SED_triples} in respective sections.

\subsection{Method to interpret the UV properties}
\label{sec:4_What can we say from the results}

We present the analysis of individual sources in the following manner. We compiled all relevant information from the literature (such as X-ray detections, periods, eccentricities, temperatures, and radii) and discussed the UV flux in the light of the above properties.
Then, we analysed the changes in positions of the sources in various CMDs and their implications. The results of the SED fits are summarised thereafter.

The nature of the UV flux is discussed in accordance with the classification of the source as given below:

\textbf{Comparison with WD models:} 
We compare the parameters of hotter companions to DA and He-core models in Fig.~\ref{fig:4_WD_mass_radius}. The sources that deviate from the WD models are mostly active binaries/triples: WOCS2003 (SB2, RS CVn), 2015 (likely an evolving YSS), 4003 (W Uma), and 5007 (SM). Among the stars that are not already classified, if the companion parameters deviate significantly from the models, we propose the source of the UV flux not to be a WD. 

\textbf{X-ray detection:} 
X-rays indicate the presence of some surface activity on the stars (hot spots, chromospheric or coronal activity). This can contaminate the UV flux to some degree. Thus, the residual UV flux can result from these activities or a hotter companion. Even if there is a hotter companion, the $T_{eff}$ and radius obtained via SED fitting may not be accurate. Thus, we cannot comment on the presence of a WD.

\textbf{Number of detections in the UVIT:} 
The number of data points is an essential variable in the SED fitting. Fewer UV data points lead to multiple SED fits with a relatively similar $\chi^2$. Thus, the fit parameters of stars detected in only 1 filter may not be entirely accurate and hence can only suggest the possibility of a hotter companion depending on the previously known information.

\textbf{SB2/Triples:} 
In this case, the single fits over the optical and IR points are not entirely trivial. Thus, fitting multiple-components SEDs to optical--IR part and its implications are explained in \S~\ref{sec:4_triples}.

\subsection{White dwarfs}
\label{sec:4_WD}

We detected 7 WDs in the UVIT images. All 7 WDs have photometry available from \citet{Yadav2008} and \citet{Williams2018}. Y1168, Y563, Y886, and Y1157 (star numbers from \citealt{Yadav2008}) are also detected by \citet{Montgomery1993}. 
We created SEDs using this photometry along with {\it Gaia} DR2 measurements.
The results of single Koester model SED fits are shown in Fig.~\ref{fig:4_SED_WD}. The U-band fluxes in KPNO and MMT filters were consistently lower than models; hence, they were not considered for the SED fitting. Including these points would create higher $\chi^{2}$ values with slightly lower $T_{eff}$.

A single star at the MSTO, with mass $\sim$1.3 \Msun, should evolve into a WD of $\sim$0.4 \Msun\ \citep{Cummings2018}. Thus, a young WD with mass $>$0.4 \Msun\ would require a heavier progenitor, i.e., a BSS.
As all the WDs we detected have cooling age $<$200 Myr, the mass alone could be an indicator of a possible BSS progenitor.
\citet{Williams2018} argued that a WD (WD29) of 0.7 \Msun\ should be a product of a 3 \Msun\ BSS, assuming that the WD mass from a BSS is similar to a WD produced by a single star. 
\citet{Landsman1998} estimated the number of WDs expected with a cooling age $<$60 Myr and $<$200 Myr as 8 and 25, respectively. We detect 2 and 7 WDs in the respective age ranges with UVIT. 
Assuming a core radius of 5.2' and tidal radius of 75', the UVIT FOV (radius of 14') should contain $\sim$80\% of the WDs. We are detecting $\sim$30\% of the expected WDs \citep{King1966}. The off-centre pointing of UVIT and unknown membership of WDs in literature are the reasons for the lower-than-expected WDs.
The luminosity of WDs increases with mass for a particular age (Fig.~\ref{fig:4_WD_mass_radius} (c) and (d)), which supports the higher fraction of high-mass WDs detected in this study.
We discuss each of the WD detected by UVIT below:

\textbf{Y1157 (WD30, MMJ6126):} \citet{Williams2018} identified it as DB spectral type WD. From Fig.~\ref{fig:4_WD_mass_radius} we found this to be a WD with 0.66--0.7 \Msun, formed in the past 200 Myr, demanding a BSS progenitor.

\textbf{Y1168 (WD15, MMJ5670):} 
Our estimate of $T_{eff}=$ 45000 K is considerably lower than the estimate of 68230 K by \citet{Fleming1997}. 
According to Wein's law, the peak radiation of a 60000 K star would be at 48 nm, which is not covered in our observations. Thus, the SED results, which depend on the spectral slope, may not be as accurate as the results from spectroscopy.
Our mass estimate of 0.66 \Msun\ is comparable to 0.55 \Msun\ as estimated by \citet{Williams2018}, which also demands a BSS progenitor.

\textbf{Y1487 (WD1):} \citet{Williams2018} categorised it as a DB type WD. We calculated the $T_{eff}$ as 13500--14750 K with a mass of 0.4--0.5 \Msun. The progenitor could be a BSS, as the mass is slightly greater than 0.4 \Msun.

\textbf{Y563 (WD2, MMJ5973):} \citet{Fleming1997} calculated T$_{eff}=$ 17150 K whereas our estimation is much higher, T$_{eff}=$ 24000--25000 K. In this temperature range, our observations in FUV filters is capable of producing a better estimate, as the peak of 17150 K lies at 170 nm. As we detect a rising flux in the UVIT filters, a hotter temperature estimate is preferred.
Mass calculated by \citet{Williams2018} at 0.69 \Msun\ is higher than our estimation at $\sim$0.52 \Msun. The difference in masses is likely the result of a difference in log $g$ between \citet{Williams2018} and our study. The mass again demands a BSS progenitor.

\textbf{Y701 (WD9) and Y856 (WD10):}
Both WDs were detected in one UVIT filter. The optical data points, along with a single UV data point, were not sufficient to fit the SED satisfactorily; thus, the $T_{eff}$ and radius estimations are unreliable, and therefore, we do not estimate the mass and cooling age.
\citet{Williams2018} calculated the mass of Y701 to be $\sim$0.56 \Msun. They also noted that Y856 is a possible double-degenerate WD system, thus making the single fit SED unreliable.

\textbf{Y886 (WD25, MMJ6061):} \citet{Williams2018} identified this as a DA type WD with 0.61 \Msun\ which is same as our results at 0.61--0.65 \Msun. The mass and cooling age of 85 Myr suggest that this WD also evolved from a BSS.

\begin{figure}[!hb]
  \centering
  \begin{tabular}{c c}
    \includegraphics[width=.43\textwidth]{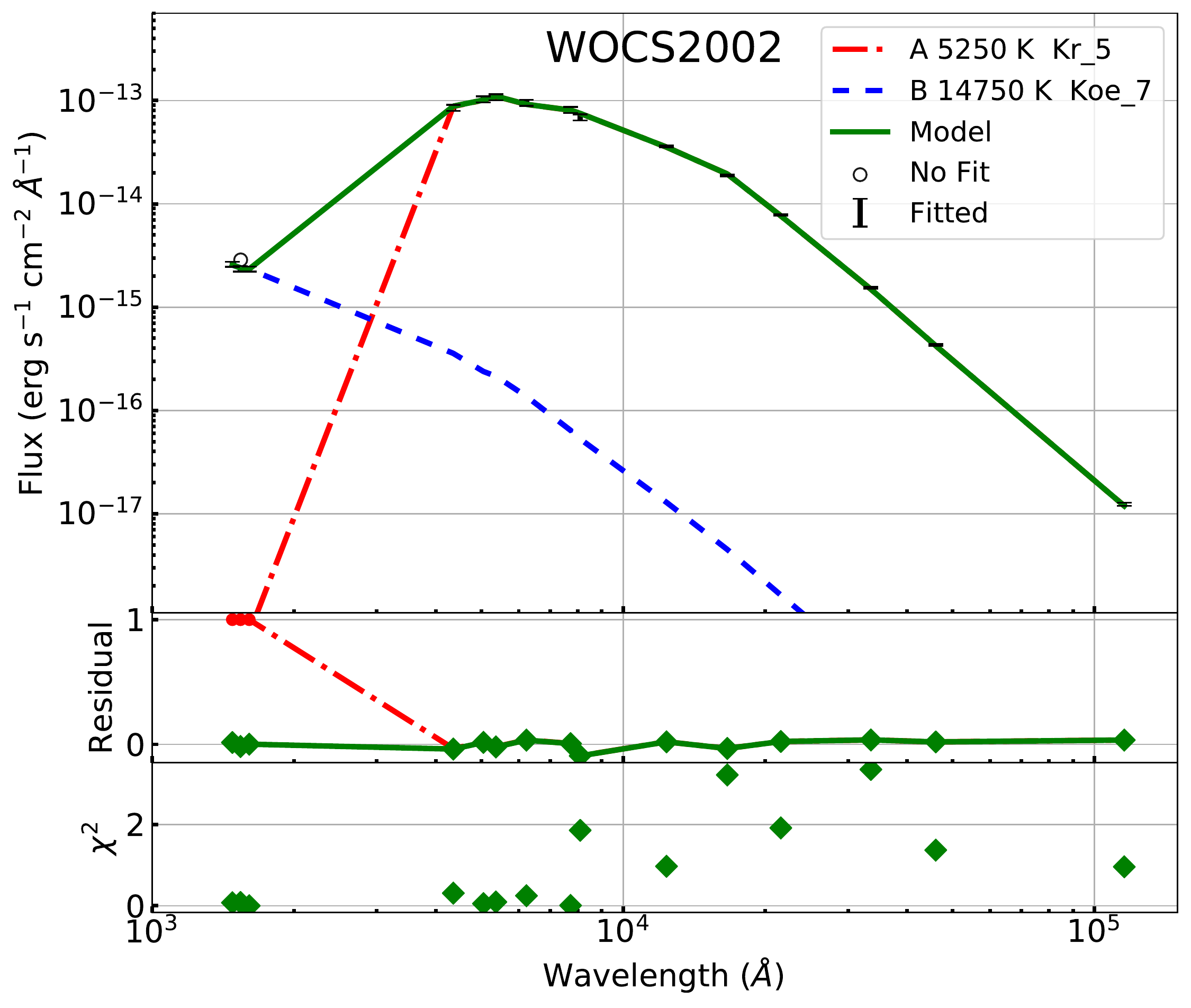} &
    \includegraphics[width=.43\textwidth]{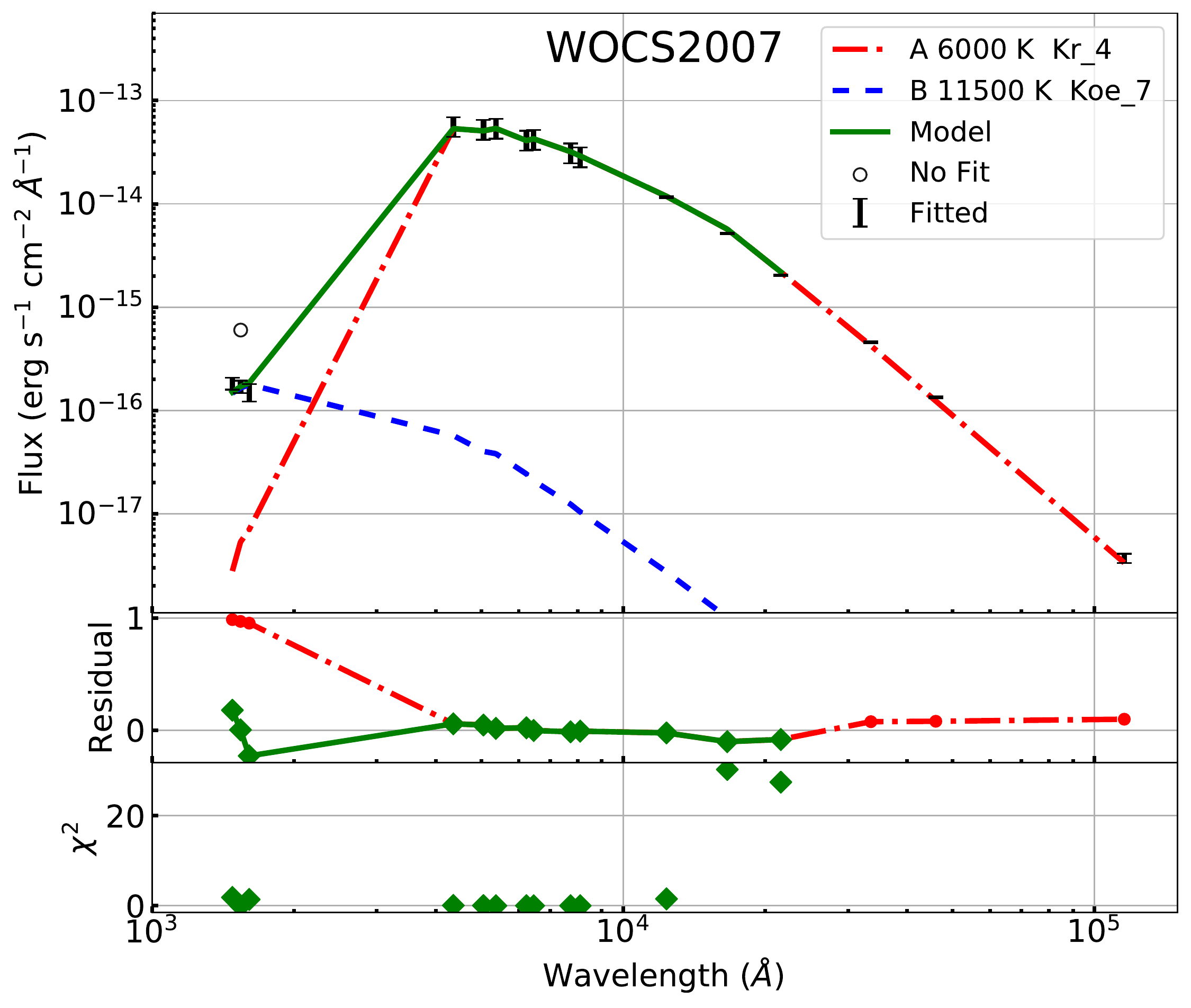} \\
    \includegraphics[width=.43\textwidth]{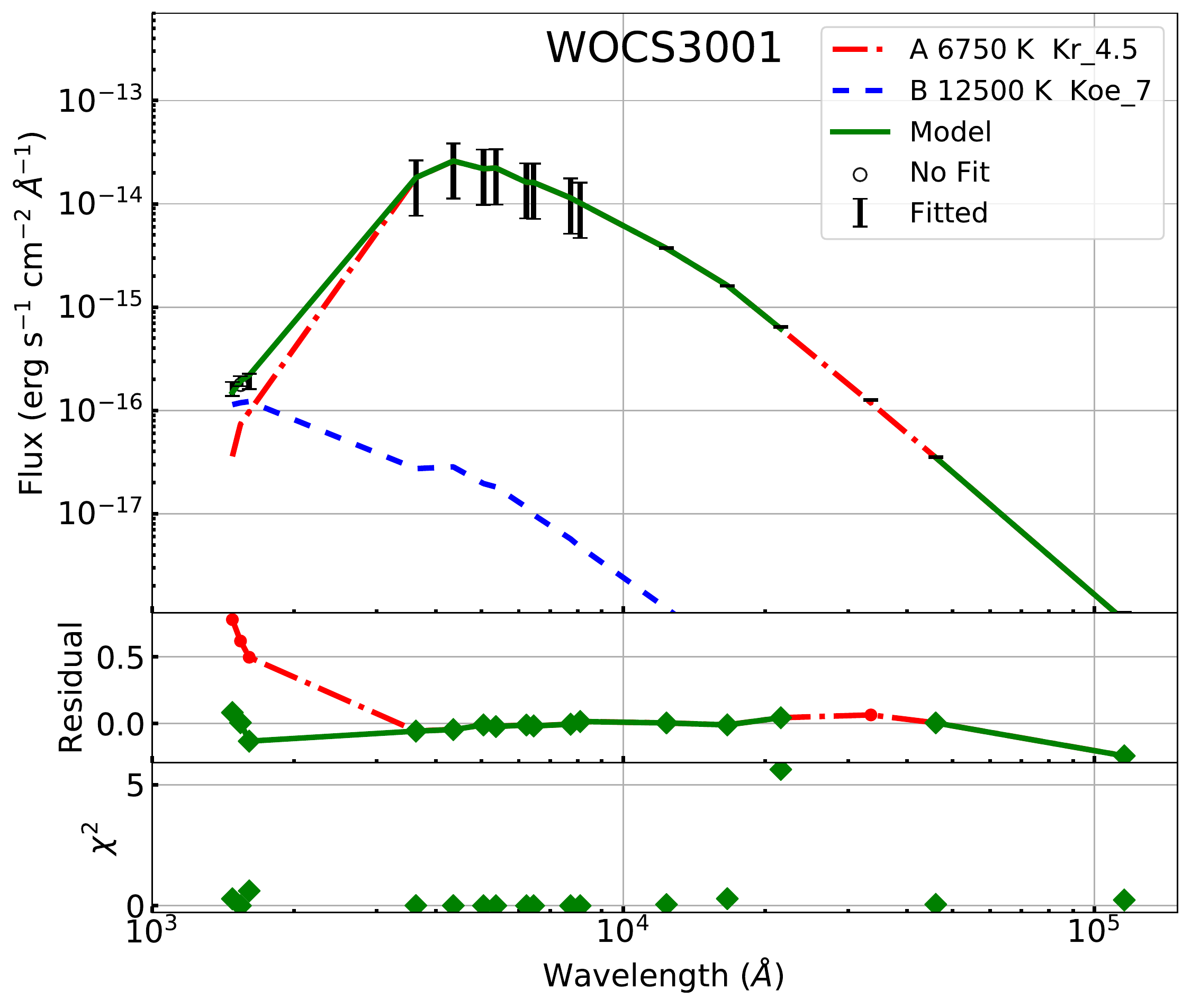}&
    \includegraphics[width=.43\textwidth]{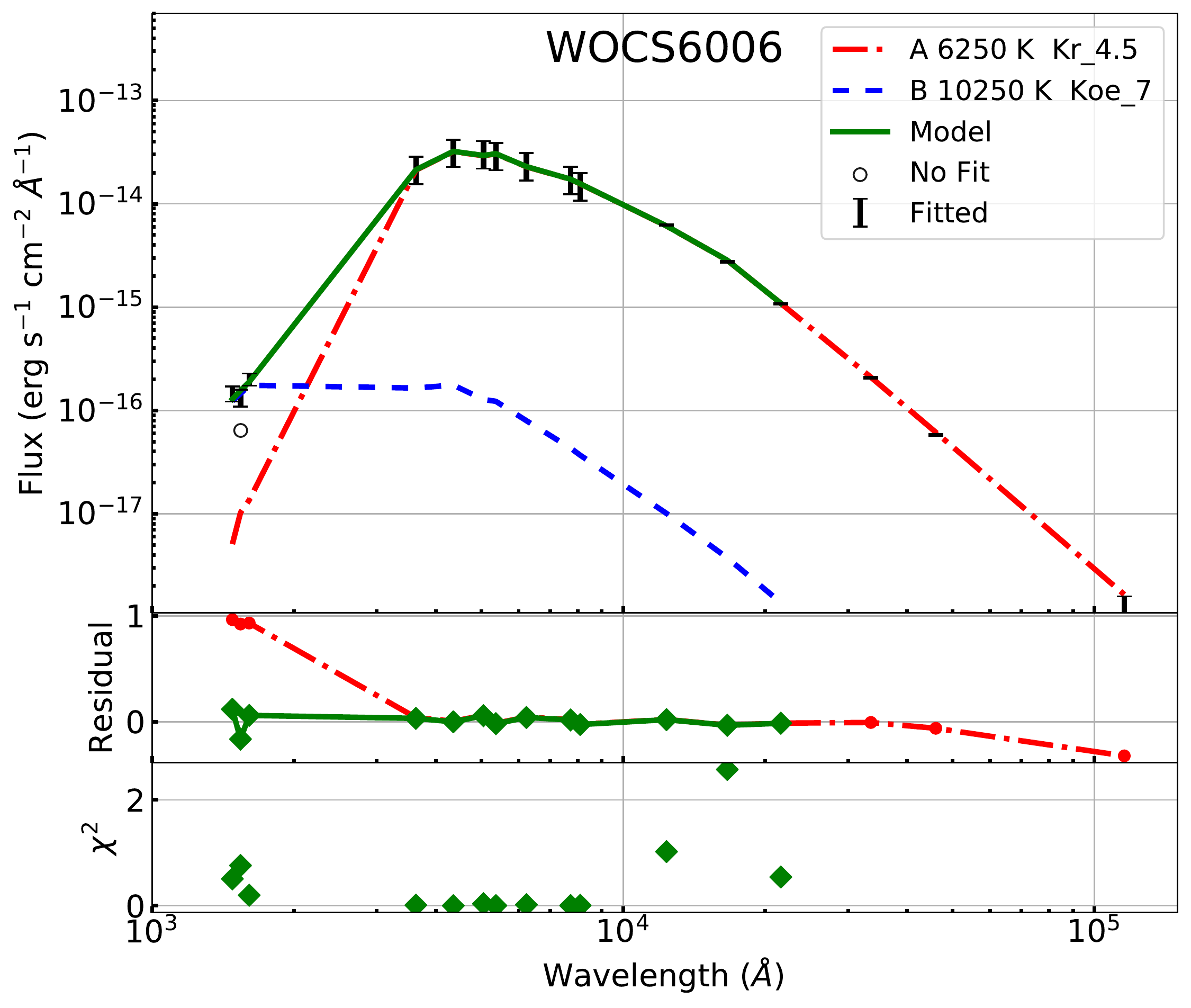} \\
  \end{tabular}
    \caption{Double SED fits of WOCS2002, WOCS2007, WOCS3001 and WOCS6006, same as \ref{fig:4_SED_3009}c} 
  \label{fig:4_SED_double_2002}
\end{figure}

\subsection{WOCS2002/S1040}
\label{sec:4_WOCS2002}
WOCS2002 contains a YSS and a WD with a circularized orbit of $P=$ 42.83 days and $e=0.027 \pm 0.028$ \citep{Latham1992}. \citet{Belloni1998} detected the star in X-rays and suggested that the X-ray emissions are due to chromospheric activity, as the WD is too cold to produce X-rays. \citet{Mooley2015} found this to be a variable in X-rays. \citet{Landsman1997} studied the source in detail using the Goddard High-Resolution Spectrograph (GHRS) and the Faint Object Spectrograph on the \textit{Hubble Space Telescope} and estimated $T_{eff}=16160 K$, log $g$ = 6.7 and M $\sim$ 0.22 \Msun. They proposed an MT scenario where the donor in a short-period ($\sim$2 days) binary began MT while on the lower giant branch, leading to the formation of a longer-period BSS + helium WD binary. \citet{Van2004} stated that their spectral fits indicate that the source of X-ray flux is coronal.

We reproduced the characteristics ($T_{eff}$, mass, and luminosity) of the WD companion to WOCS2002 using SED fits for two log $g$ values (Fig.~\ref{fig:4_SED_double_2002}). Our calculations suggest the WD component of WOCS2002 has $T_{eff}=14750$--19250 K. which is in agreement with the estimate of \citet{Landsman1997}.
We estimate the cooling age as 25--110 Myr, which suggests that the WD has formed recently. The estimated mass of 0.2--0.3 \Msun\ indicates that it is an ELM WD formed due to interaction with its close companion. As the system demands MT from the WD progenitor to the YSS progenitor, the present YSS is likely to be an evolving BSS.

\subsection{WOCS2007/S984}
\label{sec:4_WOCS2007}
WOCS2007 is classified as a BSS and an SM \citep{Geller2015}.
\citet{Shetrone2000} observed the RV variations consistent with a circular orbit with P = 1.5 days and suggested the possibility of it being a non-interacting close binary. 
They detected Li abundance similar to a single MSTO star and hence concluded that it is most likely a result of a collisional merger and has a subluminous companion.
\citet{Sandquist2003b} found no variability in the light curve. \citet{Bertelli2018} calculated the temperature of the star as $T_{eff}=6118 K$ and rotational velocity as $v$ sin $i=8 km s^{-1}$ and suggested that it is a long-period binary.

We detect significant and consistent UV excess in all three UVIT filters, while the {\it GALEX} FUV flux was found to be higher than UVIT estimates.  
The SED fits suggest the presence of a $<$150 Myr old 11500 K hotter component along with a 6000 K cooler component with luminosities of 0.01 and 8 \Lsun\ respectively (Fig.~\ref{fig:4_SED_double_2002}).

The parameters of the hot component are consistent with the WD models, and it does not have X-ray emissions. Therefore, this is likely to be a BSS + WD system. As the WD mass is $<$0.24 \Msun, it should be an ELM WD that has undergone MT.

The surface Li is reduced by both merger \citep{Lombardi2002} and MT \citep{Hobbs1991}. \citet{Shetrone2000} suggested that the high Li observed in BSS supports its formation via merger instead of MT, but the mass of the companion WD strongly indicates MT. The high Li abundance found in WOCS2007 is indeed puzzling for either merger or MT formation. The study of Li abundance variation in stellar interactions is beyond the scope of this study, but WOCS2007 could be an interesting case study for tracing the chemical signatures of BSS formation.

\subsection{WOCS3001/S1031}
\label{sec:4_WOCS3001}
\citet{Geller2015} listed this system as an SB1 binary member. It is bluer than the MS in the optical CMD. \citet{Leiner2019ApJ...881...47L} calculated $v$ sin $i=14.7 km s^{-1}$, $P=128.14$ days, $e=0.04$ and a binary mass function = 0.0143. \citet{Leiner2019ApJ...881...47L} suggested that the system is formed through MT as inferred from its rapid rotation and circularised orbit.

We detect a small amount of excess UV flux in all the 3 filters of UVIT. We could fit a cooler companion (6750 K) and a hotter companion (12500 K) of 0.3--0.45 \Msun\ (Fig.~\ref{fig:4_SED_double_2002}). 

The hotter component parameters are consistent with WD models. The mass of the possible WD is also compatible with the binary mass function estimated by \citet{Leiner2019ApJ...881...47L}. The absence of X-ray suggests a chromospherically inactive cooler component. Thus, we propose that the UV flux is indeed the result of a WD companion to WOCS3001. Although the 0.3--0.45 \Msun\ WD may or may not require MT, the circularised orbit and rapid rotation are tracers of MT in close binaries.

\subsection{WOCS6006/S1271}
\label{sec:4_WOCS6006}
This single member \citep{Geller2015} was observed in all three UVIT filters. It shifts from MSTO in the optical CMD to the beginning of the BSS sequence in the UV--optical CMD. \citet{Bertelli2018} found T$_{eff}=$ 6360 K.

The large UV flux in all three filters results in a good double fit with a cooler (6500 K) and a hotter (10250--11250 K) component (Fig.~\ref{fig:4_SED_double_2002}). According to the fitted models, there is a minor IR deficiency in \textit{WISE}.W2 and \textit{WISE}.W3 filters.

The hotter companion's parameters match well with the model WD parameters in Fig.~\ref{fig:4_WD_mass_radius}. Without any X-ray detection, the large UV flux leads us to conclude that a WD companion of $<$0.2 \Msun\ to WOCS6006. The presence of an ELM WD signals an MT between the two components, where one component has evolved into an ELM WD, while the other has remained on the MS instead of jumping to the BSS sequence. Therefore, this is a post-MT system and a BL candidate \citep{Leiner2019ApJ...881...47L}. 

\subsection{Other members}
\label{sec:4_other_members}
We also analysed 7 other UVIT detected members (excluding WOCS2002) that are also detected in X-rays. The X-ray-emitting phenomenon may or may not contaminate the UV flux, but we do not confirm any WD companion to these sources as a precaution. 
Among these, WOCS2009 is a triple system, and WOCS7009 is a suspected triple system. We characterised the third component based on the known flux of two components.

Similarly, there are 12 sources with no X-ray detection and 1 or 2 detections in UVIT filters. As fewer data points reduce the significance of SED fits, the parameters derived from the SED fits may not be definitive. Thus, we only suggest the possibility of WD companions to these stars.

The SEDs and detailed discussions of these 19 sources are as follows. The parameters of these fits are listed in Table~\ref{tab:4_All_para}.

\begin{figure}[!ht]
  \centering
  \begin{tabular}{c c}
    \includegraphics[width=.43\textwidth]{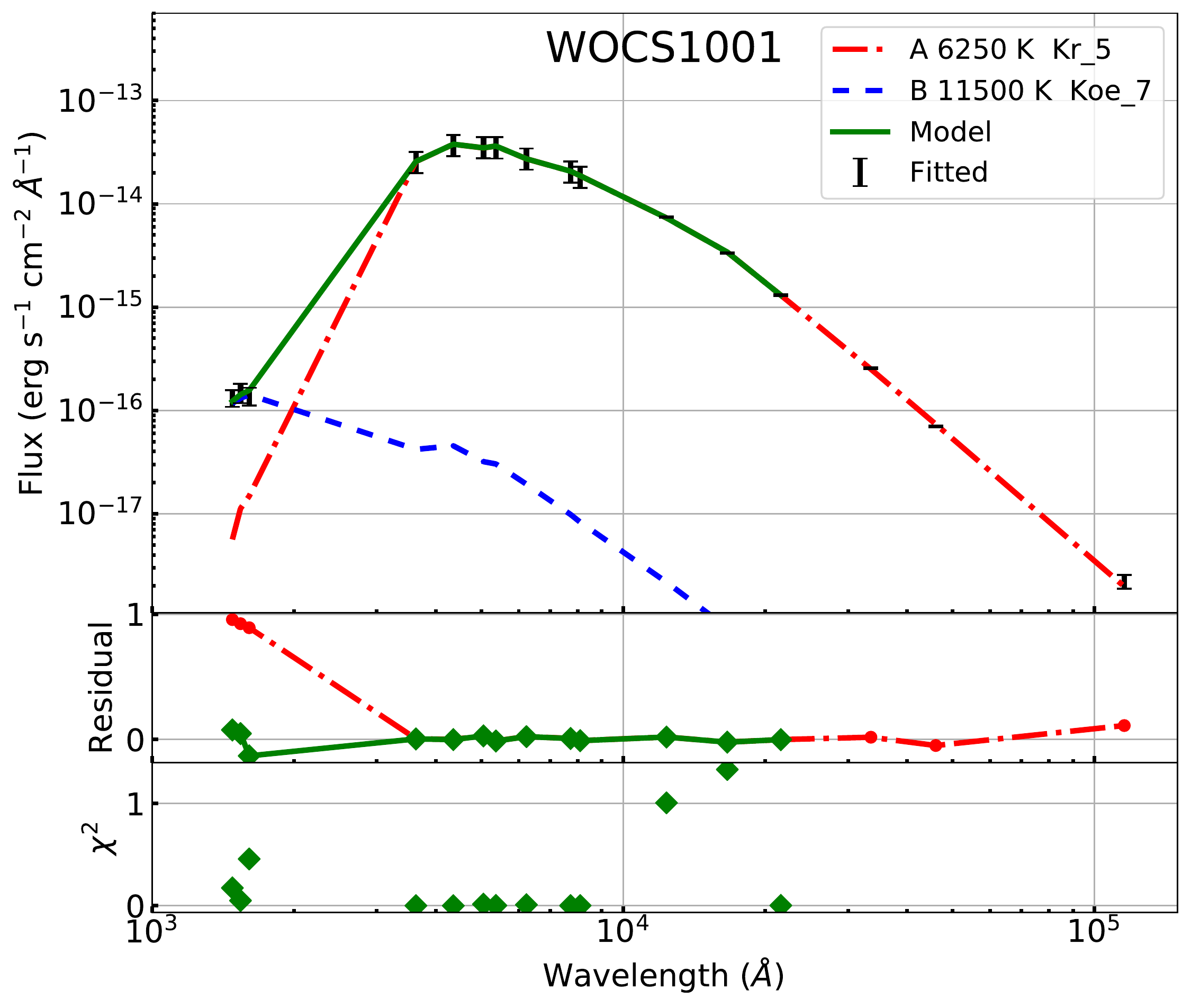} &
    \includegraphics[width=.43\textwidth]{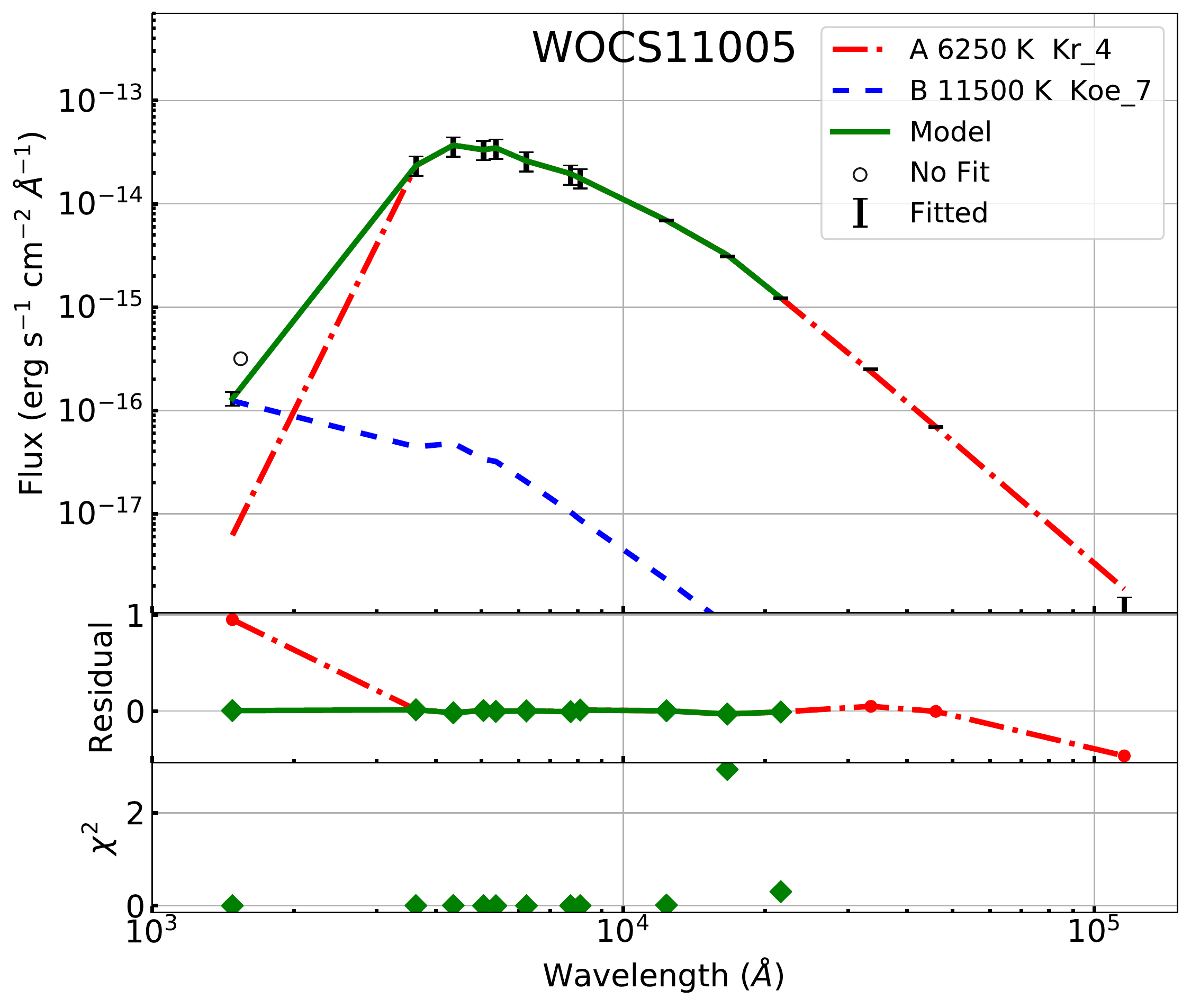} \\
    \includegraphics[width=.43\textwidth]{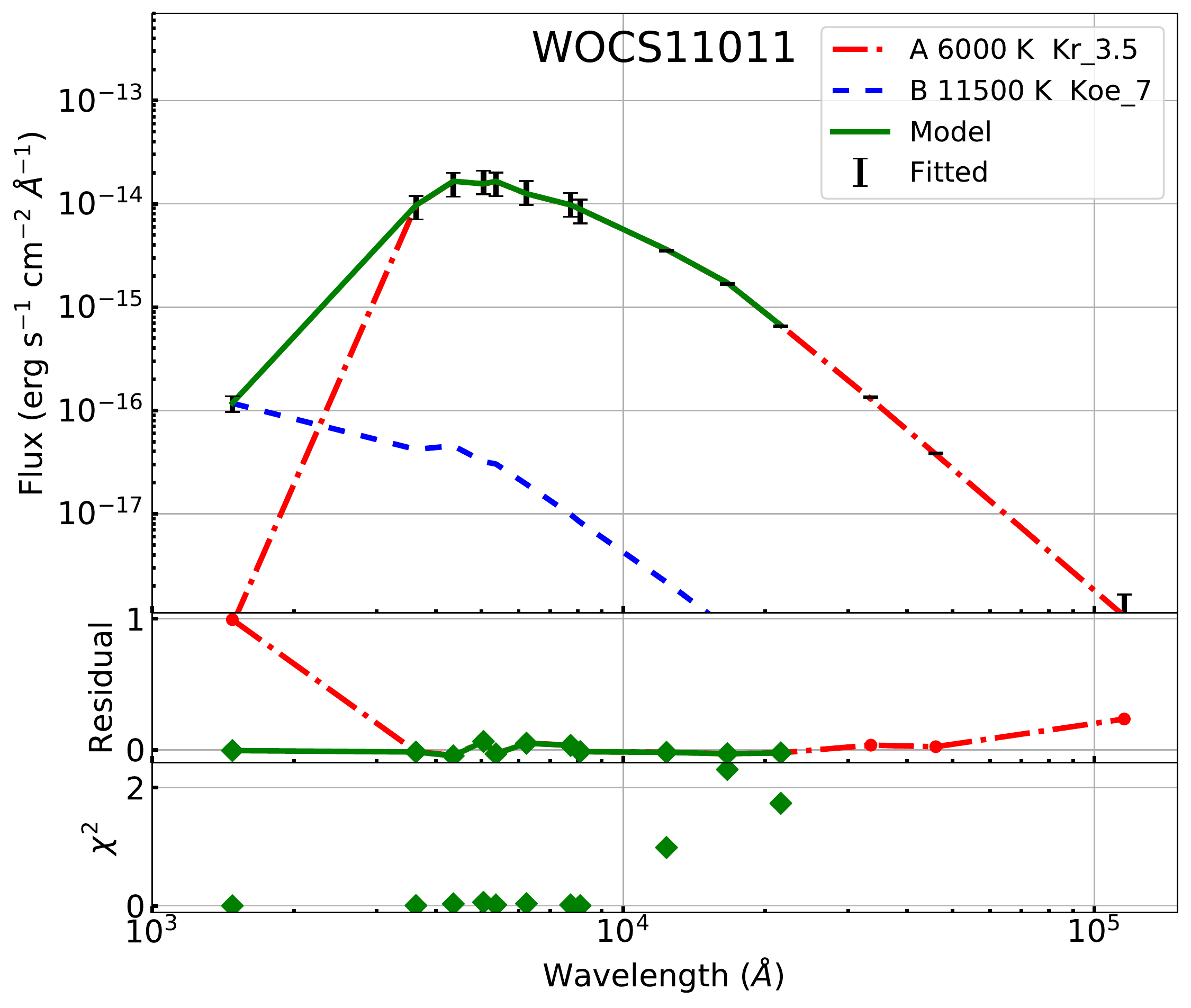} &
    \includegraphics[width=.43\textwidth]{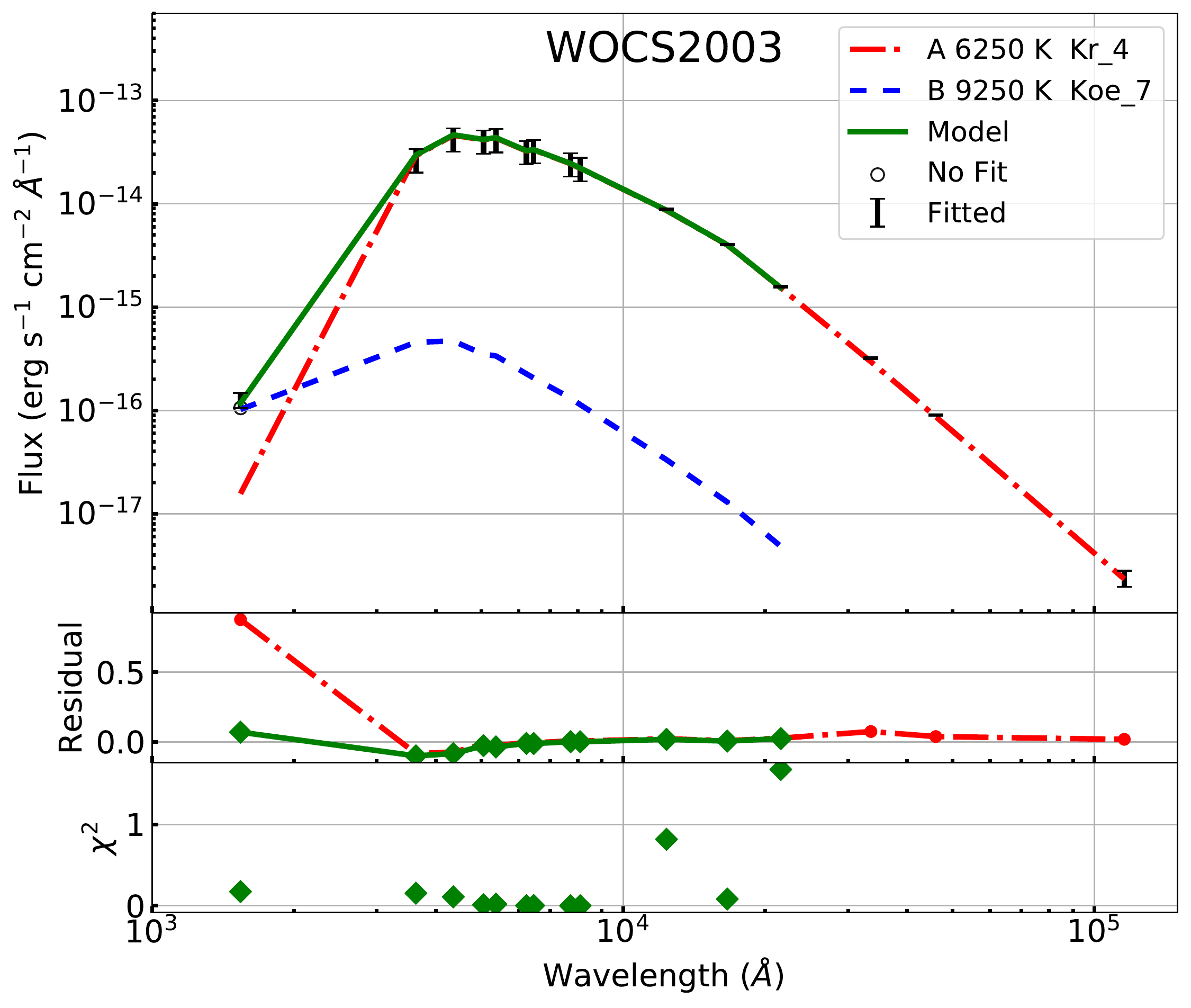} \\
  \end{tabular}
    \caption{Double SED fits of WOCS1001, WOCS11005, WOCS11011 and WOCS2003} 
  \label{fig:4_SED_double_1001}
\end{figure}

\subsubsection{WOCS1001/S1024}
\label{sec:4_WOCS1001}

This is an SB2 source with a period $P=$ 7.15961 days and eccentricity $e=$ 0.005$\pm$0.005 \citep{Latham1992}. The source was not detected in X-rays by \citet{Belloni1998}. \citet{Van2004} identified this star as the counterpart of CX111/X46. Hence, this star has X-ray emissions. \citet{Mathieu1990} find this to be a double-lined spectroscopic binary with nearly identical stars of mass 1.18 \Msun. \citet{Yakut2009} detected amplitude variations in its light curve. The orbital period and extremely circular orbit strongly suggest the possibility of an MT event in the system's past.

In the CMDs, the star shifted its location from near MSTO (Fig.~\ref{fig:4_CMD} (a)) to the beginning of the BSS branch (Fig.~\ref{fig:4_CMD} (b)) of the isochrone, with an FUV excess of about 3 mag. We calculated large $\chi^2$ for the single fit. This, along with the consistent UV flux from all three UVIT filters, led us to perform a double-component fit of one hotter (11500 K) and one cooler component (6250 K). Fig.~\ref{fig:4_SED_double_1001} shows the resultant SED fit for the Koester model hotter component with log $g$ = 7. 
We can establish one hotter and one cooler component from the SED fit. The hotter component of 11500 K is compatible with WD models in Fig.~\ref{fig:4_WD_mass_radius}. The luminosities are 5 and 0.02 \Lsun\ for cooler and hotter components, respectively.

Though the results are compatible with the presence of an optically subluminous WD companion to WOCS1001, the presence of X-ray flux and the precisely known period suggest that the UV source is not a third hotter component. Although the estimated $T_{eff}$ from the SED (11500K) is relatively high for a chromospheric activity, it is possible that there could be MT between the two stars creating a hot spot. Then, the UV and the X-ray emission would be from the hot spot on one of the binaries. It is also possible that there is a hot corona for this pair, and the detected emission could be due to coronal activity, which is usually seen in contact binaries \citep{Brickhouse1998}.

\subsubsection{WOCS11005/S995}
\label{sec:4_WOCS11005}
It is described as a single member of NGC 2682 \citep{Geller2015}. It lies near the MSTO in the optical CMD and near the beginning of the BSS sequence in the UV--optical CMD. \citet{Melo2001} estimated a slow rotation of $v$ sin $i=4.9km s^{-1}$.

The source was detected in one UVIT filter (F148W). The {\it GALEX}  FUV flux is more or less consistent with the F148W flux from the UVIT and provides support to the UVIT detection. The parameters of the hotter companion lie within the predicted values of WD models in Fig.~\ref{fig:4_WD_mass_radius}. 

The absence of detection in X-ray suggests minimal chromospheric activity, therefore, the high UV flux source could be a possible WD.
We cannot confirm the presence of a WD due to only one detection in the UVIT; thus, it is noted as `WD?' in Table~\ref{tab:4_All_para}. \footnote{After deeper observations in 2018, \citet{Subramaniam2020JApA...41...45S} refitted the star using F148W, F169M and \textit{GALEX} FUV filters. The refit resulted in an ELM WD with 12500 K, 1.29$\pm$1.16 Gyr and 0.27$\pm$0.01 \Msun\ making the primary a BL.}

\subsubsection{WOCS11011/S757}
\label{sec:4_WOCS11011}
\citet{Van2004} commented that the X-ray luminosity of the star is the result of coronal activity and is comparable to other known contact binaries of similar colour. The source is also observed by \citet{Mooley2015} in X-ray commenting that it could be a W Uma type source. \citet{Geller2015} listed this source as SB1, HS Cnc, RR, W Uma, and PV. The source lies well below the MSTO in the optical CMD and is the faintest optical source detected by UVIT. 

We detected the source in only the F148W filter in UVIT. 
A single SED fit over the optical and IR region results in a star with $T_{eff}=6000$ K and shows large UV excess flux (Fig.~\ref{fig:4_SED_double_1001}).
The double-component fit suggested a hotter component of 11500 K to compensate for the UV flux.

The hotter companion's parameters are compatible with low-mass WDs (Fig.~\ref{fig:4_WD_mass_radius}). The active nature of the binary could be the reason for the excess flux in the UV and X-ray. We cannot confirm the presence of a hotter WD component.

\subsubsection{WOCS2003/S1045}    
\label{sec:4_WOCS2003}
This source is very similar to WOCS1001/S1024 in terms of binary properties and mass of $>$1.18 \Msun\ for each component of the SB2 system \citep{Mathieu1990}. \citet{Belloni1998} detected this RS CVn system in X-rays with $P=$ 7.65 days and $e = 0.007 \pm 0.005$ \citep{Latham1992}. They expected the system to be chromospherically active. \citet{Geller2015} described it as SB2 and PV.

We detected the source in only the F154W filter and fitted the SED (Fig.~\ref{fig:4_SED_double_1001}) with one cooler and one hotter companion. We estimated the $T_{eff}$ = 6250 K for the cooler component and 9250--10000 K for the hotter component with very low $\chi^2_{red}$. 
We suggest that this source is a binary with similar temperature stars, and the excess flux in UV can be the result of chromospheric/coronal activity or hot-spot on the RS CVn system.

\begin{figure}[!ht]
  \centering
  \begin{tabular}{c c}
    \includegraphics[width=.43\textwidth]{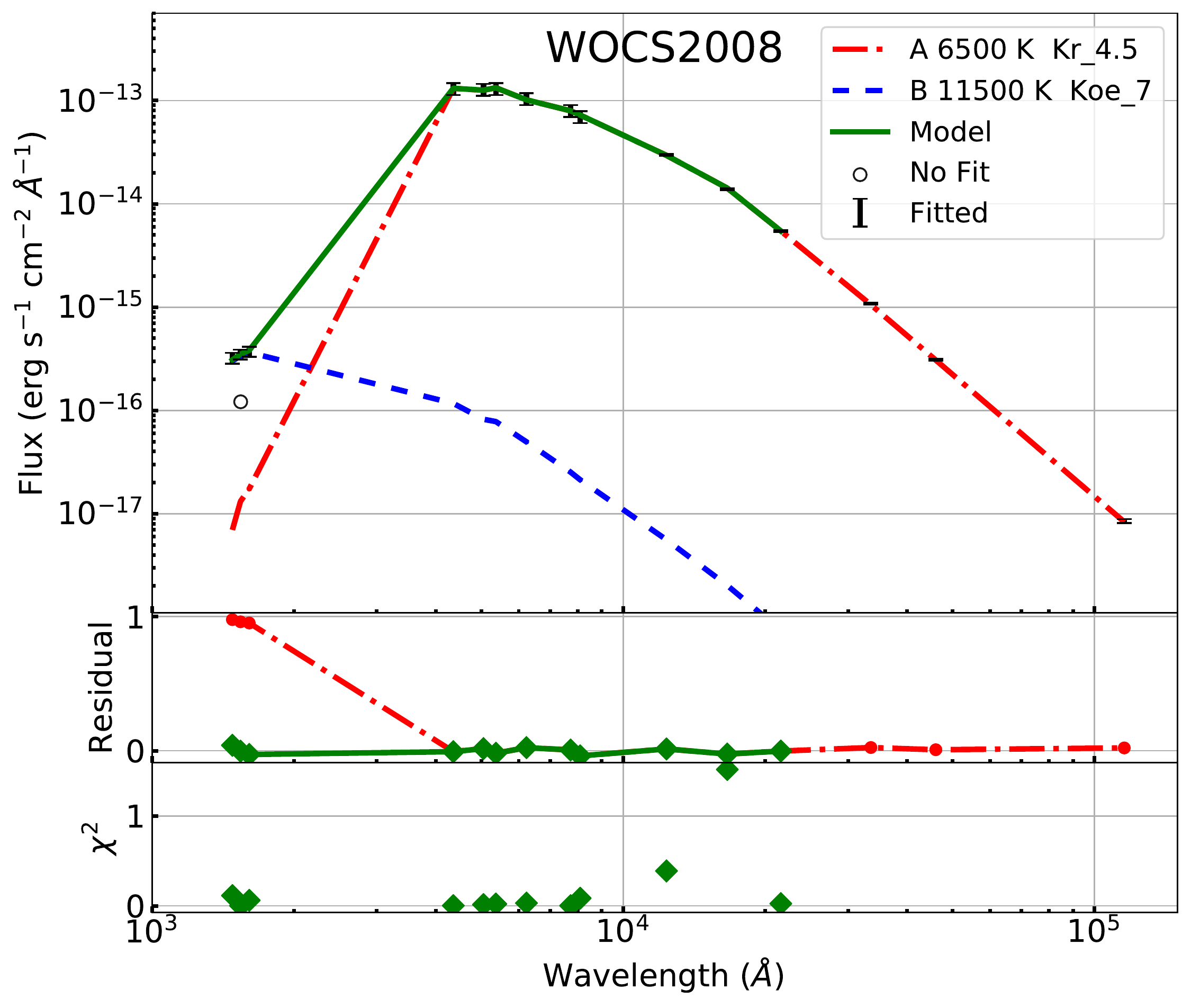} &
    \includegraphics[width=.43\textwidth]{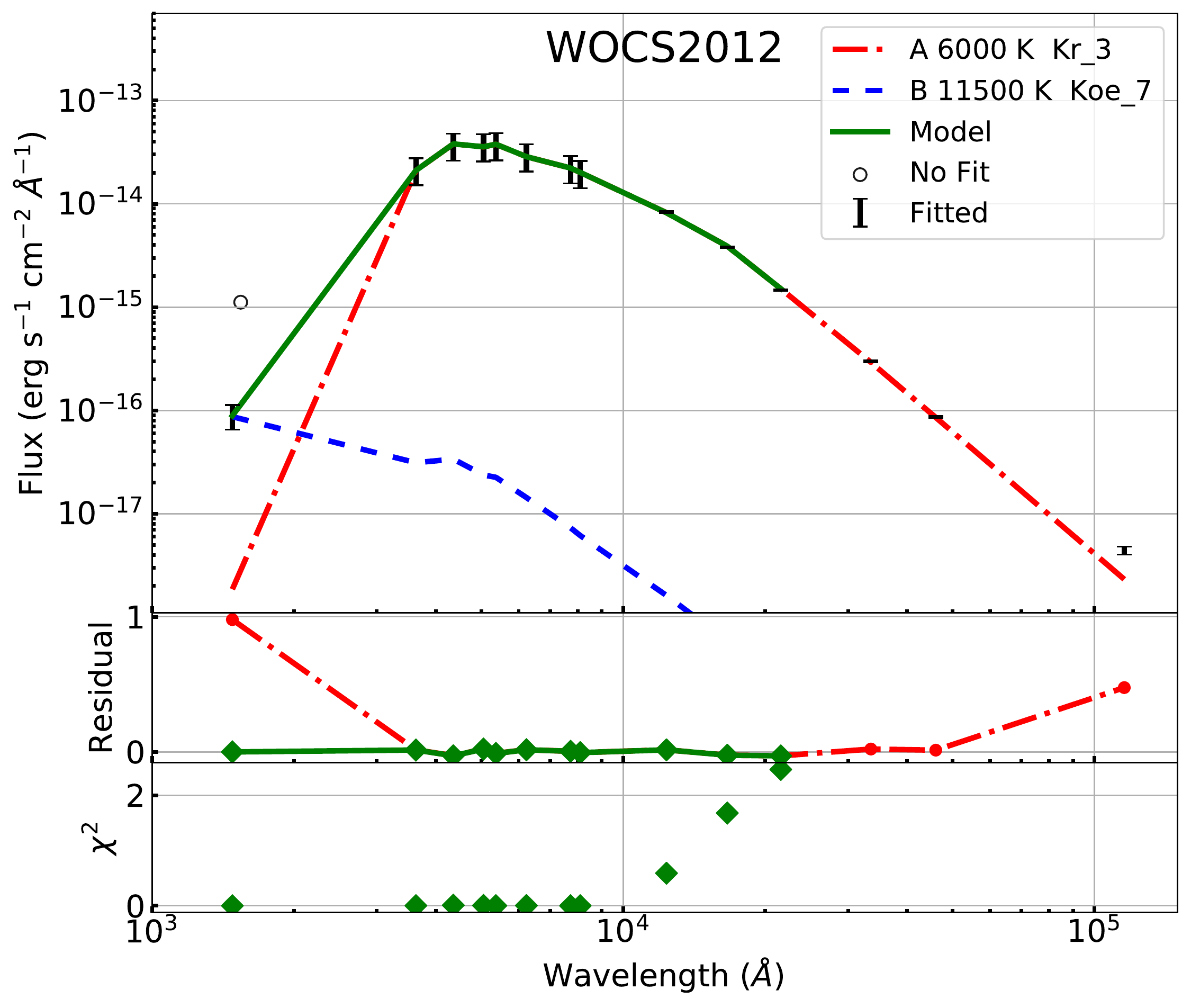} \\
     \includegraphics[width=.43\textwidth]{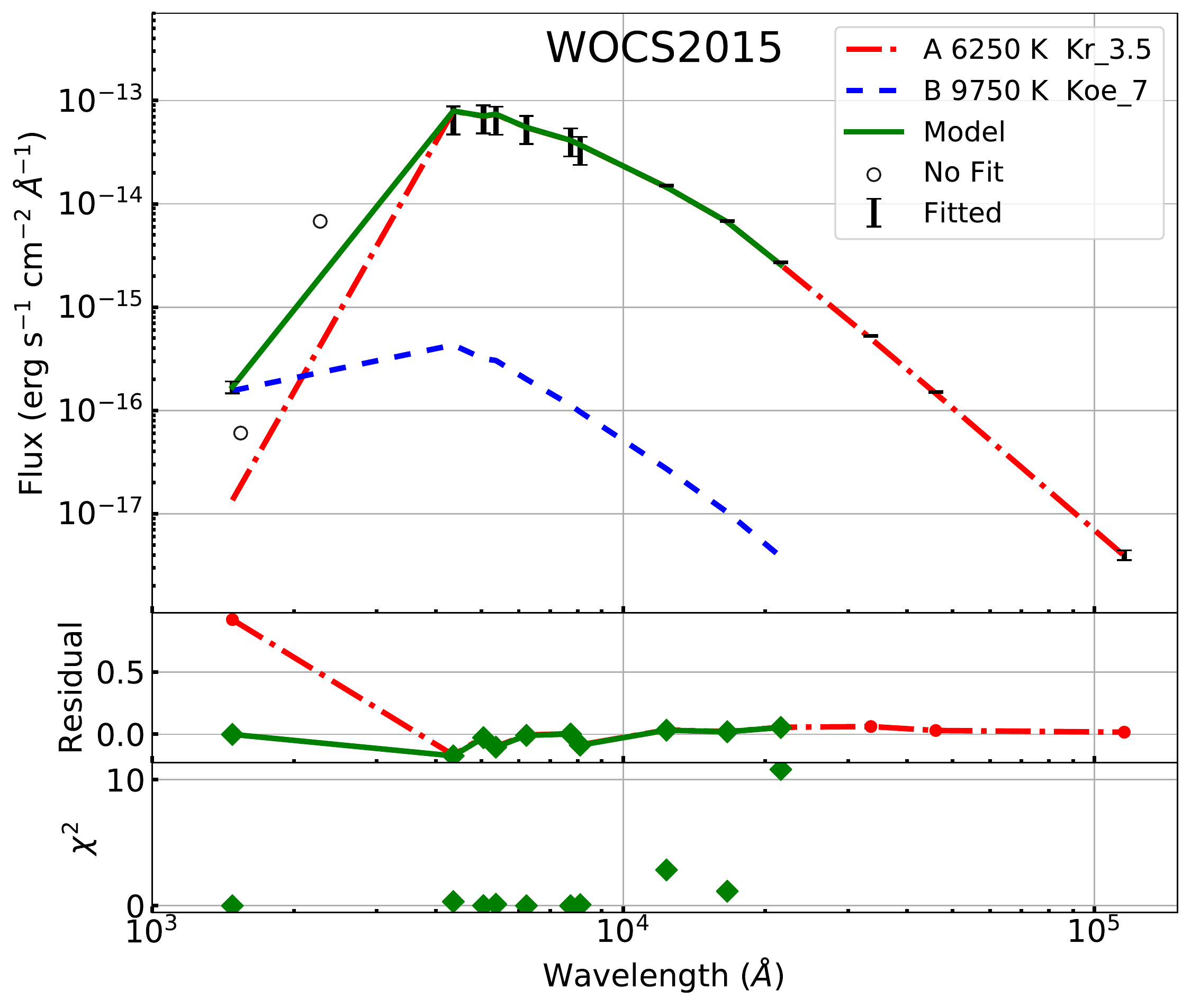} &
    \includegraphics[width=.43\textwidth]{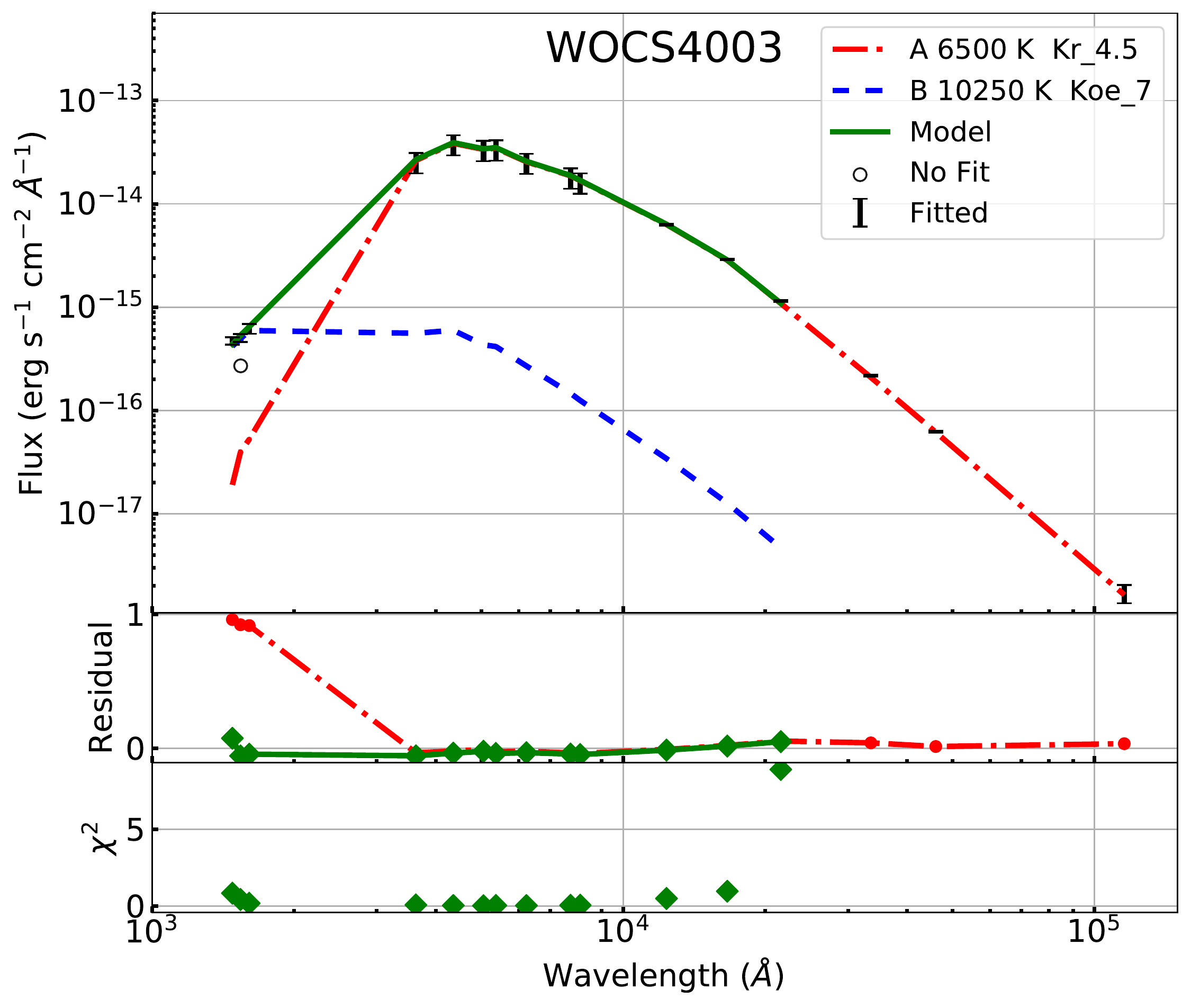} \\
  \end{tabular}
    \caption{Double SED fits of WOCS2008, WOCS2012, WOCS2015 and WOCS4003} 
  \label{fig:4_SED_double_2008}
\end{figure} 

\subsubsection{WOCS2008/S1072}
\label{sec:4_WOCS2008}
\citet{Belloni1998} detected the source in X-rays, but the X-ray emissions are not credited to any WD. The system has $P=$ 1495 days and $e=$ 0.32 \citep{Mathieu1990}. \citet{Geller2015} described the source as a YSS, SB1 and BSS candidate. \citet{Bertelli2018} calculated $T_{eff}=5915 K$ and found the chemical abundances consistent with MTO stars and no signatures of recent MT in C abundance. They suggested that it is a product of 3 stars formed by Kozai-cycle-induced merger in a hierarchical triple system \citep{Perets_2009ApJ...697.1048P}. S

We observe a large UV flux consistently in three UVIT filters.
Double-component SED fitting resulted in a hotter companion of 11500--12500 K and a cooler companion of 6500 K (Fig.~\ref{fig:4_SED_double_2008}). The hotter companion's cooling age is $<$120 Myr and mass $<$0.2 \Msun.

The WD parameters fit well within 0.18 and 0.20 cooling curves of the \citet{Panei2007MNRAS.382..779P} models (Fig.~\ref{fig:4_WD_mass_radius}). The {\it GALEX} FUV flux was lower than UVIT. The variation in FUV flux and X-ray detection could result from flares. 
The X-ray detection also means that there may be contamination in UV flux, thus making the parameters of the hotter companion unreliable. 
On the other hand, the matching WD temperatures estimated by us and \citet{Landsman1997}, in the case of WOCS2002 (YSS + WD system), points to the possibility that the X-ray flux does not contaminate the UV flux significantly. Hence, there may still be a WD companion with the estimated parameters present in the system.

\subsubsection{WOCS2012/S756}
\label{sec:4_WOCS2012}
\citet{Geller2015} listed this star as a single member. This is also one of the faintest sources observed (22.3 mag in F148W). It has not been detected in X-rays.

We detected the source in only the F148W filter. After fitting a cooler component, we observed excess flux in the UV region consistent with a WD companion of 11500 K. 
The {\it GALEX} FUV observation also showed UV flux larger than the UVIT detection (Fig.~\ref{fig:4_SED_double_2008}).
In the absence of any contradicting information and large UV flux, we suggest that the source composed of one MS star and one WD. However, the parameters of WD will not be entirely accurate due to a single UVIT data point. 

\subsubsection{WOCS2015/S792}
\label{sec:4_WOCS2015}
\citet{Geller2015} considered this as a single member and a possible BSS but suggested the possibility of a very long period binary or a BSS formed by collision. The absence of X-ray detection decreases the possibility of an interacting close binary or chromospheric activity. There was no variability detected in the light curves by \citet{Sandquist2003b}. \citet{Bertelli2018} found the APOGEE rotational velocity to be $v$ sin $i=3.63 km s^{-1}$ and $T_{eff}=5943$ K. This star lies between the MSTO and YSSs (WOCS2002, WOCS2008) in the optical CMD.

We detected the star in only the F148W filter. We could fit the observed SED with one hotter (9750-10250 K) and one cooler (6250 K) component (Fig.~\ref{fig:4_SED_double_2008}). The large $\chi^2_{red}$ value for the fit is primarily due to the small error in 2MASS.Ks filter magnitude. The {\it GALEX} detection in NUV is consistent with a single star, while {\it GALEX} FUV has smaller excess than UVIT. The hotter companion parameters differ from the models of \citet{Tremblay2009ApJ...696.1755T} and \citet{Panei2007MNRAS.382..779P}; thus, the excess UV flux may not be due to a WD. 

\subsubsection{WOCS3009/S1273}
\label{sec:4_WOCS3009}
\citet{Geller2015} listed it as a single member of NGC 2682 and a possible BSS. We did not find any X-ray detections or photometric variability in the literature.

We observed significant excess UV flux in all three filters. The resultant SED fit, in Fig.~\ref{fig:4_SED_3009} (c), shows the existence of a 10000--11000 K hotter companion. The mass and age were estimated to be $<$0.2 \Msun\ and $<$200 Myr, respectively. The estimated WD parameters are not fully compatible with the WD models. Therefore, we are not confirming the presence of a WD in the system.

\subsubsection{WOCS4003/S1036}
\label{sec:4_WOCS4003}
This source is an EV Cnc of W UMa type with $P=$ 0.44 days and $e=0.00$. \citet{Belloni1998} detected it in X-rays and related the detection to chromospheric activity and rapid rotation. \citet{Yakut2009} found the light curve to be unusual for a contact binary and estimated the temperature of two components as $T_{hotter}=6900$ K and $T_{cooler}=5200-5830$ K. The system lies slightly blueward of MSTO in the optical CMD.

We found the source to have a considerable UV flux in 3 UVIT filters, not explained by both these components' continuum flux.
Due to the unavailability of individual parameters, we performed the SED fit assuming a single cooler component (6500 K) and a hotter component (10250 K; Fig.~\ref{fig:4_SED_double_2008}).

Fig.~\ref{fig:4_WD_mass_radius} shows that the parameters of the hotter components strongly deviate from the WD models. Thus, we deduce that the UV flux results from chromospheric activity or spots.

\begin{figure}[!ht]
  \centering
  \begin{tabular}{c c}
    \includegraphics[width=.43\textwidth]{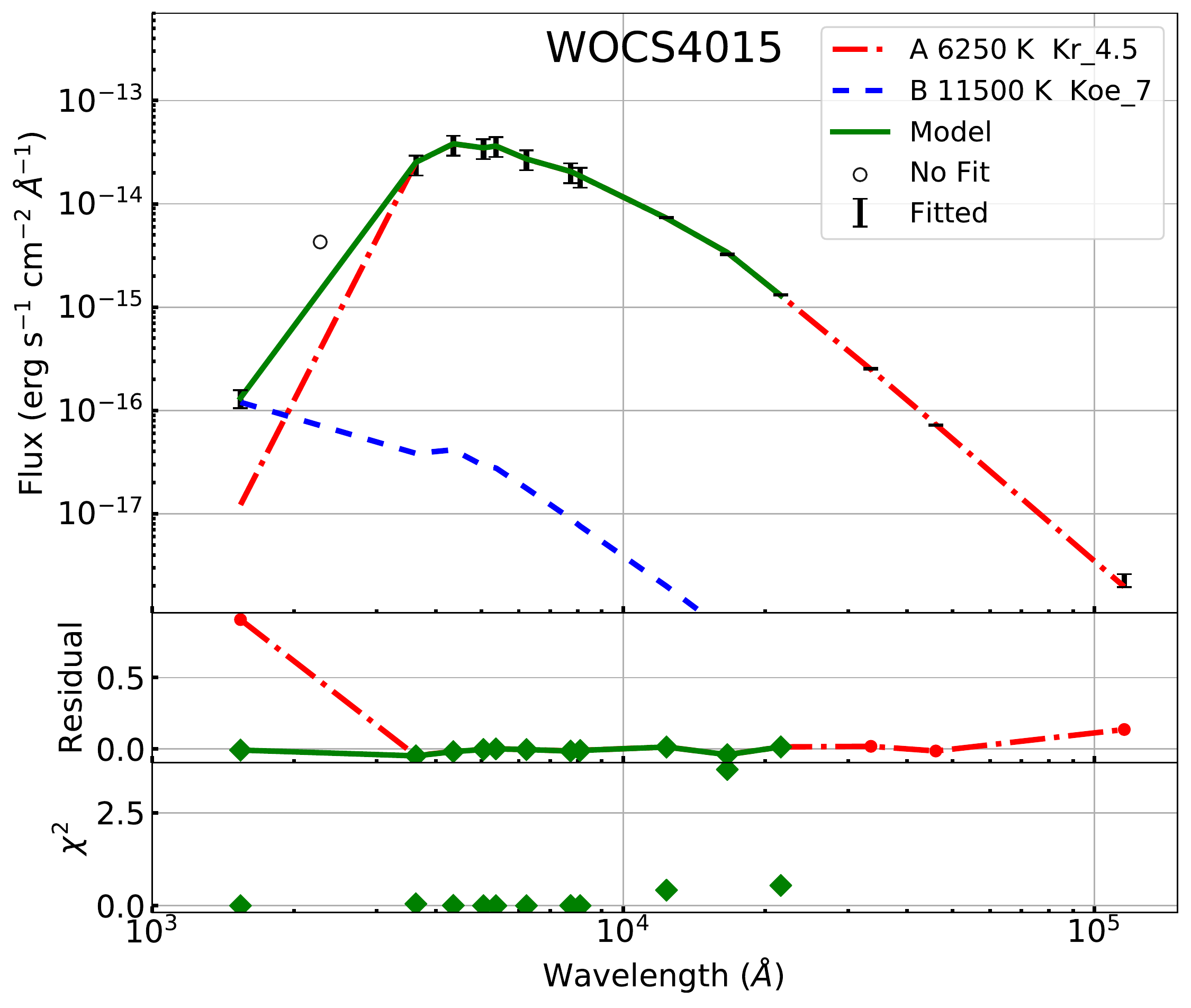} &
    \includegraphics[width=.43\textwidth]{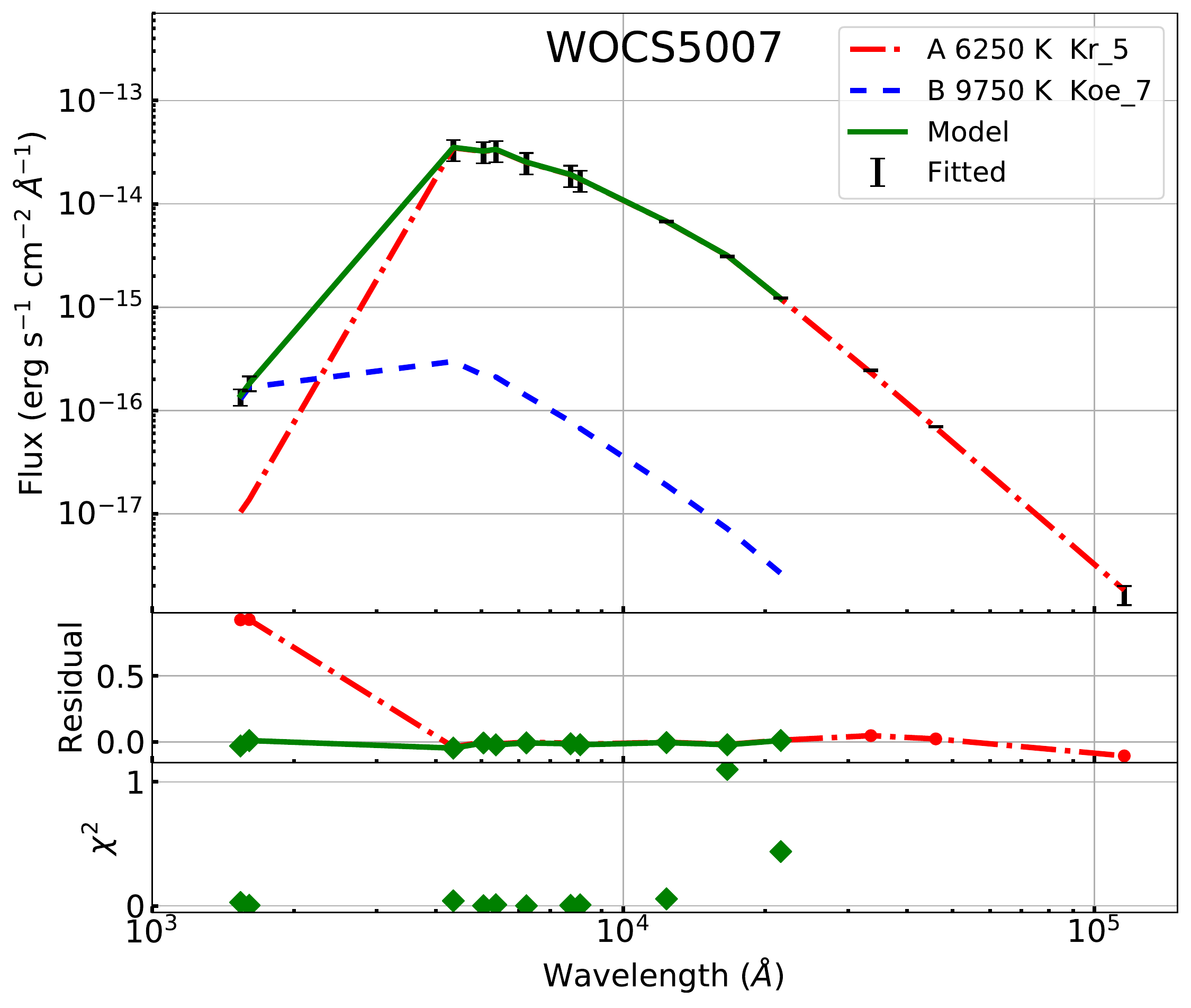} \\
    \includegraphics[width=.43\textwidth]{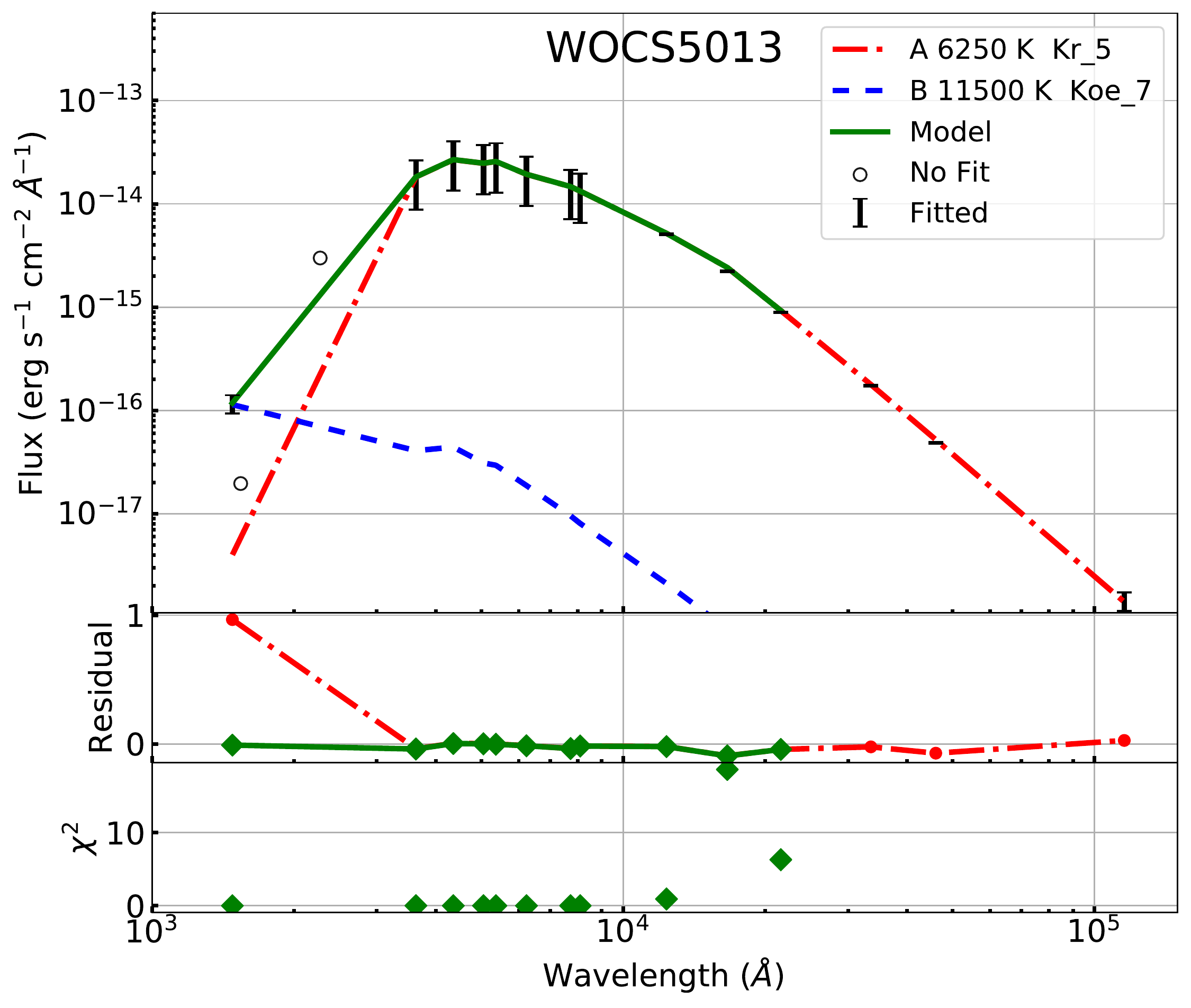} &
    \includegraphics[width=.43\textwidth]{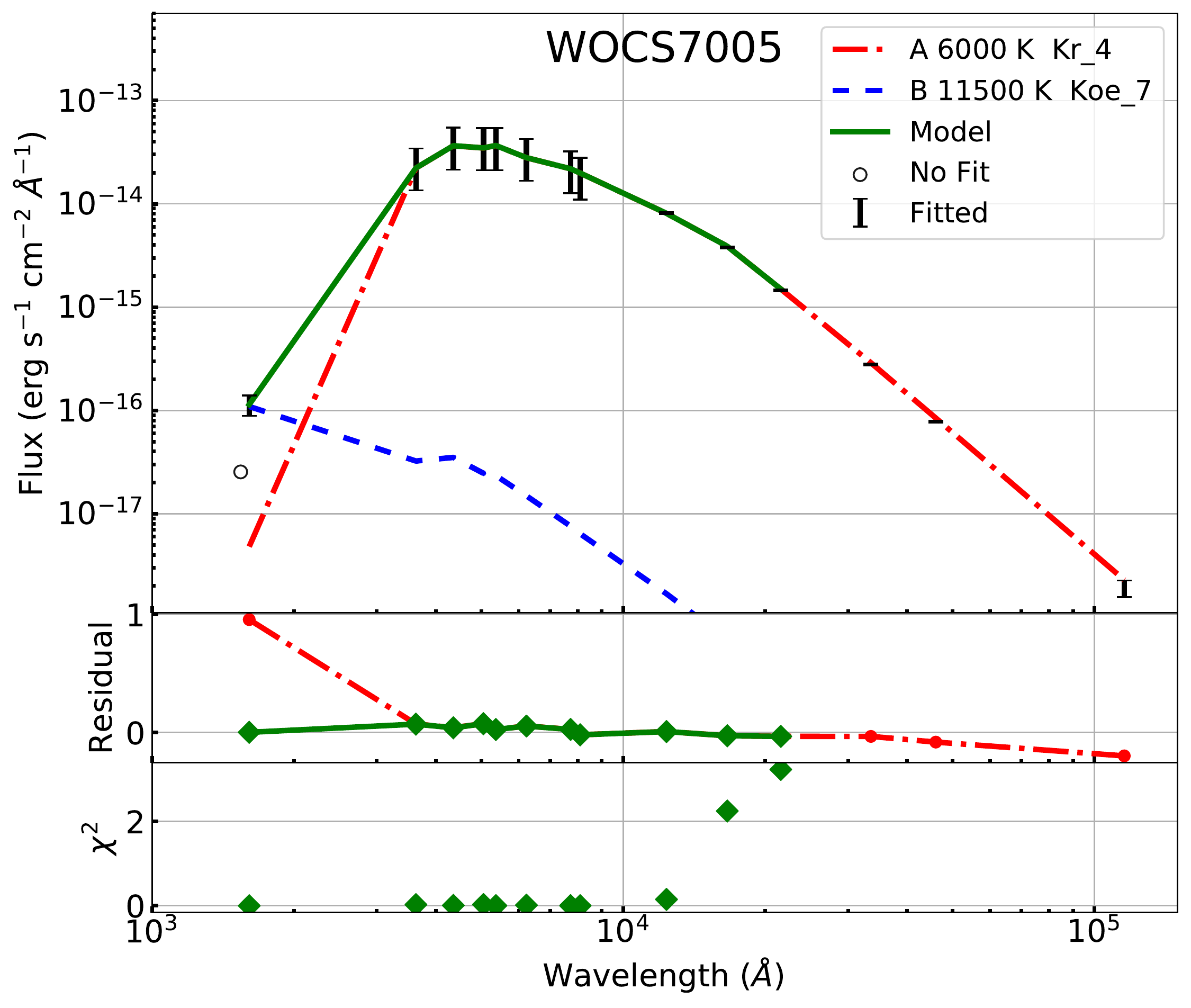} \\
  \end{tabular}
    \caption{Double SED fits of WOCS4015, WOCS5007, WOCS5013 and WOCS7005} 
  \label{fig:4_SED_double_4015}
\end{figure}

\subsubsection{WOCS4015/S1456}
\label{sec:4_WOCS4015}
This star was only categorised as a single member by \citet{Geller2015}. This source lies just below the MSTO of the optical CMD but shifts bluer in UV--optical CMD. 

We detected the star in only the F148W filter. The double fit has high $\chi^2_{red}$ (Fig.~\ref{fig:4_SED_double_4015}), but most of it can be attributed to small errors in 2MASS filter magnitudes. The WD parameters lie within the expected region in Fig.~\ref{fig:4_WD_mass_radius}, but the single detection in UVIT bars us from confirming the presence and parameters of the hotter companion.

\subsubsection{WOCS5007/S1071}
\label{sec:4_WOCS5007}
\citet{Geller2015} listed this as a single member of NGC 2682. There is a small blueward shift from the optical to the UV--optical CMD. We detected the source in F154W and F169M filters, both near the limiting magnitude. The double SED fit gives us the parameters of a hotter companion of 9750--11000 K (Fig.~\ref{fig:4_SED_double_4015}).

The absence of any X-ray detections suggests that there is insignificant chromospheric activity. The estimated parameters of the hotter companion lie not too far from the WD models (Fig.~\ref{fig:4_WD_mass_radius}). We do not confirm the presence of a WD companion due to non-detection in the F148W filter.

\subsubsection{WOCS5013/S1230}
\label{sec:4_WOCS5013}
This is a single member as described by \citet{Geller2015}. Similar to WOCS5007, there is a slight shift in the star's position from the optical to UV--optical CMD.

We detected the source in only the F148W filter. The SED fit shows high UV flux compared to a star with T$_{eff}=$ 6250 K (Fig.~\ref{fig:4_SED_double_4015}). The double fit shows the presence of a hotter companion ($T_{eff}=$ 11500 K) with temperature and mass consistent with the WD models. On the other hand, we need deeper observations in multiple filters to confirm the presence and parameters of the hotter component.

\subsubsection{WOCS7005/S1274}
\label{sec:4_WOCS7005}
\citet{Geller2015} described the source as a single member. It lies near the MSTO in the optical CMD. In the V, (F169M$-$V) CMD, it shifts near the beginning of the BSS branch of the isochrone (this CMD is not shown in Fig.~\ref{fig:4_CMD}).

We detected the source in only the F169M filter. Our double fit (Fig.~\ref{fig:4_SED_double_4015}) shows a large UV flux resulting in a hotter companion of 0.2--0.33 \Msun. According to Fig.~\ref{fig:4_WD_mass_radius}, the companion parameters are similar to WD parameters. However, the single filter detection and lower {\it GALEX} FUV flux bar us from decisively claiming the presence of a WD companion. 

\begin{figure}[!ht]
  \centering
  \begin{tabular}{c c}
    \includegraphics[width=.43\textwidth]{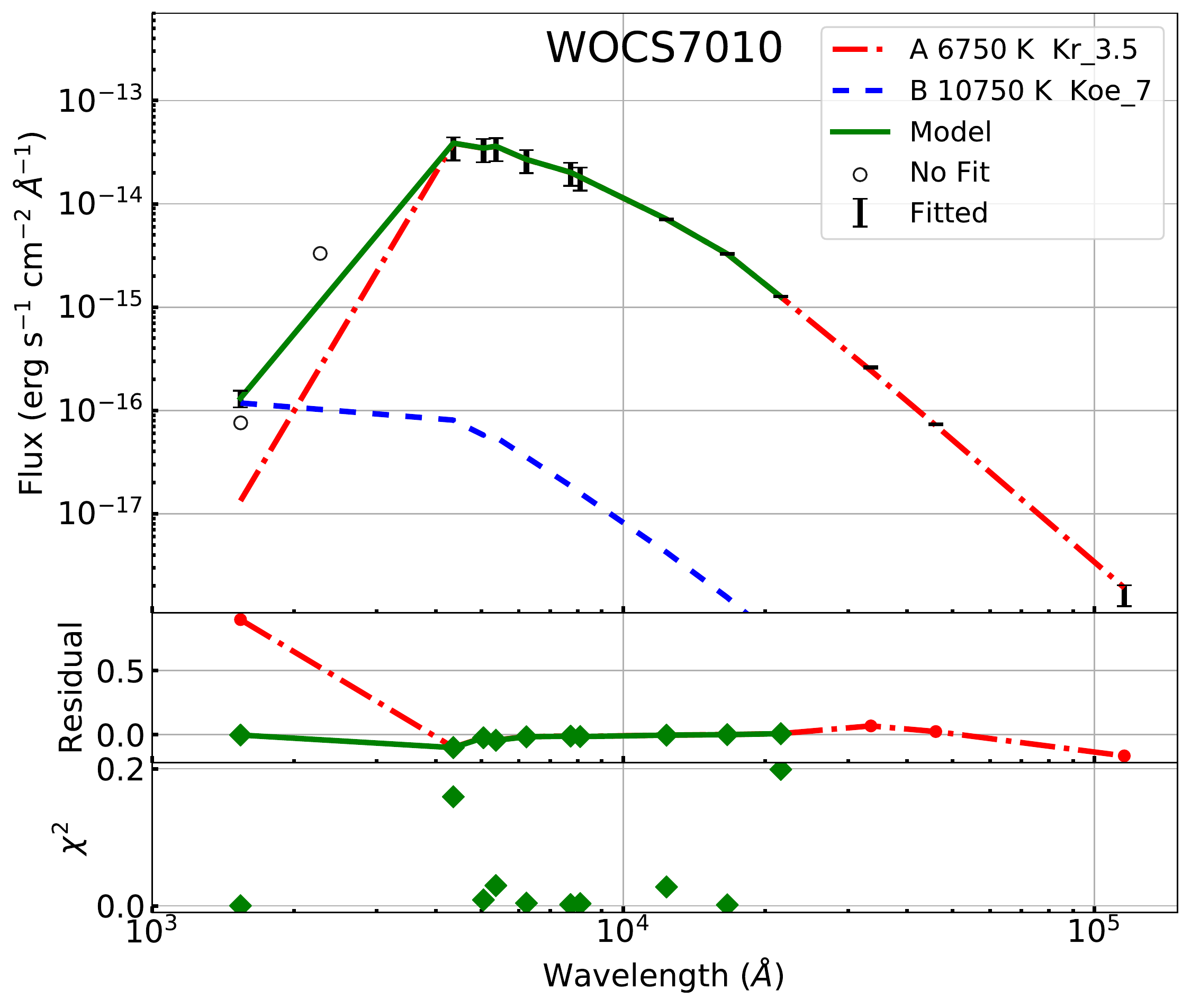} &
    \includegraphics[width=.43\textwidth]{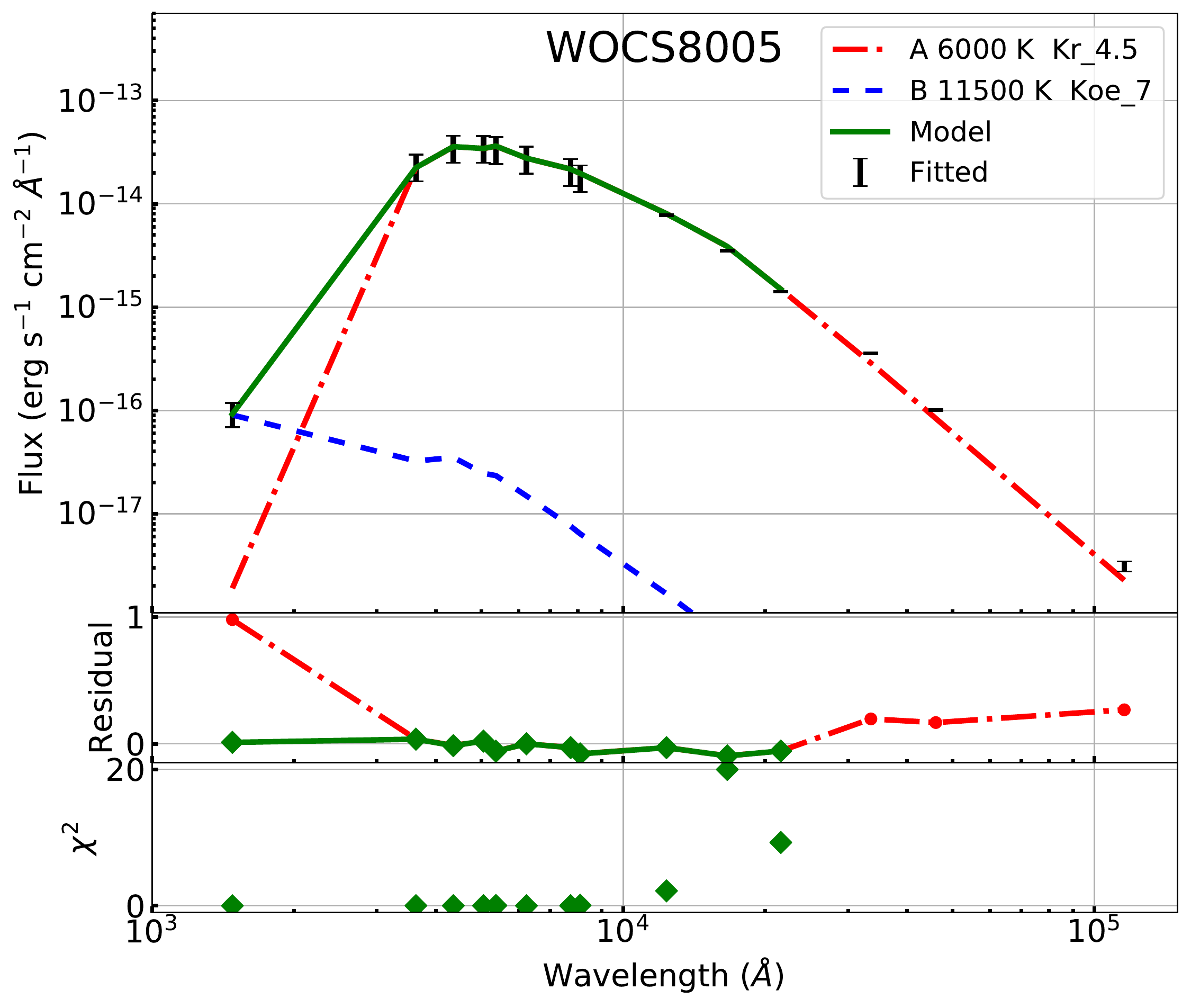} \\
    \includegraphics[width=.43\textwidth]{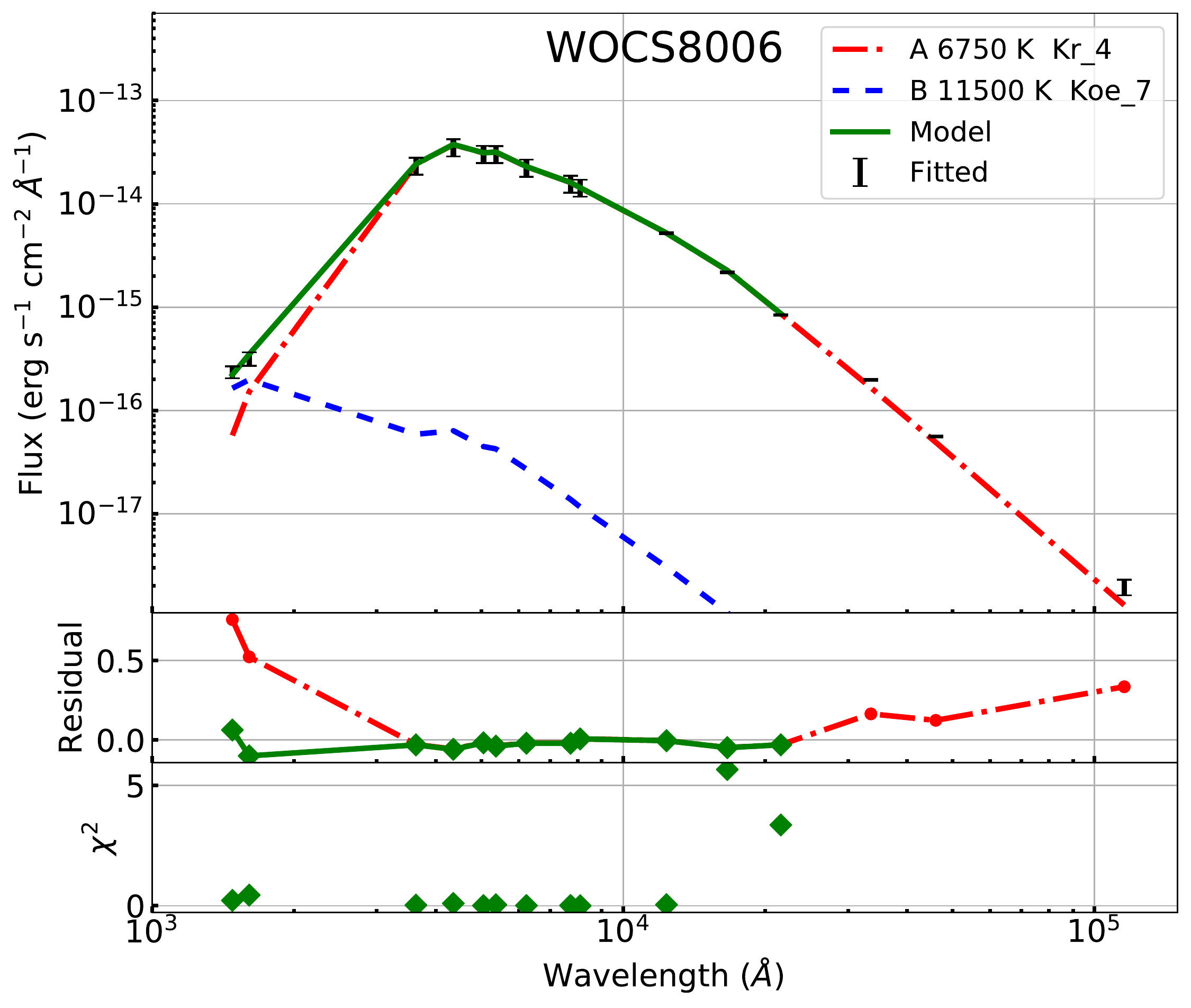} &
    \includegraphics[width=.43\textwidth]{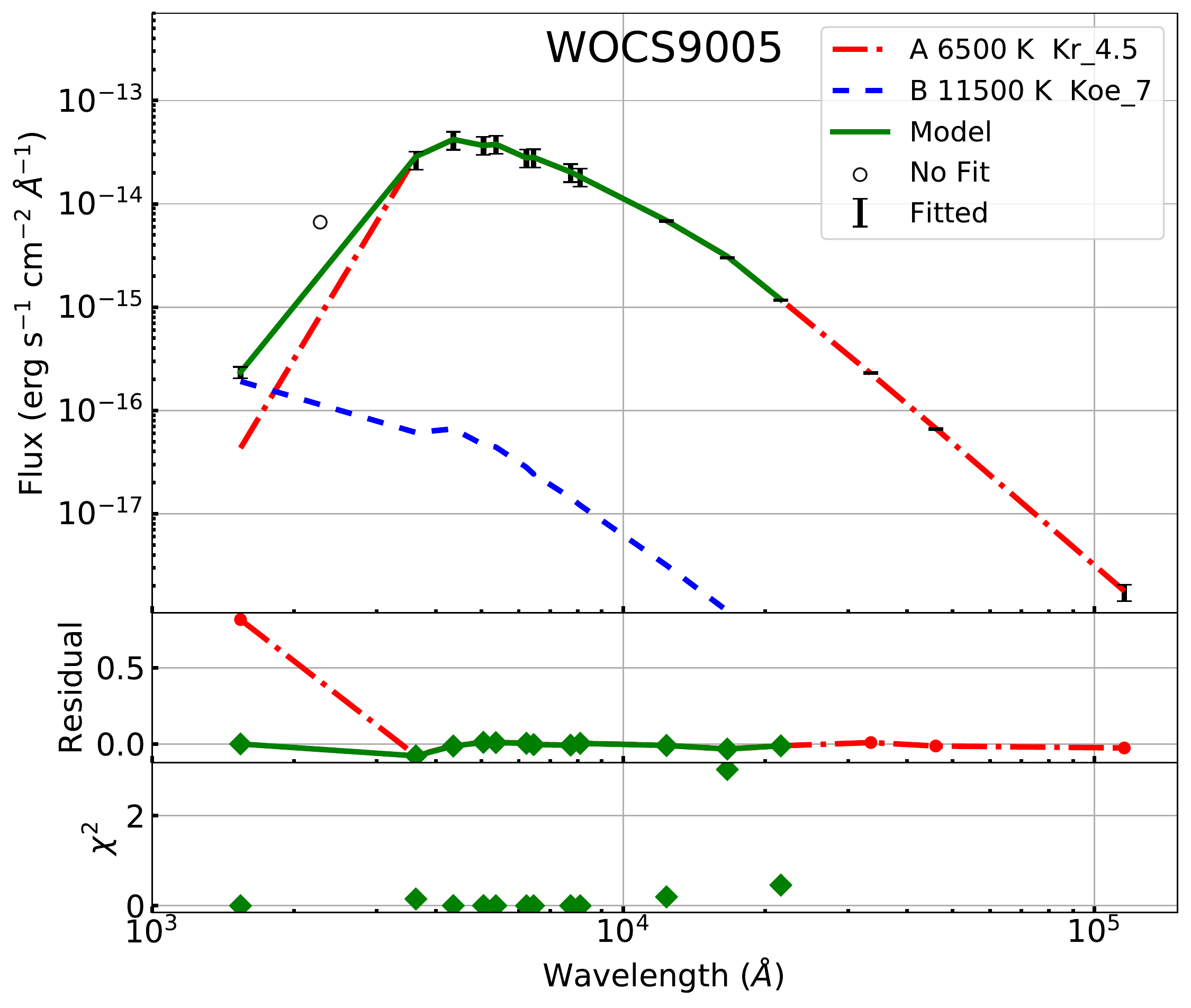} \\
  \end{tabular}
    \caption{Double SED fits of WOCS7010, WOCS8005, WOCS8006 and WOCS9005} 
  \label{fig:4_SED_double_7010}
\end{figure}

\subsubsection{WOCS7010/S1083}
\label{sec:4_WOCS7010}
\citet{Geller2015} listed this as a single member source. It lies just below MSTO in the optical CMD, but it is much bluer in the V, (F154W$-$V) CMD (not shown in Fig.~\ref{fig:4_CMD}). 

We detected the star in the F154W filter alone. The resultant single SED suggests excess UV flux (Fig.~\ref{fig:4_SED_double_7010}). {\it GALEX} observations are consistent with a double fit with a cooler component (6750 K) and a hotter component (10750--11500 K).
The mass and radius of the hotter companion are compatible with the WD models. However, due to the single filter detection, we only hint at the possibility of the presence of a WD. 

\subsubsection{WOCS8005/MMJ5951}
\label{sec:4_WOCS8005}
This is a single member \citep{Geller2015} located near the MSTO in the optical CMD but bluer in the UV--optical CMD.

We detected a large UV flux residual and a small IR residual after fitting a cooler companion SED (Fig.~\ref{fig:4_SED_double_7010}). The double fit gives the hotter companion parameters as a 120--210 Myr old WD with a mass of 0.2--0.33 \Msun. Fig.~\ref{fig:4_WD_mass_radius} shows that the obtained parameters are compatible with a WD. The only caveat is that the detection is only in the F148W UVIT filter near the limiting magnitude of the observations. Deeper observations in UV are required for further characterising the source.

\subsubsection{WOCS8006/S2204}
\label{sec:4_WOCS8006}
According to \citet{Geller2015}, this is a single-member BSS candidate. We notice a smaller blue shift from the optical to the UV--optical CMD compared to other detected members. \citet{Bertelli2018} calculated a T$_{eff}=$ 6650 K from APOGEE spectra.

We created the SED using detections in 2 UVIT filters. The final double fit shows a mild IR excess in \textit{WISE} filters (Fig.~\ref{fig:4_SED_double_7010}). Similar to WOCS8005, the fit's high $\chi^2$ value is primarily due to a small error in 2MASS filter magnitudes. Without any known activity on the surface and consistency with the models in Fig.~\ref{fig:4_WD_mass_radius}, we propose the possibility of a WD companion of $T_{eff}=$ 11500 K with a mass of $\sim$0.2 \Msun\ indicating an MT in the past 140 Myr. The cooler star's temperature of T$_{eff}=$ 6750 K matches with \citet{Bertelli2018}.

\subsubsection{WOCS9005/S1005}
\label{sec:4_WOCS9005}
\citet{Geller2015} listed the source as an SB1 BSS, but also noted that it is not a good candidate for a BSS due to its closeness to MSTO. \citet{Leiner2019ApJ...881...47L} found the orbital properties as $P= 2769$ days, $e=0.15$ and a binary mass function = 0.0368.

We detected the source in only the F154W filter. The resulting best fit suggests that the source is composed of one hotter and one cooler component (Fig.~\ref{fig:4_SED_double_7010}). The hotter component of 11500 K and $\sim$0.2 \Msun\ is a possible WD candidate due to the similarity to the WD models in Fig.~\ref{fig:4_WD_mass_radius}.

\subsection{Triple systems}
\label{sec:4_triples}
\begin{figure}[!ht]
  \centering
  \begin{tabular}{c c}
    \includegraphics[width=.45\textwidth]{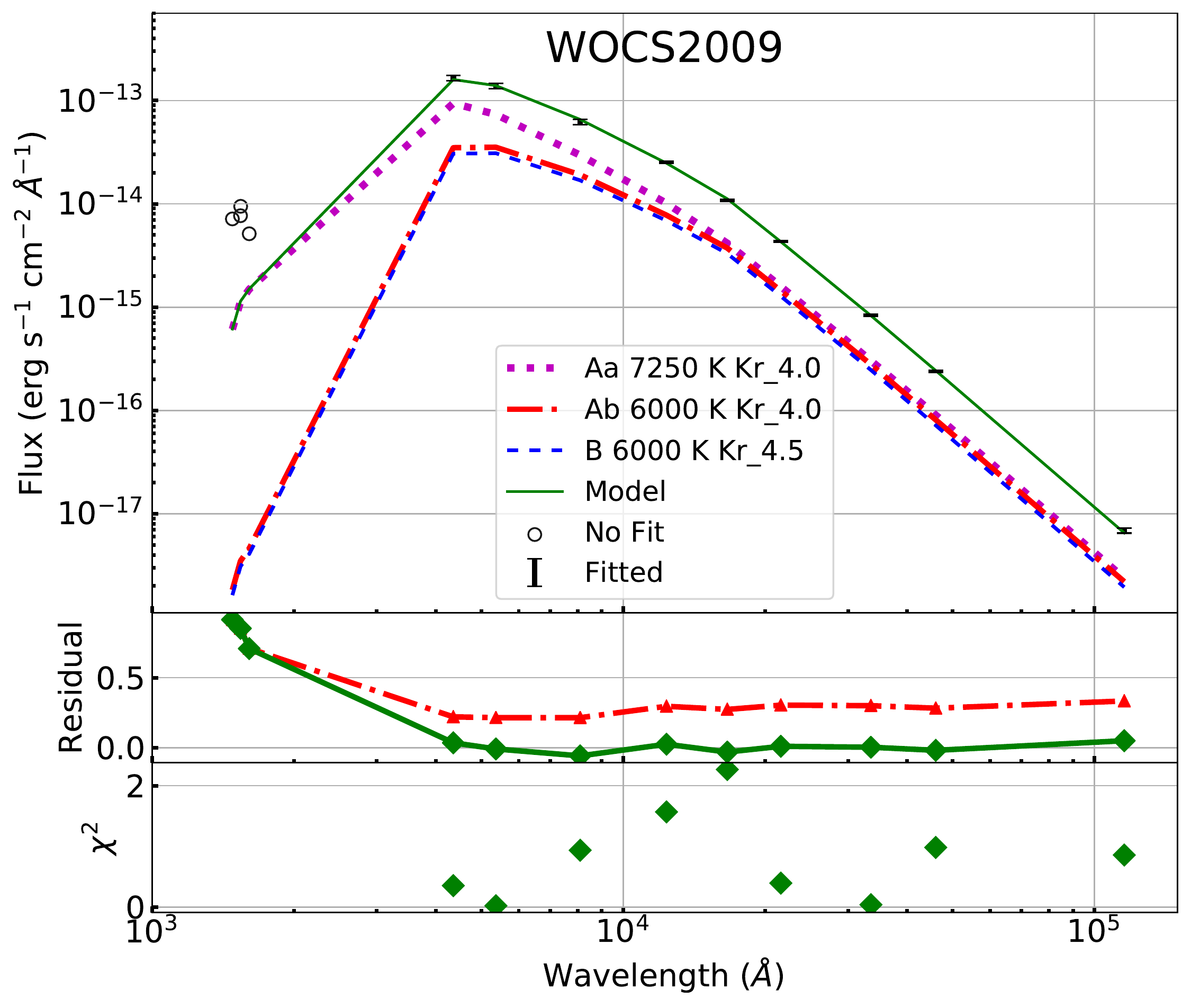} &
     \includegraphics[width=.45\textwidth]{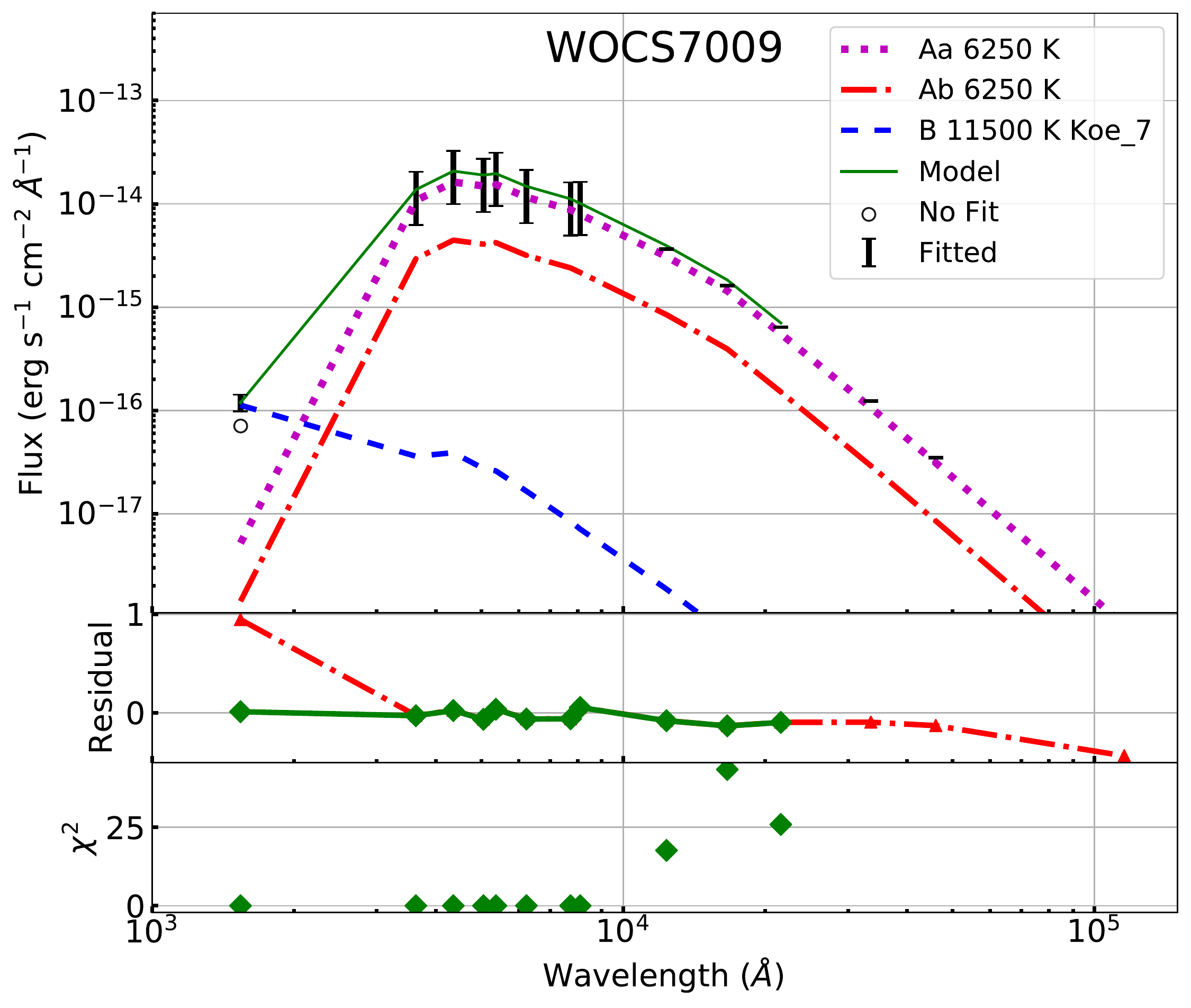} \\
 \end{tabular}  
  \caption{SEDs of triple systems, {WOCS2009:} Model SED using two known components (Aa and Ab) with third component B fitted only to the optical to IR residual. 
  {WOCS7009:} System with 2 known components (Aa and Ab) and a possible third component fitted to the UV residual.
  }
  \label{fig:4_SED_triples}
\end{figure}

\subsubsection{WOCS2009/S1082}
\label{sec:4_WOCS2009}
\citet{Goranskij1992} determined that WOCS2009 is an eclipsing close binary system with $P=$ 1.0677978$\pm$0.0000050 days. \citet{Belloni1998} detected the RS CVn system in X-ray, suggesting active regions on the surface of the close binary. \citet{Shetrone2000} found the RV variations of the eclipsing binary to be not compatible with the period of a close binary.
\citet{Van2001} studied the light curves of the eclipsing binary assuming a circular orbit (for close binary) and consisting of non-spotted stars and proposed the presence of the third component with a longer period. Further study by \citet{Sandquist2003} categorised the system as an ES Cnc, re-estimated the parameters of the close binary, and found the orbital parameters of the third component as $P=$ 1188.5$\pm$6.8 days and $e=$ 0.568$\pm$0.076. \citet{Leigh2011MNRAS.410.2370L} analysed this object using energy conservation in stellar interactions and concluded that the total mass of WOSC2009 is about 5.8 \Msun\ which demands a 3+3 encounter for the formation of the present triple system.

We detected this star in all the three UVIT filters, and the \textit{GALEX} FUV flux is similar to the UVIT fluxes.
Hereafter the close binary components will be referred to as Aa and Ab, while the third component will be referred to as B. \citet{Sandquist2003} calculated the radius and the temperature of Aa and Ab, while they only estimated the temperature of B. We calculated the flux from Aa and Ab components and found the combined SED flux to have a $\sim$0.25 residual in the optical and IR region.

We fitted a third MS star (Kurucz model) SED to the optical and IR region residual. The resultant best fit gave the temperature of the B component as 6000 K, which is consistent with the estimate of 6850 K by \citet{Sandquist2003}. 
Our estimates of L and $T_{eff}$ for the B component are found to be similar to those of the Ab component.
We detect a significant excess in the UV flux, even after fitting the three-component SED. The excess UV flux is likely to result from spot activities in the close binary, which is an RS CVn system.

\subsubsection{WOCS7009/S1282}
\label{sec:4_WOCS7009}
\citet{Belloni1998} detected the AH Cnc contact binary in X-rays. \citet{Qian2006} and \citet{Pribulla2006} suggested the existence of a third component. \citet{Yakut2009} calculated the MT rate of $9.4\times 10^{-8}$ \Msun yr$^{-1}$ for the binary with a mass ratio of 0.17, P = 0.3604360 $\pm$ 0.0000001 days. \citet{Peng2016} estimated the masses and radii of the components of the contact binary as 1.188$\pm$0.061 \Msun, 1.332$\pm$0.063 R$_{\odot}$ and 0.185$\pm$0.032 \Msun, 0.592$\pm$0.051 R$_{\odot}$ respectively, through period analysis. 
 
We used parameters from \citet{Yakut2009} to calculate the contribution from Aa ($T_{eff}$ = 6300 K, log $g = 4.31$, R = 1.40 R$_{\odot}$) and Ab ($T_{eff}$ = 6275 K, log $g = 4.17$, R = 0.68 R$_{\odot}$) components. Due to model constraints, we fitted the Aa and Ab components with $T_{eff}=$ 6250 K and log $g = 4.0$. The combined model flux of these 2 components matches with observations in the optical region suggesting that these two components are sufficient to account for the observed flux in the optical--IR SED. We detect an excess flux in F148W, which we fit with a WD model in the SED (Fig.~\ref{fig:4_SED_triples}).

The parameters of the hotter component lie within the model predictions in Fig.~\ref{fig:4_WD_mass_radius}, but the contact binary itself can be responsible for the UV flux as seen in WOCS11011, WOCS2003, and WOCS2009. The results from the SED fit do not find the presence of any cooler third component, whereas a hotter component may or may not be present.

\section{Discussion}
\label{sec:4_Discussion}

The SED analysis of 30 stars enabled us to characterise these UV-bright NGC 2682 members. We assessed the nature of UV flux in each system using the SED fit parameters, known binarity, and X-ray detections.

The SED fits of stars with multiple UVIT detections are vital to characterise any hotter component. WOCS2007, WOCS3001 and WOCS6006 have 3 UVIT detections with significant UV excess and are shifted blueward in the UV--optical CMD. The parameters of their possible hotter components more or less match with the WD models. 

WOCS2007 is a short-period BSS with a possible ELM WD companion. This makes it the second BSS to be identified to have formed via MT in NGC 2682 and also to have an ELM WD as a companion. The first one being WOCS1007, which was recently found to have an ELM WD companion \citep{Sindhu2019ApJ...882...43S}. These systems are post-MT systems where the MT should have happened while the primary is in the sub-giant phase or earlier (case A/B MT, when the core mass is still $<$0.2 \Msun).
 
We estimated the mass and age of WDs by comparing them to WD models in Fig.~\ref{fig:4_WD_mass_radius}. Among 5, 4 WDs have an estimated mass of $>$0.5 \Msun, indicating a BSS progenitor. Y1487 has a mass of 0.4--0.5 \Msun, which may or may not require a BSS progenitor. These massive WDs could be the product of single massive BSSs, or mergers in close binaries or triples (via Kozai-cycle-induced merger). Including the detection from \citet{Williams2018}, the number of WDs that demand a BSS progenitor is on the rise (5 to 6 WDs).

\begin{figure}
  \centering
    \includegraphics[width=0.45\textwidth]{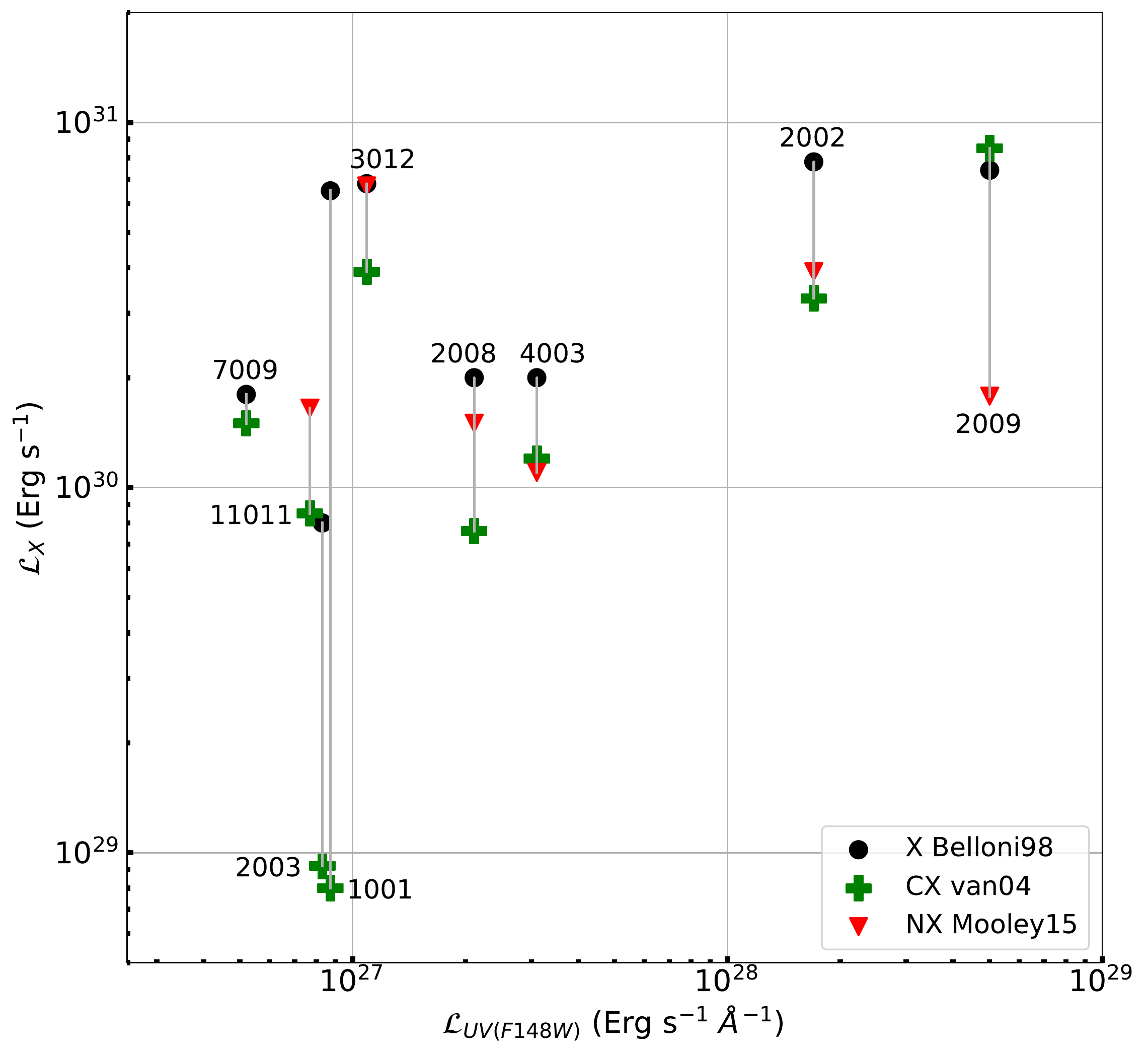} \\
  \caption{Comparison of UV and X-ray luminosity for all X-ray detected stars with X-ray luminosity taken from \citet{Belloni1998}, \citet{Van2004} and \citet{Mooley2015}}
  \label{fig:4_UV-X}
\end{figure}

Nine UVIT detected members were seen in X-ray by at least one of the missions, \citet{Belloni1998}, \citet{Van2004} or \citet{Mooley2015}. All the sources are spectroscopic binary systems. Close binaries that are spun up by tidal interactions are known sources of intrinsic X-ray emissions in old OCs \citep{Belloni1993, Van2004, van2013}. The X-ray emissions in stellar flares are also known to cause UV emissions \citep{Mitra2005}. Although we did not find any explicit correlation in UV and X-ray flux as seen in Fig.~\ref{fig:4_UV-X} (which was not expected due to non-simultaneous observations), the order of UV flux is similar to that of flares.

Acknowledging the possible UV excess flux due to X-ray activity, we fitted a hotter SED to compensate for the residual UV flux, resulting in possible companions with $T_{eff}$ ranging from 9000 to 12000 K.
Comparing the SED fit parameters to WD models, we found potential companions of 2 stars (WOCS2003, WOCS4003) to vary significantly from the WD models. Both systems are contact binaries with known periods and eccentricities, indicating that the UV flux results from binary interactions or surface activity and not due to any hotter companions. 

The SED fit of WOCS2009 shows that the third component is not a compact hot object; thus, the UV flux is due to stellar interactions/activity similar to the contact binaries. We reproduced the $T_{eff}$ of the third component of WOCS2009 consistent with the results by \citet{Sandquist2003}.
The suspected triple system, WOCS7009, showed minimal optical excess. The X-ray emissions and similarities of SED fit parameters with WD models do not conclusively comment on the possibility of a third hotter component or stellar activity.

Our estimates of ELM WD parameters in WOCS2002 are similar to \citet{Landsman1997}, even though it has X-ray emission. This suggests that the X-ray-emitting phenomenon does not necessarily contaminate the UV flux. If we extend this argument to other X-ray-emitting sources with UV excess, some of them could also have a WD companion with parameters mentioned in Table~\ref{tab:4_All_para}, such as WOCS11011, WOCS2008, and WOCS7009.

This study has increased the number of post-MT systems in NGC 2682 to at least 5 (from the previously known 2; WOCS1007 and WOCS3001). Among these, two are BSSs, one is probably an evolved BSS, and two are on the MS. It will be interesting to explore the presence of progenitors and predecessors of these types of systems. The close binary in the triple system WOCS7009 has a low-mass companion and, therefore, could be a potential progenitor of the post-MT system with a low-mass WD. Similarly, the contact binaries (WOCS1001, WOCS11011, WOCS2003, WOCS7009) could also evolve through an MT process to a BSS/MS + WD system. Y856 is suspected to be a double degenerate WD, which is a potential predecessor of a system with a BSS. All these point to an increasing amount of evidence suggesting a significant number of systems in NGC 2682 follow the binary evolution pathway through MT, potentially of case A/B type, in close binaries. Since our study is limited to stars near and brighter than the MSTO, there could be many more such fainter sources on the MS. Therefore, NGC 2682 is likely to have a relatively large number of post-MT systems. 

Some short-period post-MT systems may also merge with the evolution of the secondary, resulting in single stars with a relatively large mass. Such a formation scenario may explain the existence of massive BSSs (e.g., WOCS1010) in NGC 2682. Our finding of massive WDs with a range of cooling age (3--200 Myr) requiring BSS progenitors indicates that the BSSs have been forming and evolving from the MS of NGC 2682 in the recent past as well. This is in agreement with the suggestion by \citet{Sindhu2018} that this cluster has been producing BSSs more or less continuously. 

ELM WDs are thought to be the products of common-envelope binary evolution \citep{Marsh1995MNRAS.275..828M} and are the signposts of gravitational wave and/or possible supernova progenitors. 
Most of the low-mass WDs are in double degenerate systems or compact binaries \citep{Brown2013, Istrate2014}.
NGC 2682 contains at least 2 ELM WD + BSS systems.
The detection of a large number of ELM WDs as companions to MS and BSSs will open up a new window to understand the formation pathways of these WDs. In practice, the ELM WDs may be formed through either a Roche lobe overflow or a common-envelope ejection event.
Thus, detecting ELM WDs in binaries is also an essential tracer for MT. Binaries in OCs provide further constraints, which help determine the evolution of close binaries.

\section{Summary and conclusions}
\label{sec:4_Summary}

Our observations of NGC 2682 using UVIT detected 41 members, including MS stars, WDs, BSSs, and YSSs. The UV--optical and UV CMDs overlaid with isochrones indicate many members having UV excess. We used the SEDs to characterise 30 members (including 7 WDs) by fitting double-component SEDs to 21 members and a single-component SED to the WDs.

We detect ELM WD companions to WOCS11005, WOCS2007 and WOCS6006. Hence, these are post-MT systems. WOCS3001 also has a WD companion and most likely has undergone MT. We also estimate the mass of an ELM WD companion to WOCS2002 suggesting that it is also a post-MT binary.
WOCS2012, WOCS3009, WOCS4015, WOCS5013, WOCS7005, WOCS7010, WOCS8005, WOCS8006, and WOCS9005 require further observations to confirm the presence of hotter companions.
NGC 2682 is therefore likely to have a relatively large number of post-MT systems.
9 sources show X-ray flux and excess flux in UVIT filters and are therefore classified as sources with activity (chromospheric/ hot-spots/ coronal/ ongoing-MT). 
5 out of 7 are WDs characterised by SED fitting, and 4 of them have mass $>$0.5 \Msun\ and a cooling age of less than 200 Myr, thus demanding BSS progenitors. The massive WDs detected in NGC 2682 require BS progenitors. These massive WDs could be the product of single massive BSSs, or mergers in close binaries or triples (via Kozai-cycle-induced merger).
The SED confirms the presence of the third component in WOCS2009. It is comparable to the cooler star in the inner binary. 

\subsection{Conclusions}
\begin{itemize}
\item The UV--optical and UV--UV CMD of NGC 2682 are not as elementary as the optical CMD. The position of stars is heavily impacted by intrinsic and extrinsic factors such as surface activities and binarity. X-ray detections play an important role in identifying stellar activity.

\item This study brings out the importance of deep imaging in the UV to detect and characterise WDs and WD/ELM WD companions in non-degenerate systems.

\item As many as 12 sources need deeper UV imaging to confirm the presence of a WD companion. Spectroscopic analysis in the FUV region is generally necessary to confirm the existence of all optically sub-luminous low-mass WD companions and determine log $g$, mass, and age with more certainty. We plan to make deeper observations in FUV filters in UVIT to identify more sources with potential WD companions, and confirm the candidates identified in this study.

\item The detection of ELM WD companions to BSSs (WOCS1007 and WOCS2007) and a YSS (WOCS2002) shines a light on the formation pathways of these systems. The low mass of WDs signifies that the MT happened before the donor reached the core mass of 0.3 \Msun, indicating a case A/B MT. Such systems require close binaries as progenitors. Contact binaries (WOCS11011, WOCS4003) and close binaries (WOCS1001, WOCS2003) are likely progenitors of such BSS + ELM WD systems. 

\item Similarly, ELM WD detections along with MS stars (WOCS6006 and WOCS11005) indicate that other MT systems can be present in the MS of NGC 2682 and probably other similar clusters. They will masquerade as an MS star whose MT will be only decipherable via the presence of an ELM WD or unusually high rotation. Both MS + ELM WD and BSS + ELM WD systems can evolve to form double degenerate WD systems that could remain strong emitters of mHz gravitational waves for gigayears \citet{Brown2016}. 

\item This study demonstrates that UV observations are essential to detect and characterise the ELM WDs in non-degenerate systems. The presence of systems such as ELM WD + BSS, ELM WD + MS, WD + MS and massive hot WDs and evidence of MT on the MS show that constant stellar interactions are going on in NGC 2682, which is likely the case for more similar OCs.
\end{itemize}

\begin{savequote}[100mm]
No star is ever lost we once have seen, We always may be what we might have been
\qauthor{Adelaide Anne Proctor}
\end{savequote}

\chapter[Blue Stragglers in King 2]{Blue Stragglers in King 2 \\ \large{\textcolor{gray}{Jadhav et al., 2021, JApA, 42, 89}}}
\label{ch:UOCS4}
\begin{quote}\small
\end{quote}

\section{Introduction} \label{sec:5_intro}

The evolution of binary systems strongly depends on the initial orbital parameters and their further evolution, where any change in their orbits can lead to a widely different evolution. If one of the stars evolves and fills its Roche lobe, the system will undergo MT. The details such as duration and rate of MT will depend on the orbits and masses of the binary stars. If such a binary is present in a star cluster, and the secondary has a mass similar to the MSTO mass, the secondary will become brighter than the MSTO and appear as a BSS.
Depending on the evolutionary status of binary components, the binary system can be observed as MS+MS, contact binaries \citep{Rucinski1998AJ....116.2998R}, common envelope, MS+HB (\citealt{Subramaniam2016ApJ...833L..27S}), MS+EHB (\citealt{Singh2020ApJ...905...44S}), MS+sdB \citep{Han2002MNRAS.336..449H}, MS+WD \citep{Jadhav2019ApJ...886...13J}, WD+WD \citep{Marsh1995MNRAS.275..828M} and many more combinations. The binary evolution also depends on external factors such as collisions in a high-density environment  (which can decouple the binary; \citealt{Heggie1975MNRAS.173..729H}) and a tertiary star (which can expedite the MT/merger by reducing the orbital separation; \citealt{Kozai1962AJ.....67..591K}). 

UV imaging of the binary systems reveals the presence of hotter companions in the binary system, given that the hotter companion is luminous in UV. Old OCs such as NGC 188 and NGC 2682 are rich with BSSs, binary stars and contain many such optically sub-luminous UV-bright companions \citep{Subramaniam2016ApJ...833L..27S, Sindhu2019ApJ...882...43S, Jadhav2019ApJ...886...13J}. Similar companions have been identified to BSSs in the outskirts of GCs \citep{Sahu2019ApJ...876...34S, Singh2020ApJ...905...44S}. \citet{Subramaniam2020JApA...41...45S} provided a summary of the BSSs and post-MT systems in star clusters identified using UVIT observations.

\begin{table}
    \centering
    \caption{Age, distance, reddening (E(B$-$V)) and metallicity of King 2 estimated by various investigators are listed. [1] \citet{Dias2002A&A...389..871D}, [2] \citet{Aparicio1990A&A...240..262A}, [3] \citet{Tadross2001NewA....6..293T}, [4] \citet{Kaluzny1989AcA....39...13K}.}
    \begin{tabular}{cc ccr}
    \toprule
	Age	&	Distance	&	E(B$-$V)	&	Metallicity	&	Ref.	\\
	(Gyr)	&	(pc)	&	(mag)	&		&		\\ \hline
	6.02	&	5750	&	0.31	&	-0.42	&	[1]	\\
	6	&	5690$\pm$65	&	0.31$\pm$0.02	&	-0.5 to -2.2	&	[2]	\\
	& & &	-0.32	&	[3]	\\
 4 to 6	&$\sim 7000$&0.23 to 0.5 &	&	[4]	\\	
\bottomrule
    \end{tabular}
      \label{tab:5_info}
\end{table}

King 2 is one of the oldest cluster in Milky Way, with an age of $\sim$6 Gyr and distance of $\sim 5700$ pc (Table~\ref{tab:5_info}). However, it has been poorly studied due to its considerable distance and unknown membership information. 
For identifying and characterising hot BSSs and their possible companions, we obtained UVIT observations of the rich cluster ($\alpha_{2000} = 12\farcdeg{75};\ \delta_{2000} = + 58\farcdeg{183};\ l = 122\farcdeg{9}$ and $b = - 4\farcdeg{7}$) under \textit{AstroSat} proposal A02\_170.
\citet{Kaluzny1989AcA....39...13K} presented the first optical CMD study of this distant cluster using \textit{BV} CCD photometric data. This yielded a range of plausible ages and distances for different assumed reddenings and metallicities. The galactocentric distance of the cluster was estimated to be $\sim$14 Kpc. \citet[A90 hereafter]{Aparicio1990A&A...240..262A} did a comprehensive study of the cluster using \textit{UBVR} photometry and derived an age of 6 Gyr and a distance of 5.7 kpc for solar metallicity. They also indicated the presence of a good fraction of binaries in the MS. \citet{Tadross2001NewA....6..293T} estimated a value of [Fe/H] $= - 0.32$ using the ($U - B$) colour excess from the literature data, while \citet[WC09 hereafter]{Warren2009MNRAS.393..272W} derived a value of [Fe/H] $= - 0.42\pm0.09$ using spectroscopic data. These metallicity estimates are significantly sub-solar and inconsistent with the finding of A90. WC09 found a distance of 6.5 kpc and a slightly younger age, $\sim$4Gyr, better fitted the optical CMD and 2MASS.Ks, red clump if the reddening is adopted as E(B$-$V) = 0.31 mag. This distance puts King 2 at a galactocentric distance of 13 kpc, where its metallicity falls close to the trend of the galactic abundance gradients derived in \citet{Friel2002AJ....124.2693F}. There has been no PM study available for this cluster till {\it Gaia} DR2 \citep{Gaia2018A&A...616A...1G}. 
\citet{Cantat2018A&A...618A..93C} provided a membership catalogue of King 2 with 128 members with \textit{Gaia} DR2, and \citet{Jadhav2021MNRAS.503..236J} provided kinematic membership of 1072 stars (and 340 probable members) using kinematic data taken from {\it Gaia} EDR3.

The above discussed optical photometric studies indicate a good number of post-MS hot stars in King 2. In fact, \citet{Ahumada_2007A&A...463..789A} have identified 30 BSS candidates based on the location of these stars in the cluster. We present the UVIT and the archival data used in this study in the next section, followed by analyses, results, and discussion.

\section{UVIT and archival data}
\label{sec:5_data}

\begin{figure}
    \centering
    \includegraphics[width=0.47\textwidth]{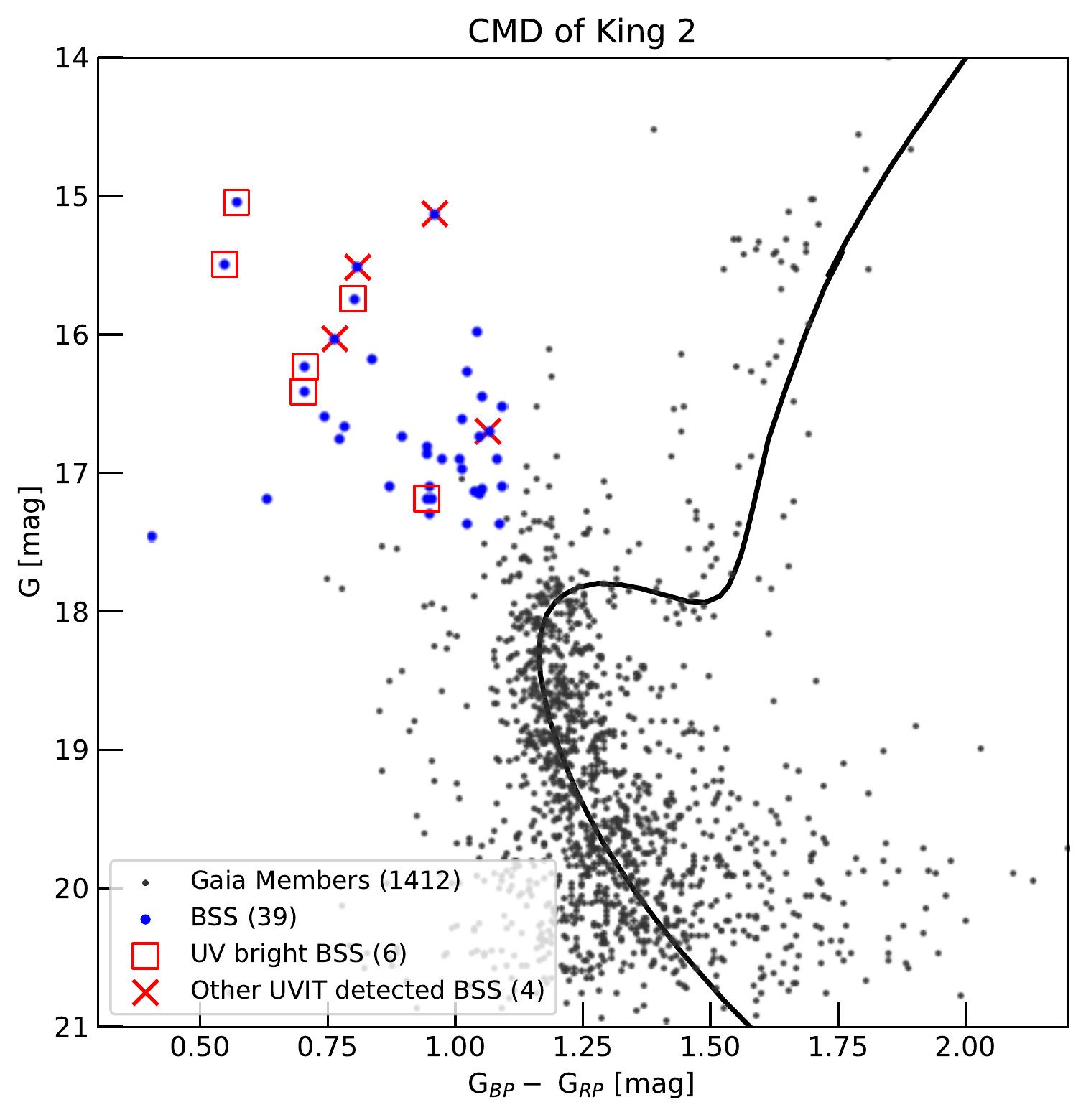}
    \caption{CMD of King 2 cluster candidates using {\it Gaia} EDR3 data. 
    All BSS members are shown as blue circles. The UV bright BSS (see \S~\ref{sec:5_sed_fitting}) are shown as red squares, and other UVIT detected BSS are shown as red X's.
    The {\it Gaia} members are shown as black dots along with the PARSEC isochrone of log $age$ = 9.7, [M/H] = -0.4, DM = 13.8 and E(B$-$V) = 0.45.}
    \label{fig:5_CMD}
\end{figure}

We observed King 2 with UVIT on 17 December 2016, simultaneously in two filters.
The cluster was observed in one FUV (F148W, limiting magnitude $\approx$ 23 mag) and one NUV (N219M, limiting magnitude $\approx$ 22 mag) filter for exposure times of $\sim$2.7 ksec. The FWHM of PSF in F148W and F219N images is 1\farcsec33 and 1\farcsec35, respectively.
We have detected ten member stars in either F148W and/or N219M filter.

We obtained archival optical (\textit{UBVR}) photometry data from A90 catalogue (Calar Alto observatory; CAHA) and cross-matched with UVIT data using {\sc topcat} \citep{Taylor2005ASPC..347...29T}. The cluster was observed with \textit{GALEX} under All-sky Imaging survey in NUV filter (exp. time $\sim$100 sec).
All the detected member stars were further cross-matched with photometric data from UV--IR wavelength bands obtained from \textit{GALEX} \citep{Bianchi2000MmSAI..71.1123B}, PAN-STARRS PS1 \citep{Chambers2016arXiv161205560C}, {\it Gaia} EDR3 \citep{Gaia2021A&A...649A...1G}, 2MASS \citep{Skrutskie2006AJ....131.1163S}, \textit{WISE} \citep{Wright2010AJ....140.1868W} using virtual observatory tools in \textsc{vosa} \citep{Bayo2008A&A...492..277B}.

\section{SED fitting and colour-magnitude diagrams} \label{sec:5_sed_fitting}

The data were corrected for reddening (E(B$-$V) = 0.31$\pm$0.02) using \citet{Fitzpatrick1999PASP..111...63F} and \citet{Indebetouw2005ApJ...619..931I} and calibrated with the cluster distance of 5750$\pm$100 pc (we have overestimated the error to cover distance estimates from \citealt{Dias2002A&A...389..871D} and A90). We have adopted the metallicity of [Fe/H]= $-$0.5 for all the stars and used Kurucz model spectrum \citep{Castelli1997A&A...318..841C} for comparison. The SED fitting was done as follows:

\begin{figure}[!ht]
    \centering
    \includegraphics[width = 0.9\textwidth]{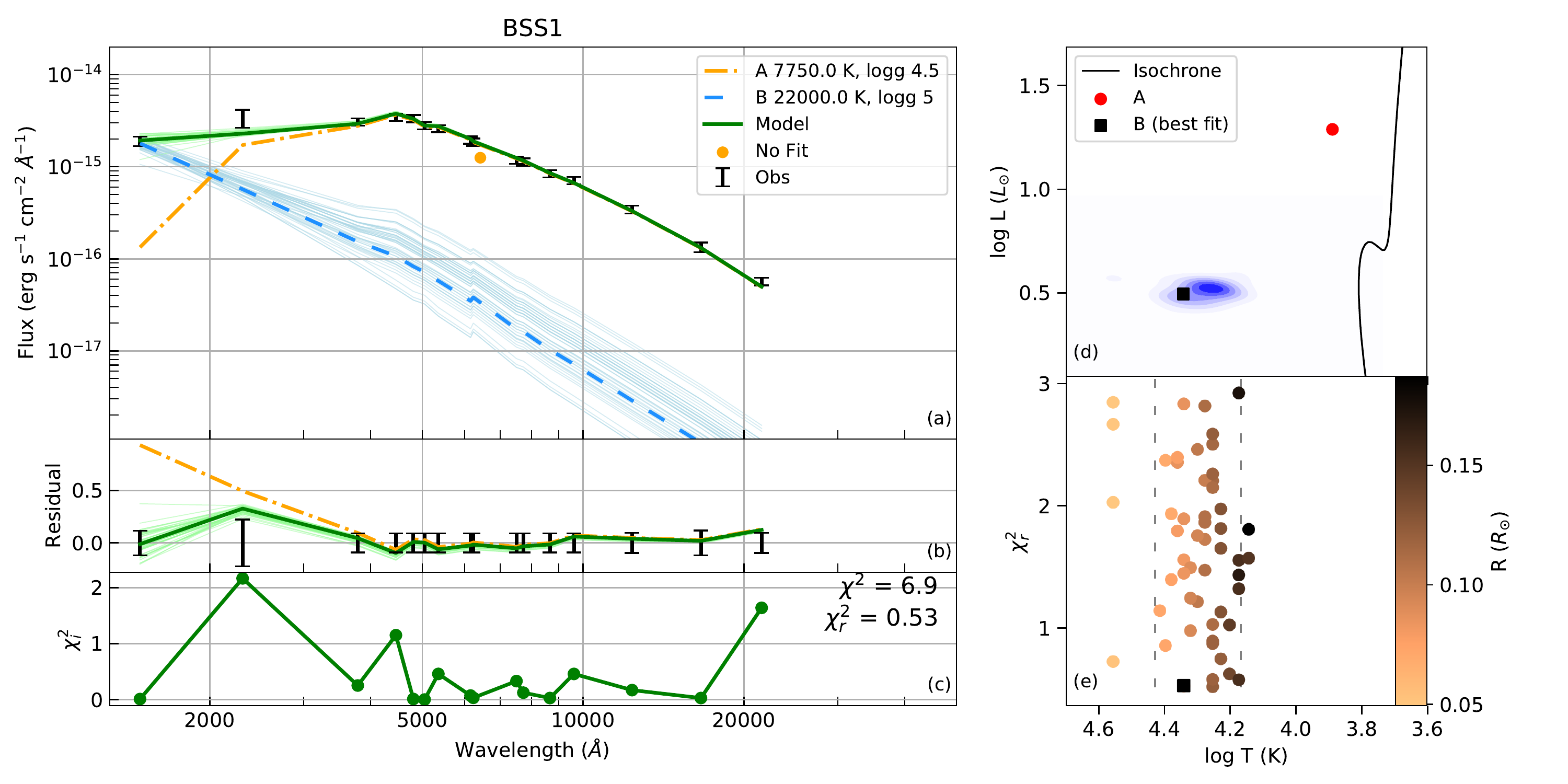}
    \caption{Two-component SED of BSS1. (a) The composite SED (green curve) is shown along with the observed flux (as black error bars). The unfitted point (in this case: CAHA.R) is shown as an orange dot. The cooler (BSS, orange dot-dashed curve) and hotter (blue dashed curve) components are also shown with their T$_{eff}$ and log $g$. The model, B component and residuals of noisy iterations are also shown as light coloured lines. (b) The fractional residual is shown for single component fit (orange dot-dashed curve) and composite fit (solid green curve). The fractional observational errors are also indicated on X-axis. (c) The $\chi_i^2$ of each data point. (d) HRD of the two components along with the isochrone for reference. The density distribution of the noisy B component fits is plotted in blue. (e) T$_{eff}$--$\chi^2$ distribution for the noisy B component fits coloured according to their radii. The dashed lines are the quoted limits of the temperature.}
    \label{fig:5_SEDs_Kurucz_1}
\end{figure} 

\begin{enumerate}
    \item We constructed the observed SEDs for all stars using the UV--IR wavelength data, as mentioned above.
    \item Kurucz models \citep{Castelli1997A&A...318..841C} of log $g \in (3.0,5.0)$ were fitted to optical and IR points (above 3000 \AA) using \textsc{vosa}\footnote{\url{http://svo2.cab.inta-csic.es/theory/vosa/index.php}} \citep{Bayo2008A&A...492..277B}. Some sources showed UV excess in multiple UV points compared to the model fit. We selected such stars to be fitted with a two-component SED. Otherwise, the single-component fits are deemed satisfactory and are stated in the lower part of Table~\ref{tab:5_parameters}.
    \item We used the cooler component parameters from above fits and then fitted a hotter component to the residual using \texttt{Binary\_SED\_Fitting}\footnote{\url{https://github.com/jikrant3/Binary_SED_Fitting} \texttt{Binary\_SED\_Fitting} is a python code which uses $\chi^2$ minimisation technique to fit two-component SEDs.}. In preliminary double component fits, the hotter components were found to be compact objects; hence they were fitted with log $g = 5$ Kurucz models.
    \item Very small errors in PAN-STARRS PS1, \textit{Gaia} EDR3 and A90 photometry led to ignoring relatively high error UV data-points; hence they were replaced with mean errors for better residual across all wavelengths. A few data points were removed to achieve better fits and lower $\chi^2$ (see Fig.~\ref{fig:5_SEDs_Kurucz_1} a).
    \item The best-fit parameters for single stars or cooler components are taken from the \textsc{vosa} fits. The hotter component parameters are taken from the least $\chi^2$ model in the two-component fitting. 
    \item The errors in cooler component parameters are fairly low and are taken as the grid values. The errors in secondary components are derived using the process mentioned in \S~\ref{sec:2_binary_SED_fitting}.
\end{enumerate}

\begin{table*}
    \centering
\resizebox{0.8\textwidth}{!}{ 
\footnotesize
   
    \begin{tabular}{lccccc cc r}
    \toprule
    \multicolumn{9}{c}{Double Fits}\\ 
Name	&	Comp.	&	log $g$	&	Teff			&	R			&	L			&	Scaling Factor	&	N$_{fit}$	&	$\chi_r^2$	\\
	&		&		&	[K]			&	[R$_{\odot}$]			&	[L$_{\odot}$]			&		&		&	($\chi_{r,single}^2$)	\\ \hline
BSS1	&	A	&	4.5	&	7750	$\pm$	125	&	2.44	$\pm$	0.04	&	19.3	$\pm$	1.8	&	8.99E-23	&	16	&	0.5 (10.5)	\\
	&	B	&	5	&	22000	$^{+4842}	_{-7269}$	&	0.122	$^{+0.150}	_{-0.037}$	&	3.1	$^{+0.7}	_{-0.4}$	&	2.23E-25	&		&		\\ 
BSS2	&	A	&	3.5	&	8250	$\pm$	125	&	3.72	$\pm$	0.06	&	57.6	$\pm$	3.4	&	2.09E-22	&	16	&	3.9 (1.2)	\\
	&	B	&	5	&	24000	$^{+6802}	_{-4996}$	&	0.234	$^{+0.131}	_{-0.089}$	&	16.4	$^{+1.5}	_{-1.2}$	&	8.27E-25	&		&		\\ 
BSS3	&	A	&	4.5	&	7250	$\pm$	125	&	3.56	$\pm$	0.06	&	31.6	$\pm$	2.1	&	1.92E-22	&	16	&	1.0 (16.1)	\\
	&	B	&	5	&	24000	$^{+3229}	_{-10186}$	&	0.094	$^{+0.191}	_{-0.016}$	&	2.7	$^{+0.8}	_{-0.3}$	&	1.34E-25	&		&		\\ 
BSS4	&	A	&	5	&	8000	$\pm$	125	&	2.16	$\pm$	0.04	&	17.2	$\pm$	1.5	&	7.06E-23	&	17	&	2.3 (9.6)	\\
	&	B	&	5	&	26000	$^{+1912}	_{-13323}$	&	0.089	$^{+0.304}	_{-0.011}$	&	3.3	$^{+0.8}	_{-0.3}$	&	1.20E-25	&		&		\\ 
BSS5	&	A	&	3.5	&	6500	$\pm$	125	&	2.29	$\pm$	0.04	&	8.4	$\pm$	0.8	&	7.90E-23	&	12	&	0.4 (19.9)	\\
	&	B	&	5	&	14000	$^{+12479}	_{-5493}$	&	0.237	$^{+0.376}	_{-0.150}$	&	1.9	$^{+2.1}	_{-0.5}$	&	8.47E-25	&		&		\\ 
BSS7	&	A	&	3	&	8500	$\pm$	125	&	2.90	$\pm$	0.05	&	39.6	$\pm$	1.4	&	1.27E-22	&	11	&	0.6 (1.4)	\\
	&	B	&	5	&	19000	$^{+8088}	_{-2901}$	&	0.270	$^{+0.102}	_{-0.134}$	&	8.6	$^{+1.2}	_{-0.8}$	&	1.11E-24	&		&		\\

\bottomrule
\end{tabular}
}
\resizebox{0.75\textwidth}{!}{ 
\begin{tabular}{lcccc ccccc r}
\multicolumn{11}{c}{Single Fits}\\
Name	&	log $g$	&	Teff	&	e\_Teff	&	R	&	e\_R	&	L	&	e\_L	&	Scaling Factor	&	N$_{fit}$	&	$\chi_r^2$	\\
	&		&	[K]	&	[K]	&	[R$_{\odot}$]	&	[R$_{\odot}$]	&	[L$_{\odot}$]	&	[L$_{\odot}$]	&		&		&		\\ \hline
BSS6	&	4	&	6500	&	125	&	2.32	&	0.04	&	8.74	&	0.41	&	8.15E-23	&	11	&	3.4	\\
BSS8	&	3	&	7000	&	125	&	4.19	&	0.07	&	38.17	&	2.05	&	2.66E-22	&	11	&	3.2	\\
BSS9	&	3.5	&	7500	&	125	&	2.81	&	0.05	&	22.58	&	1.14	&	1.20E-22	&	11	&	4.2	\\
BSS10	&	3	&	6250	&	125	&	2.91	&	0.05	&	11.67	&	0.54	&	1.28E-22	&	11	&	3.9	\\
BSS11	&	4	&	6750	&	125	&	5.28	&	0.09	&	52.55	&	2.64	&	4.22E-22	&	11	&	1.6	\\
BSS12	&	3.5	&	7000	&	125	&	1.97	&	0.03	&	8.66	&	0.43	&	5.85E-23	&	12	&	47.9	\\
BSS13	&	3.5	&	6750	&	125	&	2.32	&	0.04	&	10.17	&	0.51	&	8.16E-23	&	11	&	3.2	\\
BSS14	&	3	&	6750	&	125	&	2.38	&	0.04	&	10.63	&	0.48	&	8.54E-23	&	11	&	2.5	\\
BSS15	&	3	&	6500	&	125	&	2.44	&	0.04	&	9.71	&	0.49	&	9.02E-23	&	11	&	3.6	\\
BSS16	&	3	&	6750	&	125	&	1.99	&	0.03	&	7.42	&	0.37	&	6.00E-23	&	11	&	3.1	\\
BSS17	&	3.5	&	7000	&	125	&	2.32	&	0.04	&	11.60	&	0.55	&	8.12E-23	&	11	&	4.3	\\
BSS18	&	4	&	6500	&	125	&	2.22	&	0.04	&	8.08	&	0.34	&	7.48E-23	&	15	&	12.3	\\
BSS19	&	4.5	&	7250	&	125	&	2.89	&	0.05	&	20.92	&	1.17	&	1.26E-22	&	8	&	13.3	\\
BSS20	&	5	&	6500	&	125	&	2.23	&	0.04	&	8.05	&	0.32	&	7.53E-23	&	11	&	3.4	\\
BSS21	&	5	&	6500	&	125	&	3.08	&	0.05	&	15.32	&	0.61	&	1.43E-22	&	11	&	4.9	\\
BSS22	&	3.5	&	6250	&	125	&	3.28	&	0.06	&	14.88	&	0.62	&	1.63E-22	&	11	&	1.5	\\
BSS23	&	3.5	&	6500	&	125	&	2.84	&	0.05	&	13.00	&	0.59	&	1.22E-22	&	11	&	3.7	\\
BSS24	&	3	&	7250	&	125	&	2.12	&	0.04	&	11.24	&	0.59	&	6.77E-23	&	11	&	2.1	\\
BSS25	&	5	&	6250	&	125	&	2.64	&	0.05	&	9.73	&	0.39	&	1.06E-22	&	15	&	11.7	\\
BSS26	&	3	&	7250	&	125	&	2.28	&	0.04	&	13.06	&	0.67	&	7.89E-23	&	15	&	11.3	\\
BSS27	&	4.5	&	6250	&	125	&	2.17	&	0.04	&	6.47	&	0.26	&	7.10E-23	&	11	&	4.3	\\
BSS28	&	4	&	7250	&	125	&	2.23	&	0.04	&	12.57	&	0.68	&	7.54E-23	&	15	&	9.8	\\
BSS29	&	4	&	6500	&	130	&	2.27	&	0.04	&	8.40	&	0.41	&	7.81E-23	&	5	&	125.1	\\
BSS30	&	4.5	&	6250	&	125	&	2.45	&	0.04	&	8.35	&	0.34	&	9.08E-23	&	11	&	5.3	\\
BSS31	&	5	&	8250	&	127	&	1.40	&	0.02	&	8.09	&	0.35	&	2.98E-23	&	15	&	9.0	\\
BSS32	&	4	&	6250	&	125	&	2.57	&	0.04	&	9.27	&	0.39	&	9.98E-23	&	15	&	19.5	\\
BSS33	&	4	&	6500	&	125	&	2.04	&	0.04	&	6.79	&	0.28	&	6.29E-23	&	15	&	10.9	\\
BSS34	&	4	&	6500	&	125	&	3.80	&	0.07	&	23.42	&	1.01	&	2.19E-22	&	11	&	4.2	\\
BSS35	&	3	&	6250	&	125	&	2.41	&	0.04	&	8.07	&	0.38	&	8.75E-23	&	11	&	4.4	\\
BSS36	&	3.5	&	6500	&	125	&	2.37	&	0.04	&	9.13	&	0.44	&	8.51E-23	&	11	&	3.3	\\
BSS37	&	5	&	6500	&	125	&	2.25	&	0.04	&	8.23	&	0.34	&	7.65E-23	&	11	&	3.6	\\
BSS38	&	5	&	6500	&	125	&	2.68	&	0.05	&	11.70	&	0.48	&	1.09E-22	&	11	&	3.4	\\
BSS39	&	3.5	&	6500	&	125	&	3.31	&	0.06	&	17.67	&	0.80	&	1.65E-22	&	11	&	3.9	\\

\bottomrule
    \end{tabular}
}
    \caption{Fitting parameters of the best fit of the double and single-component fits of BSSs with the hotter component. The scaling factor is the value by which the model has to be multiplied to fit the data, N$_{fit}$ is the number of data points fitted, and $\chi_r^2$ is the reduced $\chi^2$ for the composite fit. The $\chi_r^2$ values of single fits of the cooler components are given in brackets. \textit{ Note: the log $g$ values are imprecise due to the insensitivity of the SED to log $g$.}}
    \label{tab:5_parameters}
\end{table*}

Fig.~\ref{fig:5_CMD} shows the CMD of 1412 cluster candidates identified from {\it Gaia} EDR3 with a probability of over 50\% \citep{Jadhav2021MNRAS.503..236J} and are marked as grey points.
Among these stars, we have selected 39 member stars brighter ($G<17.5$ mag) and bluer ($G_{BP}-G_{RP} < 1.1$ mag) than the MSTO as BSSs.
Seven were detected in F148W and seven in N219M (four in both filters). An isochrone of log $age$ $=$ 9.7 is over-plotted on the CMD. 20 of the BSSs are detected in the NUV filter of \textit{GALEX}. We fitted Kurucz model SEDs to all BSSs and found excess UV flux in 15 BSS (BSS 1, 2, 3, 4, 5, 7, 8, 9, 10, 19, 26, 28, 29, 33 and 36). Among these, BSS1, 2, 3, 4, 5 and 7 have multiple UV data points from UVIT or \textit{GALEX} or both. Only these six were fitted with double component SEDs because a hot component fit can be more reliable if the number of UV data points is more than one. Hereafter, these six BSS will be referred to as `UV bright BSSs', and others will be referred to as `UV faint BSSs'. The four BSS detected in UVIT but not fitted with the hotter component are shown as red X's in the CMD.

We have shown an example of a double component SED fit of BSS1 in Fig.~\ref{fig:5_SEDs_Kurucz_1} (a).
The BSS1-A component is a BSS with 7750 K, while the BSS1-B component has T$_{eff}$ of 22000 K. The reduction in residual after including the hotter component is visible in Fig.~\ref{fig:5_SEDs_Kurucz_1} (b). The $\chi_i^2$ for individual points is shown in Fig.~\ref{fig:5_SEDs_Kurucz_1} (c) with $\chi_r^2$ of 0.53.
Fig.~\ref{fig:5_SEDs_Kurucz_1} (d) shows the HRD of A and B components. The density distribution of noisy \& converging iterations is also shown to get an idea of degeneracy in temperature and luminosity. Fig.~\ref{fig:5_SEDs_Kurucz_1} (e) panel shows the best fit and the noisy \& converging iterations in T$_{eff}$--$\chi^2$ phase-plane.
The double component fits of BSS2, 3, 4, 5 and 7, and single-component fits of BSS10 and BSS15 are shown in Fig.~\ref{fig:5_SEDs_Kurucz_2}. The fitting parameters are mentioned in Table~\ref{tab:5_parameters}.

\begin{figure}
    \centering
    \includegraphics[width = 0.8\textwidth]{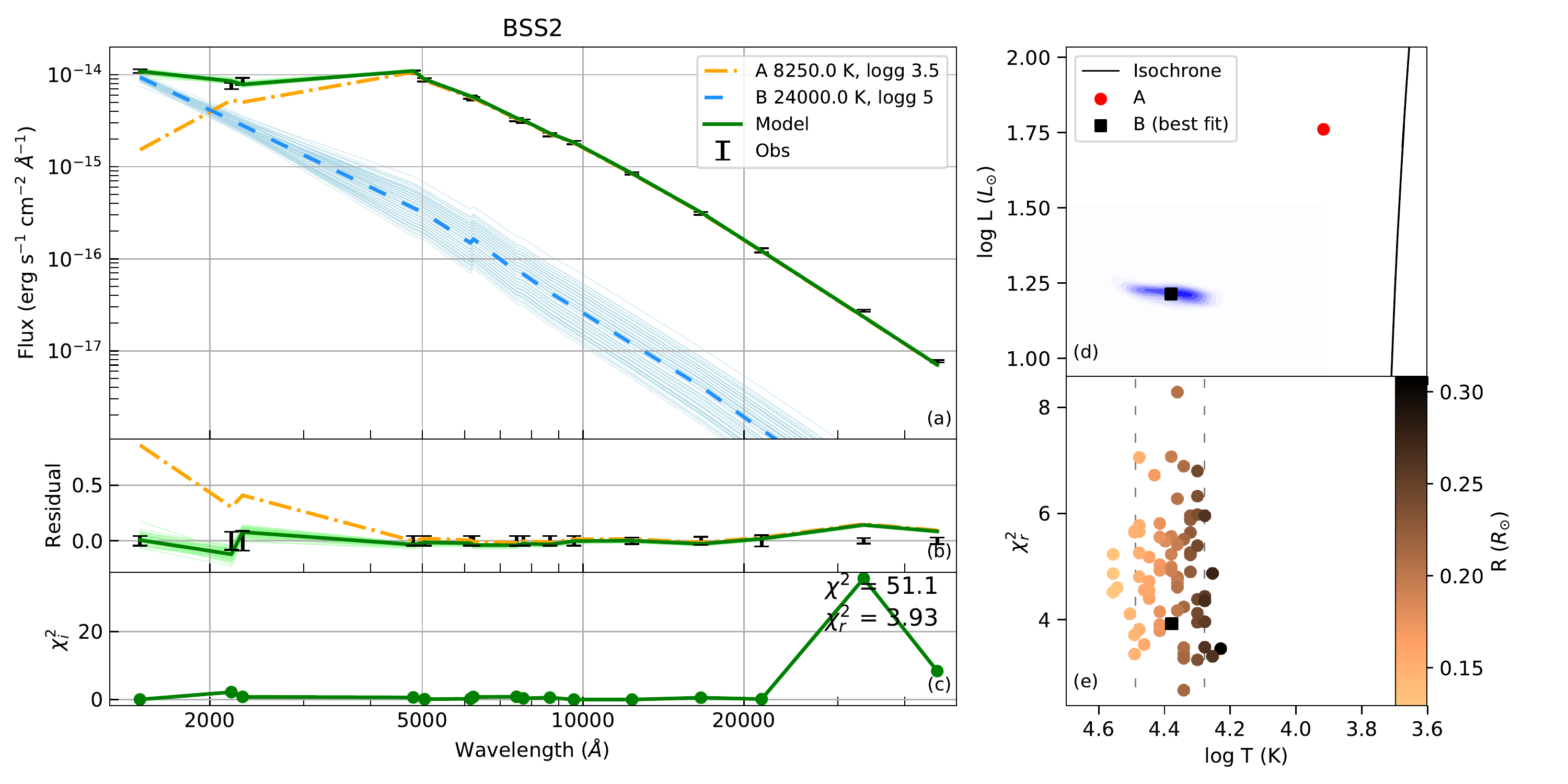}
    \includegraphics[width = 0.8\textwidth]{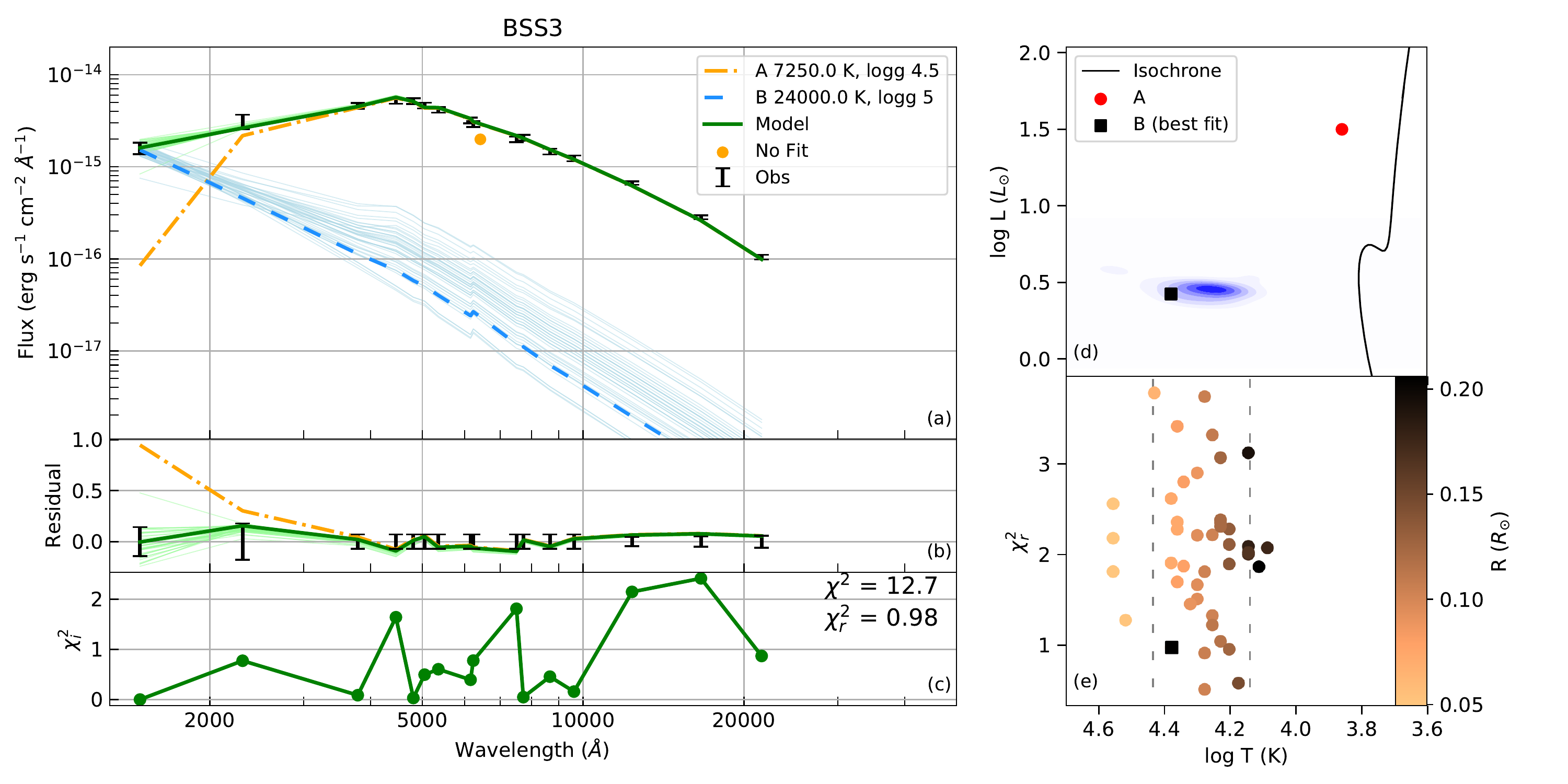}
    \includegraphics[width = 0.8\textwidth]{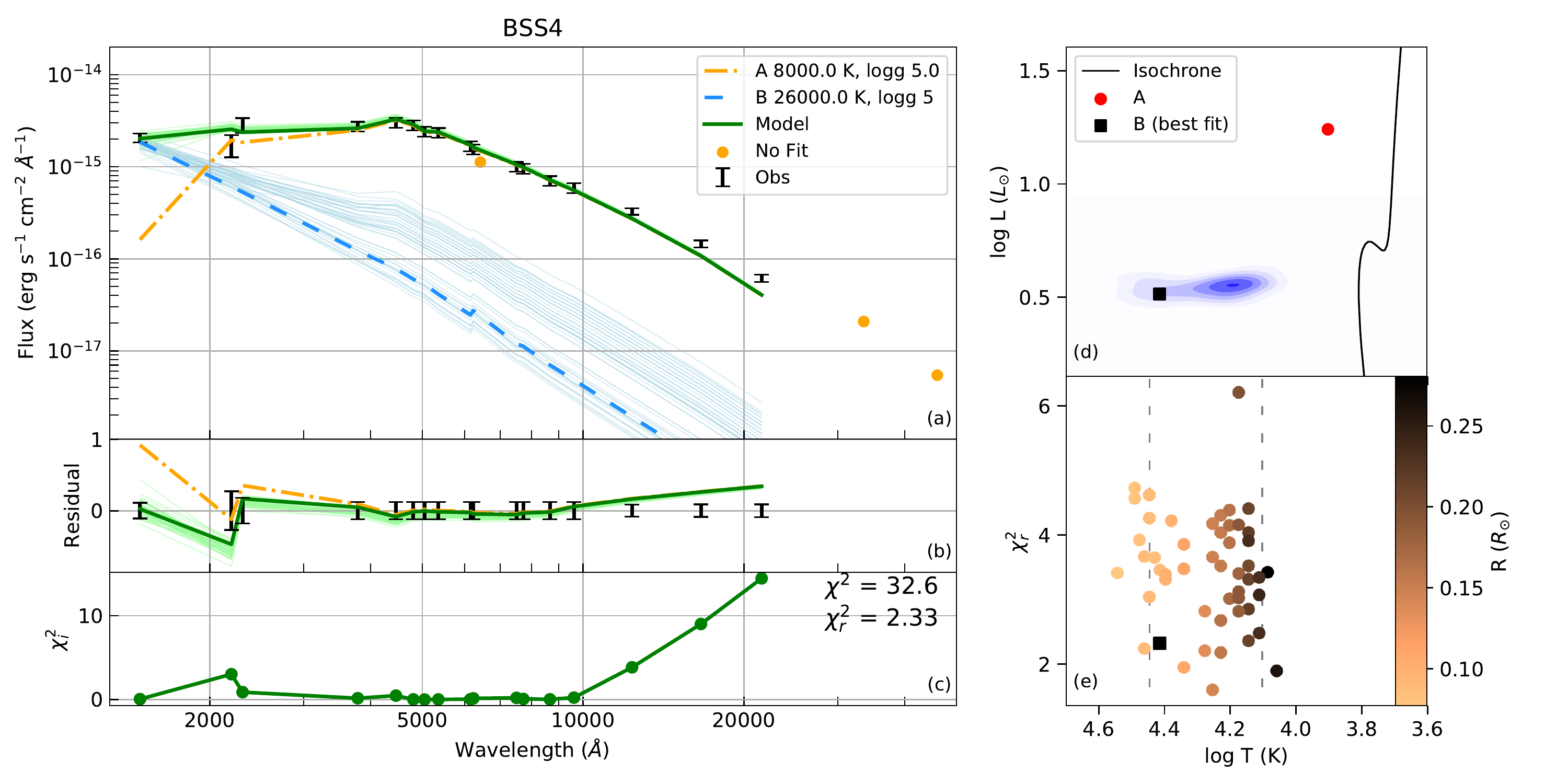}
    \caption[]{Double component fits of BSS2, BSS3 and BSS4. The descriptions of double component fits are the same as Fig.~\ref{fig:5_SEDs_Kurucz_1}. (\textit{Continued}...)}
    \label{fig:5_SEDs_Kurucz_2}
\end{figure}

\begin{figure}
    \ContinuedFloat
    \centering
    \includegraphics[width = 0.8\textwidth]{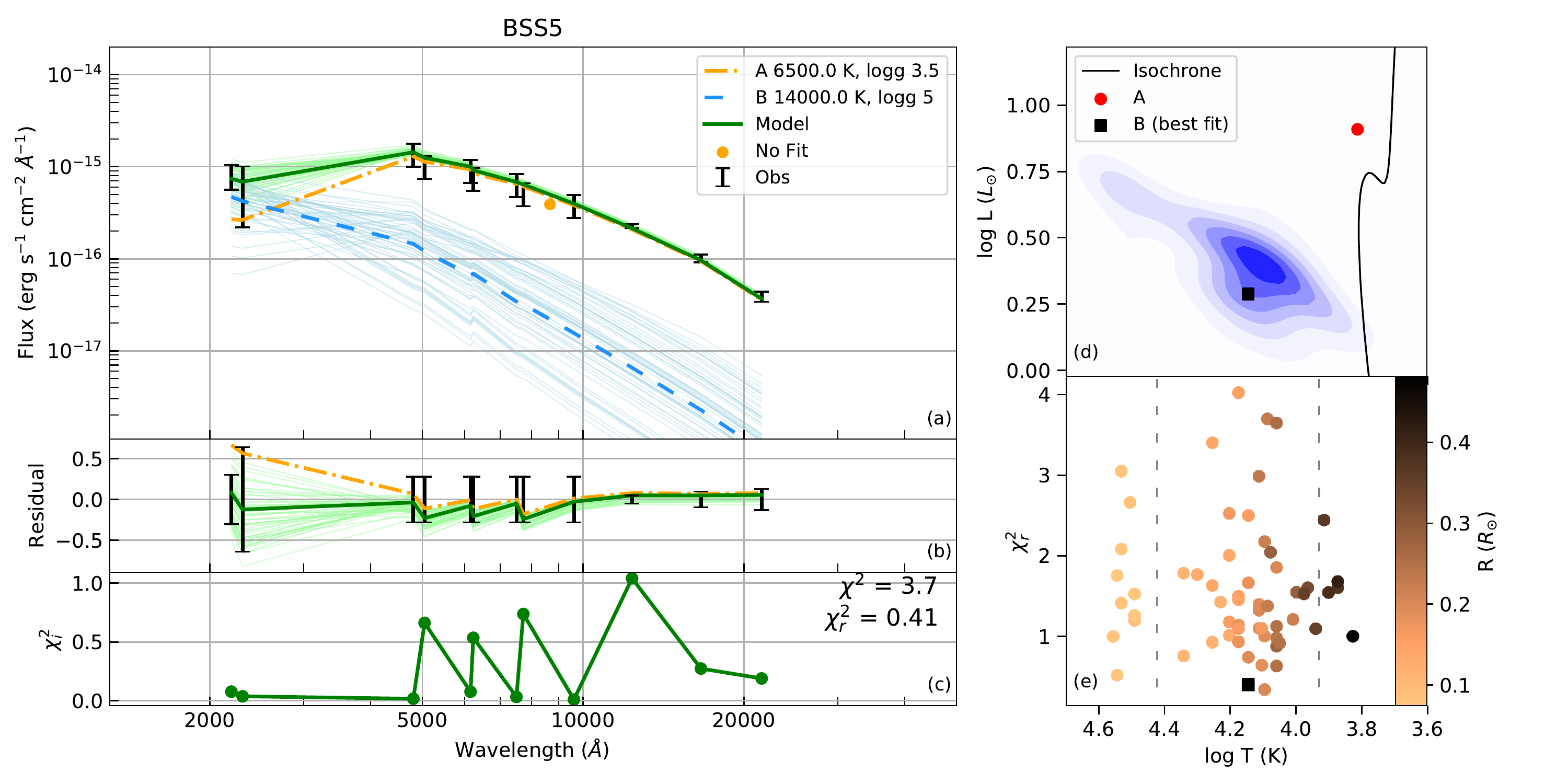}
    \includegraphics[width = 0.8\textwidth]{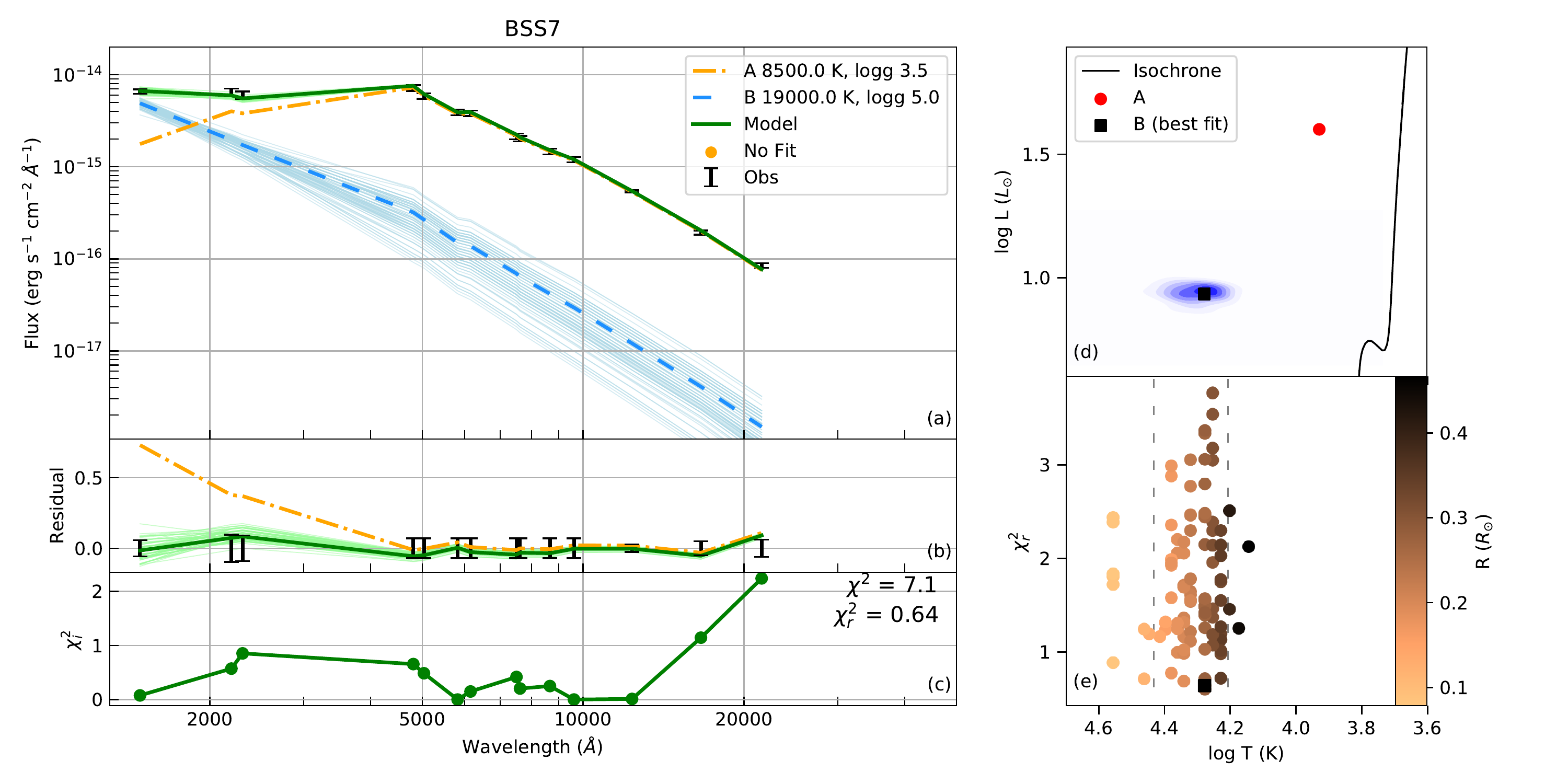}
    \includegraphics[width=0.49\textwidth]{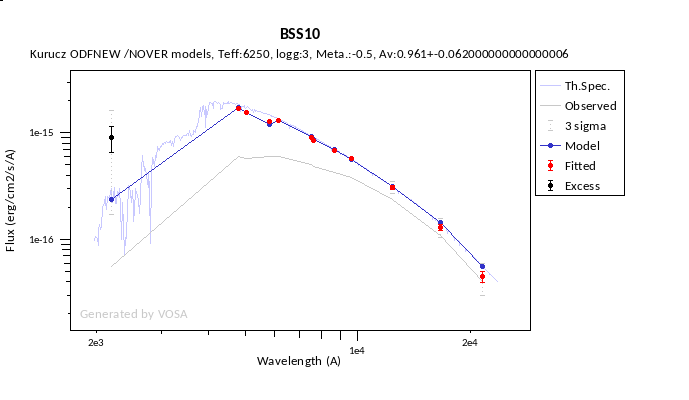}
    \includegraphics[width=0.49\textwidth]{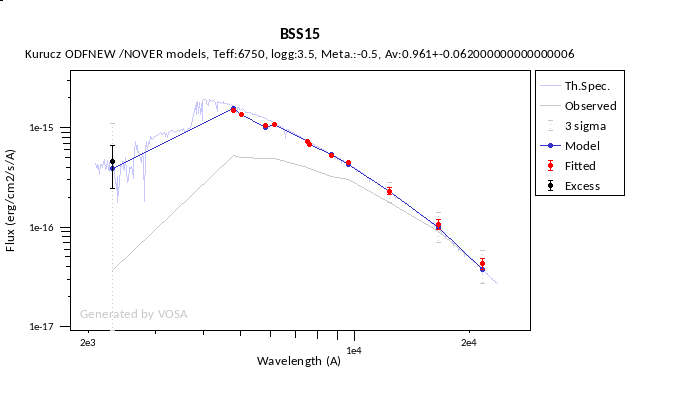}
    \caption[(Double component fits of BSS2, BSS3, BSS4, BSS5 and BSS7. And the single component fits of BSS10 and BSS15 are shown as an example with model fit (blue curve), fitted data points (red points) with 1 $\sigma$ and 3 $\sigma$ errors as solid and dashed lines. The theoretical spectra (in grey) is added for reference. The observed (reddening affected) SED is shown in grey below the corrected data-points. The title mentions the T$_{eff}$, log $g$, metallicity and A$_V$ of the model fit.]{(\textit{Continued}...) Double component fits of BSS5 and BSS7. And the single component fits of BSS10 and BSS15 are shown as an example with model fit (blue curve), fitted data points (red points) with 1 $\sigma$ and 3 $\sigma$ errors as solid and dashed lines. The theoretical spectra (in grey) is added for reference. The observed (reddening affected) SED is shown in grey below the corrected data-points. The title mentions the T$_{eff}$, log $g$, metallicity and A$_V$ of the model fit.
    }   
\end{figure}

\begin{figure}
    \centering
    \includegraphics[width=0.48\textwidth]{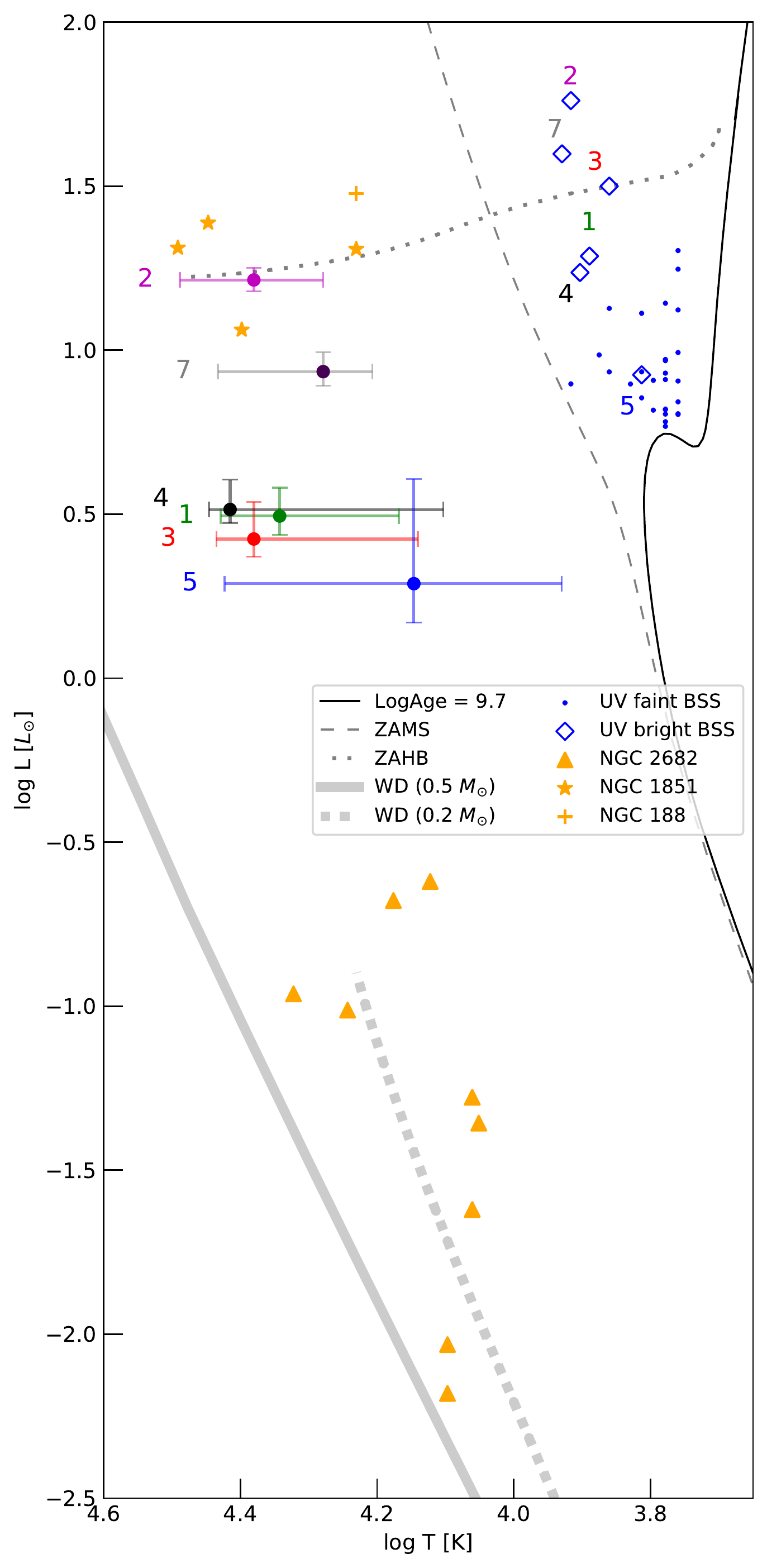}
    \caption{HRD of locations of components of binaries in King 2, NGC 2682 \citep{Sindhu2019ApJ...882...43S, Jadhav2019ApJ...886...13J, Pandey2021MNRAS.507.2373P}, NGC 188 \citep{Subramaniam2016ApJ...833L..27S} and EHB stars in NGC 1851 \citep{Singh2020ApJ...905...44S}. In King 2, the UV faint BSS (blue dots), UV bright BSS (blue diamonds with their ID) and hotter components in BSSs (coloured filled circles with numbers and error bars) are shown. Hotter components in NGC 2682 and NGC 188 are shown as orange triangles and crosses. EHB stars in NGC 1851 are shown as stars.
    The PARSEC isochrone (black curve), PARSEC zero-age MS (ZAMS; dashed grey line), WD cooling curves (thick grey curves) and BaSTI ZAHB (dotted grey curve) are shown for reference.}
    \label{fig:5_comparison_CMD}
\end{figure}

The HRD of the BSSs detected in King 2 is shown in Fig.~\ref{fig:5_comparison_CMD}.
We have included the positions of hotter companions of BSSs in NGC 188 \citep{Subramaniam2016ApJ...833L..27S} and NGC 2682 \citep{Sindhu2019ApJ...882...43S,Jadhav2019ApJ...886...13J,Pandey2021MNRAS.507.2373P}. The EHB stars in the globular cluster NGC 1851 \citep{Singh2020ApJ...905...44S} are also included.
The PARSEC\footnote{\url{stev.oapd.inaf.it/cgi-bin/cmd}} isochrone of log $age$ $=$ 9.7 is over-plotted \citep{Bressan2012MNRAS.427..127B}, along with the WD cooling curves\footnote{\url{www.astro.umontreal.ca/~bergeron/CoolingModels/}} (\citealt{Tremblay2009ApJ...696.1755T}) and BaSTI\footnote{\url{basti-iac.oa-abruzzo.inaf.it/isocs.html}} ZAHB (\citealt{Hidalgo2018ApJ...856..125H}). 

\section{Results and discussion} \label{sec:5_Discussion}

\textbf{BSS and their companions in literature:}

The BSSs have T$_{eff}$ range of 5750--8500 K and radii of 1.4--5.21 \Rsun. By comparison to isochrones, they have mass in the range of 1.2--1.9 \Msun, the brightest BSS being 3 mag brighter than the MSTO. 
The majority of BSSs in King 2 have T$_{eff}$ similar to the older NGC 188 (6100--6800 K; \citealt{Gosnell2015ApJ...814..163G}), but are cooler than NGC 2682 (6250--9000 K; \citealt{Pandey2021MNRAS.507.2373P}), which is expected due to its slightly younger age.
The BSSs in NGC 188 \citep{Geller2011Natur.478..356G, Gosnell2014ApJ...783L...8G, Subramaniam2016ApJ...833L..27S} and NGC 2682 \citep{Sindhu2019ApJ...882...43S} are known to have evolved companions. According to their luminosity and temperature, the companions were classified as WD, ELM WDs, post-AGB/HB. 

BSS2-A, BSS3-A and BSS7-A lie above/on the ZAHB in Fig.~\ref{fig:5_comparison_CMD}. There is a degeneracy in this region of the HRD where one could find both massive BSS and ZAHB stars. Stars in these two evolutionary phases will have different masses (HB mass $<$MSTO; BSS mass $>$MSTO) that could be used to lift the degeneracy.
\citet{Bond1971PASP...83..638B} measured the masses of stars in this region of the NGC 2682 CMD and determined that they are indeed more massive than MSTO i.e., BSSs. One star is found in this region of the NGC 188 CMD, and it is classified as a blue-HB \citep{Rani2021JApA...42...47R}, this star is significantly brighter than the rest of the BSSs. In the case of King 2, the BSSs show a continuous distribution up to the brightest BSS; hence BSS2-A/BSS3-A are most likely normal BSSs. However, their mass estimations (via log $g$ measurements or asteroseismology) are required before confirming their evolutionary status.

\noindent \textbf{What are the hotter companions?}

The hotter companions in UV bright BSSs have T$_{eff}$ of 14000--26000 K (spectral type B) and radii of 0.09--0.27 \Rsun. 
Fig.~\ref{fig:5_comparison_CMD} shows the density distributions of the best 100 fits for the hotter companions. Fig.~\ref{fig:5_comparison_CMD} also shows the location of various companions to BSS in NGC 188 and NGC 2682, and EHB stars in NGC 1851, one of which has a BSS as its companion \citep{Singh2020ApJ...905...44S}.  

The hotter companion to BSSs in NGC 2682 are all fainter and near the WD cooling curves. While those for NGC 1851 and NGC 188 lie closer to the ZAHB region. In King 2, the limiting magnitude of UVIT observations is 23 and 22 mag in F148W and N219M, respectively. According to WD cooling models (corrected for distance and extinction; \citealt{Tremblay2009ApJ...696.1755T}), only WDs younger than 0.7 and 16 Myr would be detectable in FUV and NUV, respectively. As seen in Fig.~\ref{fig:5_comparison_CMD}, the hotter companions are well above the WD cooling curves, and these are not WDs.

The hotter companions are likely to be hot HB stars which are also known as EHB stars or sdB stars, as inferred from their T$_{eff}$, radii, and luminosity. These are core-helium burning stars with T$_{eff}$ in the range of 20,000--40,000 K and are compact (0.15--0.35 \Rsun; \citealt{Heber2016PASP..128h2001H,Sahoo2020MNRAS.495.2844S}). As these stars are hot and not as small as WDs, they appear bright in the UV. These stars are thought to contribute to the UV upturn seen in elliptical or in early-type galaxies \citep{Brown1997ApJ...482..685B}. 
The sdB stars have a very thin hydrogen envelope and are thought to be the stripped core of an RG star \citep{Heber2016PASP..128h2001H}. \citet{Maxted2001MNRAS.326.1391M} found a good fraction of the sdB stars in detached but short-period binary systems. sdB stars are thought to provide important clues to common envelope evolution in tight binaries.

BSS2-B lies on the blue end of ZAHB and is very similar to the EHBs in the outskirts of the globular cluster NGC 1851. Similarly, a hot and bright post-AGB/HB candidate was found as a companion to a BSS in NGC 188 \citep{Subramaniam2016ApJ...833L..27S}. Hence, BSS2-B could be an EHB star. 
BSS7-B is near a known EHB from NGC 1851 and is slightly fainter than the ZAHB; hence, it is likely to be an EHB.
BSS5-B lies slightly above the WDs. Hence, it can be a very young He-WD \citep{Panei2007MNRAS.382..779P} or an sdB star.
The rest of the King 2 BSS companions are slightly fainter than the ZAHB, and they are likely sdBs.

\noindent \textbf{Formation pathways of BSSs and EHBs/sdBs:}

The BSS formation mechanism involves mass gain, while the EHB/sdB formation involves stripping the envelope of a post-MS star.
The detection of EHB/sdBs confirms binary MT as the formation mechanism for BSS1--BSS5 and BSS7. As the cooler companions are BSSs that are supposed to have gained mass, we can infer that the detected EHBs/sdBs have transferred mass to the BSSs companions. Therefore, the BSS+EHB/sdB systems in King 2 represent stars on both sides of the mass exchange. We see a range in their temperature and luminosity, suggesting a diversity among the hotter companions.

The lifespan of sdB stars is expected to be between 100 and 200 Myr (\citealt{Bloemen2014A&A...569A.123B, Schindler2015ApJ...806..178S} and references therein), after which they descend the WD cooling curve. When the current BSS expands, the MT is expected to start again due to the short orbit, which already allowed the previous instance of MT. Then the system can begin stable/unstable MT and become a WD+WD system. Alternatively, it can merge through a common-envelope phase and become a massive WD. The exact evolution will depend on the orbital parameters, MT efficiency, and mass loss. 

King 2, one of the oldest OC, lies in the outskirts of the galactic disk. It is metal-poor compared to the Galactic disc OCs. The environment is quite similar to the outskirts of GCs, which are also metal-poor, old and of comparable density. While most of the BSSs in GCs lie in the core and are formed via mergers \citep{Chatterjee2013ApJ...777..106C}, BSSs in the outskirts of GCs can form through MT as seen in EHB-4 of NGC 1851 \citep{Singh2020ApJ...905...44S}. Our study suggests that at least 15\% of the BSSs in King 2 are formed via the MT formation pathway.

We have seen sdB companions to BSSs in NGC 188 and NGC 1851 (both are older systems); however, none in NGC 2682 (which is younger). NGC 6791 of slightly younger age also has sdB stars \citep{Kaluzny1992AcA....42...29K, Reed2012MNRAS.427.1245R}. This might suggest that there is an upper age limit of $\sim$5 Gyr for the formation of sdB stars in OCs.

\section{Conclusions and summary} \label{sec:5_conclusion}
\begin{itemize}
    \item The old OC King 2 has a large population (39) of BSSs, spreading up to 3 mag brighter than the MSTO. We constructed SEDs of all the BSS using UV--IR data. The BSSs have T$_{eff}$ in the range of 5750--8250 K, luminosity in the range of 5.6--57.5 L$_{\odot}$ and mass in the range of 1.2--1.9 M$_{\odot}$.
    \item Six of the UV bright BSS showed excess UV flux and were successfully fitted with double component SEDs. The hotter components have T$_{eff}$ of 14000--26000 K and R/R$_{\odot}$ of 0.09--0.27, suggesting a range of properties. Two of the hotter companions to the BSS are likely EHB stars, while four are likely sdB stars.
    \item EHB/sdB companions imply that these 6 (out of 39) BSSs have formed via binary MT. The SED fits show that sdB stars can be created in old OCs such as King 2 (similar to old OC NGC 188 and globular cluster NGC 1851). 
\end{itemize}

\begin{savequote}[100mm]
Ah--but I was so much older then; I’m younger than that now
\qauthor{Bob Dylan}
\end{savequote}

\chapter[Blue Stragglers in Galactic Open Clusters]{Blue Stragglers in Galactic Open Clusters \\ \large{\textcolor{gray}{Jadhav \& Subramaniam, 2021, MNRAS, 507, 1699}}}
\label{ch:BSS_catalogue}
\begin{quote}\small
\end{quote}

\section{Introduction} \label{sec:6_6_Introduction}
BSSs stand out among the cluster members due to their bluer colour and luminous nature. Chapter \ref{ch:intro} mentions the different mechanisms responsible for BSS formation. However, it is not easy to pinpoint the exact pathway without information about their companions, chemical signatures and orbital parameters.
There are many dedicated studies on the BSS population of individual clusters, but only a few on BSS population across different clusters (e.g., \citealt{Leiner_2021ApJ...908..229L}). 
We require a homogeneous catalogue to study the BSSs in a larger cluster sample. \citet{Ahumada_2007A&A...463..789A} had provided the pre-\textit{Gaia} catalogue of BSSs and recently \citet{Rain_2021arXiv210306004R} have produced a \textit{Gaia} based BSS catalogue (though it is not available at the time of this writing).
A homogeneous catalogue of BSSs will help analyse the dependence of the BSS population on the cluster properties and occurrence in them. Furthermore, such a catalogue can be further used to make targeted observations of exotic BSS populations in the multi-wavelength regime, revealing and constraining the formation scenarios and stellar parameters.

Multi-wavelength studies of BSSs using \textit{Hubble Space Telescope} \citep{Gosnell2014ApJ...783L...8G, Gosnell2019ApJ...885...45G} and the UVIT \citep{Subramaniam2016ApJ...833L..27S, Sindhu2019ApJ...882...43S, Jadhav2021arXiv210213375J} have detected hot companions to BSSs. The hot companions of the BSSs were found to differ from cluster to cluster and found to show a large variety, such as ELM WD, HB stars and sub-dwarfs. These studies suggest that remnants of the donors in binary BSSs occupy a fairly large parameter space that is yet to be adequately explored. One of the aims of this study is to identify potential clusters with BSSs to be followed up with UV imaging using \textit{AstroSat}. 

In this study, we have created a catalogue of BSSs in OCs (\S \ref{sec:6_Data_and_Method}), presented and discussed the properties of BSSs and their dependence on cluster parameters (\S \ref{sec:6_Results} and \S \ref{sec:6_discussion}).

\section{Data and classification} \label{sec:6_Data_and_Method}

As BSSs lie above the MSTO of the clusters, where stars are normally not expected to be found, accurate PM information is necessary to confirm their membership. \citet{Cantat2018A&A...618A..93C,Cantat2020} provided membership and cluster parameters of OCs using \textit{Gaia} DR2 \citep{Gaia2016A&A...595A...1G, Gaia2018A&A...616A...1G}. Their catalogue provides the age, distance, extinction, and other aggregate cluster properties. We selected 670 clusters with log($age$) over 8.5 ($\sim$300 Myr) as our primary sample. The lower age cutoff was chosen to avoid confusion between BSSs and blue giants in younger clusters. Though it is known that the BSSs like systems are more prevalent in relatively older clusters, it is also required to explore their existence in younger clusters. The clusters in the age range 300 Myr--10 Gyr will serve both purposes.

\begin{figure}[!ht]
    \centering
    \includegraphics[width=0.48\textwidth]{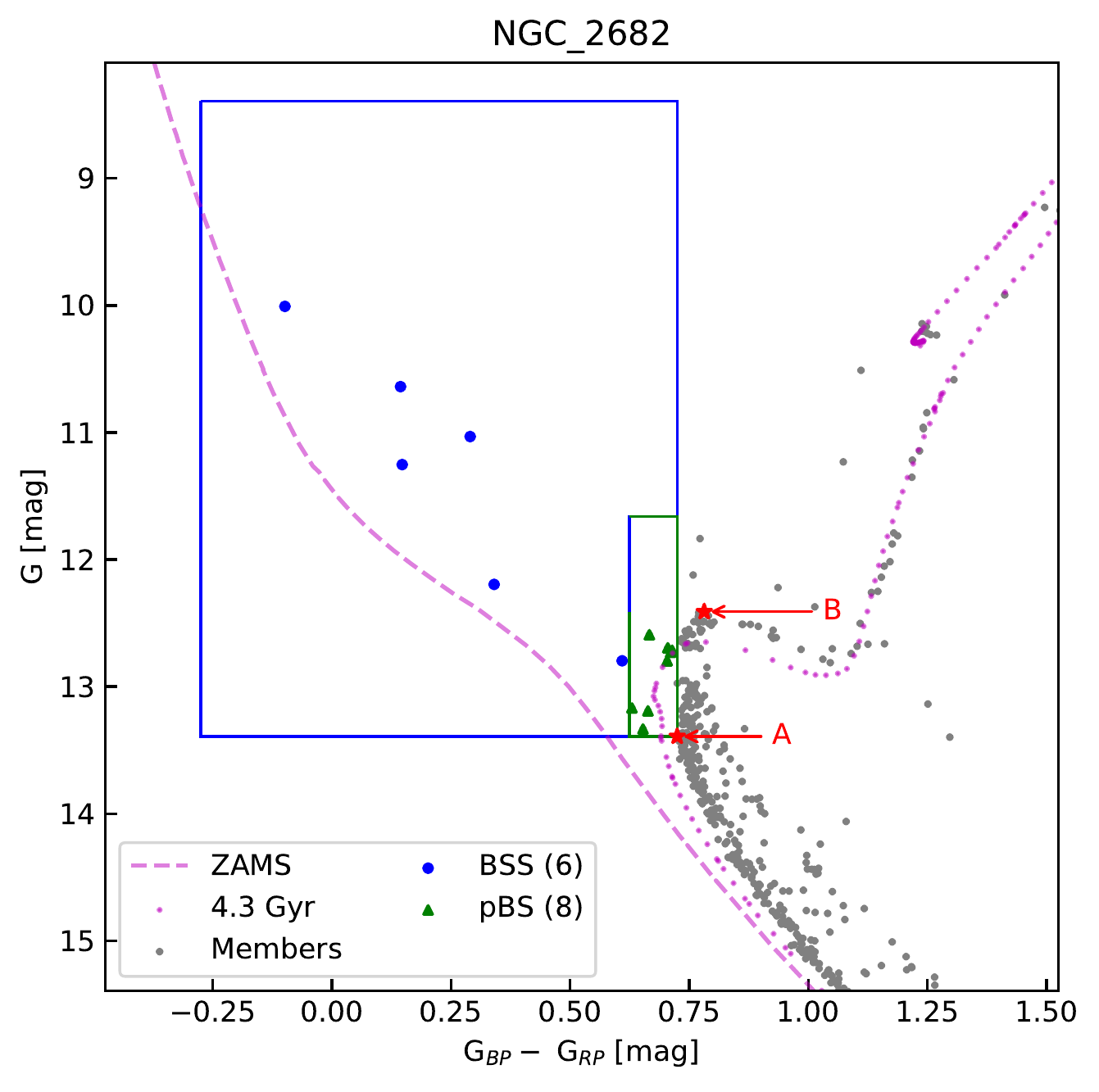}
    \includegraphics[width=0.48\textwidth]{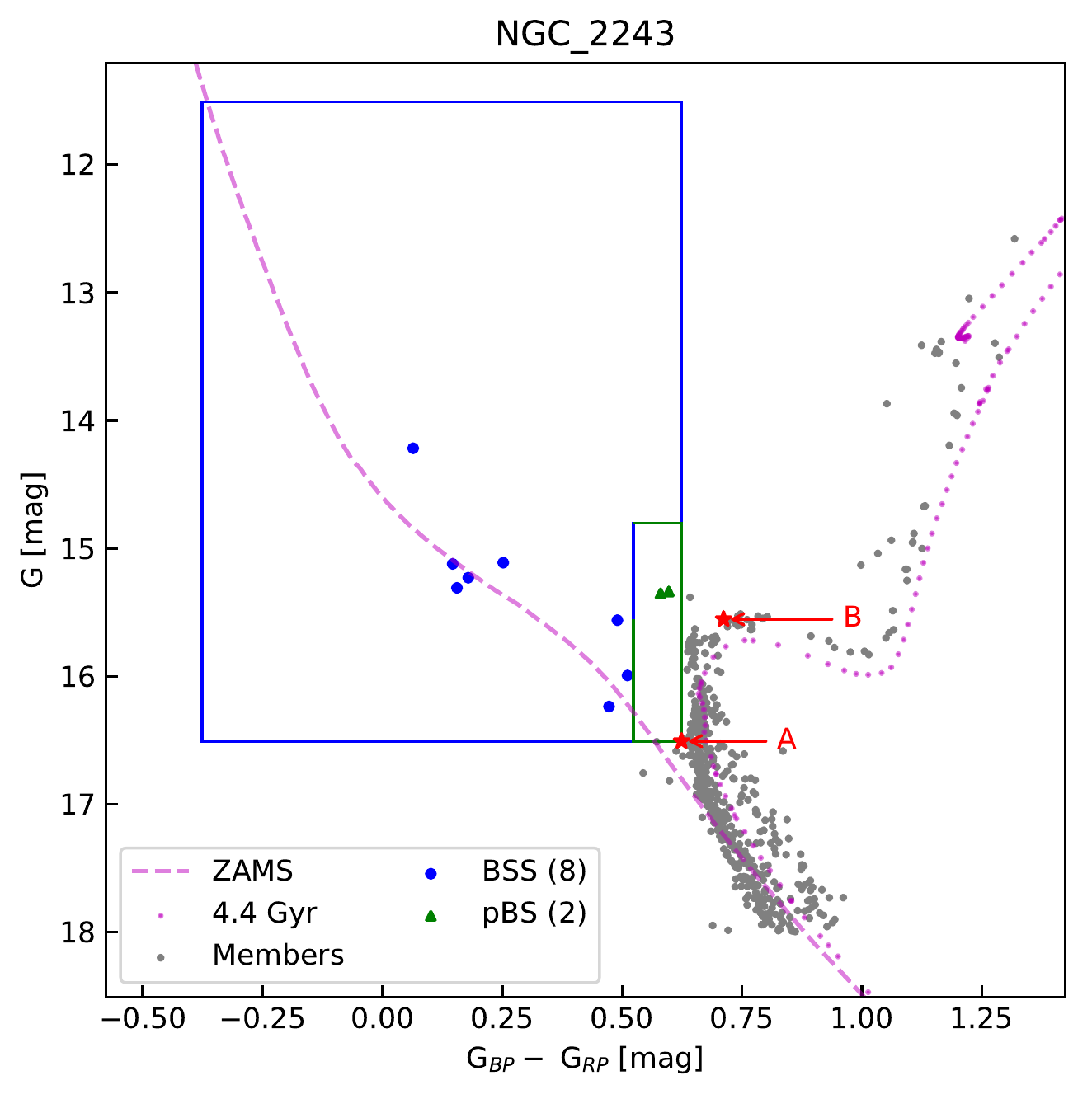}
    \includegraphics[width=0.48\textwidth]{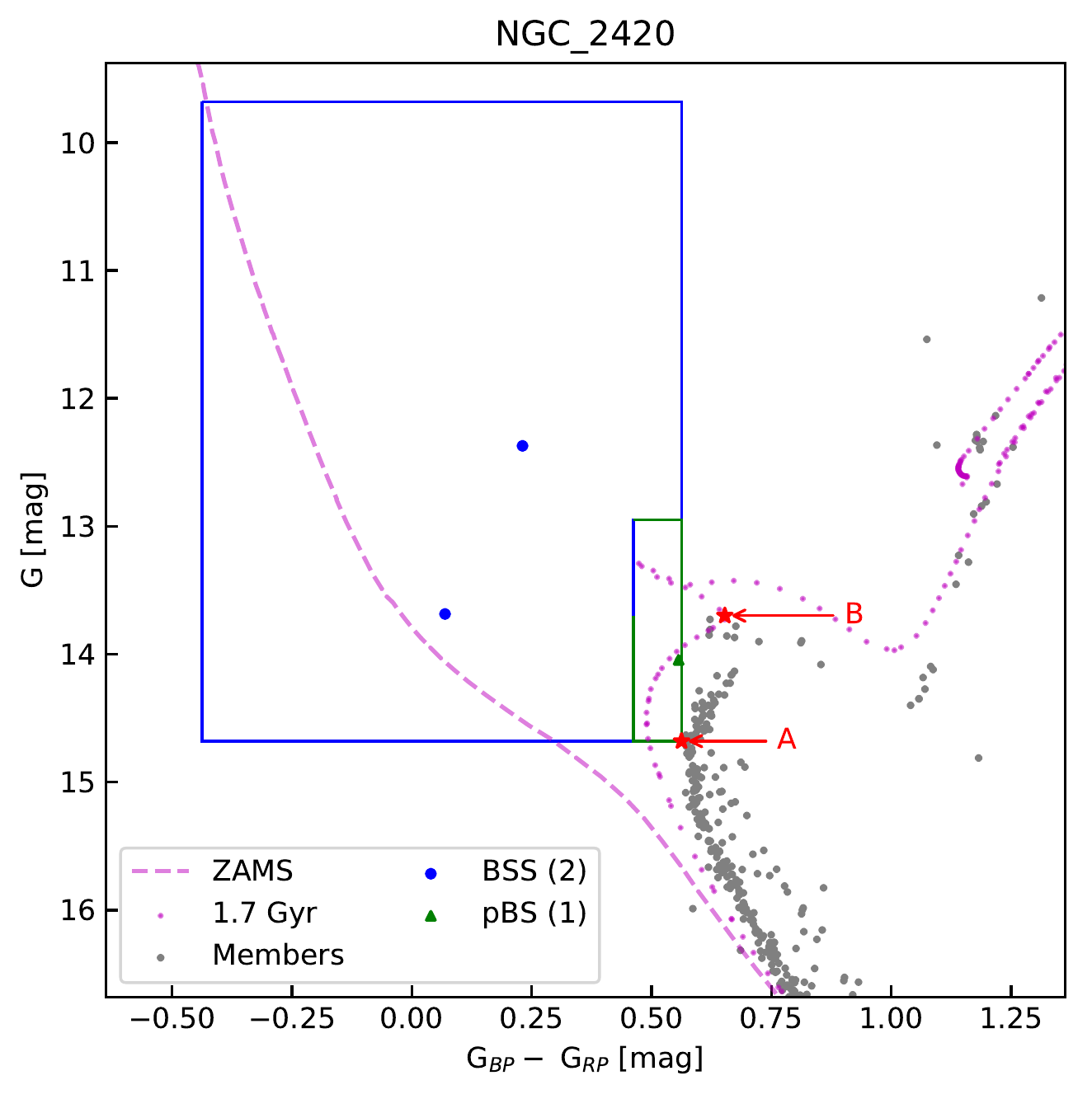}
    \includegraphics[width=0.48\textwidth]{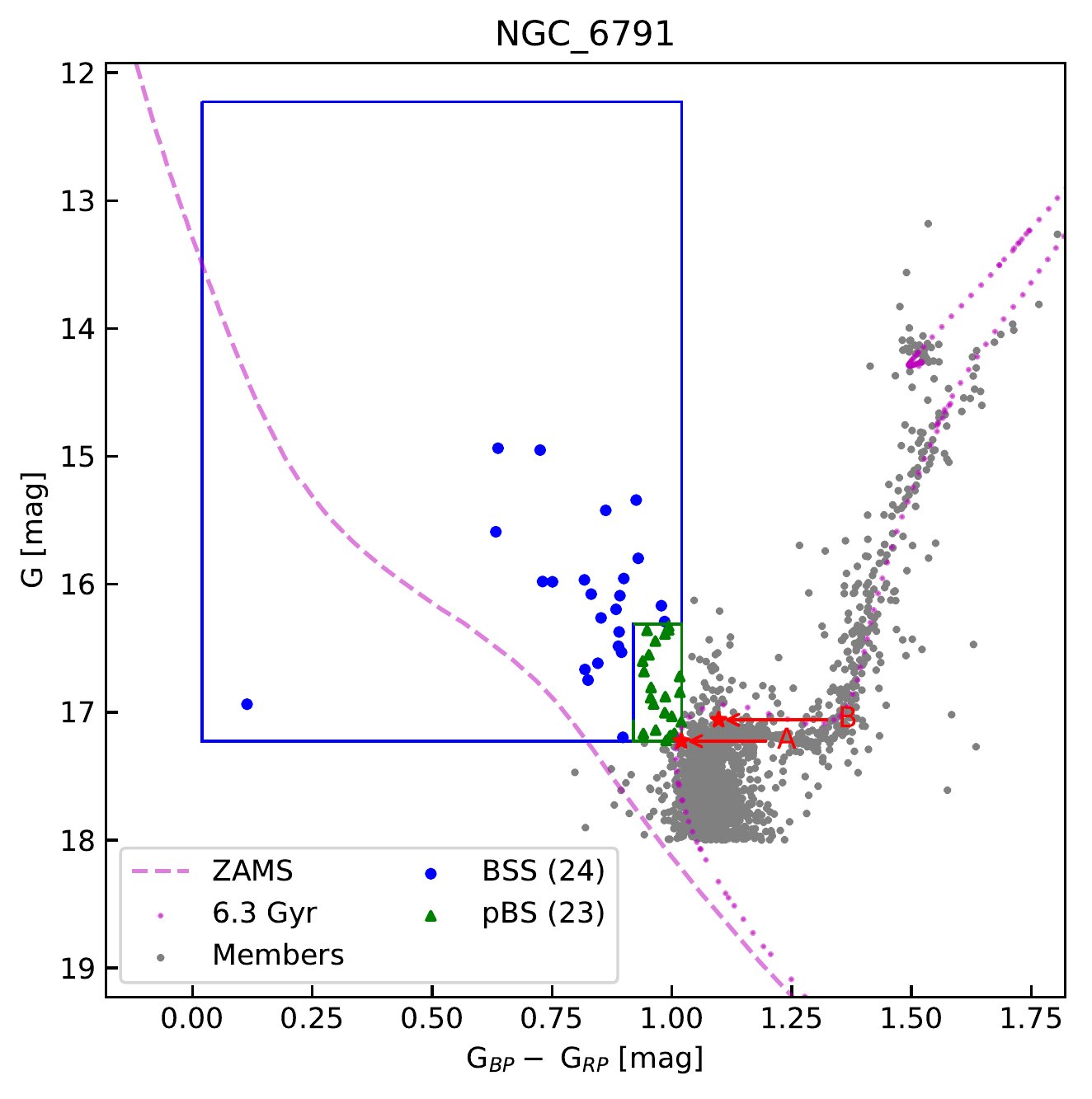}
    \caption{Schematic of classifying BSSs and pBSs in the colour-magnitude plane for OC NGC 2682, 2243, 2420 and 6791. Stars in blue and green boxes are classified as BSS and pBS respectively. The isochrone and ZAMS are shows for comparison. The manually identified A and B points are shown as red stars.}
    \label{fig:6_demo_class}
\end{figure}

Detection of BSSs requires a clear identification of the cluster MSTO. Although \citet{Cantat2020} listed the cluster parameters, identifying accurate MSTO is not trivial due to differences between isochrone models and the neural network used by \citet{Cantat2020}. The same can be seen for NGC 2682 in Fig.~\ref{fig:6_demo_class}, where the cluster data and the solar metallicity PARSEC\footnote{\url{http://stev.oapd.inaf.it/cgi-bin/cmd}} isochrone \citep{Bressan2012MNRAS.427..127B} differ by $\sim$0.1 mag in the colour axis near the MSTO. Fig.~\ref{fig:6_demo_class} also shows that the isochrone fit of NGC 2420 is not good enough to locate the turn-off point, while isochrone MSTO of NGC 2243 and 6791 are quite similar to the actual data. Another approach with an automated search of the bluest MS point in the CMD works well in rich clusters, but not particularly well in poorer clusters. Hence, we have manually selected the MSTO point (A; bluest point on the cluster CMD) and the brightest point in the MS (B; brightest point on the CMD, before starting of the sub-giant branch).

BSSs are expected to be brighter and bluer than the MSTO with a significant range. 
Near the MSTO, there is an inherent scatter in the data, sometimes shifting the MSTO stars to the bluer side. Similarly, some stars in the tip of the MS (e.g., stars with G mag $\sim$12.8 for NGC 2682 in Fig.~\ref{fig:6_demo_class}) are bluer and brighter than the MSTO. Hence, we created two regions in the CMD to select BSSs and probable BSSs (pBSs) as shown in the blue and green boxes in Fig.~\ref{fig:6_demo_class}. The blue box (to select BSSs) spans about 1 mag in colour axis and 5 mag in magnitude axis, with respect to point A. 
The green box (to select pBSs) spans 0.1 mag in the colour axis and has a bright limit at 0.75 mag above point B (to avoid any unresolved binaries brighter than the MS tip). In Fig.~\ref{fig:6_demo_class}, the fainter pBSs in NGC 2682 are genuine BSSs, while brighter pBSs are MSTO stars in the MS tip. All CMDs are manually checked, and points A and B are identified. The CMDs shown in Fig.~\ref{fig:6_demo_class} demonstrate the need for a manual selection. It also demonstrates how A and B points are identified uniformly across all the clusters studied here. We believe that this method will greatly help identify the BSSs (in the blue box) and the pBSs (in the green box) similarly across the clusters, resulting in a reliable sample. Overall, the BSS class consist of bona fide BSSs, while pBS class consists of some BSSs and some MS stars. We, therefore, mainly use only the BS population for further analysis to study their properties and dependency on cluster parameters.


\section{Calculations of stellar and cluster parameters} \label{sec:6_appendix_A}

\subsection{Absolute magnitudes of the stars} \label{sec:6_absolute_mag}
Although the isochrones are not a perfect fit for the cluster, the distances and extinction given in \citet{Cantat2020} are accurate enough for statistical analysis. Hence, we calculate the absolute magnitudes using the cluster distance modulus (DM) and Av. Similarly, the absolute magnitudes of the MSTO point are also calculated.

\subsection{Mass of the stars} \label{sec:6_star_mass}
We defined a ZAMS using a combination of three isochrones. log($age$) = 8.00 for stars fainter than 2.5 G mag, log($age$) = 7.50 for stars within 2.5 to 1.5 G mag and log($age$) = 7.00 for stars within 1.5 to $-$4.0 G mag. These magnitudes cutoffs ensure a smooth transition in the CMD and G--Mass relation. The BSSs are assumed to be single stars, and their mass is estimated by comparing their absolute magnitudes to the ZAMS. The isochrones do not completely overlap with the cluster CMD, so there are errors associated with the mass estimation. We do not recommend this method for calculating the mass of an individual BSS. However, over the large sample, this method gives a rough estimate of the mass of the stars. Furthermore, the mass hierarchy within a cluster is unaffected by the fitting of the isochrone. 

\subsection{Mass and LF of the cluster} \label{sec:6_cl_mass}
The cluster mass is calculated by comparing the LF of the cluster with a model of the same age, distance, and extinction and mass of 1 M$_{\odot}$. All clusters in the \citet{Cantat2020} catalogue are limited to 18 G mag, which gives the number of visible members ($LF_{18,\ cluster}$).

\begin{equation}
    cl\_mass = \frac{LF_{18,\ cluster}}{LF_{18,\ model}}\ [M_{\odot}]
\end{equation}
where, $LF_{18,\ model}$ is the LF of model cluster limited to apparent 18 G mag.

The total number of cluster members are derived using the complete model LF ($LF_{\infty,\ model}$) as,
\begin{equation}
    total\_stars = \frac{LF_{18,\ cluster} \times LF_{\infty,\ model}}{LF_{18,\ model}}
\end{equation}

The $total\_stars$ is dominated by very faint and small stars, which are not much important in the context of BSS evolution. Hence, we also calculated the number of stars near the turn-off using a similar technique. Finally, the $TO\_stars$ are defined as the number of MS stars within 0 to 3 mag of the MSTO. 
\begin{equation}
    TO\_stars = \frac{LF_{TO+3,\ cluster}}{LF_{\infty,\ model}} \times total\_stars
\end{equation}

We also calculated the cluster number density using the half member radius ($r_{50}$) of the cluster and the number of $TO\_stars$ as follows:
\begin{equation}
    density = \frac{TO\_stars}{4/3\ \pi r_{50}^3} [pc^{-3}]
\end{equation}

\subsection{Effective radius} \label{sec:6_eff_radius}
The effective radii of the cluster are calculated as follows: 

\begin{equation}
        R_{eff}^{cluster} = \frac{\Sigma r_{cluster}}{N_{cluster}}
\end{equation}
\noindent where $r_{sample}$ are the individual radii and $N_{sample}$ is the total number of stars in the sample. The normalised radius for individual stars is calculated as:
\begin{equation}
        r_{norm}^{sample} = r_{sample} \big/ R_{eff}^{cluster}
\end{equation}
\noindent And the normalised effective radius of the population (e.g., BSSs) is calculated as:
\begin{equation}
        R_{norm}^{sample} = \frac{\Sigma r_{sample}}{N_{sample}} \bigg/ R_{eff}^{cluster}
\end{equation}

\subsection{Dynamical relaxation time of the cluster} \label{sec:6_relaxation}
The segregation of BSSs has been linked to the number of relaxation periods passed during the cluster age. We calculated the dynamical relaxation time for the cluster from eq.~\ref{eq:1_relaxation}. We have used the radius containing half members as a substitute for the half mass radius, and $\langle m \rangle = cl\_mass/total\_stars$.


\section{Results} \label{sec:6_Results}

\begin{table}[!ht]
    \centering
\resizebox{0.98\textwidth}{!}{
\begin{tabular}{ccccc ccccc cc}
    \toprule
Cluster	&	log($age$)	&	AV	&	DM	&	r	&	cl\_mass	&	density	&	$N_{relax}$	&	$N_{BSS}$	&	$N_{pBS}$	&	$R_{eff}^{cluster}$	&	$R_{eff}^{BSS}$	\\
	&		&		&		&	[pc]	&	[\Msun]	&	[pc$^{-3}$]	&		&		&		&	[\arcmin]	&	[\arcmin]	\\ \hline
UBC\_199	&	9.06	&	0.93	&	10.46	&	2.9	&	300	&	4.37	&	16.13	&	0	&	0	&	10.77	&		\\
Skiff\_J0507+30.8	&	9.39	&	0.98	&	13.92	&	8.0	&	2626	&	1.81	&	3.70	&	2	&	2	&	4.48	&	9.54	\\
Czernik\_18	&	8.72	&	1.34	&	10.61	&	1.2	&	256	&	56.28	&	29.66	&	0	&	0	&	3.32	&		\\
COIN-Gaia\_11	&	8.9	&	1.25	&	9.13	&	3.9	&	304	&	1.81	&	7.05	&	1	&	0	&	20.58	&	26.72	\\
Berkeley\_69	&	8.9	&	1.61	&	12.66	&	2.3	&	1211	&	37.57	&	10.21	&	0	&	0	&	2.34	&		\\
\bottomrule
    \end{tabular}
    }
    \caption{Example list of all clusters with log($age$) $>$ 8.5. Full table is available online.}
    \label{tab:6_cluster_list}
\end{table}

\begin{table}[!hb]
    \centering
\resizebox{0.98\textwidth}{!}{
    \begin{tabular}{cccc cccc c}
    \toprule
Cluster	&	RA\_ICRS	&	DE\_ICRS	&	sourceID\_GaiaDR2	&	Gmag	&	BP-RP	&	$M_{BSS}$	&	\Me	&	class	\\
	&	[deg]	&	[deg]	&		&		&		&	[\Msun]	&		&		\\ \hline
Skiff\_J0507+30.8	&	76.96513	&	30.70442	&	156450304087516928	&	15.71	&	0.60	&	2.74	&	0.44	&	BSS	\\
Skiff\_J0507+30.8	&	76.67498	&	30.79202	&	157204805285086464	&	15.42	&	0.73	&	3.00	&	0.58	&	BSS	\\
COIN-Gaia\_11	&	68.60678	&	39.25254	&	178951225434386944	&	10.73	&	0.64	&	3.20	&	0.33	&	BSS	\\
Czernik\_21	&	81.65730	&	36.01864	&	184098241227893248	&	15.56	&	1.36	&	3.00	&	0.82	&	BSS	\\
Teutsch\_2	&	85.51016	&	39.12918	&	190160398586969984	&	14.83	&	0.72	&	3.40	&	0.62	&	BSS	\\

\bottomrule
    \end{tabular}
}
    \caption{Example list of all BSSs in our sample. A complete list of all BSSs and pBSs is available online.}
    \label{tab:6_catalogue}
\end{table}

\begin{figure}[!ht]
    \centering
    \includegraphics[width=0.6\textwidth]{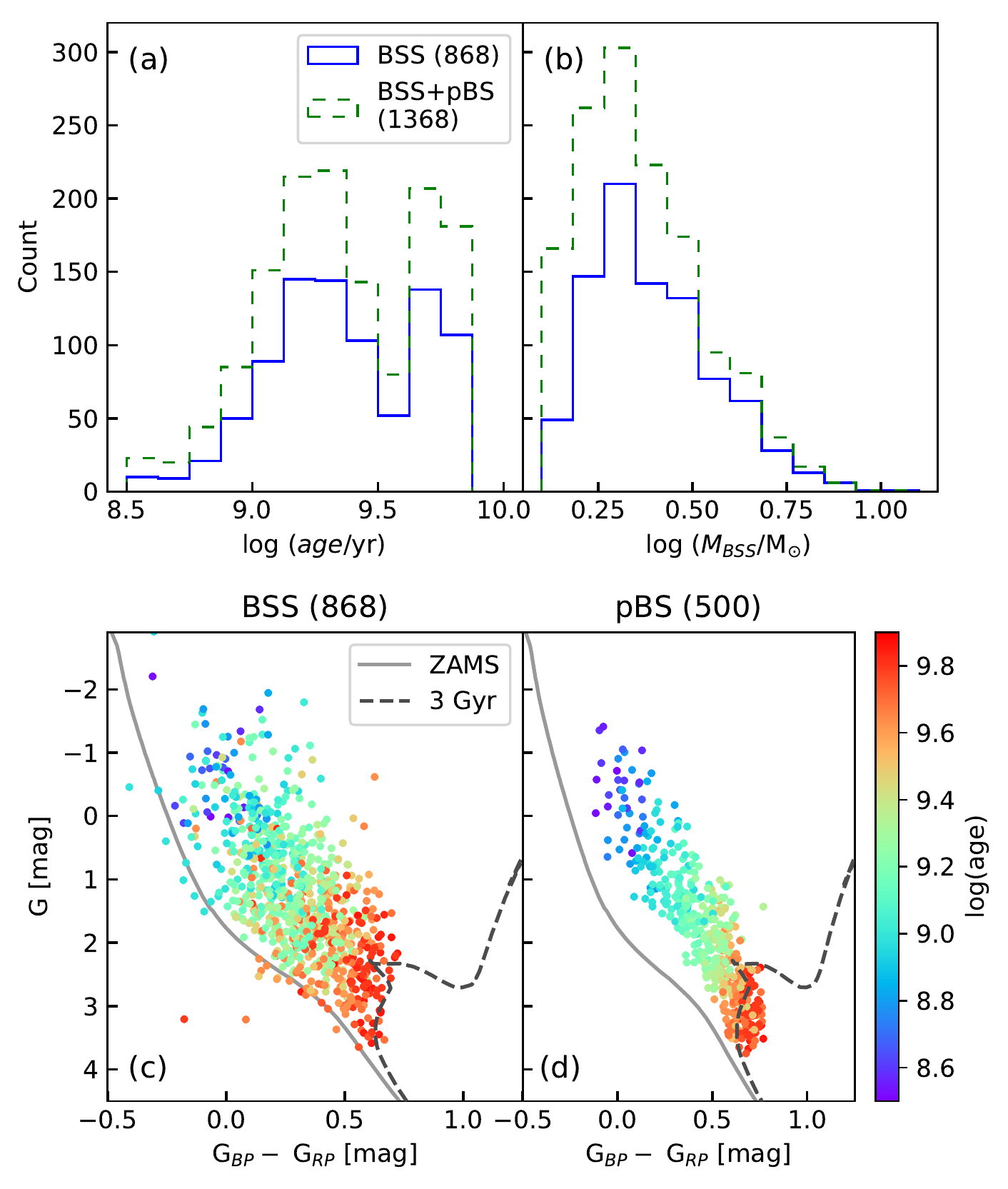}
    \caption{(a) Age and (b) mass distribution of the BSSs as blue histograms (and BSS+pBS as dashed green histograms). CMDs of (c) BSSs and (d) pBSs coloured according to the cluster age. All stars are corrected for distance and reddening using cluster parameters in \citet{Cantat2020}. An isochrone of 3 Gyr and ZAMS are shown for reference.}
    \label{fig:6_BSS_CMDs}
\end{figure}

\begin{figure}[!ht]
    \centering
    \includegraphics[width=0.98\textwidth]{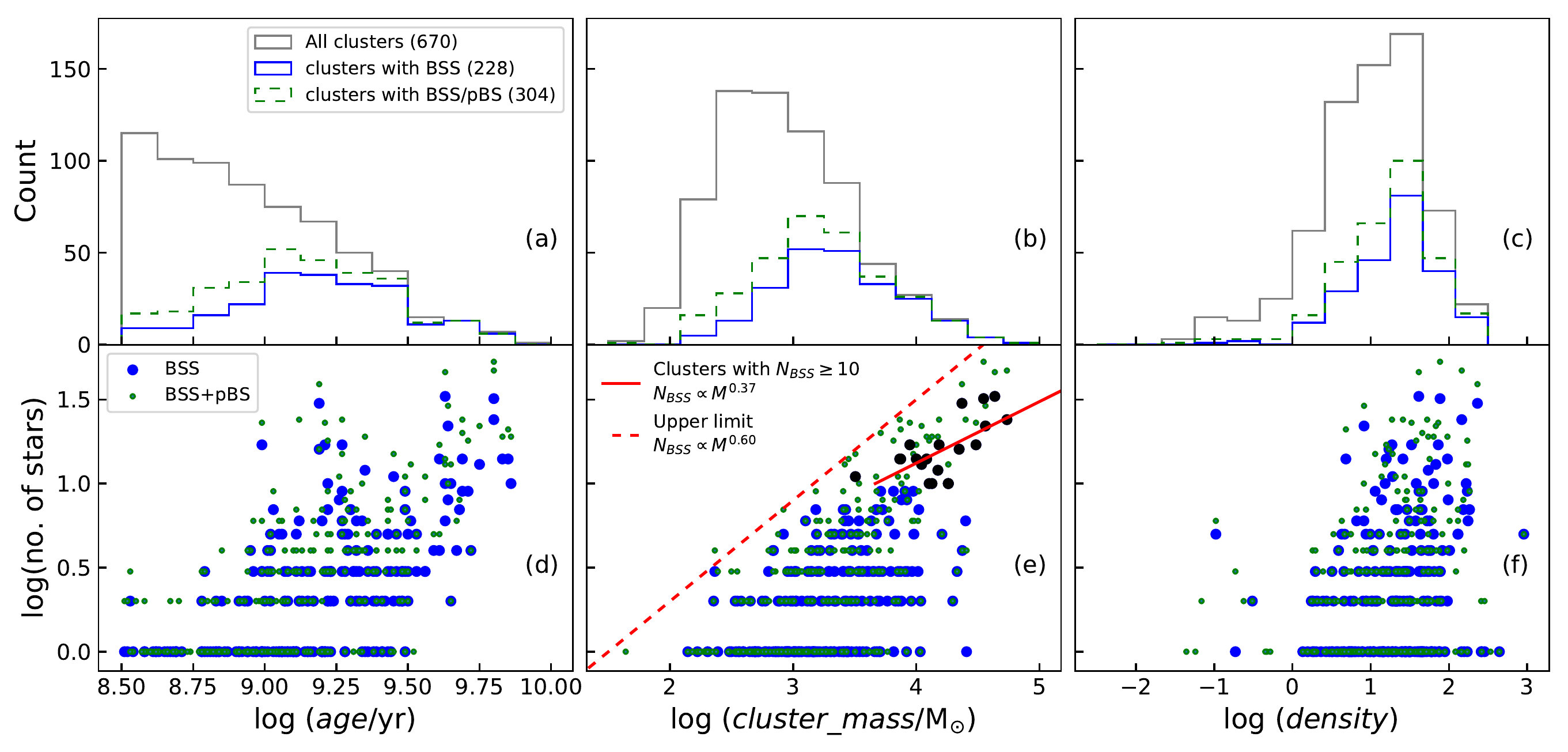}
    \caption{\textit{Upper panels} show the histograms of cluster (a) age, (b) mass and (c) density for all clusters (grey), clusters with BSSs (blue) and clusters with BSS/pBS (dashed green). \textit{Bottom panels} show the dependence of number of BSS (as blue circles) and BSS+pBS (as green dots) on the cluster (d) age, (e) mass and (f) density. The panel (e) shows a linear fit to clusters with at least 10 BSS (black filled circles) as a solid red line. The dashed red line represents the empirical upper limit on the number of BSSs in a cluster based on cluster mass.}
    \label{fig:6_cl_params}
\end{figure}

Among the 670 clusters older than 300 Myr, we found 868 BSSs in 228 clusters and 500 pBSs in 208 clusters. Overall, 304 clusters have 1368 BSS candidates (BSS+pBS).
Table~\ref{tab:6_cluster_list} shows the first five rows of the table of all 670 clusters along with the cluster properties and the number of BSSs and pBSs. Table~\ref{tab:6_catalogue} shows the example of the list of all BSSs and pBSs. The full tables are available online and in CDS\footnote{\url{https://cdsarc.cds.unistra.fr/viz-bin/cat/J/MNRAS/507/1699}}.

The age-wise distribution of 228 clusters with 868 BSSs is given in Table~\ref{tab:6_binwise} and is shown in Fig.~\ref{fig:6_BSS_CMDs} (a). {Among the given bins, the largest number of clusters (77) in this study lies in the log(\textit{age}) 9.00--9.25 bin, whereas the largest number of BSSs (247) lies in the 9.25--9.50 bin.} 
Fig.~\ref{fig:6_BSS_CMDs} (c) \& (d) shows the absolute CMDs of selected BSSs and pBSs, coloured according to the cluster age. The majority of the BSSs are located redder than the ZAMS.
The pBSs are found as a neat sequence, mainly due to the selection criteria used. 
We discuss the properties of the BSSs, pBSs and the cluster parameters below.

\subsection{Relation between BSSs and cluster properties}
The age, mass, and density distributions of clusters in this study are shown in Fig.~\ref{fig:6_cl_params} (a)--(c). The figure shows the distribution of all clusters (grey), those with BSSs (blue) and those with BSSs+pBSs (dashed green), as a function of three parameters, to understand their influence. In all the plots, the blue and the green distributions closely resemble each other. 
Figure (a) shows that only a small fraction of clusters younger than 1 Gyr have BSSs, while almost all clusters above 3 Gyr have BSSs. We notice an increasing trend in the fraction of clusters with BSSs, in the range of 300 Myr to 3 Gyr. We also notice that the number of clusters with BSSs (and pBSs) increases up to 1 Gyr, showing a flat peak in the 1--3 Gyr age range among the sample studied here. 

In this sample studied here, we have most of the clusters in the log(\textit{mass}) range 3.0--3.5, whereas the majority of the BSSs are found in the log(\textit{mass}) range of 3.5--4.5. Only a small fraction of clusters lighter than 1,000 \Mnom\ have BSSs, while almost all clusters massive than 10,000 \Mnom\ have. Clusters with mass in the range of 100--10,000 \Mnom\ show an increasing fraction. The number of clusters with BSSs (and pBSs) increases up to 1,000 \Mnom and has a flat peak between 1000--3000 \Mnom. Any such definite trend is not reflected in the density plot, where clusters with a wide range in density show the presence and absence of BSSs in the clusters. The distribution of clusters with BSSs seems to be a subset of the overall sample distribution, where denser clusters are likely to have more BSSs. 

Fig.~\ref{fig:6_cl_params} (d)--(f) shows the relation of the cluster age, mass, and density with the number of BSSs (blue) and BSSs+pBSs (green).
Here we have shown all clusters (with BSSs and pBSs) such that each blue dot denotes the number of BSSs in the cluster and green dot denotes the total number (BS+pBSs) against the cluster parameter, where the green dots can be considered as the upper limit.  
As expected, the older clusters, massive clusters and more dense clusters have more BSSs. This is apparent from the rising trend seen in panels (d) to (f). 
Looking at the table~\ref{tab:6_binwise}, we find an interesting doubling trend in the BSS frequency with age. We find that, on an average, clusters in log(\textit{age}) = 8.5--9.0 range, have 1.6 BSS/cluster. This doubles to 3.0 BSS/cluster in the range 9.0--9.25, which stays more or less similar (3.8 BSS/cluster) in the 9.25--9.5 range. This again doubles to 7.9 BSS/cluster in the 9.5--9.75 range and again doubles (and maybe a bit more, 17.8 BSS/cluster) in the 9.75--10.0 range. We get 3.8 BSS/cluster across all age ranges for the entire sample.

Interestingly, the dashed line in Fig.~\ref{fig:6_cl_params} (e) shows that there is an upper limit on the number of BSSs present in a cluster depending on the cluster mass, as follows: 
\begin{equation}
\label{eq:6_NBSS_mass}
    \textrm{log} (N_{BSS,max}) = 0.6\,\textrm{log}(cl\_mass/M_{\odot}) - 0.9
\end{equation}
This relation is estimated incorporating both the BSSs and the pBSs. This relation seems to work well for clusters more massive than 500 \Mnom. From the table~\ref{tab:6_binwise}, we find a large range in the number of BSS per cluster from 1.5 BSS/cluster in the log(\textit{mass}) range 2.0--3.0, which increases to 2.6 (3.0--3.5 range), 4.6 (3.5--4.0), 9.25 (4.0--4.5), reaching finally to 26.75 BSS/cluster (4.5--5.0). The upper limit is much higher than the older clusters (i.e., 17.8). 

\begin{figure*}
    \centering
    \includegraphics[width=0.98\textwidth]{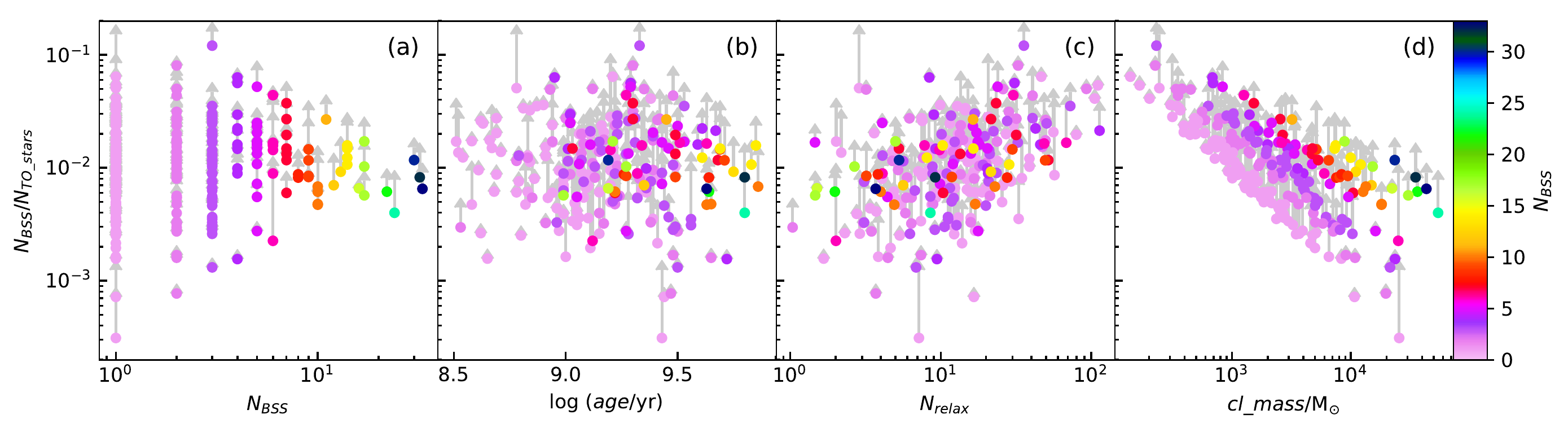}
    \caption{Variation of BSS fraction with number of BSSs, cluster age, relaxation periods passed and mass. The increase in fraction after including pBS is indicated by arrows. The clusters are coloured according to the number of BSSs.}
    \label{fig:6_fractional_BSS}
\end{figure*}

In panel (f), though we do detect a tentative rising trend with respect to density, such that the more dense clusters produce more BSSs, the observed scatter stops us from deriving a relation.
To compare the BSS frequency across different clusters, we calculated the BSS fraction by normalising with the MSTO population ($N_{BSS}/N_{TO\_stars}$).
Fig.~\ref{fig:6_fractional_BSS} (a) shows that the fraction has a large range in clusters with fewer BSSs, and it stabilises for clusters with more BSSs with a median fraction of 1\%. There is no correlation between the presence of BSSs and the cluster age. 
The clusters with fewer BSSs show linear dependence between the BSS fraction and the mass, but that is an artefact due to the range of values available for the numerator ($N_{BSS}$) and denominator ($N_{TO\_stars}$). For massive clusters, the BSS fraction tends towards the median value.
Fig.~\ref{fig:6_fractional_BSS} (c) shows the relaxed clusters tend to have higher BSS fraction.
This could be partly due to the evaporation of lower mass stars in relaxed clusters or more likely due to the inverse relation between cluster mass and relaxation.
Furthermore, we found no correlation between the presence of BSSs and the radius of the cluster, cluster distance, Galactocentric position or number of stars near the turn-off.

\subsection{Fractional mass excess}\label{sec:6_mass_excess}

\begin{figure}[!ht]
    \centering
    \includegraphics[width=0.98\textwidth]{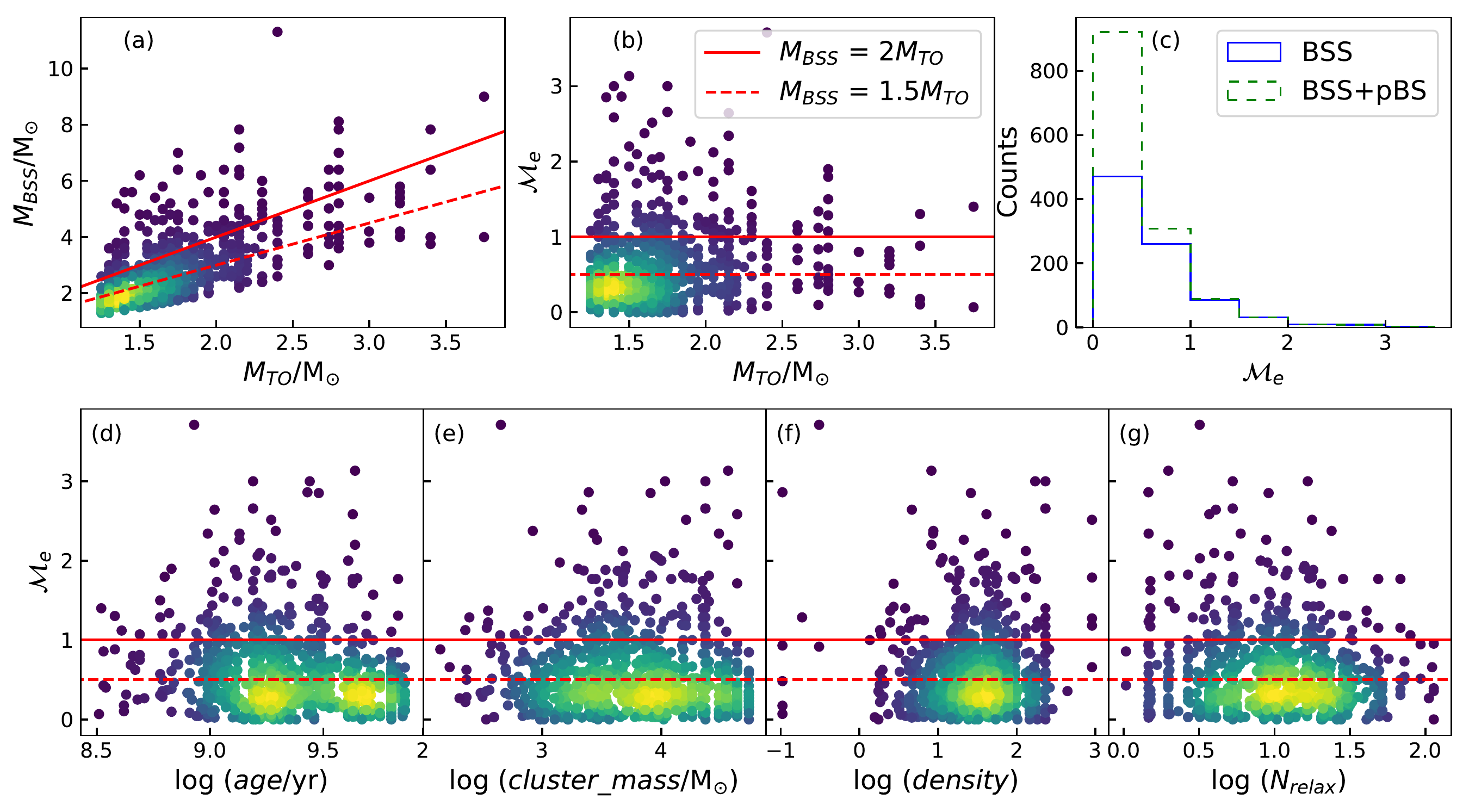}
    \caption{(a) 2-D density plots for visualising the variation of mass of the BSSs with respect to the cluster MSTO mass, where the points are coloured according to the crowding (yellow means crowding). The dashed, and solid red lines represent the BSSs with 1.5$M_{TO}$ and 2$M_{TO}$ mass, respectively. (b) Variation of \Me\ with cluster MSTO mass. (c) \Me\ distribution of the BSSs and pBSs. \textit{Bottom panels:} Variation of \Me\ with (d) cluster age, (e) cluster mass and (f) cluster density and (g) dynamical relaxation.}
    \label{fig:6_me_dist}
\end{figure}

The BSSs identified here are substantial in number across a fairly large age range, and therefore the sample has the potential to throw light on their formation mechanisms.
As BSSs have gained mass during their MS lifetime, the excess mass they have gained could reveal their formation pathway. To achieve this, we need first to estimate the mass of the identified BSSs and then make an estimate of the mass possibly gained by a BSS. The method used here to estimate the mass of the BSSs using ZAMS is given in \S \ref{sec:6_star_mass}. Fig.~\ref{fig:6_BSS_CMDs} (b) shows the distribution of mass of the BSSs. Combining the age (a) and the mass (b) plots, we find that the majority of the BSSs are older ($>$1 Gyr) and low mass ($<$3 \Mnom), but the overall range in age and mass is quite large (0.3--7 Gyr; 1--15 \Mnom).

The next step would be to understand how different formation pathways increase the mass of a star. 
In the case of a binary MT, there is a wide range in MT efficiencies: $\leq$ 0.2 in case-B MT, 0.2--0.7 in case-A MT, $\sim$1 in conservative MT scenarios \citep{Shao2016ApJ...833..108S}. Typically, wider binaries have non-conservative MT (efficiency $<$ 0.5) and can leave a remnant of the donor. On the other hand, close binaries can have more conservative MT (efficiency $>$ 0.5) and lead to mergers. 
Although it is not enough to use a single value of MT for describing all binary systems \citep{Mink2007A&A...467.1181D}, we can get an idea about the different pathways by comparing the mass of the BSSs to the mass of the cluster MSTO. As the host clusters have a range in age, they also have a range in the stars that can gain mass and become a BSS. Therefore, it is required to use fractional mass excess, based on a quantity that can be used for normalising, such as the mass at the MSTO (M$_{TO}$). We use the mass of a ZAMS star of the same magnitude as the MSTO (a conservative estimate of the MSTO mass) to determine M$_{TO}$. More discussion on the normalising mass is given in \S \ref{sec:6_reference_mass}.

Fig.~\ref{fig:6_me_dist} (a) shows the variation of $M_{BSS}$ with $M_{TO}$. Here, $M_{TO}$ is defined as the mass of the ZAMS star with the same magnitude as the MSTO.
The dashed, and solid red line represents the $M_{BSS} = 1.5M_{TO}$ and $M_{BSS} = 2M_{TO}$, respectively. For example, the solid line can be considered the position of BSSs formed by adding up the mass of two MSTO stars. It is clear from the plot that the majority of the stars follow the dashed line. Also, most of the BSS population is within the solid line. 

The $M_{BSS}$ and $M_{TO}$ have a large range (1--4 \Mnom) in our cluster sample. To identify the BSSs that a binary MT could have formed, we need to estimate the fractional mass gain of BSSs to correlate with the efficiency of MT. Hence, we defined a new parameter called `fractional mass excess' (\Me), which normalises the $M_{BSS}$ with the cluster $M_{TO}$:
\begin{equation}
    \mathcal{M}_{e} = \frac{M_{BSS}-M_{TO}}{M_{TO}}
\end{equation}
By definition, \Me\ is equivalent to the MT efficiency in the case where both accretor and progenitor were MSTO stars. 
Fig.~\ref{fig:6_me_dist} (b) shows the relation between \Me\ and  $M_{TO}$ and \Me\ can directly give information about the MT efficiency in clusters with any $M_{TO}$. The plot shows that typically $M_{TO}$ is less than 2.0 \Mnom. Moreover, most BSSs are below \Me\ $<$ 0.5.
Fig.~\ref{fig:6_me_dist} (c) shows the histogram of \Me\ for all BSSs and pBSs in our sample. The distribution peak is in the 0--0.5 range, with most of the BSSs having \Me\ $<$ 1, and a few having large \Me\ in the range of 1--3.5. We also note that the large \Me\ values are mostly present in clusters with $M_{TO}$ $<$ 2.5 \Mnom. 

Hereafter we classify BSSs as low-\Me\, high-\Me\ and extreme-\Me\ BSSs for \Me\ $<$ 0.5,  0.5 $<$ \Me\ $<$ 1.0 and \Me\ $>$ 1.0 respectively. Overall, there are 471 (BSSs)--921 (BSS+pBS) (54--67\%) low-\Me\ BSSs, 260--308 (30--22\%) high-\Me\ BSSs and 137--139 (16--10\%) extreme-\Me\ BSSs. As expected, most of the BSSs fall within the low to high-\Me. However, we find that there is a non-negligible fraction of extreme-\Me\ BSSs, that requires attention.

Fig.~\ref{fig:6_me_dist} (d)--(g) present the variation of \Me\ with the cluster age, mass, density, and number of relaxation periods passed. We find the low-\Me\ BSSs to be the majority in all clusters. The very young and very old clusters appear to have less extreme-\Me\ BSSs. In the (d) panel, two yellow blobs identify two peaks in the BSSs (similar to the distribution in Fig.~\ref{fig:6_BSS_CMDs} (a)), though likely to be a selection effect, suggests that the low-\Me\ BSSs dominate. There appears to be an increase in the presence of extreme-\Me\ BSSs as the cluster mass increases till $\sim$40000 \Mnom, as suggested by the panel (e). Irrespective of cluster mass, the low\Me\ BSSs is the dominant population. The density parameter (panel (f)) shows no correlation with the maximum possible \Me.

\subsection{BSSs properties in age and mass binned clusters} \label{sec:6_agewise}

\begin{table}[!ht]
    \centering
    \begin{tabular}{cccc|ccc}
    \toprule
\multicolumn{2}{c}{Binning type}			&	No. of 	&	No. of	&	\multicolumn{3}{c}{No. of BSSs in \Me\ classes}					\\	\cmidrule(lr){5-7}
	&		&	Clusters	&	BSSs	&	low	&	high	&	extreme	\\	\hline
\multirow{5}{*}{\rotatebox[origin=c]{90}{\parbox[c]{2cm}{\hfill log($age$)}}}	&	8.50--9.00	&	56	&	90	&	43	&	27	&	20	\\	
	&	9.00$-$9.25	&	77	&	234	&	109	&	71	&	54	\\	
	&	9.25$-$9.50	&	65	&	247	&	121	&	84	&	42	\\	
	&	9.50$-$9.75	&	24	&	190	&	123	&	49	&	18	\\	
	&	9.75$-$10.0	&	6	&	107	&	75	&	29	&	3	\\	\hline
\multirow{5}{*}{\rotatebox[origin=c]{90}{\parbox[c]{2cm}{\hfill log($cl\_mass$)}}}	&	2.0$-$3.0	&	55	&	85	&	44	&	28	&	13	\\	
	&	3.0$-$3.5	&	88	&	207	&	112	&	61	&	34	\\	
	&	3.5$-$4.0	&	60	&	272	&	140	&	83	&	49	\\	
	&	4.0$-$4.5	&	21	&	193	&	101	&	56	&	36	\\	
	&	4.5$-$5.0	&	4	&	111	&	74	&	32	&	5	\\	\hline
	&	Total	&	228	&	868	&	471	&	260	&	137	\\
	\bottomrule
    \end{tabular}
    \caption{The distribution of BSSs in different \Me\ classes for clusters binned by age and mass. The first column shows the limits of cluster age and mass for each bin. Second and third columns have the number of clusters and BSSs in the respective bins. The last three columns show the number of BSSs divided into low-\Me, high-\Me\ and extreme-\Me\ BSSs. The last row shows the \Me\ class distribution of all BSSs.}
    \label{tab:6_binwise}
\end{table}

\begin{figure}[!ht]
    \centering
    \includegraphics[width=0.7\textwidth]{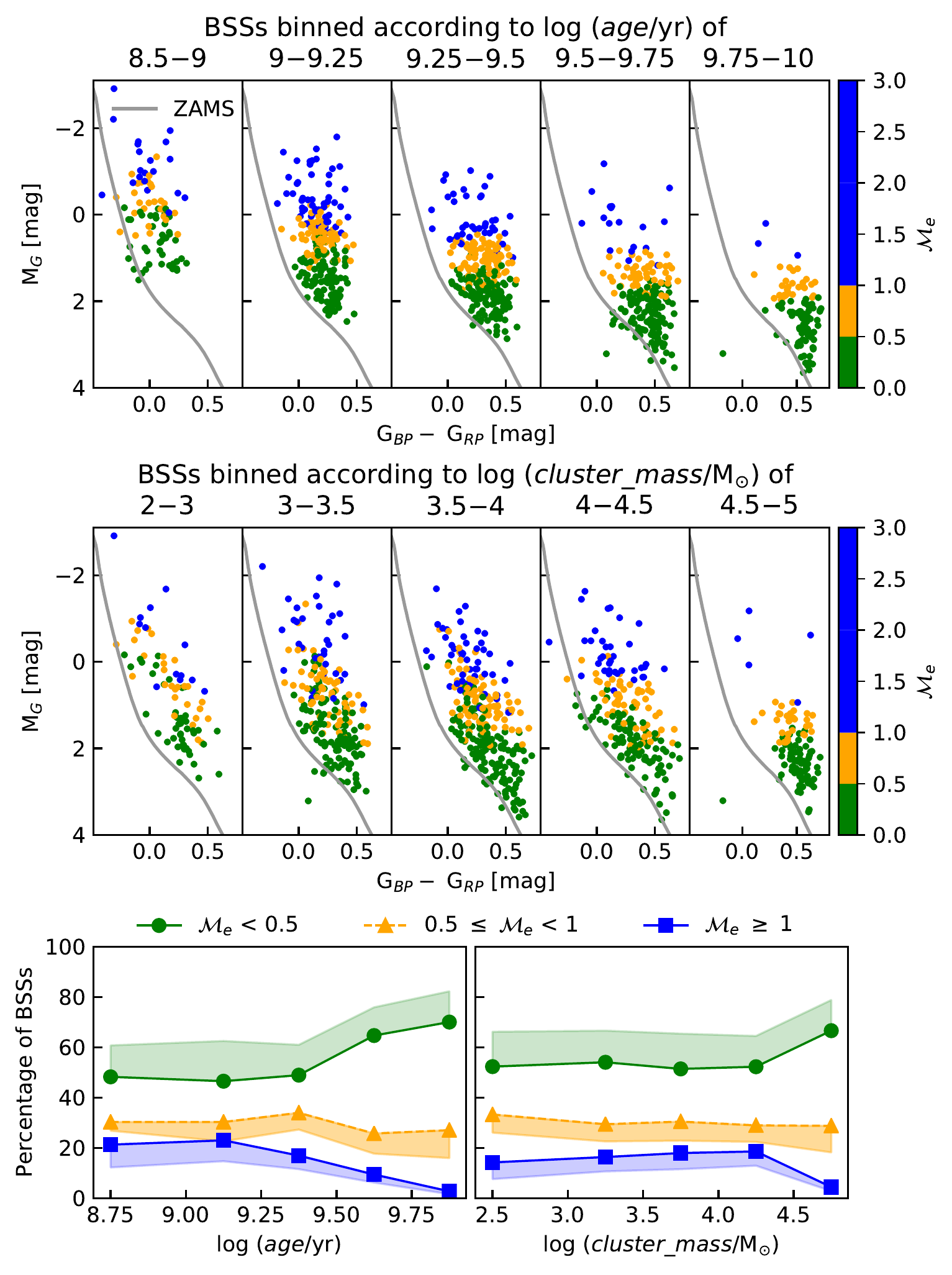}
    \caption{\textit{Top panels}: CMDs of BSSs across various age bins. \textit{Middle panels:} CMDs of BSSs across various cluster mass bins. The ZAMS are shown in the CMDs for reference. The BSSs are coloured according to \Me. \textit{Bottom panels:} The variation of percentage of stars in the three BSS classes for the above age and cluster mass bins. The shaded regions indicate the possible percentages if pBSs were included in the tally.}
    \label{fig:6_age_wise}
\end{figure}

For a more in-depth and quantitative analysis across cluster ages and mass, we divided the BSSs in age and mass bins as mentioned in Table~\ref{tab:6_binwise}.
The top panels in Fig.~\ref{fig:6_age_wise} show the CMDs of BSSs binned according to the cluster age. The middle panels show the CMDs of BSSs binned according to the cluster mass. All the BSSs are coloured according to their fractional mass excess and divided into the above-mentioned three classes. As seen in Fig.~\ref{fig:6_me_dist}, the most populous age range and mass range are 1--3 Gyr and 1000--30000 \Mnom\ respectively. In the upper panels, we see that the brightest BSSs are in the younger clusters, as expected, and becomes progressively fainter in older clusters. When we bin the clusters by mass, the BSSs have a similar range of brightness across the group, except for the most massive bin. There is tentative evidence for an increase in the low-\Me\ BSSs as a function of age, whereas no such trend is visually noticed with respect to mass. 

To quantitatively assess the presence of the above trend, the percentage of binned BSSs in the three classes for the age/mass groups are tabulated in Table~\ref{tab:6_binwise} and shown in the bottom panels of Fig.~\ref{fig:6_age_wise}. The panels also include the possible change in each class if pBSs were included in the calculations. From the bottom panels, it becomes apparent that the fraction of extreme-\Me\ BSSs decreases with age beyond 1 Gyr and has a maximum fraction near 1 Gyr age. On the other hand, the fraction of low-\Me\ BSSs increases steadily with the cluster age, beyond 1-2 Gyr. The high-\Me\ fraction follows the trend of the extreme-\Me\ BSSs, as a function of age, though with a lesser degree of reduction. The trends seen here might suggest that the high and extreme-\Me\ BSSs might belong to similar formation pathways and are similarly affected by cluster age. In contrast, the low-\Me\ BSSs, which are likely to have a different formation pathway, are affected oppositely with age.
The cluster mass seems to have less impact on the fractional mass excess of the BSSs present. The low and extreme-\Me\ BSSs show a constant fraction in most bins, with a deviation in the most massive bin. The high-\Me\ BSSs do not show any trend at all. 

\subsection{BSS segregation} \label{sec:6_segregation}

\begin{figure}[!ht]
    \centering
    \includegraphics[width=0.6\textwidth]{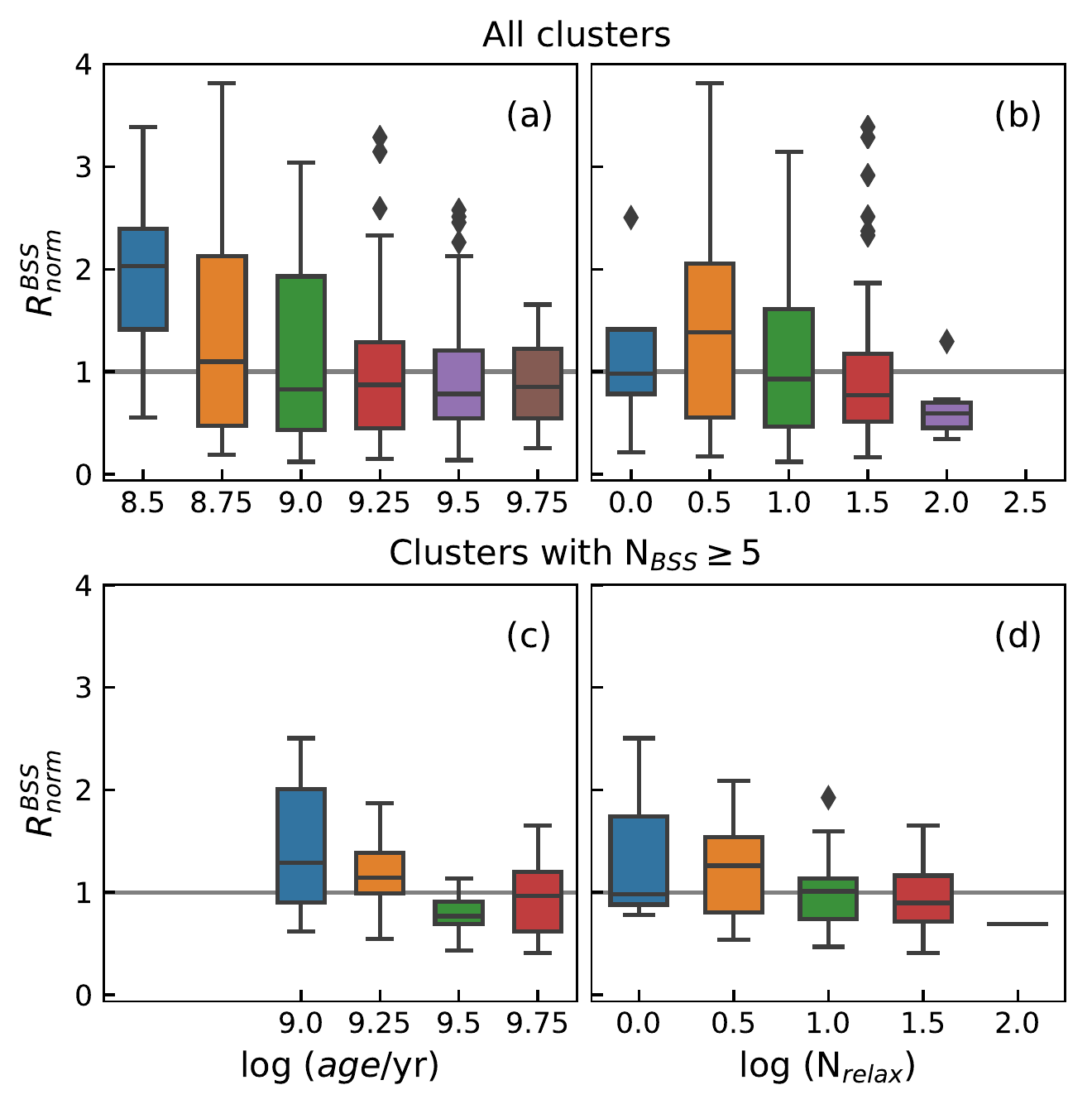}
    \caption{Box plots of the relationship of normalised effective radii of BSSs with the cluster age and dynamic relaxation time. The upper panels show the relationship for all clusters, while the lower panels only show for clusters with 5 or more BSSs.}
    \label{fig:6_segregation}
\end{figure}

BSSs are the most massive stars in a cluster; hence they are known/expected to be segregated towards the centre. To analyse the radial segregation, we calculated the effective radii of the cluster and BSSs (see \S \ref{sec:6_eff_radius} for definitions). 
Fig.~\ref{fig:6_segregation} (a) shows the variation of normalised BSS radii ($R_{norm}^{BSS}$) with cluster age. If the normalised radius has a value less than 1, there is radial segregation. On average, the BSSs in clusters older than 1 Gyr are segregated, though there is quite a large range in the values of the $R_{norm}^{BSS}$ for clusters of all ages. Note that there can be large errors in $R_{norm}^{BSS}$ due to the low number of BSSs in a cluster. The trend of decreasing $R_{norm}^{BSS}$ with age is more prominent in clusters with at least 5 BSSs (Fig.~\ref{fig:6_segregation} (c)).

Fig.~\ref{fig:6_segregation} (b) \& (d) both show the variation of $R_{norm}^{BSS}$ as a function of the dynamical relaxation. The increasing segregation in clusters with larger $N_{relax}$ is more clear in these plots when compared to Fig.~\ref{fig:6_segregation} (a) and (c). Unfortunately, there are only 3 BSSs per cluster on average. Hence, the individual effective radii are not statistically significant. The global results from Fig.~\ref{fig:6_segregation} (d) demonstrate that the segregation of BSSs, though a function of cluster age, is related to the dynamical evolution of the cluster. 

We also compared the radial distance of individual BSS with its \Me. On average, BSSs with \Me\ $<$ 1.5 are segregated. However, contrary to expectations, BSSs with \Me\ $>$ 1.5 are not segregated. These are massive BSSs and therefore are expected to be segregated. This would then indicate that these stars are not forming from the already segregated stars but from the non-segregated systems. It is possible that these are short-lived and hence may evolve before getting radially segregated. The (g) panel of Fig.~\ref{fig:6_me_dist} shows the relation between \Me\ and log(N$_{relax})$. The figure shows that the extreme-\Me\ BSSs decreases with the dynamical relaxation of the clusters.
The decline of extreme-\Me\ BSSs in relaxed clusters supports that they have not had enough time to segregate. The above points could provide some constraints/pointers to the formation pathways of extreme-\Me\ BSSs.

\subsection{Probable BSSs and data limitations}
The probable BSS (pBS) classification was devised to select all possible BSSs which can be very close to the MSTO while not contaminating the larger BS sample with MS-tip or MSTO stars. This group has both genuine BSSs and some MS stars. Therefore, one needs to be cautious while using the pBSs data in our catalogue.

Here we point out the potential issues which lead to incompleteness in this BS catalogue: These are either issues with the catalogue itself, such as calibration issues in G $<$ 12 mag, faintness cutoff of membership catalogue at G = 18 mag, missing PM and hence missing cluster members, or reduction of the cluster sample due to missing clusters near Galactic centre. A more detailed discussion on the \textit{Gaia} DR2 data can be found in \S 2.6 of \citet{Rain_2021arXiv210306004R}.

\subsection{Contamination in the blue straggler regime} \label{sec:6_contamination}
The BSS region has been known to overlap with the expected location of HB \citep{Bond1971PASP...83..638B, Jadhav2021arXiv210213375J}. We can estimate the number of HB stars in these clusters assuming the ratio of BSSs and HB stars in OCs is similar to globular clusters \citep{Leigh2009MNRAS.399L.179L}:
\begin{equation}
    \textrm{log}\frac{N_{HB}}{N_{BSS}} \approx 0.44\ \textrm{log} \frac{N_{TO+1}}{1000} - 0.36
\end{equation}
where $N_{TO+1}$ are the stars within 0 to 1 mag fainter than MSTO. HB stars have $M_G\sim1$ mag (from BASTI HB models; \citealt{Hidalgo2018ApJ...856..125H}). Among our sample, 157 clusters with BSS older have an MSTO fainter than 1 mag. Based on the above equation, there can be $\sim$49 HB stars in these 157 clusters. However, only blue HB (BHB) stars can be confused with BSSs. Assuming $B_{BHB}/B_{HB}$ of 0.25 \citep{Arimoto1981Ap&SS..76...73A}, there can be $\sim$12 BHB stars in our sample of 878 BSSs (1.4\%). As this is a very small fraction, the general results presented here are not affected.

There are two BSSs (one each in NGC 6791 and Haffner 5) that are quite bluer than the ZAMS (bottom left corner of Fig.~\ref{fig:6_BSS_CMDs} (c)). The star in Haffner 5 is very close to another bright star, while the star in NGC 6791 is likely a background B-type star \citep{Reed2012MNRAS.427.1245R}. These two stars do not impact the overall analysis presented here, but more caution is advised while studying individual BSSs in clusters.

The mass estimation assumed the BSSs are single stars. However, a large percentage of BSSs are known to be in binaries: 16/21 in NGC 188 \citep{Geller2008AJ....135.2264G,Mathieu2009Natur.462.1032M}; 11/14 in NGC 2682 \citep{Geller2015ApJ...808L..25G}; 6/12 in NGC 6791 \citep{Tofflemire2014AJ....148...61T}; 7/17 in NGC 6819 \citep{Milliman2014AJ....148...38M}; 4/13 in NGC 7789 \citep{Nine2020AJ....160..169N}. However, most of the binaries are single-lined spectroscopic binaries (SB1). Among the 5 clusters, the fraction of double-lined spectroscopic binaries (SB2s) is 4/77 (5.2\%). The shift in the CMD will be significant for SB2 systems, leading to the overestimation of BSS mass. Assuming the SB2 percentage of BSSs in these 5 clusters is similar to the complete sample, 5.2\% (45 BSSs and 26 pBS) of the BSS candidates have overestimated mass.


\section{Discussion} \label{sec:6_discussion}
Out of 670 clusters in the age range 300 Myr to 10 Gyr, 228 OCs have bona fide BSSs, while 76 more clusters have probable BSSs. The number of BSSs roughly increases with the age and mass of the cluster. Here we discuss possible formation pathways suggested by this sample of BSSs and compare our findings with those in the literature.

\subsection{Reference mass for calculating \texorpdfstring{\Me}{Me}}
\label{sec:6_reference_mass}
Understanding the relation between the mass of BSSs and the cluster turn-off is essential to put boundaries on the formation pathways. The most massive stars present in the cluster at the formation time of BSS tell about the possible pathways. However, the best estimation we can have regarding the most massive stars in the cluster is the cluster MSTO. This includes the assumption that the BSSs are not too old and the MSTO has not changed significantly during the lifetime of the BSSs. 
However, there are two ways to calculate the MSTO mass: (i) We can use the isochrone corresponding to the cluster age and get the MSTO mass (ii) We can use ZAMS and estimate the mass corresponding to the absolute G magnitude of the MSTO. The MSTO mass from the second method is larger than the first method.
Using the first method will lead to overestimating the \Me\ values. Hence, to be conservative, we have used the second approach. 

We would like to note another possibility, that is, using the ZAMS mass corresponding to the MS-tip. In general, for clusters older than 1 Gyr, the mass as per the isochrone of cluster age at MSTO and MS-tip will be very similar. However, the mass from ZAMS will be much higher for MS-tip when compared to MSTO. Therefore, using ZAMS mass at MS-tip will be the most conservative approach, whereas the isochrone mass at the MSTO will be the least conservative approach. We also note that BSSs that are bluer but with similar brightness as MSTO will have negative excess mass if we consider the ZAMS mass at MS-tip as a reference. Therefore, we decided to use the moderately conservative ZAMS mass at MSTO as the reference. This decision does not impact the results much, as we do not attempt to pick each BSS and identify its formation pathway individually. On the other hand, we only attempt some statistics and trends concerning the fractional mass excess, \Me.

\subsection{BSS formation pathways}
As \Me\ is a proxy to the MT efficiency, we can roughly divide the BSSs into mergers and MT products. 
The low-\Me\ BSSs encompass less-efficient MT binaries and merger products of low mass secondary. If we consider the typical mass-ratios of binaries to be $\sim0.7$, the mergers systems will contribute only a small portion in low-\Me\ BSSs. High-\Me\ BSSs are likely results of conservative MT and hence can be considered merger-dominated. The extreme-\Me\ BSS have more mass than two MSTO stars, and therefore, they are either formed earlier when the MSTO mass was larger, or they are a product of more than 2 MSTO stars. The first scenario is less likely due to a very slow shift in $M_{TO}$ and the evolution of such BSSs away from the MS, particularly for M$_{TO}$ $<$ 2.0 clusters where these BSSs are identified. In summary, the majority of low-\Me\ BSSs are likely to be MT products, whereas the majority of high-\Me\ BSS are merger products, and extreme-\Me\ BSS are likely to be multiple merger/MT products. We also found that the fraction of high and extreme-\Me\ BSSs decreases with age. It is, therefore, possible that this is indicative of a reduction in the BSS formation through merger processes in clusters older than 1--2 Gyr. Moreover, the rising low-\Me\ fraction could suggest a steadily increasing occurrence of BSSs formation through MT in clusters older than 1--2 Gyr. 

The extreme-\Me\ BSSs apparently require more than twice the TO mass to be formed. They can result from unresolved binaries, gas accretion, multiple MT/merger events. As mentioned in \S~\ref{sec:6_contamination}, there is a 5.2\% chance that the BSSs are bright due to being part of unresolved binaries. This still leaves $\sim$130 extreme-\Me\ BSSs, which are likely single stars (or partners of a sub-luminous companion).
They have unlikely to be gained mass via gas accretion, as the clusters older than 5 Myr rarely have molecular gas \citep{Lada2003ARA&A..41...57L, Leisawitz1989ApJS...70..731L}.
\citet{Leigh2011MNRAS.410.2370L} have shown that 2+2, 2+3, 3+3 encounters are possible in the lifetimes of OCs. Moreover, multiple encounters can develop into mini-clusters of 5--7 stars, which can significantly increase the chance of collisions \citep{Geller2015ApJ...808L..25G}. Such encounters can also interrupt the MT for 20--40\% of the cases \citep{Leigh2016ApJ...818...21L}. Overall, the large percentage of extreme-\Me\ BSSs indicates that multiple encounters are quite abundant in OCs.

\subsection{Comparison with literature}
The maximum number of BSSs have a power-law dependence on cluster mass (N$_{BSS,max}$ $\alpha$ M$_{clus}^{0.6}$). Similar mass dependence was found in globular clusters by \citet{Knigge_2009Natur.457..288K} (N$_{BSS}$ $\alpha$ M$_{core}^{0.38\pm0.04}$), where they show a correlation between the number of BSSs and the core mass. The authors concluded that most BSSs, even those found in cluster cores of the globular clusters, come from binary systems. Similar to this, we found a power-law fit to clusters with at least 10 BSSs (Fig.~\ref{fig:6_cl_params}). Also, the relation between the number of BSSs and the mass of the cluster is found to have a very similar value for the exponent (N$_{BSS}$ $\alpha$ M$_{clus}^{0.37 \pm 0.10}$). This would also suggest a binary origin for BSSs in OCs, which is indeed true, as we find about 84\% of BSSs to be of binary origin. Therefore, the BSS formation and evolution is quite similar in open and globular clusters, both dominated by binaries.

\citet{Ahumada_2007A&A...463..789A} produced the most comprehensive list of BSSs before the arrival of \textit{Gaia}. Their list of 1887 candidate BSSs in 470 clusters can be considered an upper limit on the number of BSSs. Higher precision astrometry/membership with \textit{Gaia} will remove field contamination from their list. They had identified 148 (8\%) stars in 26 old clusters as high-mass BSSs. The fraction of extreme-\Me\ BSSs (with \Me\ $>$ 1) is 10--16\% from our estimates. This difference is likely due to the better classification using \textit{Gaia} and the differences in the cluster sample.

\citet{Leiner_2021ApJ...908..229L} analysed the BS population in 16 nearby old OCs. They defined $\delta M = M_{BSS}-M_{MSTO}$, where the stellar masses are calculated by comparison with isochrones. They analysed 16 clusters in the age range 1--10 Gyr; in comparison, this study used 165 clusters in the same age range. \citet{Leiner_2021ApJ...908..229L} stated that stars above $\delta M$ = 1 \Mnom\ are rare. Our estimates show that 12--20\% of the BSSs in the same cluster sample are extreme-\Me\ BSSs. This discrepancy is the result of the definition of $M_{TO}$. We could replicate their Fig. 3 using the ZAMS mass of the MS-tip as the $M_{TO}$. As the MS-tip stars are roughly 1.2 mag brighter than the MSTO (for the old clusters), $M_{TO}$ used by \citet{Leiner_2021ApJ...908..229L} is significantly larger than our $M_{TO}$, resulting in a conservative estimate of the excess mass of BSSs. As mentioned before, their method produces negative excess mass for fainter (than the MS-tip) and bluer BSSs, making it difficult to classify them as BSSs and understand their formation pathways.

\citet{Rain_2021arXiv210306004R} recently produced a BSS catalogue using the parent membership data from \citet{Cantat2020}, although for clusters of all ages. They found 899 BSS candidates in 408 OCs. They also found that BSSs are largely absent in clusters younger than 500 Myr. The difference between their catalogue and this work can come from the adopted age criteria, selection method and different MP cutoffs used in the two studies (50\% in \citealt{Rain_2021arXiv210306004R} and 70\% in this work).

We also note that though we have estimated the number of pBS stars, we do not use them in the analysis to find correlations with cluster properties. This is because these are only BSS candidates. The pBSs generally have low excess mass and are likely to be the product of MT. If they are indeed BSSs, then the fraction of BSSs formed via binary interactions will increase to 90\% and reduce the fraction requiring multiple mergers/MT. 

We plan to carry out detailed analyses of BSSs in the catalogue to obtain their stellar properties in the future. Furthermore, interesting clusters and BSSs identified in this study will be followed with UV imaging (UVIT) and spectroscopic observations (3.6 m Devasthal Optical Telescope).

\section{Summary}  \label{sec:6_Conclusions}
This study aimed to explore the properties of BSSs as a function of cluster parameters and identify potential clusters for further study. The data of 868 BSSs in 228 clusters in the 0.3--10 Gyr age and 10$^2$--10$^4$ \Mnom\ mass range are used to explore their properties. The conclusions we derive from this study are the following:
\begin{enumerate}

\item The number of BSSs found in a cluster is dependent on the cluster age. We derive the average BSS/cluster for different age ranges, which shows an increasing trend with age: 1.6 BSS/cluster (log(\textit{age}) 8.5--9.0); 3.4 (9.0--9.5); 7.9 (9.5--9.75) and 17.8 (9.75--10.0).

\item The maximum number of BSSs found in a cluster is related to the cluster mass by a power-law (N$_{BSS,max}$ $\alpha$ M$_{clus}^{0.6}$). The clusters with at least 10 BSSs show a power-law relation (N$_{BSS}$ $\alpha$ M$_{clus}^{0.37 \pm 0.10}$) similar to globular clusters, indicating binary dominated BSS formation in both types of clusters.

\item The number of BSS found in a cluster is not dependent on the density, radius, distance, Galactocentric distance or number of stars near the turn-off. The BSS fraction does not correlate with age but has a positive correlation with respect to cluster relaxation.

\item We introduced the term \textit{fractional mass excess} for a BSS (\Me\ $\sim$ BSS mass normalised to MSTO) to differentiate various formation pathways. We divided the BSSs into 3 groups: low-\Me\ ($<$ 0.5), high-\Me\ ($0.5 < \mathcal{M}_e < 1.0$) and extreme-\Me\ ($>$ 1.0). We suggest that the low-\Me\ group is likely to be formed via MT, high-\Me\ to be formed via binary mergers and extreme-\Me\ to be through multiple mergers/MT. 
 
\item Majority of BSSs are formed by interactions in binaries, either by MT (as suggested by low-\Me\ = 54\%) or merger in binaries (high-\Me\ = 30\%). In addition, a not-so-big but significant fraction of BSSs is likely to be formed through multiple mergers/MT, as suggested by 16\% of extreme-\Me\ BSSs. 

\item The percentage of high and extreme-\Me\ BSSs show a similar decreasing trend with age beyond 1--2 Gyr, possibly indicating a reduction in the BSS formation through merger processes in older clusters. The rising low-\Me\ fraction beyond 1--2 Gyr could suggest a steadily increasing occurrence of BSS formation through MT beyond 1--2 Gyr.

\item The radial segregation of the BSSs is not found to be a direct function of age but rather a direct function of the dynamical relaxation of the cluster. On average, BSSs with \Me\ $<$ 1.5 are found to be segregated. Contrary to expectations, those with larger \Me\ are comparatively less segregated, suggesting that they are not formed from the already segregated stars.
\end{enumerate}

\begin{savequote}[100mm]
Our destiny is in the stars, so let's go and search for it
\qauthor{Dr. Who}
\end{savequote}

\chapter{Conclusions and Future Work}
\label{ch:summary}
\begin{quote}\small
\end{quote}

\section{Summary and conclusions}
This thesis is focused on understanding the binary stars and their evolutionary products in OCs. The results of the study are presented in Chapters \ref{ch:UOCS3}--\ref{ch:BSS_catalogue}.

\begin{itemize}[leftmargin=*]
    \item Cluster Membership and UV Catalogues
    \begin{itemize}[leftmargin=*]
        \item In \textbf{Chapter \ref{ch:UOCS3}}, we started with analysing cluster stars to confirm their membership in the cluster. We developed a machine learning based method to determine the individual stellar membership using \textit{Gaia} EDR3 data. We provide the \textit{Gaia} EDR3 membership catalogues along with the classification of the stars as members, candidates, and field in six OCs observed by UVIT/\textit{AstroSat}.
        The classification was further utilised by creating UV-optical catalogues of the six OCs: Berkeley 67, King 2, NGC 2682, NGC 2420, NGC 2477 and NGC 6940. 
        
        \item The combination of the \textsc{gmm} and the \textsc{prf} techniques was powerful in identifying cluster members and accounting for statistical flags in the \textit{Gaia} EDR3 data. 
        From UV--Optical CMDs, we found that majority of the sources in NGC 2682 and a few in NGC 2420, NGC 2477 and NGC 6940 showed excess UV flux. 
        
        \item Chapter \ref{ch:UOCS2} and \ref{ch:UOCS4} present the comprehensive study of NGC 2682 and King 2. The 92 (576) members in the FUV (NUV) image of NGC 2477 provide the data necessary for future studies of the UV properties of stars in the extended turn-off and various evolutionary stages from MS to red clump.
    \end{itemize}
    
    \item UVIT Open Cluster Study
    \begin{itemize}[leftmargin=*]
        \item In 2018, we started a long-term \textit{UVIT Open Cluster Study} (UOCS) to understand the UV bright stellar population in the OCs using UVIT. \textbf{Chapter \ref{ch:UOCS2}} presents the detailed UV analysis of the MS stars in NGC 2682. The UV and UV-optical CMDs and overlaid isochrones are presented for the member stars, including BSSs, triple systems, WDs, and spectroscopic binaries (SB). We analysed the UV-optical-IR SEDs of MS stars along with BSSs and found many stars with excess UV flux. I could confirm 4 MS+WD systems \citep{ Jadhav2019ApJ...886...13J, Subramaniam2020JApA...41...45S}. 
        \textbf{Chapter \ref{ch:UOCS4}} presents the study of BSSs in old OC King 2. Using \textit{Gaia} EDR3 astrometry and multiwavelength photometry, we fitted single and double component SEDs to 33 and 6 BSSs, respectively. The hotter companions alongside BSSs were found to be sdB and/or EHB type stars. As the sdB/EHB formation requires mass loss and BSS formation requires mass gain, it is most likely that these systems are post-MT systems similar to NGC 2682. 
         
        \item This study found various types of post-mass-transfer systems and end products in NGC 2682: (i) there are massive WDs likely to be evolved from BSSs, (ii) the ELM WDs which are the end products of the donor in a binary system, (iii) 3 BLs which are accretors present on the MS. 
        Overall, NGC 2682 is a large melting pot of a variety of stellar interactions.
        This study demonstrates that UV observations are vital to detecting and characterising the ELM WDs in non-degenerate systems, which are ideal test beds to explore the formation pathways of these peculiar WDs. 
        
        \item The discovery of sdBs in King 2 makes it one of the youngest OC with sdB members. In this study, we identify rare BSS+sdB systems and may put a new lower limit ($\sim$6 Gyr) on the age of clusters with sdBs. 
        
        \item Overall, there is an extensive variety in the hot companions to BSSs and BLs, depending on the cluster age and environment. More multiwavelength analysis of various OCs is warranted before we completely understand the variety of companions to BSSs and BLs and their connection to the cluster properties. The search to build up a post-MT binary sample in other clusters is ongoing and part of the UOCS.

    \end{itemize}
    
    \item Blue Stragglers in Galactic Open Clusters
    \begin{itemize}[leftmargin=*]
        \item \textbf{Chapter \ref{ch:BSS_catalogue}} present a homogeneous catalogue of BSSs in Galactic OCs. This study was only possible due to the precise astrometry from \textit{Gaia} which has led to accurate cluster membership and cluster parameters. 
        
        \item This study showed that the BSSs are most abundant in old and massive clusters. Furthermore, there exists a power-law relation between the maximum BSSs present in OCs with the cluster mass, which was previously seen in globular clusters. The radial segregation of BSSs is found to be correlated with the dynamical relaxation of the cluster rather than the age. 
        The formation mechanism of BSSs is dominated by binary MT (54--67\%) though there exists a 10--16\% chance of BSSs forming through more than 2 stellar interactions.
        
        \item Though the exact nature of such formation scenarios is difficult to pin down; we provide observational constraints on the different possible mechanisms.
        The statistics and trends presented here are expected to constrain the BSS formation models in OCs.
    \end{itemize}
\end{itemize}

\section{Ongoing and future work}

The thesis gave me a sneak peek into the stellar evolution in clusters and binary systems. A few exciting offshoots of the thesis work are summarised below.

\begin{itemize}[leftmargin=*]
    \item 
    As binary stars were integral to most of my work, we estimated binary fractions and binary segregation in 23 OCs \citep{Jadhav2021AJ....162..264J}. We found a large range in values of binary fractions across different clusters, but overall the brighter stars have more binary fraction than their fainter counterparts. The high mass binary stars were segregated with respect to low mass stars in unrelaxed clusters but not in relaxed clusters. This clearly showed that the segregation of most massive systems in OCs happens on a faster timescale than the complete cluster relaxation.
    
    \item As part of the UOCS, I am analysing the UV bright population of NGC 6791, including WDs, BSSs and MS stars. I am also involved in multiwavelength studies of NGC 7789 \citep{Vaidya2022arXiv220108773V} and NGC 2506 (Panthi et al. in prep), focusing on BSSs.
    
    \item We are analysing H$\alpha$ emitting stars in the Magellanic Bridge (MB) using SALT spectra to understand the ongoing star formation. The analysis of the 14 spectra is in progress. We are planning to study the spectral features, estimate kinematic information and age of emission-line stars in the MB and use them to understand the nature of in-situ star formation in the MB.
    
    \item Since 2019, I have been a part of the science team for the proposed 1-m UV-optical space telescope, named as, \textit{Indian Spectroscopy and Imaging Space Telescope} (\textit{INSIST}), which is the follow-up mission for UVIT. My primary contribution is in developing science cases and white papers (OCs, WDs and planetary nebulae) for the mission. I also participate in the discussions of spectrograph design, optical-mechanical trade studies and planning, which gives me an insight into the steps that go into the initial stages of a space mission. 
    The experience I gained by working with the \textit{INSIST} team has manifested an interest in planning and executing a mission, including survey strategies and data pipelines.
\end{itemize}

The upcoming decade will provide ample panchromatic wide-field data through following surveys: \textit{Gaia}, LSST, \textit{Euclid} and \textit{Roman}, SDSS-V, 4MOST, WEAVE etc. Here, I present some ideas for future projects based on the available/upcoming data and work done during the thesis period.

\begin{itemize}[leftmargin=*]
\item Stellar binarity

    \begin{itemize}[leftmargin=*]
        \item The observed binary fraction varies with the luminosity of the primary star. The binary fraction of FGK, B and O type stars is 34\%, 56--58\% and 42--69\% respectively (\citealt{Luo2021arXiv210811120L} and references therein). 
        The multiplicity fraction is also known to vary with metallicity \citet{Badenes2018ApJ...854..147B} and age \citep{Jaehnig2017ApJ...851...14J}. However, no explicit dependencies are established between the binary properties and binary fractions. I plan to expand our previous study, \citet{Jadhav2021AJ....162..264J}, using the latest astrometric, photometric (\textit{Gaia} DR3, LSST) and spectroscopic (LAMOST, APOGEE, RAVE) data to increase the sample of known binary and higher-order systems. The combination of wide-field multiwavelength photometric surveys with time-domain and spectroscopic follow-ups is a critical tool to analyse multiple systems.
    
        \item A major open problem in our present understanding of binary populations is the mass-ratio distribution.
        The expected mass-ratio distribution changes depending on the formation scenario: (i) peaked towards unity for cloud fragmentation \citep{Bate2000MNRAS.314...33B} (ii) flat for tidal capture \citep{Kroupa2003MNRAS.346..354K}. The distribution is also known to be different for intermediate-mass stars and brown dwarfs \citep{Goodwin2013MNRAS.430L...6G, Thies2015ApJ...800...72T}. In reality, the observed mass-ratio distribution will be a combination of the formation mechanisms and set by the techniques used to identify binaries and estimate the mass ratios. I plan to use clusters and field stars to get the mass-ratio distribution of binary stars. This is a fundamental constraint on the physics of star formation.
    \end{itemize}

    \item  BSSs and post-MT systems
    \begin{itemize}[leftmargin=*]
        \item The relation of stellar interactions and chemical peculiarities in the post-interaction systems is not yet understood. We plan to study the surface abundances of Li, C, N, and O in isolated and post-interaction (BSSs/BLs) systems to evaluate the abundance patterns and their relationship with stellar age, mass, cluster metallicity and orbital parameters. Wide-field spectroscopic surveys and necessary targeted observations will be used to get abundance parameters. 
        
        \item BSSs are complicated systems formed via MT or mergers. The orbital parameters of binary progenitors significantly impact the formation and detectability of present-day binary companions. We plan to use the simulations to identify the frequency of BSSs and present-day binary properties. The lifetime of BSSs (after mass accretion) is also not entirely known. We plan to use N-body simulations to get a statistical bearing on the longevity of the BSSs and use stellar simulation codes to evolve BSSs after the MT individually.

    \end{itemize}
\end{itemize}




\bibliographystyle{mnras}
\bibliography{references}
\printthesisindex

\end{document}